%% file: zwanenburg-rmp-arxiv.tex
\documentclass[rmp,twocolumn,aps]{revtex4-1}
\usepackage{graphicx,amsmath,amssymb}
\pdfoutput=1
\begin{document}
\input{titlepage}
\tableofcontents

\section{Introduction and motivation}\label{introduction}
\input{sqe-FZ}

\section{Quantum confinement}\label{confinement}
\input{confinement-FZ}

\section{Physics of Si nanostructures}\label{physics}
\input{physics}

\section{Quantum dots in Si and SiGe}\label{quantumdots}
\input{quantumdots}
\section{Dopants in silicon}\label{dopants}
\input{dopants_v11}
\section{Relaxation, coherence and measurements}\label{timeresolved}
\input{timeresolved}

\input{acknowledgements}

\end{document}

%% file: titlepage.tex
\title{Silicon Quantum Electronics}

\author{Floris A. Zwanenburg} \email{f.a.zwanenburg@utwente.nl}
\affiliation{NanoElectronics Group MESA+ Institute for Nanotechnology University of Twente, Enschede The Netherlands}
\affiliation{Centre of Excellence for Quantum Computation and Communication Technology, The University of New South Wales Sydney  Australia}
\author{Andrew S. Dzurak, Andrea Morello, Michelle Y. Simmons}
\affiliation{Centre of Excellence for Quantum Computation and Communication Technology, The University of New South Wales Sydney Australia}
\author{Lloyd C. L. Hollenberg}
\affiliation{Centre of Excellence for Quantum Computation and Communication Technology, University of Melbourne Melbourne Australia}
\author{Gerhard Klimeck}
\affiliation{School of Electrical and Computer Engineering Birck Nanotechnology Center, Network for Computational Nanotechnology Purdue University West Lafayette Indiana USA}
\author{Sven Rogge}
\affiliation{Centre of Excellence for Quantum Computation and Communication Technology, The University of New South Wales Sydney Australia}
\affiliation{Kavli Institute of Nanoscience Delft University of Technology Delft The Netherlands}
\author{Susan N. Coppersmith, Mark A. Eriksson}
\affiliation{University of Wisconsin-Madison Madison Wisconsin USA}

\date{\today}

\begin{abstract}
This review describes recent groundbreaking results in Si, Si/SiGe and dopant-based quantum dots, and it highlights the remarkable advances in Si-based quantum physics that have occurred in the past few years. This progress has been possible thanks to materials development of Si quantum devices, and the physical understanding of quantum effects in silicon. Recent critical steps include the isolation of single electrons, the observation of spin blockade and single-shot read-out of individual electron spins in both dopants and gated quantum dots in Si. Each of these results has come with physics that was not anticipated from previous work in other material systems. These advances underline the significant progress towards the realization of spin quantum bits in a material with a long spin coherence time, crucial for quantum computation and spintronics.
\end{abstract}

\maketitle

%% file: sqe-FZ.tex
\subsection{Silicon Quantum Electronics}\label{sqe}

The exponential progress of microelectronics in the last half century has been based on silicon technology. After decades of progress and the incorporation of many new materials, the core technological platform for classical computation remains based on silicon. At the same time, it is becoming increasingly evident that silicon can be an excellent host material for an entirely new generation of devices, based on the quantum properties of charges and spins. These range from quantum computers to a wide spectrum of spintronics applications. Silicon is an ideal environment for spins in the solid state, due to its weak spin-orbit coupling and the existence of isotopes with zero nuclear spin. The prospect of combining quantum spin control with the exquisite fabrication technology already in place for classical computers has encouraged extensive effort in silicon-based quantum devices over the past decade.

While there are many proposed physical realizations for quantum information processors \cite{LloydScience93,LaddNature10,bulutaOther11}, semiconductor-based quantum bits (qubits) are extremely interesting, in no small part because of their   commonalities with classical electronics  \cite{LossPRA98,KaneNature98}. Electron spins in quantum dots have received considerable attention, and significant experimental progress has been made since the original \textcite{LossPRA98} proposal. Experiments on lithographically defined quantum dots in GaAs/AlGaAs heterostructures have shown qubit initialization, single-shot single-electron spin read-out \cite{elzermanNature04}, and coherent control of single-spin \cite{koppensNature06} and two-spin \cite{pettaScience05} states. One of the major issues in AlGaAs/GaAs heterostructures is the inevitable presence of nuclear spins in the host material, leading to relatively short spin relaxation and coherence times.

A way to increase the coherence time is to use materials with a large fraction of non-magnetic nuclei. Natural silicon consists of $95\%$ non-magnetic nuclei (92\% $^{28}$Si and 3\% $^{30}$Si) and can be purified to nearly $100\%$ zero-nuclear-spin isotopes. Various proposals have been made for electron spin qubits based on donors  in Si \cite{vrijenPRA00, desousaPRA04, HillPRB05, HollenbergPRB06} and Si quantum dots \cite{friesenPRB03}. The key requirement for spin quantum bits is to confine single electrons to either a quantum dot or a donor, thus posing a scientific challenge. In contrast with the technological maturity of classical field-effect transistors, Si quantum dot systems have lagged behind GaAs systems, which were historically more advanced because of the very early work in epitaxial growth in lattice-matched III-V materials. \textcite{Kouwenhoven:1997p1788} studied the excitation spectra of a single-electron quantum dot in a III-V material.  Even though Coulomb blockade in Si structures was observed very early~\cite{paulAPL1993,aliAPL1994}, it took another 5 years before regular Coulomb oscillations were reported \cite{simmelPRB1999}. Silicon systems needed nearly ten years to achieve single-electron occupation in quantum dots \cite{simmonsAPL07,zwanenburgNL09,limAPL09-2} and dopants \cite{SellierPRL06,fuechsleNatureN12}. For quantum dots this has laid the foundation for spin filling in valleys in few-electron quantum dots \cite{borselliAPL2011, LimNanotech11}, tunnel rate measurements in few-electron single and double quantum dots \cite{thalakulamAPL10}, Pauli spin blockade in the few-electron regime \cite{borselliAPL11}, and very recently Rabi oscillations of singlet-triplet states \cite{mauneNature12}. In the case of dopants valley excited states \cite{FuechsleNnano10}, gate-induced quantum-confinement transition of a single dopant atom \cite{LansbergenNphys08}, a deterministically fabricated single-atom transistor \cite{fuechsleNatureN12} and single-shot read out of an electron spin bound to a phosphorus donor \cite{morelloNature10} have been reported. The importance of deterministic doping has recently been highlighted in the 2011 ITRS Emerging Research Materials chapter, where a remaining key challenge for scaling CMOS devices towards 10 nm is the control of the dopant positions within the channel \cite{ITRS11}. All these results underline the incredible potential of silicon for quantum information processing.

It is tempting to project the achievements in integrated-circuit technology onto a supposed scalability of quantum bits in silicon. Even though current silicon industry standards, with 22 nm features, have higher resolution than typical quantum devices discussed in this review, superb patterning alone does not guarantee any sort of `quantum CMOS' (Complementary Metal-Oxide-Semiconductor). As one example, interface traps have a very different effect on classical transistors (where they serve as scattering centers or shift threshold voltages) than in quantum dots (where they also affect spin coherence).  Nonetheless, a fully-integrated CMOS foundry has been used for many steps in the fabrication of silicon quantum devices~\cite{Nordberg:2009p115331}.

While silicon-based devices generate special interest for quantum computation, because of zero nuclear spin isotopes and low spin-orbit coupling, they also face some special challenges and display physics that, until recently, has been little explored in the context of quantum computation.  Examples of the challenges include the relatively large effective mass in silicon and the large difference in lattice constant between silicon and germanium. An example of the unexplored physics is the presence of multiple conduction band valleys in silicon.

As described in this review, there have been rapid advances addressing the challenges and exploring the new physics available in silicon-based quantum devices.  The extent to which these advances will lead to larger-scale quantum systems in silicon is an exciting question as of this writing.

\subsection{Outline of this review}
This review covers the field of electronic transport in silicon and focuses on single-electron tunneling through quantum dots and dopants. We restrict ourselves to experiments and theory involving electrons confined to single or double (dopant) quantum dots, describing the development from the observation of Coulomb blockade to single-electron quantum dots and single dopant atom transistors. Ensembles of quantum dots or dopants are beyond the scope of this article. Also, the review is strictly limited to electron transport experiments, and does not cover optical spectroscopy measurements. Optical spectroscopy on quantum dots and ensembles of dopants is a very active and emerging field, see for example the recent work by \textcite{stegerScience12,greenlandNature10} and references therein.

\textsl{Section \ref{confinement} Quantum Confinement} starts with a general introduction to transport through quantum-confined silicon nanostructures. The silicon bandstructure is described in \textsl{Section \ref{physics} Physics of Silicon nanostructures} with specifics such as the valley degeneracy and splitting in bulk and quantum dots, and wave function control and engineering of dopant states.  \textsl{Section \ref{quantumdots} Quantum dots in Si and SiGe} explains the development from the discovery of Coulomb blockade in 1990 to  single-electron occupancy in single and double quantum dots in recent years. Analogously, dopant transport in silicon has evolved from tunneling through 1980's MOSFETs to current-day single-atom transistors, see \textsl{Section \ref{dopants} Dopants in silicon}. The remarkable advances of Sections \ref{quantumdots} and \ref{dopants} have lead to the relaxation and coherence measurements on single spins in \textsl{Section \ref{timeresolved} Outlook: relaxation, coherence and measurements}.

%% file: confinement-FZ.tex
This section introduces quantum electronic experiments in silicon, starting with the quantum mechanical confinement of electrons in silicon, which can be achieved by a combination of electrostatic fields, interfaces between materials, and/or placement of individual atoms. All of these approaches lead to single-electron tunneling devices consisting of a silicon potential well coupled to source, drain and gate electrodes.

\subsection{From single atoms to quantum wells}\label{3D}
\begin{figure*}[t]
\includegraphics[width=\textwidth, clip=true]{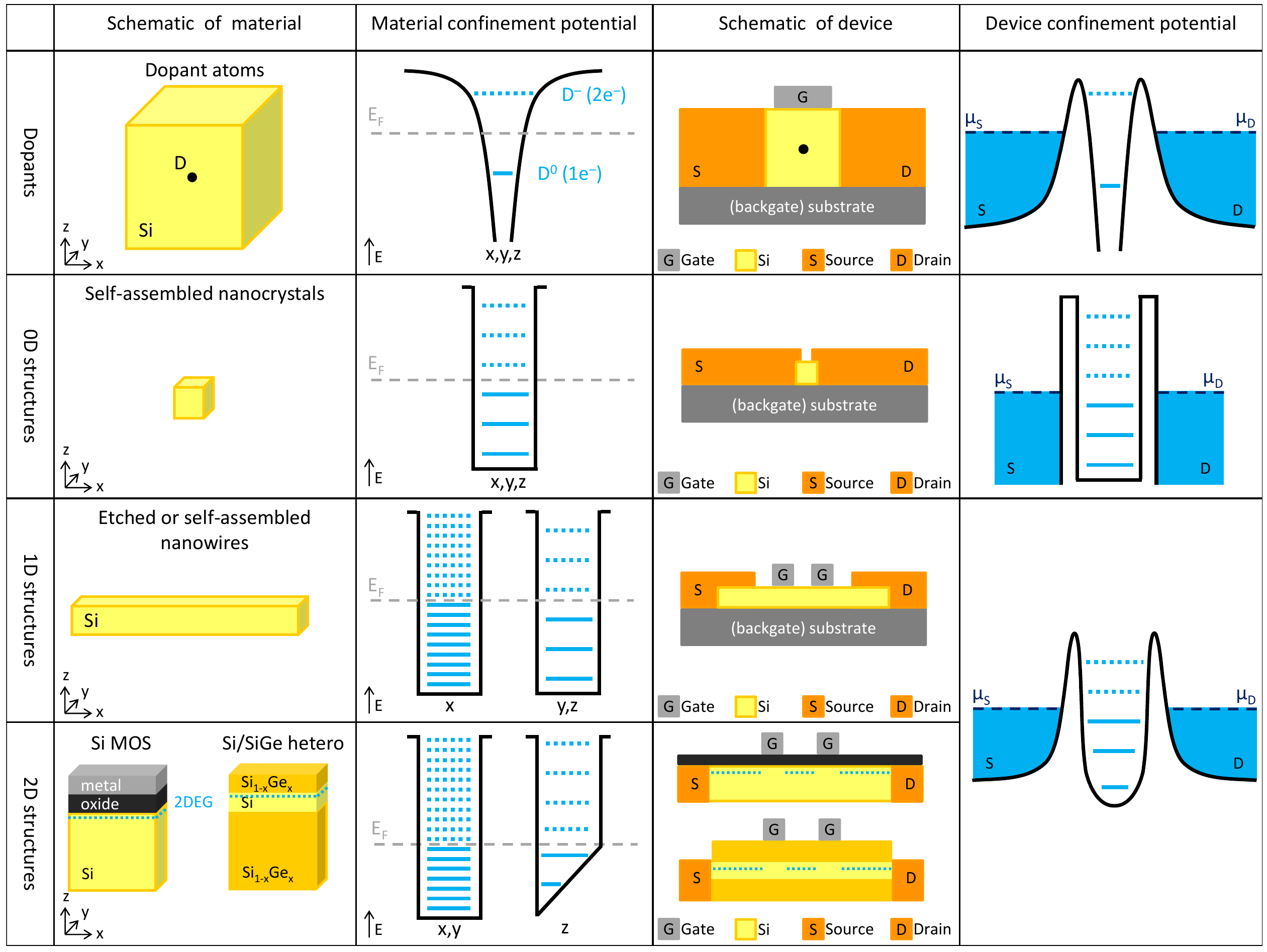}
\caption{\label{confinement_fig3} (Color online) \textbf{Combining material and electrostatic confinement to create single-electron transistors.} \textsl{First column}: schematic of dopants, 0D-, 1D- and 2D-structures. \textsl{Second column}: in the corresponding confinement potentials in x-, y- and z-directions  electron states are occupied up to the Fermi energy $E_F$ (dashed grey line). Occupied and unoccupied electron states are indicated as straight and dashed lines respectively. \textsl{Third column}: Schematic of the silicon nanostructure integrated into a transport device with source, drain and gate electrodes. \textsl{Fourth column}: The potential landscape of the single-electron transistor is made up of a potential well which is tunnel-coupled to source and drain reservoir and electrostatically coupled to gates which can move the ladder of electrochemical potentials, as described in Section \ref{qt}.}
\end{figure*}

Electrons in Si nanostructures are confined using a combination of material and electrostatic potentials. The shape and size of nanostructured materials provide natural confinement of electrons to 0, 1 or 2 dimensions. The exact confinement potential of the structure in x, y and z-directions sets the additional requirements in terms of additional electric fields. Figure \ref{confinement_fig3} gives an overview of materials of different dimensionality and their integration into single-electron tunneling devices.\\

\noindent \textbf{Dopants}\\
The electrostatic potential of a single dopant atom is radially symmetric, resulting in the same steep potential well in all directions, as shown in the first row of Fig.~\ref{confinement_fig3}. The Bohr radius $a_B$ is the mean radius of the orbit of an electron around the nucleus of an atom in its ground state, and equals for example 2.5 nm for phosphorus in silicon. A dopant atom has three charge states: the ionized $D^+$ state, the neutral $D^0$ state (one electron bound to the dopant) and the negatively charged $D^{-}$ state (two electrons bound to the dopant). Because the $D^+$ state corresponds to an empty dopant it does not appear as an electron state in the potential well. Measuring electron transport through a single atom has been a great challenge, as described in Section \ref{dopants}, but the single-dopant regime as sketched in the third column has been reached by several groups. Depending on the architecture, the source and drain reservoirs can be made up of highly-doped Si \cite{SellierPRL06,PierreNNano10,fuechsleNatureN12}, or of a two-dimensional electron gas \cite{TanNL10}. The same goes for the gates, but they can also be metallic \cite{TanNL10}. The resulting single-electron transistors consist of a steep dopant potential well connected to source and drain reservoirs.\\

\noindent \textbf{0D structures}\\
Like dopants, self-assembled nanocrystals  provide confinement to zero dimensions, but the confinement is better described by a hard-wall potential well in x, y and z-directions and is much wider (Fig.~\ref{confinement_fig3}). The energy levels of an electron in a quantum well of size $L$ are quantized according to basic quantum mechanics, see for example \textcite{cohen-tannoudji}. The corresponding level spacing $\Delta E$ is on the order of $h^2/m_{\textrm{eff}}L^2$, where $m_{\textrm{eff}}$ is the electron effective mass. The separation between energy levels thus decreases quadratically with the well width: as a result, the discrete levels of e.g.\ a 30 nm size nanocrystal are expected to have energy spacings 2 orders of magnitude smaller than those of a dopant with a 3 nm Bohr radius. Making source and drain contacts requires very precise alignment by means of electron-beam lithography. The tunnel coupling of these devices relies on statistics; creating tunable tunnel coupling to self-assembled dots is very challenging. A highly-doped substrate can be used as a global backgate and metallic leads on a dielectric as a local gate.\\

\noindent \textbf{1D structures}\\
The high aspect ratio of nanowires (NWs) implies a large level spacing in the transverse directions, and a small level spacing in the longitudinal direction ($L_x \gg L_{y,z}$), creating a (quasi) 1-dimensional channel with few subbands in the transverse direction (see second row of Fig.~\ref{confinement_fig3}). Within this channel a zero-dimensional well can be created by local gates on the nanowire, or by Schottky tunnel barriers to source and drain contacts. In the latter case the barrier height is determined by the material work functions and hardly tuneable in-situ --- the tunnel coupling will generally decrease as electrons leave the well and the wave function overlap with source and drain shrinks. Local gates, however, can tune the tunnel barriers since the applied gate voltage induces an electric field which locally pulls up the conduction band. Electrons tunnel from the quantum well into reservoirs which are part of the nanowire itself. The metallic leads connecting the nanowire to the macroscopic world must be ohmic; i.e., the contacts should have high transparency, to prevent the formation of multiple quantum dots in series (particularly if the contacts are very close to the quantum dot).\\

\noindent \textbf{2D structures}\\
A 2-dimensional electron gas (2DEG) can be created in Si MOSFETs (Metal-Oxide-Semiconductor Field Effect Transistors) and in Si/SiGe heterostructures. Electrons are unconfined in the $x$-$y$-plane and are confined by a triangular potential well perpendicular to the plane, as sketched in in Fig.~\ref{confinement_fig3}. More realistic band diagrams are drawn in Fig.~2 in the review by \textcite{andoRMP82} for Si MOS and Fig. 11 in the review by \textcite{schafflerSSCT97} for Si/SiGe heterostructures. In a 2DEG-based quantum dot, the lateral confinement is a soft-wall potential  defined by top gate electrodes, enabling tunnel-coupling to source and drain reservoirs in the 2DEG. Those reservoirs are connected to macroscopic wires via ohmic contacts, which are often highly doped regions at the edge of the chip. The resulting potential landscape is highly tunable thanks to local electrostatic gating via the top gates.

\subsection{Transport regimes}\label{qt}
Having introduced quantum-confined devices, we now cover the basics of quantum transport through single-electron transistors (SETs), which are made up of a zero-dimensional island, source and drain reservoirs, and gate electrodes.

\begin{figure}[h]
\includegraphics[width=0.48\textwidth]{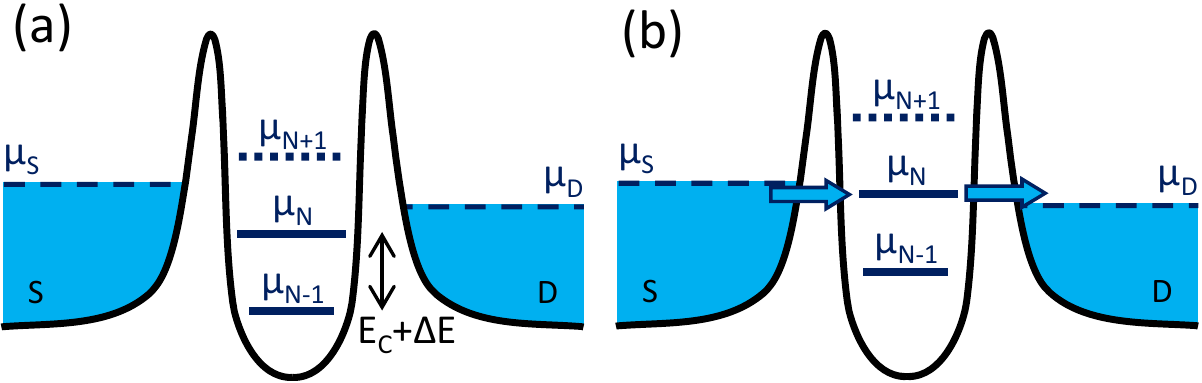}
\caption{\label{confinement_fig1}  (Color online)  \textbf{Schematic diagrams of the electrochemical potential of a single-electron transistor.} (a) There is no available level in the bias window between $\mu_S$ and $\mu_D$, the electrochemical potentials of the source and the drain, so the electron number is fixed at $N$ due to Coulomb blockade. (b) The $\mu_N$ level aligns with source and drain electrochemical potentials, and the number of electrons alternates between $N$ and $N-1$, resulting in a single-electron tunneling current.}
\end{figure}
Electronic measurements on single electrons require a confining potential which is tunnel coupled to electron reservoirs in source and drain leads, see Fig.~\ref{confinement_fig1}. The SET-island is also coupled capacitively to one or more gate electrodes, which can be used to tune the electrostatic potential of the well. The discrete levels are spaced by the addition energy $E_{\textrm{add}}(N) = E_C + \Delta E$, which consists of a purely electrostatic part, the charging energy $E_C$, plus the energy spacing between two discrete quantum levels, $\Delta E$. $\Delta E$ is zero when two consecutive electrons are added to the same spin-degenerate level. The charging energy $E_C=e^2/2C$, where $C$ is the sum of all capacitances to the SET-island\footnote{We refer to other review articles on quantum dots and single-electron transistors for more background and details: \textcite{Beenakker:1991p1, GrabertOther93, KouwenhovenNato97, Kouwenhoven:2001p701, ReimannRMP02, VanDerWiel:2003p1382, hansonrmp07}}.

\begin{figure}[!ht]
\includegraphics[width=0.4\textwidth]{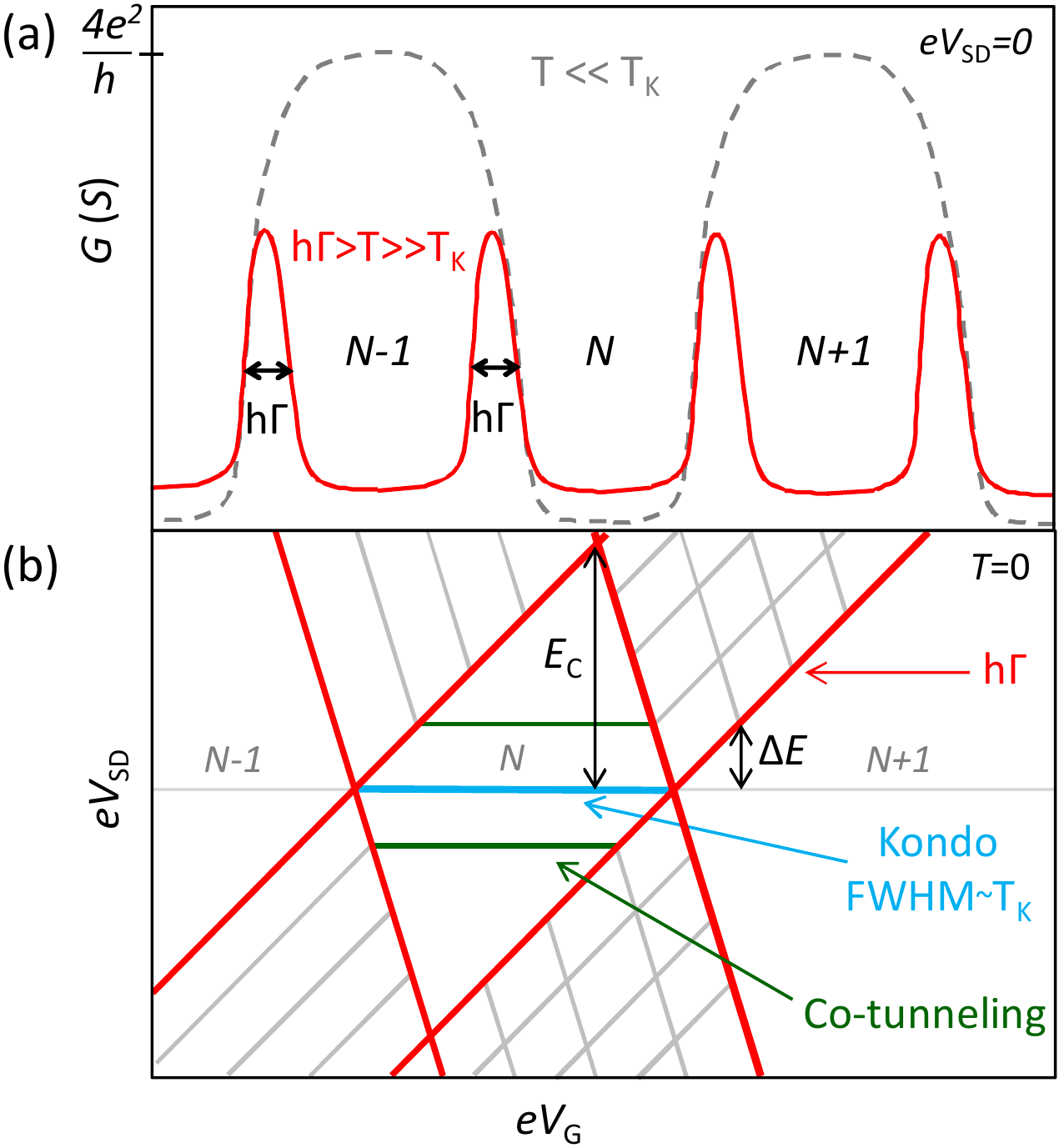}
\caption{ (Color online) {\bf Zero-bias and finite-bias spectroscopy.} (a) Zero-bias conductance $G$ of transport versus gate voltage $V_G$ both at $T\gg T_K$ (solid line) and $T \ll T_K$ (dashed line). In the first regime, the full width at half maximum (FWHM) of the Coulomb peaks corresponds to the level broadening $h\Gamma$. In the Kondo regime ($T \ll T_K$), Coulomb blockade is overcome by coherent second-order tunneling processes (see main text). (b) Stability diagram showing Coulomb diamonds in differential conductance, $dI/dV_{SD}$, versus $eV_{SD}$ and $eV_G$ at $T = 0 K$. The edges of the diamond-shaped regions (red) correspond to the onset of current. Diagonal lines of increased conductance emanating from the diamonds (gray) indicate transport through excited states. The indicated internal energy scales $E_C$, $\Delta E$, $h\Gamma$ and $T_K$ define the boundaries between different transport regimes. Co-tunneling lines can appear when the applied bias exceeds $\Delta E$ (see main text). Adapted from \onlinecite{LansbergenTHE10}.} \label{fig2_TRcd}
\end{figure}

In the limit of low temperature, if we only consider sequential tunneling processes, energy conservation needs to be satisfied for transport to occur. The electrochemical potential $\mu_N$ is the energy required for adding the $N$th electron to the island. Electrons can only tunnel through the SET when $\mu_N$ falls within the bias window (see Fig. ~\ref{confinement_fig1}(b)), i.e. when $\mu_S \geq \mu_N \geq \mu_D$. Here $\mu_S$ and $\mu_D$ are the electrochemical potential of the source and the drain respectively. Current cannot flow without an available level in the bias window, and the device is in Coulomb blockade, see Fig.~\ref{confinement_fig1}(a). A gate voltage can shift the whole ladder of electrochemical potential levels up or down, and thus switch the device from Coulomb blockade to single-electron tunneling mode. By sweeping the gate voltage and measuring the conductance, one obtains Coulomb peaks as shown in Fig.~~\ref{fig2_TRcd}(a).

Usually, one measures the conductance versus source-drain voltage $V_{SD}$ and gate voltage $V_{G}$ in a bias spectroscopy, as shown in Fig.~\ref{fig2_TRcd}(b). Inside the diamond-shaped regions, the current is blocked and the number of electrons is constant. At the edges of these Coulomb diamonds a level is  resonant with either source or drain and single-electron tunneling occurs. When an excited state enters the bias window a line of increased conductance can appear parallel to the diamond edges. These resonant tunneling features have other possible physical origins, as described in detail by \textcite{EscottN10}. From such a bias spectroscopy one can read off the excited-states and the charging energy directly, as indicated in Fig.~\ref{fig2_TRcd}(b).

The simple model described above explains successfully how quantization of charge and energy leads to effects like Coulomb blockade and Coulomb oscillations. Nevertheless, it is too simplified in many respects.
Up until now we only worried about the electronic properties of the localized state but not about the physics of the electron transport through that state. In this section, based on \onlinecite{LansbergenTHE10}, we will describe the five different regimes of electron transport through a localized stated in a three-terminal-geometry. How electrons traverse a quantum device is strongly dependent on the coherence during the tunneling process and thus depends strongly on $eV_{SD}$ and $k_{B}T$. These external energy scales should be compared to the internal energy scales of the tunneling geometry that determine the transport regime, namely the charging energy $E_C$, the level spacing $\Delta E$, the level broadening $h\Gamma$ and the Kondo temperature $T_K$. Here, $\Gamma$ is the total tunnel rate to the localized state which can be separated into the tunnel coupling to the source electrode $\Gamma_S$ and to the drain electrode $\Gamma_D$, i.e. $\Gamma = \Gamma_S + \Gamma_D$. The internal energy scales are all fixed by the confinement potential, and the external energy scales reflect the external environment, namely the temperature $T$ and the applied bias $V_{SD}$.

\begin{figure}[!ht]
\includegraphics[width=0.4\textwidth]{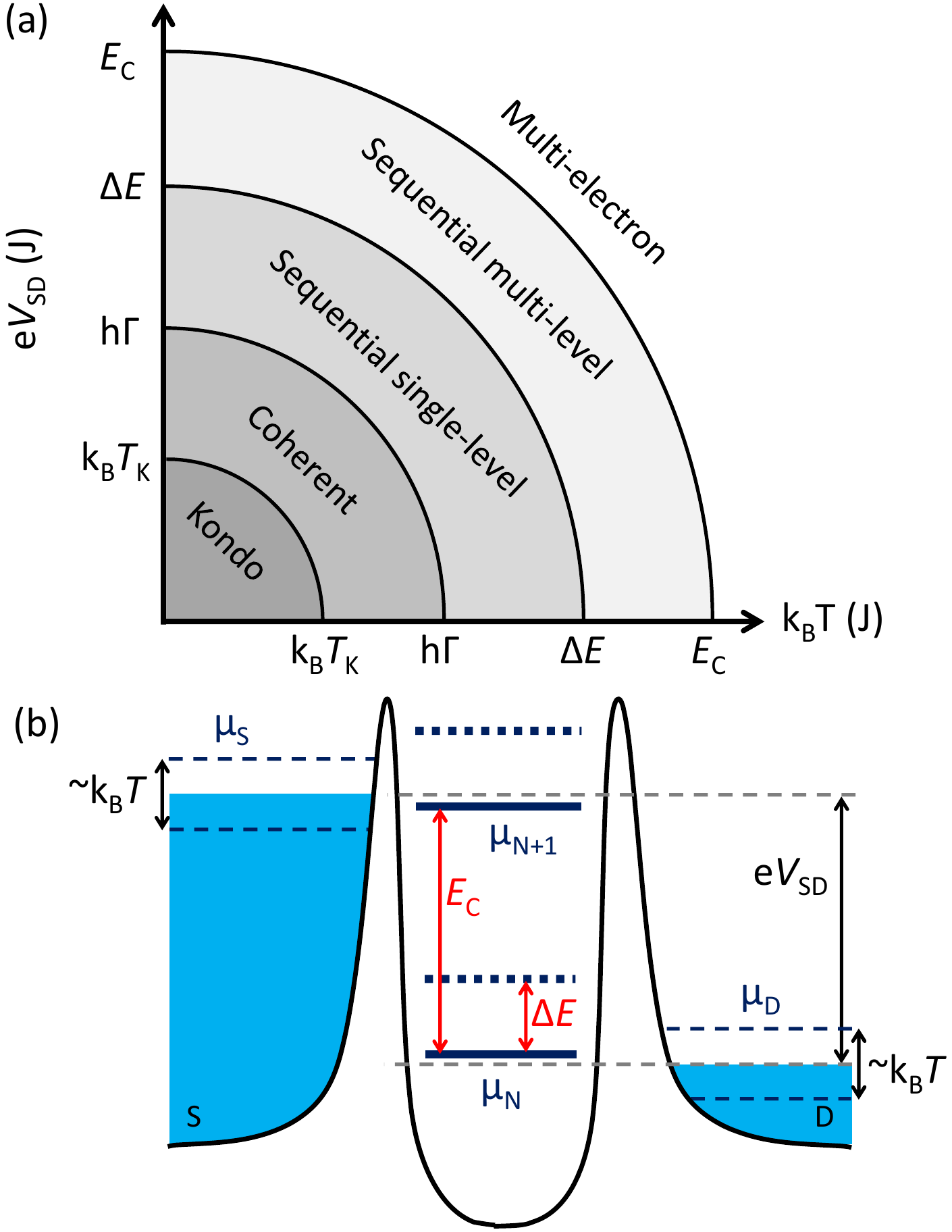}
\caption{ (Color online) \textbf{The five separate transport regimes in a three-terminal quantum device.} (a)  Schematic depiction of the regimes in which transport through a localized takes place as a function of the external energy scales $k_B T$ and $V_{SD}$. The transitions between regimes take place on the order of the internal energy scales $E_C$, $\Delta E$, $h\Gamma$ and $T_K$. (b) Potential landscape of the three terminal geometry, where the quantum states and the electrochemical potential of the leads are shown together with $k_B T$, $V_{SD}$ and $E_C$, $\Delta E$.} \label{fig2_TR}
\end{figure}

Much literature  describes the electronic transport in all possible proportionalities of these energy scales with each other \cite{ButtikerIBM88, BeenakkerPRB91, AlhassidRMP00}. The internal energy scales are typically related to each other by $T_K \ll h\Gamma \ll \Delta E \ll E_C$, and occasionally by $T_K \ll \Delta E < h\Gamma \ll E_C$, limiting the number of separate transport regimes that we need to consider. Fig.\,\ref{fig2_TR}(a) is a schematic depiction of transport regimes as a function of $eV_{SD}$ and $k_{B}T$. It should be noted that the boundaries between transport regimes are typically not abrupt transitions. For clarity, internal and external energy scales (except $T_K$ and $h\Gamma$) are indicated in a schematic representation of our geometry, see Fig.\,\ref{fig2_TR}(b).

Here, we will not make a distinction between the external energy scales $k_BT$ and $eV_{SD}$ when we compare them to internal energy scales, as indicated by Fig.\,\ref{fig2_TR}(a). The reason behind this equality is that both these external energy scales have a very similar effect on the transport characteristics. Their only relevant effect is that they introduce (hot) phonons to the crystal lattice, either directly by temperature or by inelastic tunneling processes induced by the non-equilibrium Fermi energies of the source/drain contacts.

Next, we will describe the five separate tunneling regimes and their corresponding expressions for the source/drain current $I$ shortly. These regimes are the so-called multi-electron regime, the sequential multi-level regime, the sequential single level regime, the coherent regime and the Kondo regime, see Fig.~\ref{fig2_TR}(a).

\subsubsection{The multi-electron regime}
Firstly there is the multi-electron regime ($E_C \ll k_B T , eV_{SD}$) where Coulomb blockade does not occur, as mentioned in the start of this chapter. This regime is not relevant for this review.

\subsubsection{The sequential multi-level regime}
At $\Delta E \ll k_B T, eV_{SD} \ll E_C$ the system is in the sequential multi-level regime. The transport is given by \cite{BeenakkerPRB91, vanderVaartPhysicaB93} \begin{equation}
  I = e \frac{\left( \Gamma_{\textrm{in}}^1 + \Gamma_{\textrm{in}}^2 + ... + \Gamma_{\textrm{in}}^n \right) \Gamma_{\textrm{out}}^1}{\Gamma_{\textrm{in}}^1 + \Gamma_{\textrm{in}}^2 + ... + \Gamma_{\textrm{in}}^n + \Gamma_{\textrm{out}}^1} \label{vgl2_1},
\end{equation}
where the subscript denotes the direction of transport, into or out of the localized state, and the superscript indicates the level, where $1$ refers to the ground state and $n$ indicates the highest orbital within the energy window set by $eV_{SD}$. The current thus depends on the ingoing rates of all levels in the bias window and the outgoing rate of \textsl{only} the ground state. Physically, electrons can enter any orbital state that is energetically allowed. Once a single electron is transferred to the localized state, Coulomb blockade prevents another electron from entering. For dopants, the bound electron will relax back to the ground state before it has a chance to tunnel out of the localized state, since the orbital relaxation times ($\sim$\,ps-ns \cite{LansbergenPRL11}) are typically much faster than the outgoing tunnel rates ($\sim$1 ns). For quantum dots the physics is similar but tunnel rates and orbital relaxation rates are slower, e.g. $\sim$\,1-10ns`in GaAs quantum dots  \cite{FujisawaScience98}. The inelastic nature of the relaxation prohibits  coherent transfer of electrons from the source to the drain electrode.

\subsubsection{The sequential single-level regime}
The next transport regime is the sequential single-level regime, roughly bounded by $h\Gamma \ll k_B T, eV_{SD} \ll \Delta E$, where only a single level resides inside the bias window. This regime is a transition between phase-coherent and phase-incoherent transport between source- and drain -electrodes and the tunneling current depends vitally on $k_B T$. For V$_{SD} = 0$ the conductance is given by \cite{BeenakkerPRB91}
\begin{equation}
G = \frac{e^2}{4 k_B T} \frac{\Gamma_{\textrm{in}}^1 \Gamma_{\textrm{out}}^1}{\Gamma_{\textrm{in}}^1 + \Gamma_{\textrm{out}}^1}
\label{vgl2_2},
\end{equation}
where $\Gamma_{\textrm{in}}$ is the tunnel rate into the localized state and $\Gamma_{\textrm{out}}$ is the tunnel rate out. Note that $\Gamma_{\textrm{in}} = \Gamma_S, \Gamma_{\textrm{out}} = \Gamma_D$ for $V_{SD} > 0$ and $\Gamma_{\textrm{in}} = \Gamma_D, \Gamma_{\textrm{out}} = \Gamma_S$ for $V_{SD} < 0$.

If the localized state is strongly coupled to the contacts higher-order transport processes become apparent in the Coulomb blocked region, \textsl{i.e.} the so called co-tunneling lines indicated in Fig.\,\ref{fig2_TRcd}(b). This is the case when $E_C / \Gamma$ approaches unity in the open regime. There is an elastic and inelastic component to the co-tunneling \cite{AverinPRL90,Nazarov09}. The elastic component leads to a constant background current in the Coulomb diamond. The inelastic component leads to a step in the current when the applied bias exceeds $\Delta E$. The current is given by
\begin{equation}
I_{\textrm{el}} = \frac{\rho^2e^2}{8 \pi^2 h} \Gamma_{\textrm{in}}\Gamma_{\textrm{out}} \frac{1}{\Delta E}
\label{vgl2_3},
\end{equation}
\begin{equation}
I_{\textrm{in}} = \frac{\rho^2e^2}{6h} \Gamma_{\textrm{in}}\Gamma_{\textrm{out}} \left(\frac{k_BT}{E_e}+\frac{k_BT}{E_h}\right)
\label{vgl2_4},
\end{equation}
for the elastic and inelastic co-tunneling respectively with $E_e+E_h=E_C$, where the energies $E_e$ and $E_h$ denote the distance to the Fermi energy of the filled and empty state and $\rho$ is the density of states. The complex co-tunneling line shape is discussed in depth in \onlinecite{wegewijsArxiv01}.

\subsubsection{The coherent regime}
As soon as the external energy scales are much smaller then $h\Gamma$ ($ T_K \ll k_B T, eV_{SD} \ll h\Gamma \ll \Delta E$) the system is in the coherent regime, where the conductance is given by \textcite{ButtikerIBM88}
\begin{equation}
G = \frac{e^2}{\hbar} \frac{\Gamma_{\textrm{in}}^1 \Gamma_{\textrm{out}}^1}{\left(\Gamma_{\textrm{in}}^1 + \Gamma_{\textrm{out}}^1\right)^2}
\label{vgl2_3}
\end{equation}
The conductance is thus given by the quantum conductance $e^2/\hbar$ multiplied by a factor that only depends on the symmetry between $\Gamma_S$ and $\Gamma_D$. It has been proven explicitly that this expression, easily derived for resonances in 1D double barrier structures \cite{RiccoPRB84}, also holds in three dimensions \cite{KalmeyerPRB87}.

\subsubsection{The Kondo regime}
The final transport regime occurs when $eV_{SD}, k_B T \ll T_K$. The Kondo temperature is the energy scale below which second-order charge transitions other than co-tunneling start to play a role in the transport \cite{MeirPRL93}. In first-order transitions, the transferred electrons make a direct transition from their initial to their final state. It should be noted that the constant interaction model only considers first-order charge transitions~\cite{KouwenhovenNato97}. In a second-order transition, the transferred electron goes from the initial to the final state via a virtual state of the atom or dot. A virtual state is an electronic state for which the number operator does not commute with the Hamiltonian of the system and therefore has a finite lifetime. The lifetime of the virtual state is related to the Heisenberg uncertainty principle, as the electron can only reside on the virtual state on a timescale $t \sim \hbar / \left( \mu_N - \mu_{S,D} \right)$, where $\mu_N - \mu_{S,D}$ is the energy difference between the virtual state and the nearest real state. The main characteristic of this transport regime is a zero-bias resonance inside the Coulomb diamond for $N$=odd, as we will explain next, see also Fig.\,\ref{fig2_TRcd}(a) and (b).

When $N$\,=\,even, the total localized spin is zero due to the (typical) even-odd filling of the (spin) states, resulting in zero localized magnetic moment. When $N$\,=\,odd, one electron is unpaired, giving the localized state a net magnetic moment. In contrast to metals doped with magnetic impurities, the conductance of double barrier structures actually \textsl{increases} due the Kondo effect. This is because the density of states in the channel at a $\mu_S, \mu_D$ (associated with the newly formed Kondo singlet state) acts as a transport channel for electrons, as if it were a ``regular" localized state in the channel. The Kondo temperature can be expressed as \cite{GlazmanKluwer03}
\begin{equation}
T_K = \sqrt{E_C \Gamma} \exp(- \pi \frac{\mu_{N} - \mu_{S,D}}{2 \Gamma})
\label{vgl2_4}
\end{equation}
assuming $\mu_N - \mu_{S,D} \ll \mu_{N-1} - \mu_{S,D}$. The zero-bias Kondo resonance is furthermore characterized by its temperature and magnetic field dependence. The conductance of the Kondo resonance has a logarithmic temperature dependence, which is described by the phenomenological relationship \cite{GoldhaberGordonPRL98}
\begin{equation}
  G(T) = \left( G \right)_0 \left( \frac{T_K^{'2}}{T^2 + T_K^{'2}} \right)^s \label{vgl2_5}
\end{equation}
where $T_K^{'}=T_K/\sqrt{2^{1/s}-1}$, $G_0$ is the zero-temperature Kondo conductance and $s$ is a constant found to be equal to 0.22 \cite{GoldhaberGordonPRL98}.

%% file: physics.tex
Here we describe the fundamental physical properties of Si nanostructures. Some of these arise from the electron confinement into a small region (tens of nanometers or less) and are similar to those of other semiconductors, but other properties are present only in Si. One example arises because Si has multiple degenerate valleys in its conduction band, described in the first section. The valleys play an important role in both dopant and quantum dot devices, although the details of the valley physics in those two systems are different. Moreover, in heterostructures, strain often plays an important role, and the interplay between strain, disorder, and the properties of the valleys are important in determining the low-energy properties of the devices.

\subsection{Bulk silicon: valley degeneracy}\label{physicselectrons}
Because silicon is used in many technical applications, methods for manufacturing extremely high purity samples are well-developed. Silicon has several stable nuclear isotopes, with $^{28}$Si, which has no nuclear spin, being the most abundant (its abundance in natural silicon is 92\%). This availability of a spin-zero silicon isotope is useful for applications in which one wishes to preserve the coherence of electron spins, since the absence of hyperfine interaction eliminates a possible decoherence channel for the electron spin, see section VI.A.4. 

\begin{figure}[h]
\label{fig:si_crystal_structure}
\includegraphics[width=0.48\textwidth]{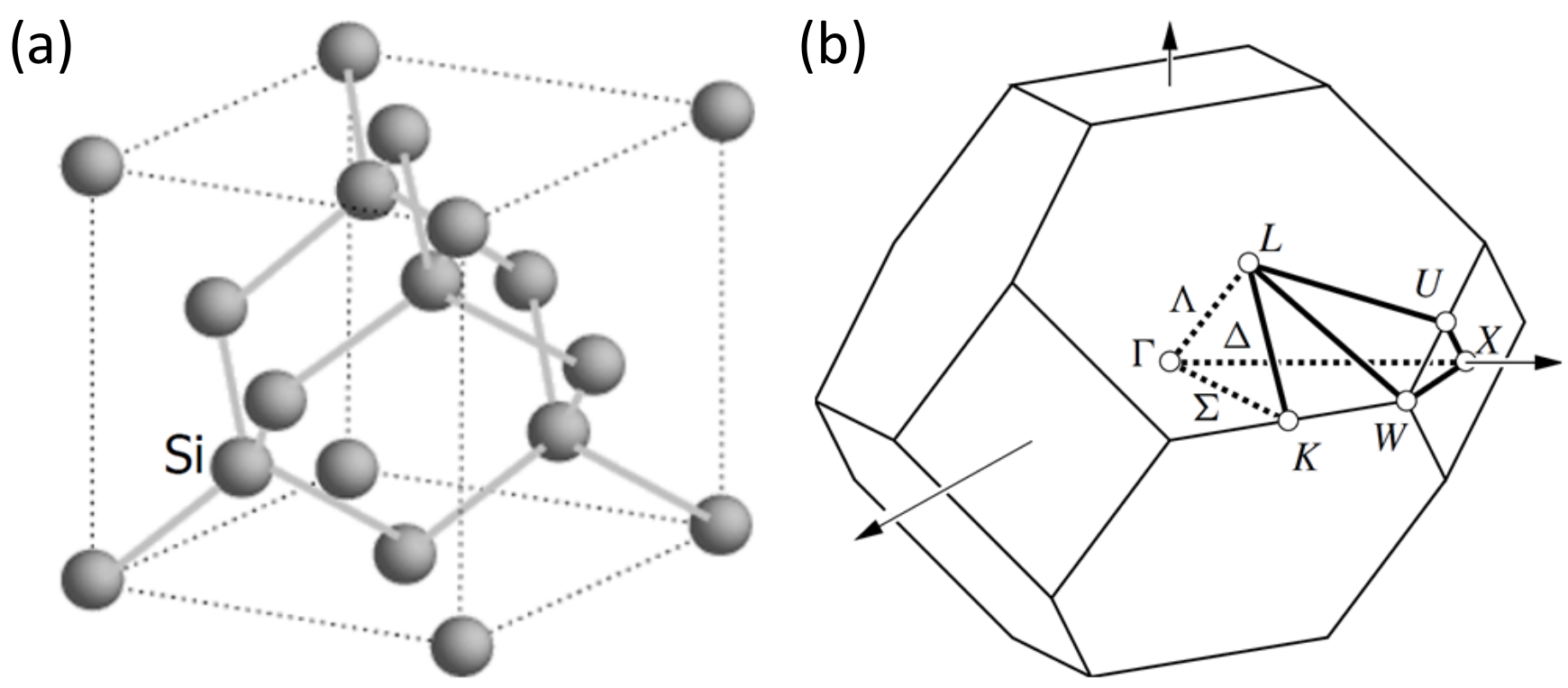}
\caption{\textbf{Silicon crystal in real and reciprocal space.} (a) 3D plot of the unit cell of the bulk silicon crystal in real space, showing the diamond or Face-Centered Cubic lattice, which has cubic symmetry. (b) Silicon crystal in reciprocal space. Brillouin zone of the silicon crystal lattice. It is the Wigner-Seitz cell of the Body-Centered Cubic lattice. $\Gamma$ is the center of the polyhedron. Figure from \textcite{DaviesBook}.}
\end{figure}

The properties of electrons in silicon have been studied in great detail for many decades~\cite{Cohen_book:1988,Yu_book:2001}. Here we review aspects of the material that will prove critical in understanding the challenges that arise as one works to create devices with desired properties on the nanoscale. One such aspect is how the effects of multiple valleys present in the conduction band in bulk silicon appear in specific silicon nanodevices. The manifestations of valley physics in quantum dots are different from those in dopant-based devices, and understanding the relevant effects is critical for manipulating the spin degrees of freedom of the electrons in nanodevices. In the following subsections, we first define and discuss the conduction band valleys in bulk silicon, and then the behavior and consequences of valley physics for quantum dots and for dopant devices.

Crystalline silicon is a covalently bonded crystal with a diamond lattice structure, as shown in Fig.~5. The band structure of bulk silicon~\cite{Phillips:1962p1931}, shown in Fig.~6, has the property that the energies of electron states in the conduction band is not minimized when the crystal momentum $k=0$, but rather at a nonzero value, $k_0$, that is 85\% of the way to the Brillouin zone boundary, as shown in Fig.~6(b). Bulk silicon has cubic symmetry, and there are six equivalent minima. Thus we say that bulk silicon has six degenerate valleys in its conduction band.

\begin{figure}[t]
\label{fig:bulk_si_band_structure}
\includegraphics[width=0.48\textwidth]{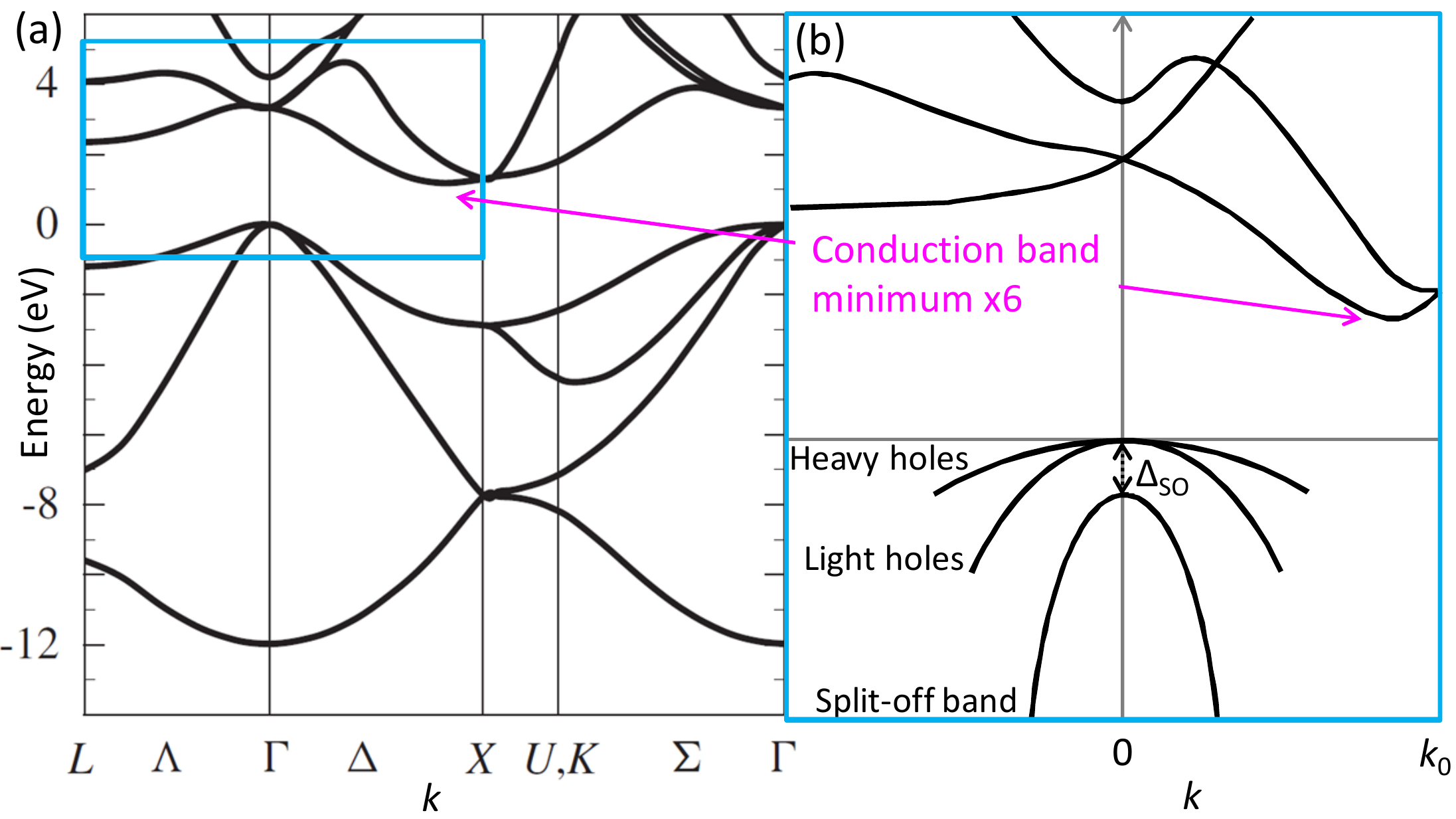}
\caption{ (Color online) \textbf{Band structure of bulk silicon.} (a) The conduction band has six degenerate minima or valleys at $0.85 k_0$.  Results kindly supplied by G.P.~Srivastava, University of Exeter. Figure from \textcite{DaviesBook}. (b) Zoom-in on the bottom of the conduction band and the top of the valence band (schematic, not exact). The bandgap in bulk Si is 1.12 eV at room temperature, increasing to 1.17 eV at 4 K \cite{greenJAP90}. The heavy and light hole bands are degenerate for $k = 0$. The split-off band is separated from the other subbands by the spin-orbit splitting $\Delta_{\textrm{so}}$ of 44 meV.}
\end{figure}

In conventional electronic devices, the presence of multiple valleys typically does not affect transport properties in a profound way. However, valley physics plays a critical role in quantum electronics because of  interference between different valleys that arises when the electronic transport is fundamentally quantum. For example, the  presence of an additional valley greatly complicates spin manipulation because it can lift Pauli spin blockade,
which is fundamental for many strategies for spin manipulation in quantum dot nanodevices~\cite{onoScience02,Johnson:2005p165308,Koppens:2005p1346,Rokhinson:2001p035321,Huttel:2003p1958}.  In pure bulk silicon, the valleys are degenerate (the energies  of the six states related by the cubic symmetry are the same), but in nanodevices this degeneracy can be and usually is broken by various effects that include strain, confinement, and electric fields. When valley degeneracy is lifted, at low temperatures the carriers  populate only the lowest-energy valley state, thus eliminating some of the quantum effects  that arise when the valleys are degenerate.

\begin{figure}
\includegraphics[width=0.48\textwidth]{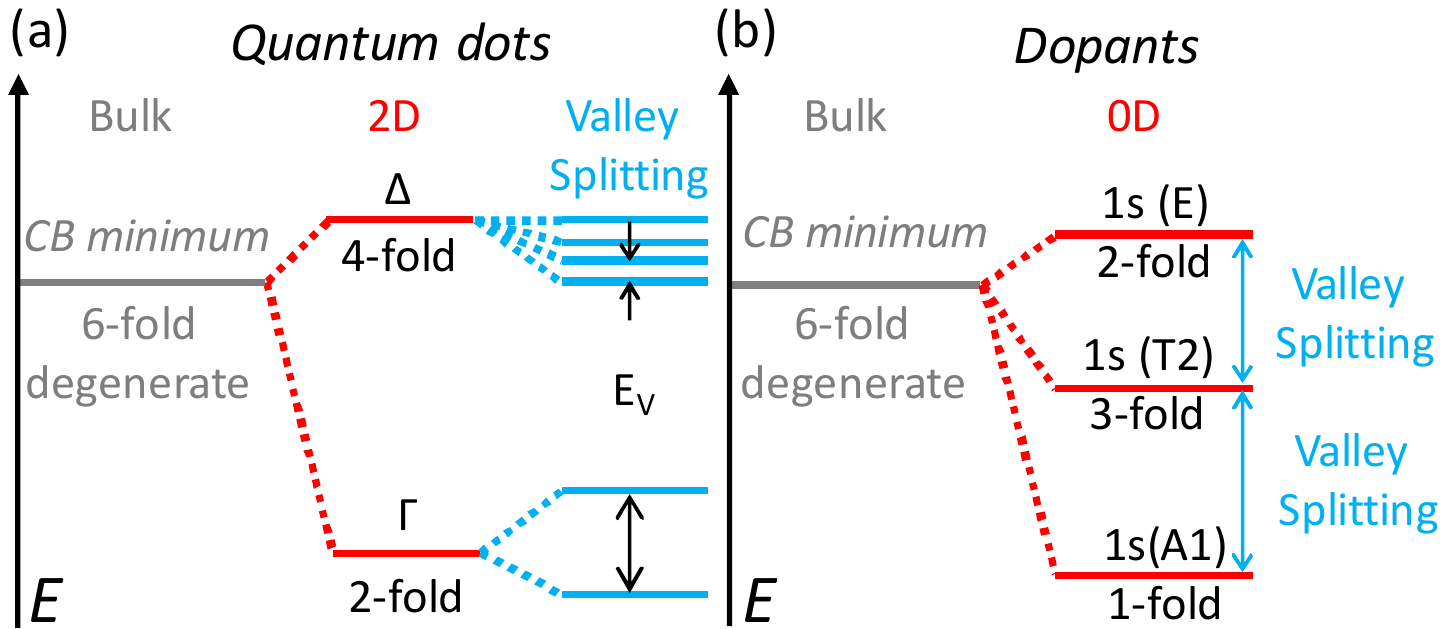}
\caption{ (Color online) \textbf{Valley splitting of of dopants and of quantum dots in silicon quantum wells.}
(a) For a quantum well, in which a thin silicon layer is sandwiched between two layers of Si$_x$Ge$_{1-x}$, with $x$ typically $\sim 0.25-0.3$, the six-fold valley degeneracy of bulk silicon is broken by the large in-plane tensile strain in the quantum well so that two $\Gamma$-levels are about ${\rm 200~meV}$ below the four $\Delta$-levels~\protect{\cite{Schaffler:1992p1515}}. The remaining two-fold degeneracy is broken by the confinement in the quantum well and by  electric fields, with the resulting valley splitting typically ${\rm \sim 0.1-1~meV}$. (b) For phosphorus dopants, strong central-cell corrections near the dopant break the six-fold valley degeneracy of bulk silicon so that the lowest-energy valley state is non-degenerate (except for spin degeneracy), lowered by an energy ${\rm 11.7~meV}$. The degeneracies of higher-energy levels are broken by lattice strain and by electric fields.}
\label{valleys-bulk-2D-dopants-alternative}
\end{figure}

Fig.~\ref{valleys-bulk-2D-dopants-alternative} shows a summary of valley splitting in heterostructures and in dopant devices. For strained silicon quantum wells, the large in-plane strain lifts the energies of the in-plane (x and y) valleys.  The remaining two-fold  degeneracy of the z-valleys is broken by electronic z-confinement induced by electric fields and by the quantum well itself, resulting in a valley splitting of order $0.1-1$ meV. The breaking of the two-fold valley degeneracy is very sensitive to atomic-scale details of the interface, and is discussed in detail in Sec.~\ref{physicsdots} and in the supplemental material.

For an electron bound to a dopant in silicon, the valley degeneracy of bulk silicon is lifted because of the strong confinement potential from the dopant atom~\cite{kohnPRB55}. For phosphorus donors in silicon, the electronic ground state is non-degenerate, with an energy gap of ${\rm \sim 11.7~meV}$ between the non-degenerate ground state and the excited states~\cite{RamdasRPP1981,Andresen:2009p169}. Thus, additional degeneracy of the electronic ground state is not a concern in dopant devices. However, the fact that the conduction band minimum in silicon is at a large crystal momentum $k_0$ that is near the zone boundary gives rise to other physical effects that are important for quantum electronic devices. One such consequence arises because the wave functions of the electronic states in dopants oscillate in space on the very short length scale $ \sim 2 \pi/k_0$, which is roughly on the scale of one nanometer. These charge oscillations differ from the electron charge variations due to Bloch oscillations because they can cause the exchange coupling to change sign, and thus have significant implications for the design of quantum electronic devices, as discussed in Section \ref{physicsdopants}.

\subsection{Quantum wells and dots}\label{physicsdots}
In the quantum well devices we discuss here, one starts with a material with  a two-dimensional electron gas (2DEG), and then lithographically patterns top gates to which voltages are applied that deplete the 2DEG surrounding the quantum dot. By carefully adjusting the gate voltages, one can achieve dots with occupancy of a single electron, see section \ref{fewelectron}. Moreover, the same gate voltages that are used to define the dot are also
used to perform the manipulations required for initialization, gate operations, and readout of charge and spin states~\cite{mauneNature12}, see section VI.C.4. 

\subsubsection{Valley splitting in quantum dots}\label{motivehetero}
Understanding the valley degrees of freedom is important for ensuring that the valley splitting is in a regime suitable for spin-based quantum computation. Even in the low-density limit appropriate to single-electron quantum dots, where electron-electron interactions~\cite{andoRMP82} are unimportant, valley splitting is complex: the breaking of the valley degeneracy involves physics on the atomic scale, orders of magnitude smaller than the quantum dot itself, so it depends on the detailed properties of alloy and interface disorder.
 Because the locations of the individual atoms in a given device are not known, statistical approaches to atomistic device modeling or averaging theories like effective mass must be utilized. Theory, modeling, and simulation provide insight into the physical mechanisms giving rise to valley splitting, so that device design and fabrication methods can be developed to yield dots with valley splitting compatible with use in spin-based quantum information processing devices.

In bulk silicon, there are six degenerate conduction band minima in the Brillouin zone (valleys) as depicted in Fig.~5. One modern strategy for fabricating Si devices for quantum electronics applications is to use a bi-axially strained thin film of Si grown
on a pseudomorphic $\rm Si_xGe_{1-x}$~ substrate.
In such devices,
the silicon quantum well is under large tensile strain, and the six-fold degeneracy is broken into a two-fold one~\cite{schafflerSSCT97}.
Confinement of electrons in the $z$-direction in a 2-dimensional electron gas lifts the remaining two-fold valley degeneracy, resulting in four $\Delta$-valleys with a heavy effective mass parallel to the interface at an energy several tens of meV above the two $\Gamma$-valleys~\cite{andoRMP82}, as shown in Fig.~\ref{valleys-bulk-2D-dopants-alternative}.
The sharp and flat interface produces a potential step in the $z$-direction
and can lift the degeneracy of the $\Gamma$-valleys in two levels separated by the valley splitting $E_V$.
Built-in or externally applied electric fields break the symmetry of the Hamiltonian and can couple the various valleys and thus lift the valley degeneracy.
Theoretical predictions for the valley splitting of flat interfaces
are generally on the order of 0.1--0.3 meV~\cite{Ohkawa:1977p917,BoykinAPL04,Culcer:2010p155312, saraivaArxiv10}. Experimental values in Si inversion layers mostly vary from 0.3--1.2 meV, but some are substantially smaller~\cite{kohlerOther79, nicholasOther80, pudalovOther85,Weitz:1996p542,koesterSST97,Lai:2006p161301}.
A giant valley splitting of 23 meV measured in a similar structure~\cite{Takashina:2006p236801} is still not completely understood theoretically~\cite{saraivaArxiv10}.

\begin{figure}[htbp]
\begin{center}
\includegraphics[width=0.48\textwidth]{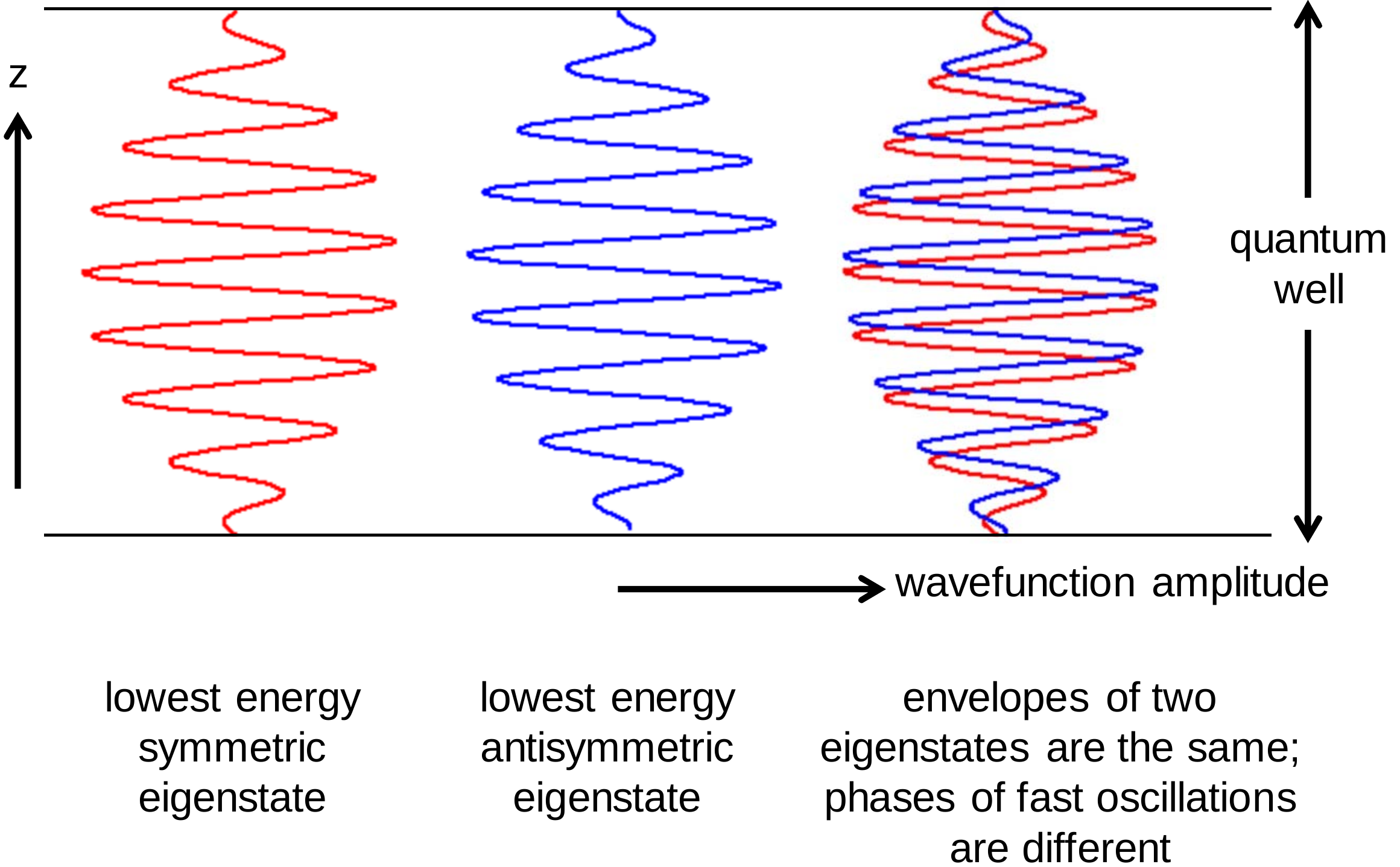}
\caption{ (Color online) \textbf{Sketch of the two lowest energy eigenstates in an infinite square well of the two-band model presented in the supplemental material.} The envelopes of the two eigenfunctions are very similar to each other and to the sine behavior obtained in the absence of valley degeneracy; the effects of the valley degeneracy give rise to  fast oscillations within this envelope.  For a square well, one eigenfunction is symmetric and the other  is antisymmetric; the symmetries are different because the fast oscillations have different phases, as measured from the quantum well boundaries. This sensitive dependence of valley splitting on the atomic-scale physics near the well boundary is the source of the sensitive dependence of the valley splitting on disorder at the quantum well interfaces.} \label{fig:wavefunction_sketch}
\end{center}
\end{figure}

The two main approaches for understanding valley splitting in silicon heterostructures are tight-binding calculations~\cite{boykinPRB04,Boykin:2005p1461,kharcheAPL07,srinivasanAPL08,boykinPRB07} and theories that use an effective mass formalism~\cite{Friesen:2007p115318,Saraiva:2009p081305,Friesen:2010p115324}. Section I in the supplemental material reviews a simple one-dimensional tight-binding model~\cite{BoykinAPL04} that illustrates some of the physical mechanisms that lead the breaking of the valley degeneracy and hence the emergence of valley splitting. A pictorial sketch of the two lowest-energy eigenstates of this one-dimensional model is presented in Fig.~\ref{fig:wavefunction_sketch}. The eigenfunctions have very similar envelopes and fast oscillations with a period very close to $2\pi/k_0$, where $k_0$ is the wavevector of the conduction
band valley minimum. The different alignments of the phases of the fast oscillations with sharp interfaces cause the energies of the two states to be different, thus giving rise to valley splitting.

Valley splitting has a complicated dependence on environmental and structural conditions.  Large-scale atomistic tight-binding calculations can incorporate realistic inhomogeneity in the atomic arrangement, both in terms of alloy disorder and in terms of disorder in the locations of interface steps, as discussed in section III of the supplemental material.
Technically well controlled interfaces in Si are buffers of either $\rm SiO_2$ or $\rm Si_xGe_{1-x}$, which are intrinsically atomistically disordered.
Some of the effects of this disorder can be understood qualitatively using effective
mass theory, but because of the importance of atomic-scale physics in
determining valley splitting, atom-scale theory is required for quantitative understanding.
For $\rm Si_xGe_{1-x}$, there are 3 critical disorder effects to consider: atom-type disorder, atom-position disorder, alloy concentration disorder.
A detailed discussion of the characterization of the effects of these different types is presented in section III of the supplemental material.

Many  features of the physics that give rise to valley splitting can be understood qualitatively and  semi-quantitatively using effective mass
theories~\cite{Kohn:1955p915,Kohn_inbook:1957}, if these theories are formulated carefully to incorporate the microscopic effects that give rise to valley splitting~\cite{Fritzsche:1962p1560,Pantelides:1978p797,FriesenPRL05,Friesen:2007p115318,Nestoklon:2006p235334}. In the envelope function or effective mass formalism, the theory is written in terms of an envelope function for the wave function, which is well-suited for describing variations on relatively long scales (such as the quantum dot confinement). The effects of the degenerate valleys are incorporated using a valley coupling parameter that is treated as a delta function whose strength is determined by the atomic scale physics~\cite{Friesen:2007p115318,Chutia:2008p193311,Saraiva:2009p081305}. The envelope function formalism has the advantage that one can obtain analytic results for valley splitting in nontrivial geometries~\cite{Friesen:2007p115318,Culcer:2010p205315,Culcer:2010p155312,Friesen:2010p115324}. However, the theory must explicitly incorporate information from the atomic scale, either as a valley coupling parameter that is fit to tight-binding results, as the output of a multiscale approach~\cite{Chutia:2008p193311,Saraiva:2009p081305}, or by explicit atomistic calculation on large scales, as embodied by the NEMO tool suite~\cite{klimeckCMES02,BoykinAPL04,klimeckNanoDM07,steigerIEEE11}. More details of effective mass theory treatment of valley splitting are  in
 the supplemental material.

\subsubsection{Mixing of valleys and orbits}\label{phys-valleyphysics}
When the valley splitting $E_V$ is much greater than the orbital level spacing $\Delta E$, electrons will occupy single-particle levels with orbital numbers 1, 2, 3, ...~and valley number $V1$, the lowest valley state (see Fig.~9(a)).
Conversely, if $\Delta E \gg E_V$ the first four electrons will occupy the valleys $V1$ and $V2$ in the lowest orbit before going to the next orbit with $n=2$, as shown in Fig.~9(b).
However, valleys and orbits can also hybridize \cite{Friesen:2010p115324}, making it inappropriate to define distinct orbital and valley quantum numbers (see  Fig.~9(c)).
Depending on the degree of mixing, the valley-orbit levels $VO1$, $VO2$ etc, behave mostly like valleys or like orbits. Instead of referring to a pure valley splitting $E_V$ the term valley-orbit splitting is used, $E_{VO} = E_{VO2} - E_{VO1}$ for the difference in energy between the first two single-particle levels, $E_{VO1}$ and $E_{VO2}$. This is referred to as the ground-state gap~\cite{Friesen:2010p115324}.

\begin{figure}[t]
\includegraphics[width=0.4\textwidth]{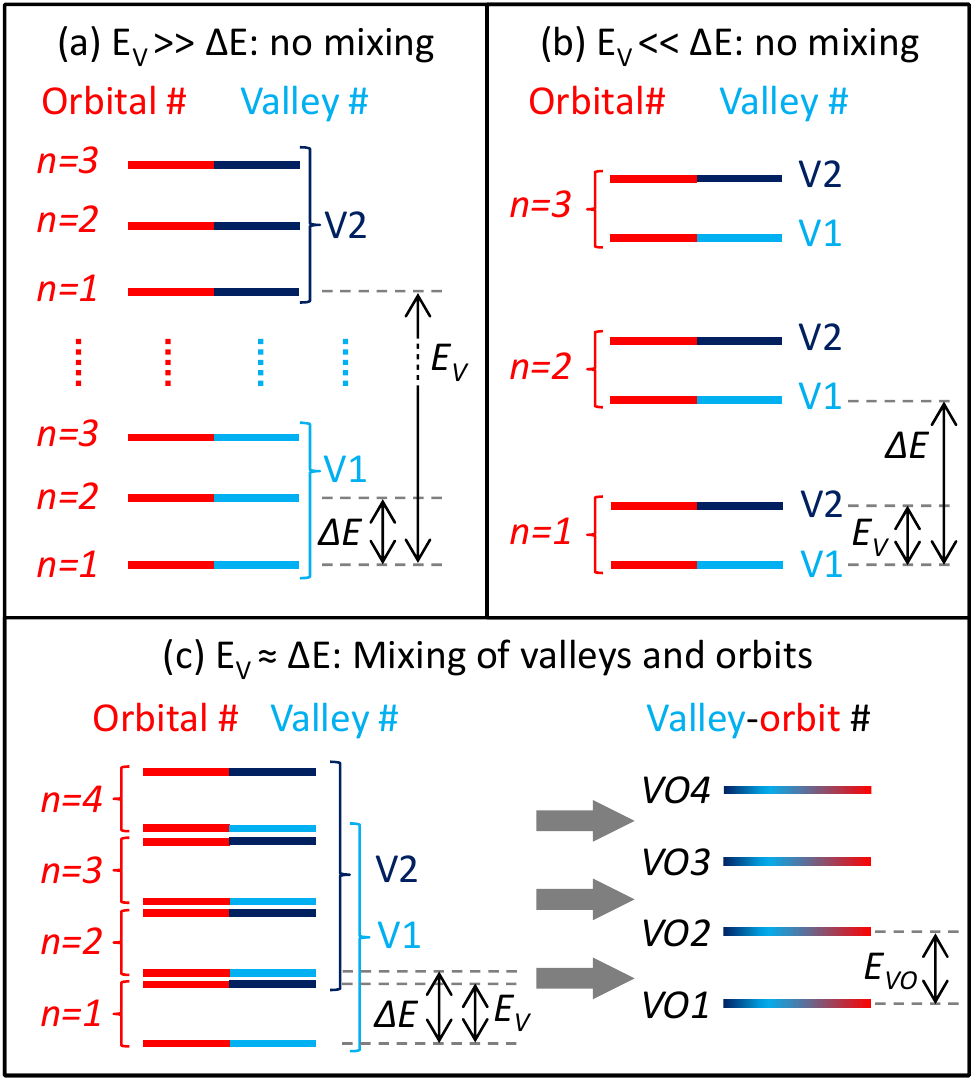}\label{valleymixing}
\caption{ (Color online) \textbf{Valley-orbit mixing.} (a,b) If the valley splitting $E_V$ and orbital level spacing $\Delta E$ have very different values, the orbital and valley quantum numbers are well-defined and there will be no mixing of orbital and valley-like behavior. (c) When $E_V$$\approx$$\Delta E$ the valleys and orbits can hybridize in single-particle levels separated by the valley-orbit splitting $E_{VO}$.}
\end{figure}

The behavior of the valley splitting in real quantum wells is complicated by the fact that  in real devices the quantum well interface is not perfectly smooth and oriented perpendicular to $\hat{z}$. The energy difference between the two lowest eigenstates depends on the relationship between the phase of the fast oscillations of the wave function with the heterostructure boundary, and a step in the interface alters this phase relationship. The lowest energy wave function minimizes the energy, and, as shown in Fig.~8, can cause the phase of the fast oscillations to become dependent on the transverse coordinates $x$ and $y$. This coupling between the $z$-behavior and the $x$-$y$ behavior is called valley-orbit coupling.

\begin{figure}[htb]
\includegraphics[width=0.4\textwidth]{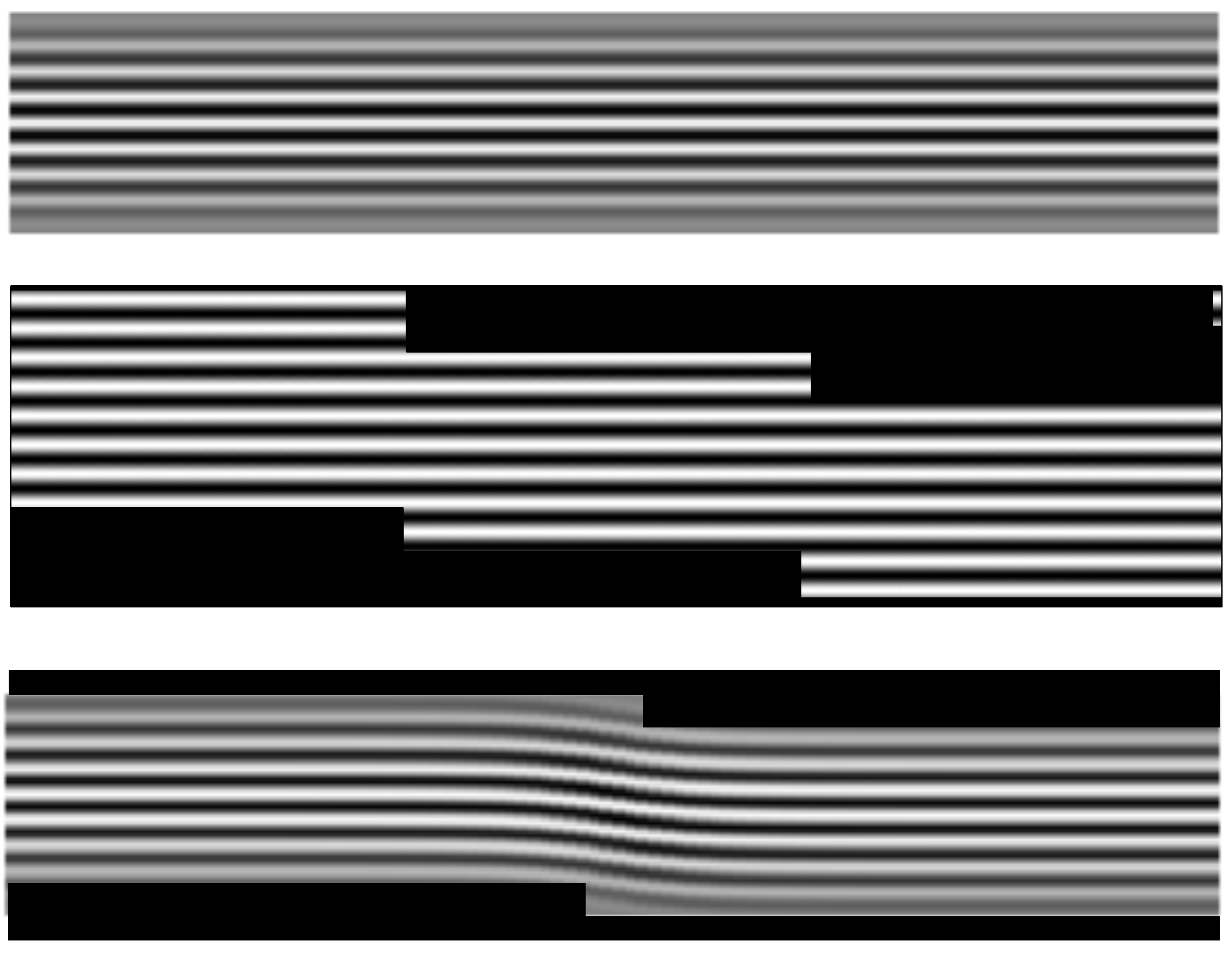}\label{fig:valleyorbit}
\caption{\textbf{Valley-orbit coupling from interface steps.} \textsl{Top}: gray-scale visualization of wave function oscillations in the presence of a perfectly smooth interface, oriented perpendicular to $\hat{z}$. \textsl{Middle}: The relationship between the phase of the wave function oscillations and the interface is different on the two sides of an interface step. When the steps are close together, the phase does not adjust to the individual steps, and the valley splitting is suppressed. \textsl{Bottom}: When steps are far enough apart, the oscillations line up with the interface location on both sides of the steps, which causes the phase of the oscillations to depend on the transverse coordinate. This coupling between the behavior of the wave function in the $z$ direction and in the $x-y$ plane, which arises even when the well is atomically thin, is known as valley-orbit coupling.}
\end{figure}

As discussed in subsection~\ref{motivehetero} above, in a silicon quantum well  under tensile strain, there are two low-lying conduction band valleys at wavevectors $+k_0 \hat{z}$ and $- k_0 \hat{z}$, whose energies are split by the effects of confinement potentials and electric fields perpendicular to $z$. In the limit of a perfectly smooth interface aligned perpendicular to $\hat{z}$, the valley splitting of a quantum well with typical width and doping is of order 0.1 meV, a magnitude that can be understood using the simple one-dimensional model presented in section I of the supplemental material.

If the step density of the quantum well interface is reasonably high, then the transverse oscillations of the charge density cannot align with the entire interface, and valley splitting is greatly suppressed~\cite{andoPRB79,Friesen:2006p202106,Friesen:2007p115318}. The physical picture that emerges from effective mass theory that incorporates valley-orbit coupling is that the envelope function for the wave function in a silicon heterostructure is qualitatively similar to typical wave functions in quantum dots, but that there are also fast oscillations with wave vector $\sim k_0$ in the z-direction. The fast oscillations of the two valley states have different phases. In the presence of interfacial disorder such as interfacial steps, the value of the valley phase that minimizes the energy becomes position-dependent, so that one fixed value of the phase cannot minimize the energy everywhere, and the energy difference between the
two different valley states decreases. This suppression explains measurements performed in Hall bars~\cite{Weitz:1996p542, koesterSST97, Khrapai:2003p113305, Lai:2004p156805} that yield very small values for the valley splitting of only $\mu$eV, and also why singlet-triplet splittings in dots with two electrons have been observed with both positive and negative values at non-zero magnetic field~\cite{borselliAPL2011} --- if the electron wave function straddles a step, then the valley splitting is small, which, together with the effects of electron-electron interactions, causes the triplet state to have lower energy than the singlet state. If an electron is confined to a region small enough that it does not extend over multiple steps, then the valley splitting is not affected by the steps. Over the past several years, measurements of valley splitting in quantum point contacts~\cite{goswamiNPhys06} and of singlet-triplet splittings in quantum dots~\cite{borselliAPL11,borselliAPL2011,simmonsPRL11,Thalakulam:2011p045307} in Si/SiGe heterostructures demonstrate that these splittings can be relatively large, of order ${\rm 1~meV}$, when the electrons are highly confined.  These splittings are large enough that valley excitations are frozen out at the relevant temperatures for quantum devices (${\rm \sim 100~mK}$).

There are two different manifestations of valley-orbit coupling.  The first, illustrated in the bottom panel of Fig.~10, occurs when the phase of the valley oscillations depends on the transverse coordinate. The second type of valley-orbit coupling can be visualized by considering an interface with a nonuniform step density.   A wave function localized in a region with few steps has larger valley splitting and hence lower energy than a wave function localized in a region with many steps~\cite{Shi:2011p233108}. Therefore, the presence of the valley degree of freedom leads to translation of the wave function in the $x$-$y$ plane. Valley-orbit coupling is important when the scale of the variations of the orbital and valley contributions to the energy are similar, a situation that occurs frequently in few-electron quantum dot devices.

Because valley-orbit coupling and valley splitting depend on interface details, the observation of valley splittings that vary substantially between devices~\cite{borselliAPL11} is not unexpected. Understanding and controlling this variability is important for being able to scale up the technology and for the development of devices that exploit the valley degree of freedom~\cite{CulcerPRB09,LiPRB10,CulcerPRL12,Shi:2012p140503}. Therefore, improved understanding of the physical mechanisms that affect valley splitting in real devices remains an important topic of active research. The valley-orbit coupling also contains phase information, which can be used for quantum computation \cite{WuPRB12}.

\subsection{Dopants in Si}\label{physicsdopants}

\subsubsection{Wave function engineering of single dopant electron states}\label{wavefunction}

\begin{figure}[t]
\centering
\includegraphics[width=0.48\textwidth]{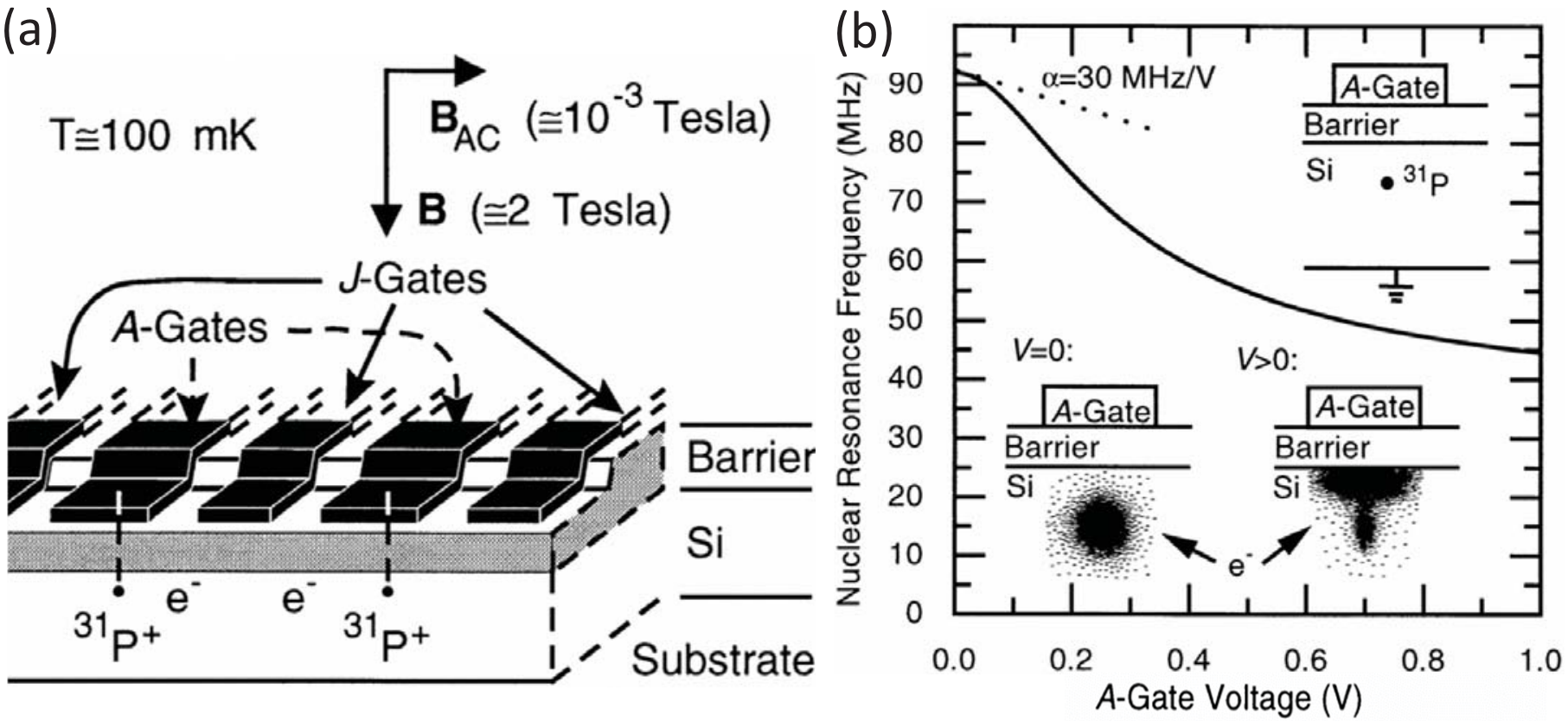}
\caption{\textbf{A silicon-based nuclear spin quantum computer} (a) Schematic of Kane's proposal for a scalable quantum computer in silicon using a linear array of $^{31}$P donors in a silicon host. $J$-gates and  $A$-gates control respectively the exchange interaction $J$ and the wave function, as shown in (b). Reproduced from \textcite{KaneNature98}.} \label{Kane1}
\end{figure}

The central theme of quantum electronics applications using single dopants is the ability to modify the dopant electron wave function using external electric fields and/or to manipulate the spin degrees of freedom using magnetic fields. In many proposals for dopant based qubits using either electron or nuclear spins as the qubit states, dopant electron wave function engineering is critical to effect single and two qubit gates. Since most work has been done on n-type dopants, this section will focus on donors. The original idea comes from the Kane proposal for a nuclear-spin based quantum computer in silicon \cite{kaneFDP00} where the single qubit operations are implemented by tuning the contact hyperfine interaction to bring the donor electron into resonance with a transverse oscillating magnetic driving field (see Fig.~\ref{Kane1}).
To see this we write the effective spin qubit Hamiltonian of a single donor nucleus-electron system in the presence of a gate potential with strength $V$ at the donor position as \cite{KaneNature98, goanIJQI05}
\begin{equation}
{ H_{1Q}} = \mu_B B_z  \sigma_e^z - g_n \mu_n B_z  \sigma_n^z  + A(V_A) {\vec   \sigma}_n.{ \vec \sigma}_e,
\label{equation:1Qham}
\end{equation}
where $\mu_B$ is the Bohr magneton, $g_n$ the Land\'{e} factor for $^{31}$P, and $\mu_n$ is the nuclear magneton. The contact hyperfine interaction strength $A$ can be tuned by an applied electric field arising from a bias $V_A$ on an $A$-gate as:
\begin{equation}
A(V_A) = {2 \over 3} |\psi(0,V_A)|^2 \mu_B g_n \mu_n \mu_0,
\end{equation}
where $\mu_0$ is the permeability of silicon and $\psi(0,V_A)$ is the donor electron wave-function evaluated at the nucleus under the $A$-gate bias $V_A$.

\begin{figure}
\label{fig:HFStark}
\includegraphics[width=0.4\textwidth]{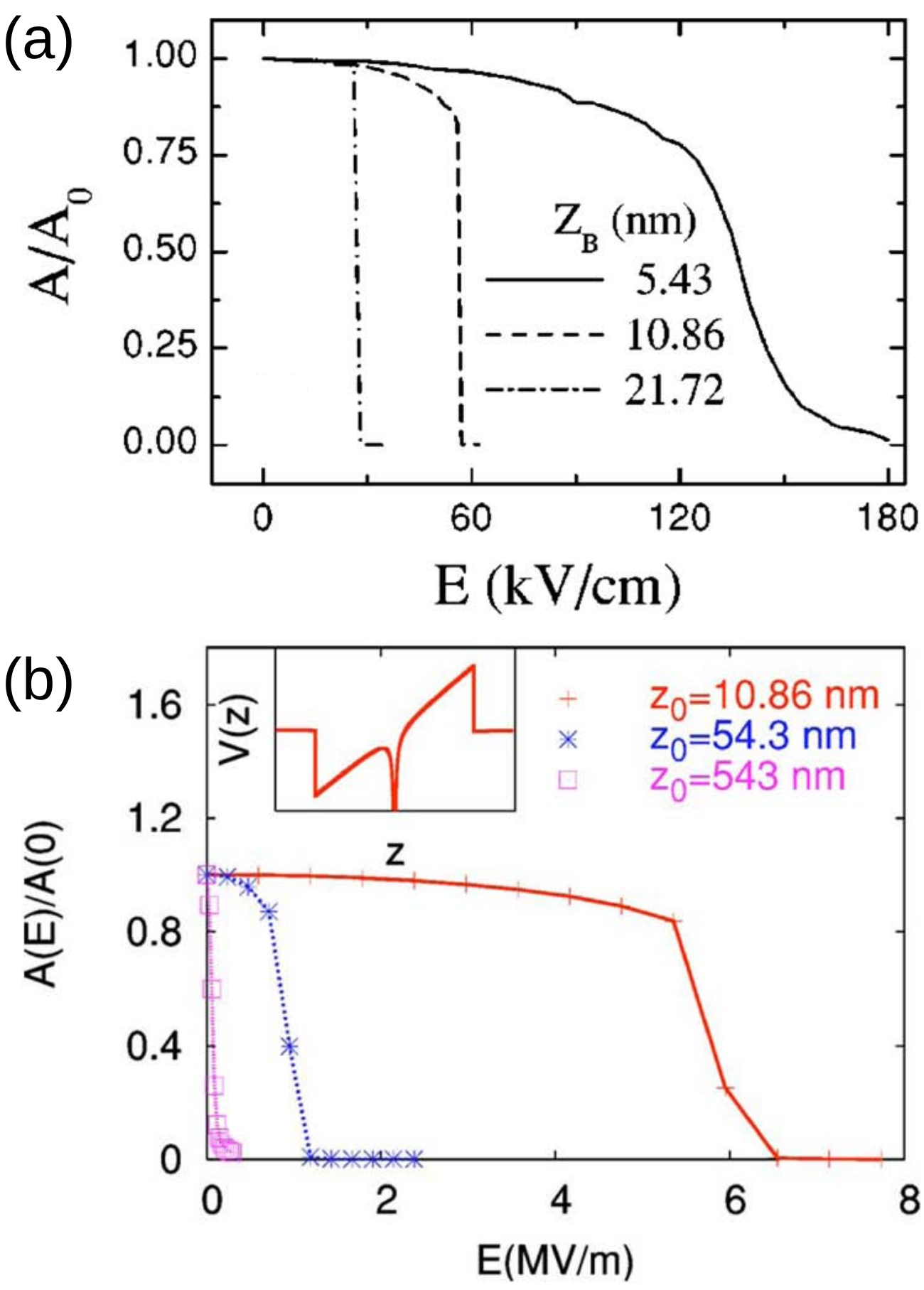}
\caption{ (Color online) \textbf{Relative Stark shift of the contact hyperfine interaction for different donor depths ($z$) calculated for a uniform field in the z direction.} (a) Using the tight-binding approach \cite{martinsPRB04}, and (b) Direct diagonalization in momentum space \cite{wellardPRB05}. Agreement in overall trends is reasonable, and for the $z=10.86$ nm case both methods predict ionization at $\sim$6 MV/m.}
\end{figure}

The change in the strength of the contact hyperfine coupling due to the application of a gate bias has been studied by several authors since Kane's proposal. To determine the change in the contact hyperfine coupling strength it is necessary to calculate the shift in the donor electron wave function at the position of the donor nucleus. Depending on the applied bias polarity, an $A$-gate control electrode will either draw the wave function toward, or away, from the gate. In either scenario the wave function at the donor nuclear position is perturbed to some extent. The resulting tuning of $A$ depends critically on device parameters such as the depth of the donor from the interface, and the gate/interface geometry. The level of sophistication of the treatment of the donor electron wave function in these devices has steadily improved since the original calculations following \textcite{KaneNature98}. The earliest approaches used fairly simple hydrogenic wave functions scaled by the dielectric constant of silicon. \textcite{larionovNANO00} treated the bias potential analytically, and the shift in the hyperfine interaction constant as a function of applied bias voltage was calculated using perturbation theory. \textcite{wellardNANO02}, again using scaled hydrogenic orbitals treated the problem using a more realistic gate potential (modeled using a commercial semi-conductor software package, with built in Poisson solver). The donor electron wave function was expanded in a basis of hydrogenic orbitals in which the Hamiltonian was diagonalized numerically. \textcite{KettlePRB03} extended these calculations using a basis of non-isotropic scaled hydrogenic orbital states. \textcite{SmitPRB03,SmitPRB04} used group theory over the valley manifold and perturbation theory to describe the Stark shift of the donor electron while \textcite{martinsPRB04,martinsPRB05} applied tight-binding theory to obtain the first description of the Stark shift of orbital states and the hyperfine interaction incorporating Bloch structure. Meanwhile, the effective mass treatment was further developed in a combined variational approach \cite{FriesenPRL05} and \cite{CalderonJAP2009}, and in \cite{DebernardiPRB06} using a Gaussian expansion of the effective-mass theory (EMT, see section II of the supplemental material) envelope functions. This was followed by the application of direct diagonalization in momentum space \cite{wellardPRB05} allowing the potential due to the $A$-gate to be included at the Hamiltonian level and gave a similar picture of the Stark shift of the hyperfine interaction as a function of external field strength and donor depth as the earlier tight-binding treatment of \textcite{martinsPRB04} (see Fig.~12). Although not optimized computationally, the momentum space diagonalization approach has served as a consistency check against larger scale real-space tight-binding calculations of the Stark shift of the donor hyperfine interaction at low fields \cite{RahmanPRL07} in the overall benchmarking against experiment \cite{BradburyPRL06} which shows the theoretical description has converged to a reasonable level in terms of internal consistency and comparison with experiment  (see Fig.~13). It should be noted that in such descriptions encompassing the overall donor electron wave function it is the relative change in the contact hyperfine interaction as a function of electric field that is computed since these approaches do not describe well the details of the electron state at the nucleus. Absolute calculations of the contact hyperfine interaction are the domain of ab-initio theories where they have had remarkable success despite the truncation of the long range part of the donor potential \cite{OverhofPRL04,gerstmannPSSB11}.

In more recent years, the effect of depth and proximity to the interface on donor orbital states \cite{CalderonPRL06,CalderonPRB08,rahmanPRB09,haoPRB09} has received more attention as key experimental measurements became available. A turning point was the measurement of donor orbital states through transport in FinFET devices. The observed donor energy levels were very different from the bulk spectrum (see section \ref{d_discussion}). Extensive tight-binding calculations were used to explore the space of electric field and donor depth on the quantum confinement conditions of the donor-associated electron, identifying Coulombic, interfacial, and hybridized confinement regimes.  These calculations provided an excellent description of the low lying donor states observed and determination of the donor species \cite{LansbergenNphys08}. It would appear that the theoretical description of electric field ``wave function engineering" of the donor electron across device dimensions is now well understood. The context of the Kane donor qubit has spurred further refinements of the theoretical description of donor states, including the site-specific contact and non-isotropic hyperfine interaction terms \cite{iveyPRB75,IveyPRB75a} for wave function mapping under electric fields \cite{ParkPRL09}, interaction with magnetic fields and gate control of the g-factor \cite{rahmanPRB09b,thilderkvistPRB94}, dynamics of molecular donor-based systems \cite{HollenbergPRB04,HuPRB05,WellardPRB06,RahmanNANO11}, cross-talk in hyperfine control \cite{KandasamyNANO06},  coherent single electron transport through chains of ionized donor chains \cite{rahmanPRB09b}, spin-to-charge readout mechanisms \cite{FangPRB02,HollenbergPRB04}, and the calculation of donor levels in the presence of STM-fabricated nanostructures providing modifications to the overall potential in a single-atom transistor, as shown in section \ref{d_dopant_STM} \cite{fuechsleNatureN12}.

\begin{figure}
\label{fig:HFStarkLowFields}
\includegraphics[width=0.48\textwidth]{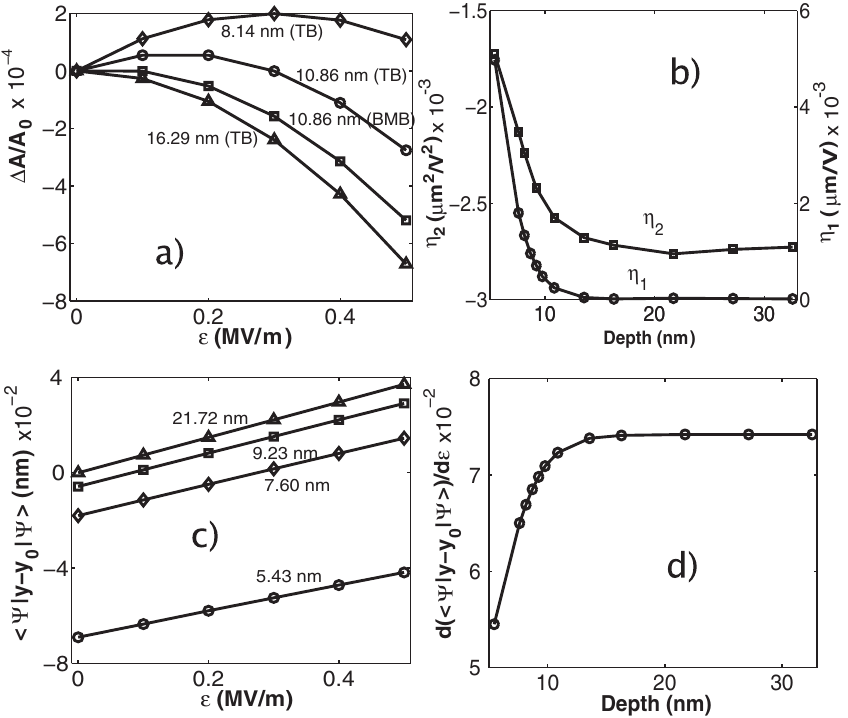}
\caption{\textbf{Low-field Stark shift of the hyperfine interaction for momentum space diagonalization (BMB) and tight-binding (TB) methods.} (a) Electric field response of hyperfine coupling at various donor depths (BMB and TB). (b) Quadratic (lefthand axis) and linear (right-hand axis) Stark coefficients as a function of donor depth (TB). (c) Shift of the ground state electron distribution (dipole moment) as a function of the electric field (TB). (d) The electric field gradient of the dipole moments as a function of donor depth (TB). From \textcite{RahmanPRL07}.}
\end{figure}

\subsubsection{Two-donor systems and exchange coupling}\label{wavefunction}

In the quantum computing context, the two main approaches to directly couple the spins of donor electrons are through the Coulomb-based exchange interaction between proximate donor electrons, or the magnetic dipole interaction. The Kane model uses gate control of the exchange interaction as per the two-qubit effective spin Hamiltonian:
\begin{eqnarray}
{ H_{2Q}} &=& \mu_B B_z  \sigma_{e_1}^z - g_n \mu_n B_z  \sigma_{n_1}^z  + A_{1}(V_{A1}) {\vec \sigma}_{n_1}.{\vec \sigma}_{e_1}\\\nonumber
&+& \mu_B B_z  \sigma_{e_2}^z - g_n \mu_n B_z  \sigma_{n_2}^z  + A_{2}(V_{A2}) {\vec \sigma}_{n_2}.{\sigma}_{e_2}\\\nonumber
&+& J(V_J) {\vec \sigma}_{e_1}.{\sigma}_{e_2}.
\label{equation:twoQ}
\end{eqnarray}
In this equation we apply equation \ref{equation:1Qham} on two dopants and add the exchange-coupling $J$ between the dopants. There have been a number of papers investigating the construction and fidelity of two-qubit gates (e.g.\ such as the controlled-NOT) from this Hamiltonian \cite{FowlerPRA03,HillPRA03,HillPRA04,fangPRB05,kerridgeJOPCM06,tsaiAIP08,tsaiPRA09}. From a microscopic physics viewpoint, in general the exchange energy $J$ is stronger than the dipole interaction for smaller separations, however it behaves as \cite{HerringPR64}
\begin{equation}
J(R) \sim (R/a^*)^{5/2}{\rm exp}(-2 R/a^*),
\label{eq:herring_flicker}
\end{equation}
where $R$ is the donor separation and $a^*$ is the effective Bohr radius of the electron wave function. The exchange coupling dominates over dipole coupling for donors that are separated by less than approximately 20-30 nm.

\begin{figure}
\label{fig:JoscillationsKHDS}
\includegraphics[width=0.4\textwidth]{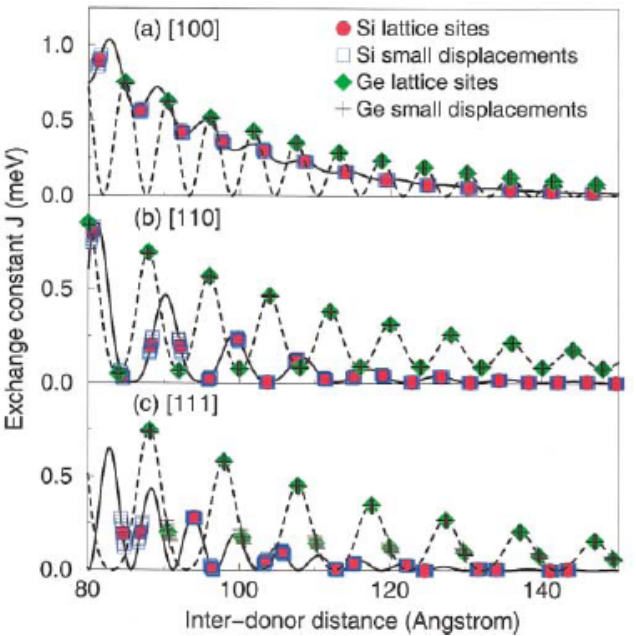}
\caption{ (Color online) \textbf{J-oscillations in the exchange coupling.} Calculated exchange coupling between two phosphorus donors in Si (solid lines) and Ge (dashed lines) along high-symmetry directions for the diamond structure. Values appropriate for impurities at substitutional sites are given by the circles (Si) and diamonds (Ge). Off-lattice displacements by 10\% of the nearest-neighbor distance lead to the perturbed values indicated by the squares (Si) and crosses (Ge). Reproduced from \textcite{KoillerPRL02}.}
\end{figure}

The valley degeneracy of the silicon conduction band gives rise to a far more complicated dependance of $J$ on the donor separation (so-called ``exchange oscillations") as noted in the early work of Cullis and Marko \cite{CullisPRB70}, and is particularly relevant in the Kane quantum computer context \cite{KoillerPRL02,KoillerPRL03,koillerOther05}  (see Fig.~14). The effect persisted in effective mass treatments in which the exchange integrals over Bloch states were carried out numerically \cite{WellardPRB03,KoillerPRB04}. For some time these ``exchange oscillations" were seen as a fundamental limitation of donor based quantum computing as it was thought that to achieve a given exchange coupling the donors would have to be placed in the lattice with lattice site precision \cite{KoillerPRL02}, although \textcite{KoillerPRB02} found that strain could be used to lift the valley degeneracy and alleviate the problem to some extent. In these treatments the exchange coupling is calculated in the Heitler-London approximation \cite{KoillerPRB04,CalderonPRB06} using effective mass wave functions containing a single Bloch component from each valley minimum, hence it is perhaps not surprising that the overlap integral results in an oscillatory behavior in the donor separation at the level of the lattice constant. When the exchange integral is computed using a more accurate wave function including many such Bloch states to reproduce the observed donor levels and valley splittings, the interference effect is somewhat smeared out \cite{wellardPRB05} over the background Herring-Flicker dependence in equation \ref{eq:herring_flicker} (see Fig.~15). Nonetheless, the issue remains that in fabricating donor devices there will be some level of imprecision in the donor atom placement and hence a variation in the (un-gated) value of $J$ between donor pairs, however, using STM fabrication these errors might be constrained to the single lattice site level.

In any case, all components of a quantum computer will need some form of characterization. For all donor qubit logic gates (single and two qubit), considerations of background noise sources and decoherence also need to be taken into account, e.g. see \textcite{wellardQCMCO01,wellardJoPD02,saikinPRB03, FowlerPRA03,HillPRA03} (the decoherence of donor electron spins is covered in Section \ref{timeresolved}). Robust control techniques have been developed specifically for the eventuality of some level of variation in the exchange coupling \cite{HillPRL07}, which in conjunction with gate characterization protocols \cite{devittPRA06,coleJPMG06} have the potential to produce high fidelity two qubit gates in the Kane scheme \cite{testolinPRB05}. \textcite{tsaiPRA09} have applied control techniques to optimize the CNOT gate in the Kane scheme. A more serious impediment to employing the exchange interaction for quantum gates is the effect of charge noise \cite{vorojtsovPRB04,huPRL06}. Because the exchange interaction is ultimately derived from an overlap of electronic wave functions, variations in the background potential such as from charge noise in the device can affect the exchange coupling and may require further development of the materials design \cite{kaneMRS05}, and/or quantum control techniques.

\begin{figure}
\label{fig:JBMB}
\includegraphics[width=0.4\textwidth]{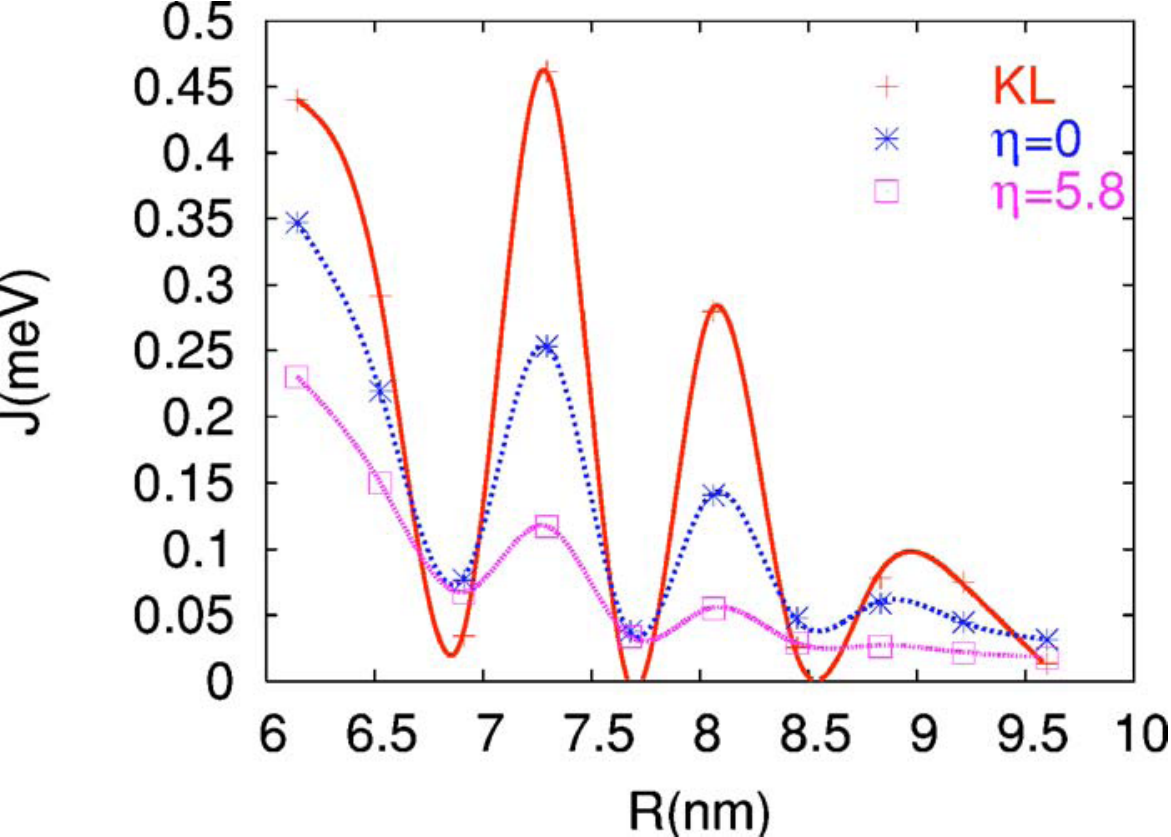}
\caption{ (Color online) \textbf{Smoothing out the exchange oscillations - the exchange coupling $J$ as a function of donor separation along [110].} Top curve: Calculation using the effective-mass wave function. Middle curve: Calculation of $J$ based on wave functions obtained using direct momentum diagonalization over a large basis of Bloch states (BMB) with no core-correction of the impurity potential ($\eta = 0$). Bottom curve: BMB calculation of $J$ with a core-correction ($\eta = 5.8$) that reproduces the donor ground-state and valley-splitting. Note that the points refer to substitutional sites in the silicon matrix. Although the donor separations are relatively small in this case, the spatial variation of the exchange interaction appears to be significantly damped compared to the effective mass treatment. All $J$ values are calculated in the Heitler-London approximation. Reproduced from \textcite{wellardPRB05}.}
\end{figure}

The control of the exchange interaction $J$ has also received considerable attention since the original Kane paper. Early calculations of the dependence of $J$ on an external $J$-gate bias were carried out by \textcite{FangPRB02} using a Gaussian expansion (see Fig.~16). Subsequent calculations of the $J$-gate control in various approaches describing the two-electron physics were carried out \cite{kettleJOPCM04, wellardJOPCM04, fangPRB05,kettlePRB06, CalderonPRB07} given further insight into the controllability of the exchange interaction. However, the gate modification of the overlap between electron states is a difficult calculation and most likely a full configuration interaction framework incorporating valley physics and Bloch structure is required to obtain quantitative results to compare with experiments once measurements are made. A related problem is the calculation of the two-electron donor state (D$^-$), notoriously difficult in the case of a hydrogen ion in vacuum, but even more so when the non-trivial valley physics is added in to complicate such simple points of reference as Hund's Rule. In the context of single donor quantum computing \textcite{FangPRB02} calculated the effect of electric fields on the $D^-$ state, which was a key component of the spin-to-charge conversion read-out scheme of Kane. In \textcite{HollenbergPRB04} time-dependent calculations of the D$^0$ D$^0 \rightarrow$ D$^+$ D$^-$ transition were undertaken in a proposal for resonant based spin-to-charge conversion. More recent calculations have focussed on the complication of valley physics in the D$^-$ bound states particularly under electric fields \cite{koillerPRB10,RahmanPRB11}, with some notable success in comparisons with recent experimental measurements \cite{LansbergenNphys08,fuechsleNatureN12}.

\begin{figure}
\label{fig:FangJV}
\includegraphics[width=0.48\textwidth]{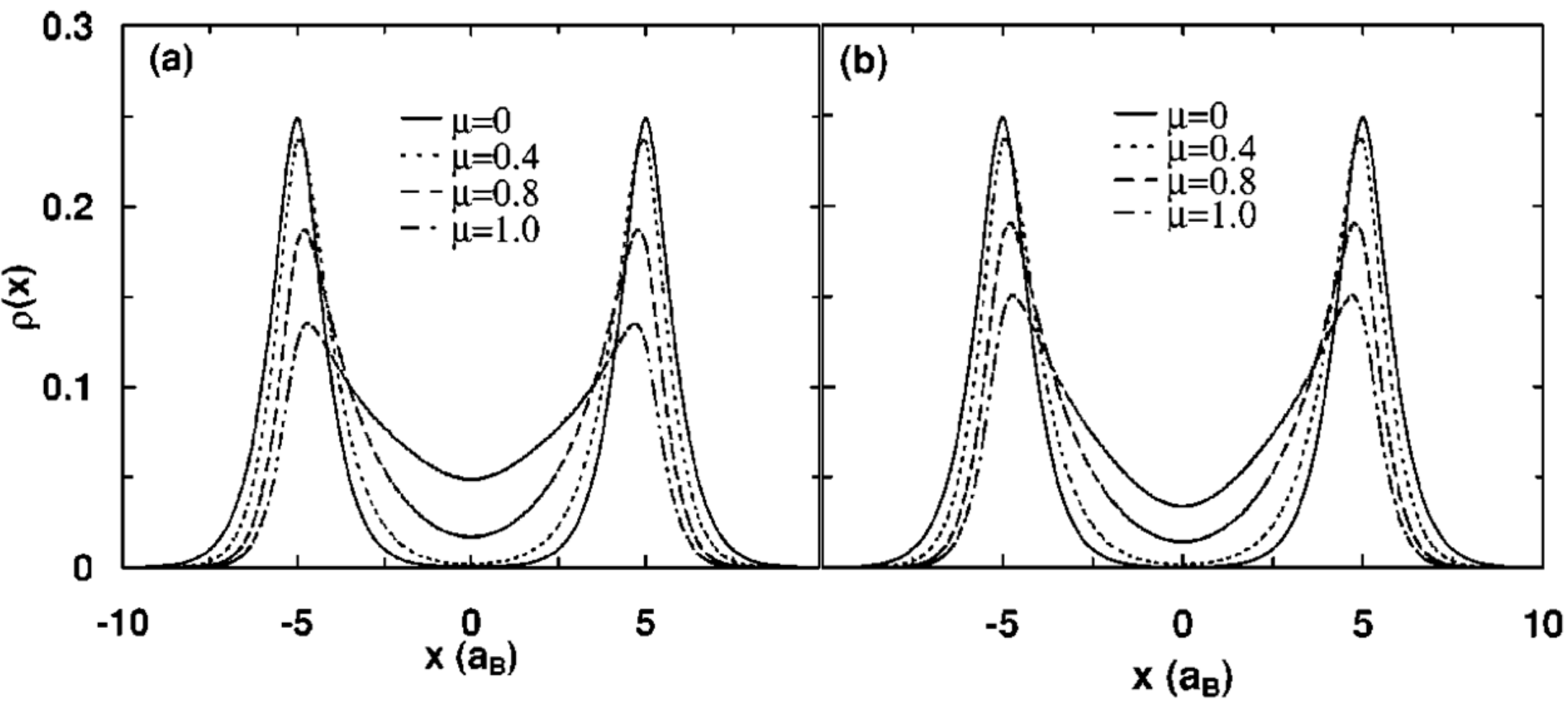}
\caption{\textbf{Gate control of  the two-donor system.} Averaged charge distribution along the interdonor axis for various strengths of the $J$-gate potential ($\mu$) for the singlet (a) and triplet (b) states (fixed donor separation at 10 $a_B$). Reproduced from \textcite{FangPRB02}.}
\end{figure}

\subsubsection{Planar donor structures: delta-doped layers and nanowires}\label{planardonor}

The atom-by-atom fabrication of monolayer donor structures using STM techniques represents the state-of-the-art in precision silicon devices (see section \ref{d_dopant_STM}). From a theoretical point of view these structures present new challenges in order to describe not just their inherent physics (band structure, Fermi level, electronic extent, valley splitting, effect of disorder etc), but their use as in-plane gates in quantum electronic devices, including quantum computing. In understanding the physics of these highly doped monolayer systems ab-initio techniques have been used to good effect. Paradoxically, ab-initio techniques whilst being severely limited to relatively small numbers of atoms can handle planar systems with a high degree of symmetry, exploiting periodic boundary conditions of the supercell in the plane of the structure with sufficient silicon ``cladding" vertically for convergence. The earliest calculations in this context were by \textcite{QianPRB05} for the infinite 2D planar (``delta-doped") ordered layer using a Wannier-based Density Functional Theory (DFT) approach (see Fig.~17(a)). \textcite{CarterPRB09} carried out an extensive DFT calculation of the same Si:P structures using a single zeta polarized basis providing a comprehensive picture of the band structure, effective potential, Fermi energy and electronic width as a function of planar doping density, finding converged results for cladding above 80 layers (see Fig.~17(b)). More recently the effect of disorder on the physics of the delta-doped layer has been investigated both in a DFT approach \cite{CarterPRB09,carterNano11}, and in a self-consistent tight-binding approach which can handle much larger supercell sizes and hence more accurately represent instances of disorder \cite{LeePRB11}. These calculations indicate that the valley spitting of the sub-Fermi bands is quite sensitive to the degree of disorder and will play an important role in eventual device applications.

\begin{figure}\label{fig:DeltaDoped}
\includegraphics[width=0.4\textwidth]{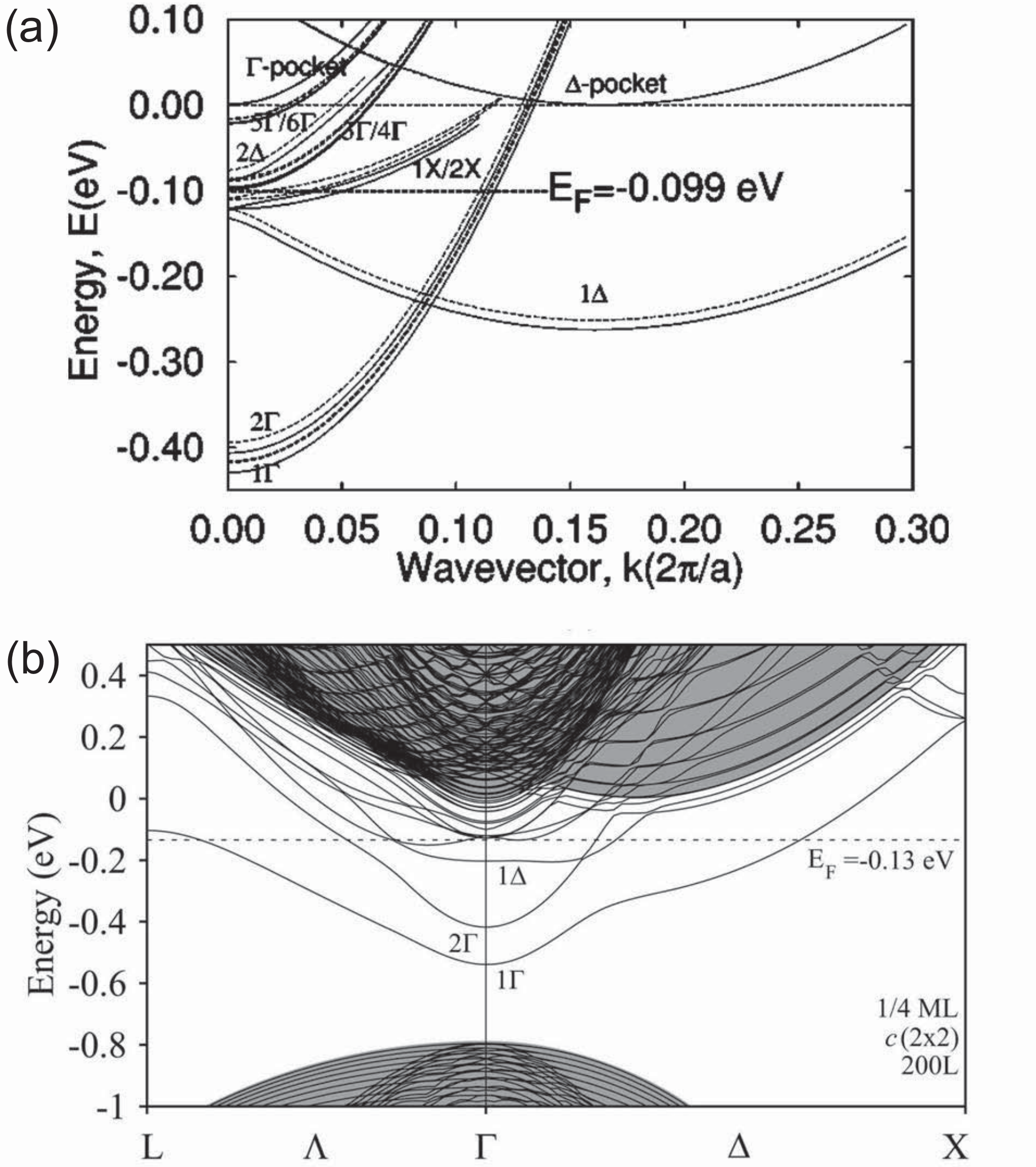}
\caption{\textbf{Band structure of the 1/4 monolayer phosphorus $\delta$-doped layer} Top: (a) The calculation by \textcite{QianPRB05}: the solid lines show the band structure without exchange-correlation and short-range effects, while the dotted lines show the band structure obtained in the full model. (b) The DFT calculation in a supercell with 200 cladding layers by \textcite{CarterPRB09}. The plane projected bulk band structure of Si is represented by the gray continuum. The Fermi level is indicated by a horizontal dashed line. Reproduced from \textcite{QianPRB05} and \textcite{CarterPRB09}.}
\end{figure}

The question of convergence between methodologies still remains on important quantities such as valley splitting. \textcite{drummArxiv12} have applied distinct DFT approaches based on localized and de-localized basis sets to calculate the properties of delta-doped layers. They obtain convergence in the description of the valley splitting and Fermi level only when the localized basis set is extended to the double zeta polarized level. The DFT calculations of the band structure have informed a self-consistent effective mass description of Si:P monolayer structures \cite{drummPRB12}, which has been effective in describing states observed in a STM fabricated 7-donor planar quantum dot \cite{FuechsleNnano10}. The self-consistent tight-binding approach has also been employed beyond the delta-doped layer to describe recent STM fabricated devices. In \textcite{weberScience12} the electronic structure of Si:P monolayer wires only four atoms wide was calculated and gave results in terms of the number of conduction modes in good agreement with experiment. The most ambitious calculation to date was a simulation of the single-atom transistor \cite{fuechsleNatureN12} where the same self-consistent tight-binding approach was used to determine the effective potential due to planar gates at the channel-donor site and subsequently coupled with a tight-binding description of the donor electronic levels. The agreement of the calculated D$^0$ and D$^-$ charge transitions with the measurements was indeed remarkable given the complexity of the device and is a strong indication that the theoretical description of donor based quantum electronic devices is well in hand.

\begin{figure}\label{fig:NNANO}
\includegraphics[width=0.50\textwidth]{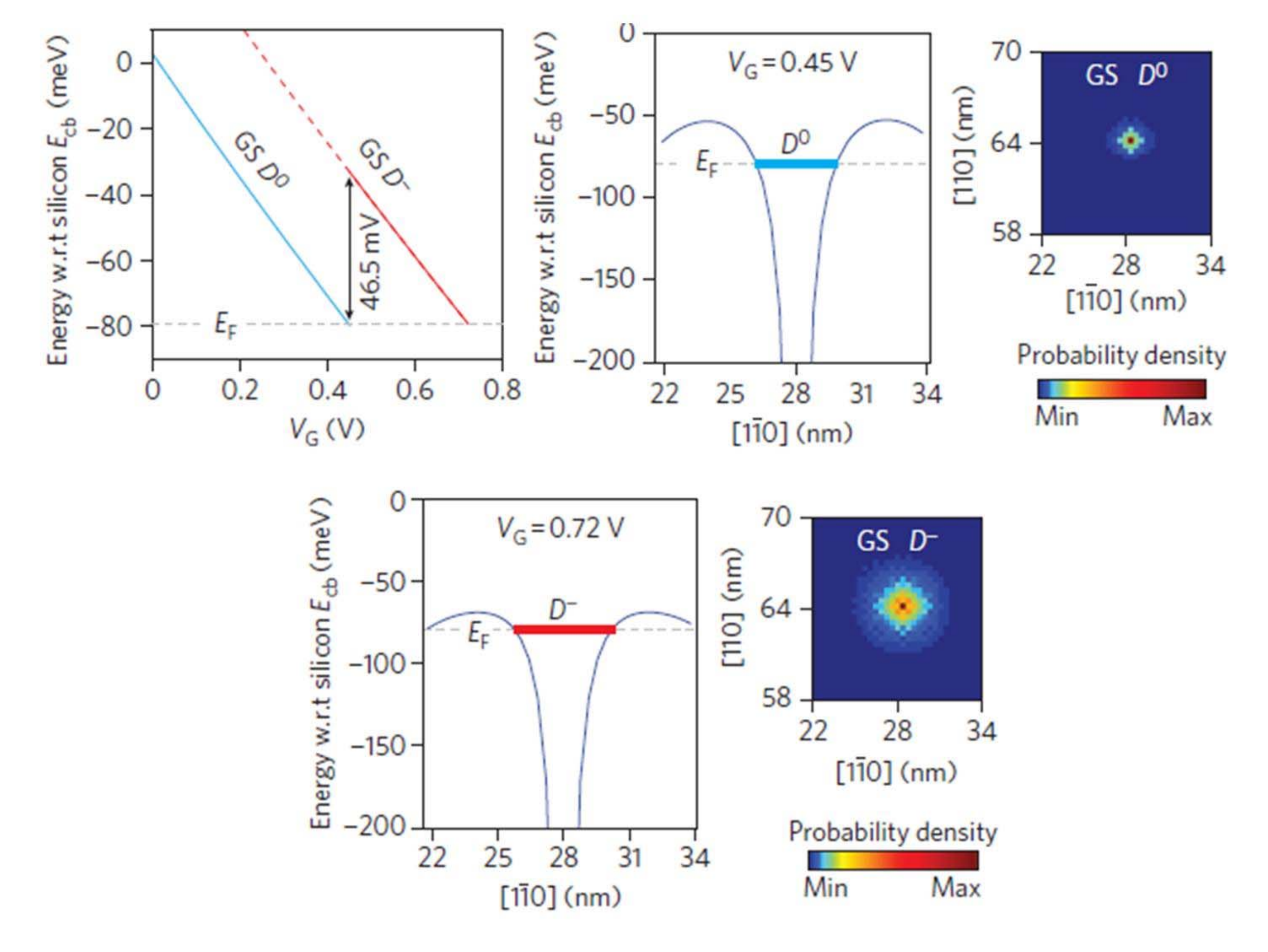}
\caption{ (Color online) \textbf{Calculated electronic spectrum of a single-atom transistor.}
Top left: Calculated energies of the $D^0$ and $D^-$ ground states (GS) as a function of the applied gate voltage $V_G$. The difference in the energy of these two ground states gives a charging energy of $E_C\approx46.5$ meV, which is in excellent agreement with the measurement in this device. Potential profiles between source and drain electrodes calculated for $V_G = 0.45$ V (top middle) and 0.72 V (bottom left). The calculated orbital probability density of the ground state for the $D^0$ potential (top right) is more localized around the donor than for the $D^-$ potential (bottom right), which is screened by the bound electron. Reproduced from \textcite{fuechsleNatureN12}.}
\end{figure}

%% file: quantumdots.tex
Quantum dots showing Coulomb blockade and displaying single-electron
physics can be created in Si and SiGe in many different ways.  In this
section we first briefly review the early work aimed at the
demonstration of Coulomb charging effects in Si and SiGe.  An emphasis
in this work was the quest to see Coulomb effects at as high a
temperature as possible.  We then discuss modern approaches to quantum
dot fabrication.  The application of charge sensing methods is shown
to enable a wide range of experiments, including calibration of the
absolute electron number, spin-state spectroscopy, and the measurement
of spin filling as a function of electron number.  We close this
section with a discussion of both transport and charge sensing
measurements in silicon-based double quantum dots.

\subsection{Early work: Coulomb blockade in silicon}\label{dotsearlywork}

In this section we discuss early experiments studying Coulomb blockade
in Si devices.  Additional background and details can be found
in~\textcite{meiravSST96,ahmedJVSTB97,likharevPIEEE99,tilkePQE01,takahashiJPCM2002,onoJAP2005}.

Experiments exploring intentional Coulomb blockade and transport
through Si/SiO$_2$ and Si/SiGe quantum dots dates to the early 1990s,
shortly after the discovery of Coulomb
blockade~\cite{fultonPRL1987,scottPRL89,meiravPRL1990,fieldPRB90}.
The primary requirements for the observation of Coulomb blockade are
to isolate a small island while maintaining a weak but nonzero tunnel
coupling to the leads.  The addition of one or more gates to control
the charge on the dot is essential for more complicated experiments.

Coulomb blockade was achieved very early in structures formed by
etching delta-doped SiGe or doped silicon-on-insulator (SOI)
structures~\cite{paulAPL1993,aliAPL1994}.  \textcite{aliAPL1994} made use of
two separate lithography and etching steps to modulate the thickness
of a patterned silicon-on-insulator layer, resulting in a weakly
coupled island between two leads.  Coulomb blockade,
which was observed in measurements of current versus source-drain
voltage that showed a Coulomb gap, persisted up to $T=3.8$~K.  The
Coulomb gap could be modulated by an integrated side gate.  In this
type of highly-doped SOI structure, current in the doped leads was
three-dimensional, as the mean free path was smaller than the lead
thickness.

Silicon nanowires formed in SOI can be transformed into a quantum dot
by pattern-dependent oxidation (PADOX), a process that makes use of
the dependence of oxidation in silicon on the exposed surface area and
strain~\cite{takahashiIEDM1994,takahashiEL1995}.  One of the features
of this process is that very small quantum dots can be formed,
enabling measurement of Coulomb oscillations at high temperatures,
with a demonstration of some modulation persisting to room temperature
as early as 1994~\cite{takahashiIEDM1994}. Fujiwara and co-workers studied the few-electron regime in similar devices using photoexcitation techniques \cite{fujiwaraPRL97}.  Electron
beam lithography can be used to help control the shape of small silicon dots that show
Coulomb effects at temperature above 100~K~\cite{leobandungAPL1995}.Very narrow triangular cross-section wires also can be formed by anisotropic etching on SIMOX,
resulting in Coulomb effects at room temperature from disorder-induced
dots along the length of each wire~\cite{ishikuroAPL1996}.

Coulomb blockade can in fact be observed in devices that are similar to production FETs, provided a small island of electrons can be isolated in the device.  Isolation of such an island of electrons can be accomplished by the use of a gate that does not overlap the source and drain, leading to Coulomb blockade in CMOS devices~\cite{BoeufOther03}.
This approach has culminated very recently in a single-electron transistor operating at room temperature~\cite{shinAPL10,shinNL11}.

In 1994 Matsuoka and co-workers proposed using ``two-story gates'' to
create single-electron devices~\cite{matsuokaAPL1994}. A single gate was used to form an inversion
layer for transport, and an upper gate was reverse-biased to generate
barriers and define a quantum dot~\cite{matsuokaAPL1995}.  While this
structure has only a single gate to control the tunnel barriers and
differs in significant ways from later work, it anticipates the use of
two layers of gates that would be used more than a decade later for
experiments on spin blockade, spin measurement, and spin manipulation
(see Sections \ref{doubledotsspin} and \ref{timeresolved}).

\subsection{Single quantum dots}\label{singledots}
This section assesses the experimental analogues of the quantum dot concepts different in silicon nanostructures as explained in section \ref{3D}.

\subsubsection{Self-assembled nanocrystals}\label{nanocrystals}
The material dimensions of nanocrystals can easily be made as small as 10 nm, resulting in large and thus easily observable level splittings, even at room temperature \cite{otobeAPL98}. On the other hand, those dimensions make electron transport measurements cumbersome because the crystals are not easily connected to source and drain reservoirs. Self-assembled silicon nanocrystals with diameters varying from 3-12 nm have been grown by chemical vapor deposition techniques \cite{BaronOther00, steimleOther07}.  Coulomb oscillations have been observed by electrostatic trapping between Al source and drain electrodes \cite{duttaOther00}. \textcite{ZaknoonNL08} showed charging energies of $\sim 50$ meV using scanning tunneling spectroscopy. Twelve resonances in the conductance versus bias voltage were attributed to the twelve-fold conduction band degeneracy owing to spin and the six-fold valley degeneracy as described in section \ref{physicselectrons}.

\begin{figure}[htb]
\includegraphics[width=0.48\textwidth]{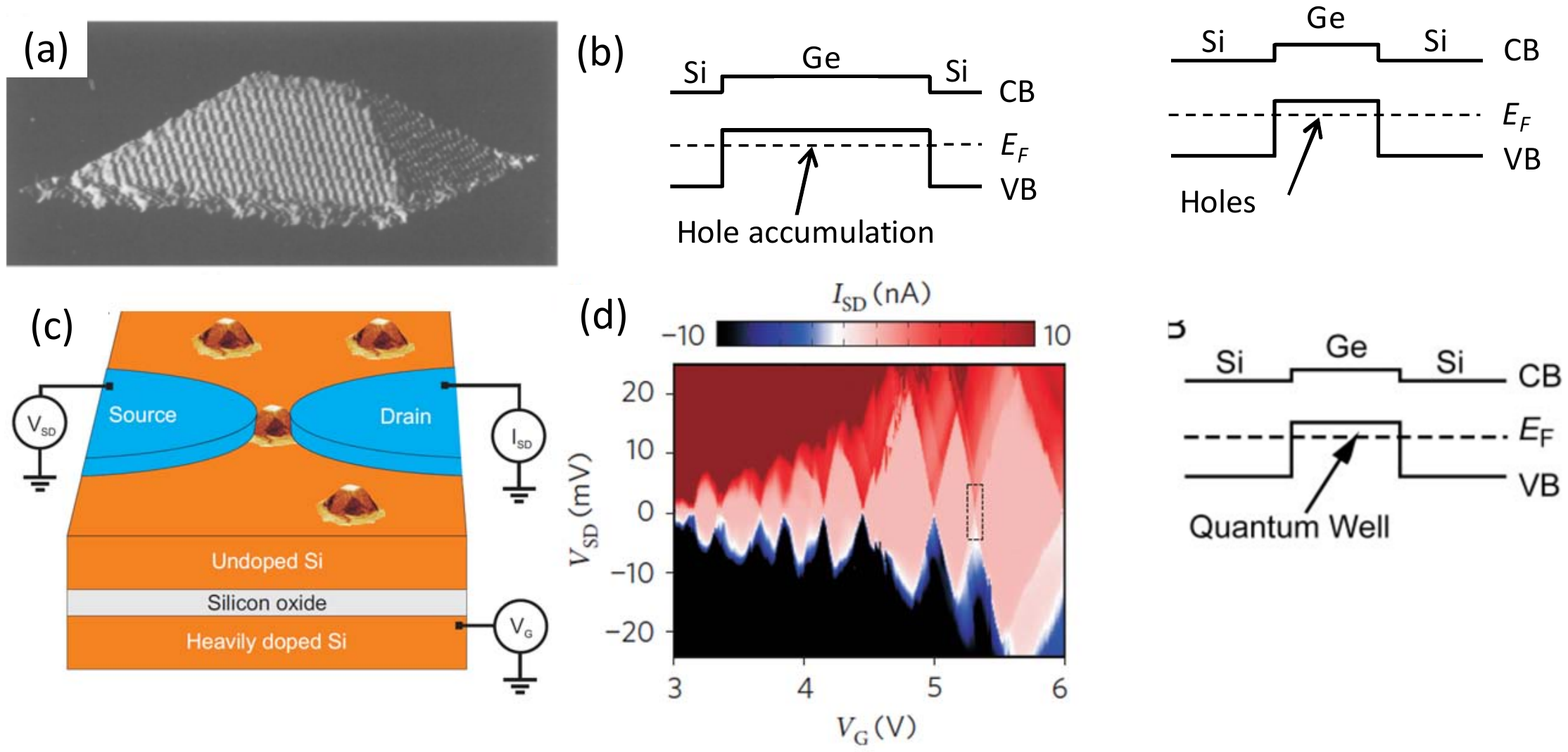}
\caption{ (Color online) \label{nanocrystals_fig1} \textbf{Self-assembled nanocrystals.} (a), STM image of a Ge/Si(001) cluster with a height of 2.8 nm. Scan area is $40 \times 40$ nm, from \cite{MoPRL90}. (b), Band diagram for a Si/Ge/Si heterostructure, showing the accumulation of holes owing to the valence band offset between Ge and Si. (c), Schematic of a quantum-dot device obtained by contacting a single SiGe nanocrystal to aluminum source/drain electrodes. The heavily doped substrate is used as a back-gate for the measurements in (d) where $I_{\textrm{SD}}$ is plotted as a function of $V_\textrm{G}$ and $V_{\textrm{SD}}$. (c,d) from \textcite{KatsarosNNano10}.}
\end{figure}

Small Ge islands can be grown on Si(001) via Stranski-Krastanov growth resulting in huts, pyramids and domes with heights of 5-70 nm and lateral dimensions varying from 20-80 nm \cite{EagleshamPRL90, MoPRL90, Medeiros-RibeiroScience98, RossScience99, StanglRMP04, KatsarosPRL08}, see Fig.~\ref{nanocrystals_fig1}(a). The group of De Franceschi in Grenoble made Al contacts to Ge domes with an additional 2 nm Si capping layer \cite{KatsarosNNano10,KatsarosPRL11}, see Fig.~\ref{nanocrystals_fig1}. In this configuration the SiGe nanocrystal acts as a confining potential for holes due to the valence band offset between Ge and Si at the heterostructure interface \cite{schafflerSSCT97, vandewallePRB86}. Free holes will accumulate in the Ge when the Fermi level lies below the valence band edge of the Ge center, see Fig.~ \ref{nanocrystals_fig1}(b). Electron transport measurements at 15 mK show Coulomb diamonds with charging energies varying from few to 20 meV as 8 holes leave the quantum dot. Due to the limited tunability reaching the few-charge regime in self-assembled nanocrystals will be a great challenge.\\

\subsubsection{Bottom-up grown nanowires}\label{nanowires}
\begin{figure}[h]
\includegraphics[width=0.48\textwidth]{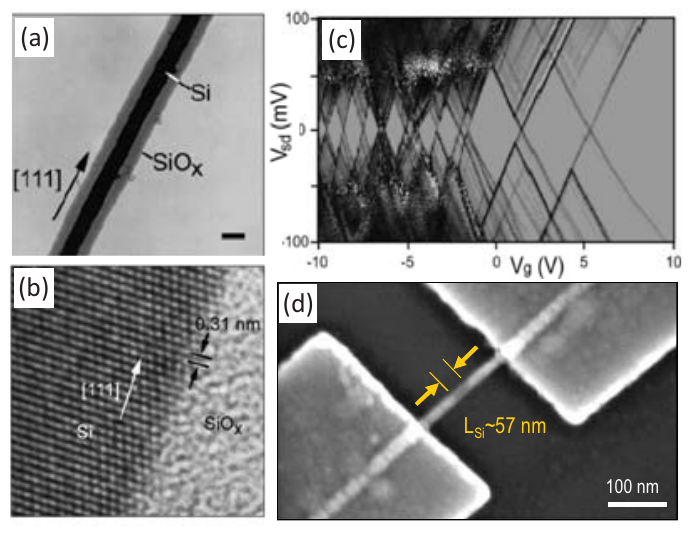}
\caption{ (Color online) \label{nanocrystals_fig2} \textbf{Bottom-up grown nanowires.} (a), TEM image of a Si nanowire; crystalline material (the Si core) appears darker than amorphous material (SiO$_x$ sheath) in this imaging mode. Scale bar, 10 nm. (b) High-resolution TEM image of the crystalline Si core and amorphous SiO$_x$ sheath. The (111) planes (black arrows) are oriented perpendicular to the growth direction (white arrow). (a) and (b) adapted from \textcite{MoralesScience98}. (c) Stability diagram of a p-Si nanowire quantum dot, from \textcite{zhongNL05} (d) SEM image of a nanowire quantum dot with NiSi Schottky contacts, taken from \textcite{zwanenburgJAP09}.}
\end{figure}

Bottom-up grown nanowires are generally synthesized by means of a vapor-liquid-solid process \cite{WagnerAPL64}, allowing for growth of single-crystal Si and Ge nanowires \cite{MoralesScience98} with diameters varying from 3-100 nm and lengths up to tens of microns, see Fig.~\ref{nanocrystals_fig2}(a,b). Both n-type and p-type dopants have been incorporated, and their location depends on the diameter \cite{XiePNAS09}. The doping can be varied during growth: such modulation doping has been used to intersect heavily-doped n-Si regions with two short lightly-doped regions, resulting in single-electron tunneling at 1.5 K \cite{YangScience05}. Within one nanowire heterostructures of different materials can be created both radially and axially, such as core/shell Ge/Si nanowires \cite{LauhonNature02}. In the latter case the valence band offset will induce hole population in the Ge core, see Fig.~\ref{nanocrystals_fig1}(b).

When metallic contacts are made to nanowires the Schottky tunnel barriers can define the quantum dot length as shown in core/shell Ge/Si nanowires \cite{LuPNAS05} and Si nanowires \cite{zhongNL05}, see Fig.~\ref{nanocrystals_fig2}(c). The Si nanowire quantum dot length can be shortened by silicidation transforming the device into e.g.\ a NiSi-Si-NiSi nanowire as shown in Fig.~ \ref{nanocrystals_fig2}(d) \cite{WeberNL06, zwanenburgJAP09, MongilloOther11}.

After the demonstration of Coulomb blockade oscillations in Ge/Si nanowires by the Lieber group from Harvard \cite{LuPNAS05}, they joined forces with the Marcus group and created double quantum dots with tuneable tunnel barriers, see Section \ref{doubledots}. Here the source and drain contacts were ohmic, while the tunnel barrier were defined by local top gates \cite{huNNano07}. \textcite{roddaroPRL08} used the same configuration to create single quantum dots and probe the hole spin states, see section \ref{spinsdots}. Ge/Si nanowires were found to have a strong spin-orbit interaction, which can be tuned by means of an electric field \cite{haoNL10}. Recent spin lifetime measurements \cite{huNNano11} indicate spin-orbit interaction as the dominant mechanism for spin relaxation.  According to the work by \textcite{kloeffelPRB11}, the unusually strong spin-orbit coupling makes them particularly attractive candidates for quantum information processing via electric-dipole induce spin resonance~\cite{golovachPRB06,Nowack:2007p1430,nadjNature10}, and for research on Majorana fermions~\cite{majoranaOther37}.

Very recently, Ge/Si nanowires with a triangular cross section and a height of just three unit cells were realized by molecular beam epitaxy \cite{Zhang2012}. These wires are directly grown on planar Si without the use of any catalyst, and preliminary low-temperature measurements show Coulomb blockade.

\subsubsection{Electrostatically Gated Si/SiGe quantum dots}\label{sigedots}
A powerful way to achieve tunability of tunnel couplings in quantum dots is to provide confinement in one or more directions through the use of electrostatic gates. Using Si/SiGe heterostructures or MOS structures, it is possible to form high-quality two-dimensional electron systems that can be partitioned into tunable quantum dots using depletion or accumulation gates, a procedure described in detail in this section and the next. In general, at least direction of confinement must be provided by a non-electrostatic method; usually a materials interface is used, the two most common being the interface between single-crystal silicon and its amorphous oxide (in MOS structures, see next section), and the epitaxial interface between single-crystal Si and Si$_{1-x}$Ge$_{x}$. When the precise composition $x$ is unimportant and no confusion will arise, we refer to these heterostructures as Si/SiGe. Both MOS devices and Si/SiGe devices have been reviewed extensively: see, for example, \cite{sze,wolf1990} for the former, and \cite{mooneyMSE1996,schafflerSSCT97} for the latter.

A convenient, if incomplete, figure of merit for two-dimensional electron systems is the mobility $\mu$.  For Si MOS, mobilities in the range $5,000-15,000~\mathrm{cm^2/Vs}$ are quite good (see e.g. \textcite{engAPL05,Eng:2007p808}), and mobilities in excess of 40,000 have been reported~\cite{kravchenkoRPP2004}. The low-temperature mobility in Si/SiGe two-dimensional electron gases is not limited by defects at the interface and has been improving rapidly in recent years.  In 1995, Ismail and coworkers reported a low-temperature mobility of $520,000~\mathrm{cm^2/Vs}$ in a modulation doped Si/SiGe heterostructure.  Even higher mobility $800,000~\mathrm{cm^2/Vs}$ was reported by a group from Hitachi in 1998~\cite{sugiiSST1998}. Very recently, Si/SiGe two-dimensional electron systems have been formed using undoped structures with a positively-biased accumulation gate.  In this approach, an intervening oxide such as Al$_2$O$_3$~\cite{laiAPL05} is used to separate the accumulation gate from the semiconductor surface to avoid injecting current into the heterostructure \cite{luAPL2007}.  The positively biased accumulation gate removes the need for any doping in the structure, removing a source of background impurities and eliminating the modulation doping layer altogether, both of which cause scattering.  Resulting mobilities as high as $\mu = 1.6\times 10^{6}~\mathrm{cm^2/Vs}$ have been reported~\cite{luAPL2009}.  Further, the removal of intentional doping appears to significantly reduce low-frequency charge noise in the devices.

Because both Si and Ge have isotopes with zero nuclear spin, the proposal by Loss and DiVincenzo to use quantum dots as hosts for semiconductor spin qubits \cite{LossPRA98} led to great interest in the development of high-quality quantum dots in Si/SiGe heterostructures \cite{vrijenPRA00,friesenPRB03}.  The challenge in
the early work in this field was to find ways to fabricate such dots with low-leakage gates, sufficient tunability, and in such a way as to yield stable, low-noise devices.  As we discuss later in this review, modern Si/SiGe quantum dots have achieved performance that rivals that of any materials system available.  In this section we discuss the materials and device research that enabled this advance.

Here we discuss a few critical materials issues relevant to Si/SiGe heterostructures. Interest in Si/SiGe arises because of the inevitability of defects at the interface between crystalline Si and its amorphous oxide. Heterostructures formed from Si and Si$_{1-x}$Ge$_{x}$ offer a natural alternative with, in principle, no interfacial traps (although other types of disorder, such as atomic steps and strain variation are certainly present).

Although both Ge and Si have the diamond structure, Ge sits one row beneath Si in Group IV of the periodic table, so that the lattice constant of Si$_{1-x}$Ge$_{x}$ increases as $x$ increases, achieving a mismatch between pure Si and Ge of approximately 4.17\%~\cite{schafflerSSCT97}.  Because of this mismatch, pure Ge will
grow epitaxially only three monolayers on Si (REF).  Beyond this critical thickness, self-assembled quantum dots or ``huts'' form \cite{MoPRL90}, as discussed in Sec.~\ref{nanocrystals}, preventing the growth of uniform quantum wells.

\begin{figure}[h]
\includegraphics[width=0.35\textwidth]{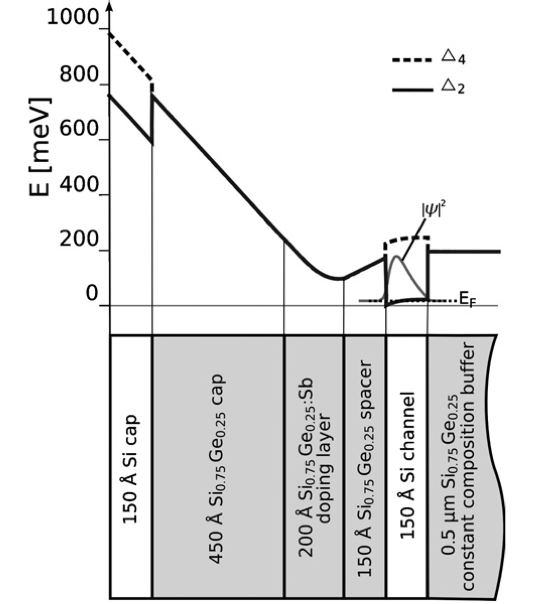}
\caption{\label{fig:sige-design} \textbf{Layer design and corresponding band diagram of a Si/SiGe modulation doped heterostructure used to form top-gated quantum dots.}  Reproduced from \textcite{bererSST07}.}
\end{figure}

Because the lattice constant of Si$_{1-x}$Ge$_{x}$ depends on $x$, a full description of a heterostructure of these two materials must include the strain of the various layers.  For the structures considered here, the layers of interest typically include a Si quantum well with Si$_{1-x}$Ge$_{x}$ barriers on either side, as shown in Fig.~\ref{fig:sige-design}; typically, $x\sim 0.3$.  If the quantum well is below the critical thickness for dislocation formation, the in-plane lattice constant will remain unchanged passing vertically from the Si$_{1-x}$Ge$_{x}$ through the Si quantum well and into the upper barrier.  The band offsets at the Si/Si$_{1-x}$Ge$_{x}$ interfaces depend on this in-plane lattice constat.  For an unstrained, relaxed Si$_{0.7}$Ge$_{0.3}$ barrier layer, the minimum in the conduction band is approximately 160~meV lower inside a Si quantum well compared with the barriers \cite{schafflerSSCT97}.

Because it is very challenging to grow bulk, relaxed Si$_{1-x}$Ge$_{x}$ with even moderately large $x$, relaxed Si$_{1-x}$Ge$_{x}$ substrates conventionally are formed by slowly increasing the Ge concentration $x$ from zero to the desired final value over a thickness of several microns.  This procedure induces the
formation of misfit dislocations, increasing the overall lattice constant, and can yield low-defect structures~\cite{mooneyMSE1996}. The relaxation process itself does result in small inhomogeneities, which can be observed with nano-beam x-ray measurements \cite{evansOther12}.

Quantum dots in Si/SiGe demonstrating Coulomb blockade were first
formed using a combination of etching and electrostatic gating.
\textcite{notargiacomoAPL2003} observed Coulomb blockade oscillations in
a gated nanowire etched into a Si/SiGe heterostructure.  This early device had a single overall
top gate used to control the number of electrons in the quantum dot.
\textcite{kleinAPL2004} formed a quantum dot with three separate
electrostatic gates.  These gates were formed of
the same two-dimensional electron gas as the quantum dot, source and
drain leads~\cite{erikssonQIP2004}.  To avoid current flowing from the gates to the dot, deep
trenches were etched between the gates and the dots; the intervening
gaps make it difficult to apply local fields and separately gate the
quantum dot and the tunnel barriers.  This drawback was partially
ameliorated by the demonstration that gates could be formed by metal
deposited into etched regions surrounding the dot \cite{sakrAPL2005},
and by the use of extremely small top gates used to break an etched
wire into a gated quantum dot~\cite{slinkerNJP2005}.  The drawback of
etching, however, is the potentially large degree of side-wall
depletion~\cite{kleinJAP2006}.

\begin{figure}[h]
\includegraphics[width=0.5\textwidth]{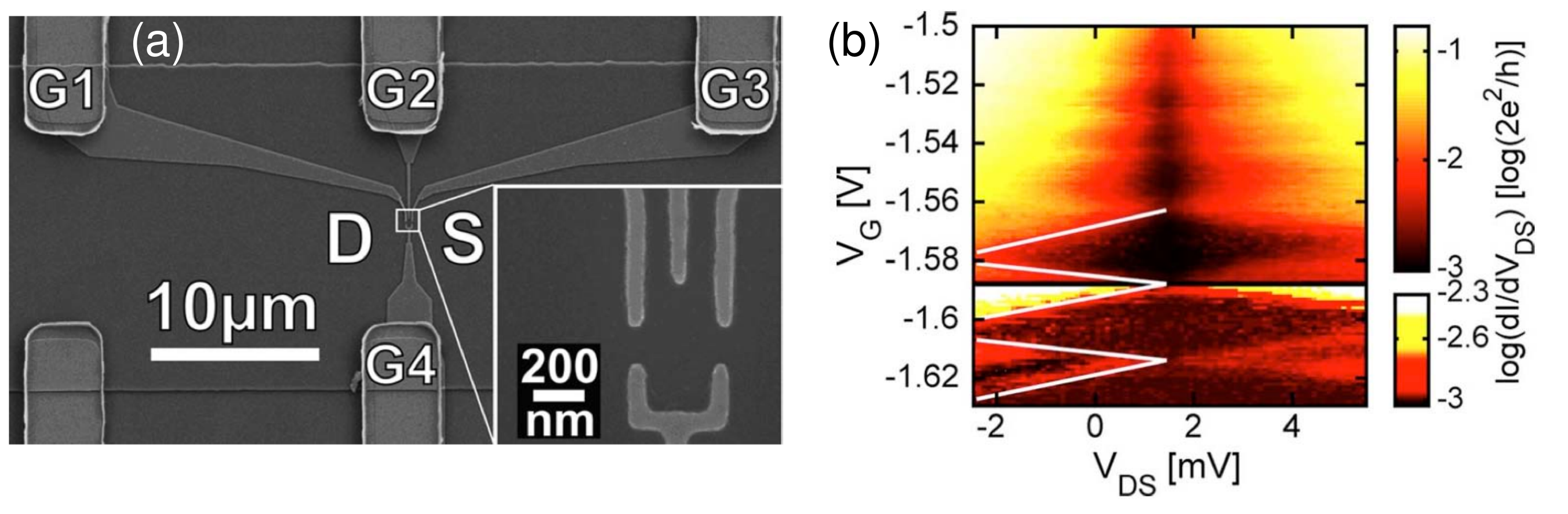}
\caption{ (Color online) \label{fig:schottky-gates} (a) Scanning electron micrograph
  of the Schottky gates used to form a gated quantum dot in
  Si/SiGe. (b) Coulomb diamonds: conductance of the dot as a function
  of the voltage $V_\mathrm{G}$ applied to gates G1 and G2 and of the
  drain-source voltage $V_\mathrm{DS}$.  Figure
  from~\textcite{bererAPL06}.}
\end{figure}

\textcite{bererAPL06} demonstrated a fully top-gate defined quantum dot
formed in a modulation-doped 2DEG, as shown in Fig.~\ref{fig:schottky-gates}.  They showed that
Pd Schottky gates, when fabricated on heterostructures like that shown
in Fig.~\ref{fig:sige-design}, in which care was taken to reduce the
dopant density near the surface, enabled low-leakage
gates~\cite{bererAPL06}.  There had been great concern about leakage
between the top gates and the electron-gas, but the Pd Schottky gate approach has proven to be very robust~\cite{kleinAPL07, wildNJP2010, payetteAPL12}.  The Schottky gate
approach has also been used successfully to gate heterostructures with
enhanced concentration of $^{28}$Si and $^{70}$Ge~\cite{sailerPSS09}.
A second approach to eliminating leakage is to use a dielectric
material beneath the gates, creating metal-oxide-semiconductor split
gates to define the quantum dot~\cite{shinSST11}.

The primary advantage of top-gated quantum dots, in which the lateral confinement is entirely provided by adjustable gate voltages, is their extreme tunability.  At zero gate voltage in most cases current can flow directly under a gate, enabling a smooth transition from a completely open two-dimensional electron gas to a fully confined quantum dot.  This tunability led both to the observation of the Kondo effect in a Si/SiGe top-gated quantum dot~\cite{kleinAPL07} and the demonstration of single-electron occupation, as shown in Fig.~\ref{chargesensing_fig1} below.

\subsubsection{Quantum dots in planar MOS structures}\label{planarmos}

The silicon MOSFET is arguably the world's most important electronic device, being the basic component of all modern microprocessor chips. Its success has been built on the ability to grow a high-quality SiO$_2$ layer on the Si(001) surface by thermal oxidation, forming a high band-gap insulator that isolates the gate from the silicon channel. In current processor chips a SiO$_2$ layer of $\sim$1 nm is sufficient to maintain gate voltages that are a significant fraction of a volt with negligible leakage. The Si/SiO$_2$ interface, which confines the electron layer in a MOSFET, can also have relatively low disorder, with reported electron mobilities as high as 40,000 cm$^2/V$s \cite{kravchenkoRPP2004}, although the imperfect lattice match between the Si and SiO$_2$ creates defects at the interface, thus constraining the electron mobilities below those attainable at Si/SiGe interfaces. Despite this, it is possible to form quantum dots in MOS structures that can be controlled down to the single electron level with high tunability.

In this section we focus on quantum dots formed at the Si/SiO$_2$ interface via the use of multiple surface gates that provide electrostatic confinement in all three dimensions. In general an upper gate is used to induce an electron layer at the interface (as in a `traditional'ù MOSFET), while two or more lower gates provide tunable tunnel barriers between the electron reservoirs and the dot. As already described in Section IV.A, one of the earliest such structures \cite{matsuokaAPL1994} exhibited Coulomb blockade oscillations, although these preliminary results were rather irregular.

\begin{figure}[h]
\includegraphics[width=0.48\textwidth]{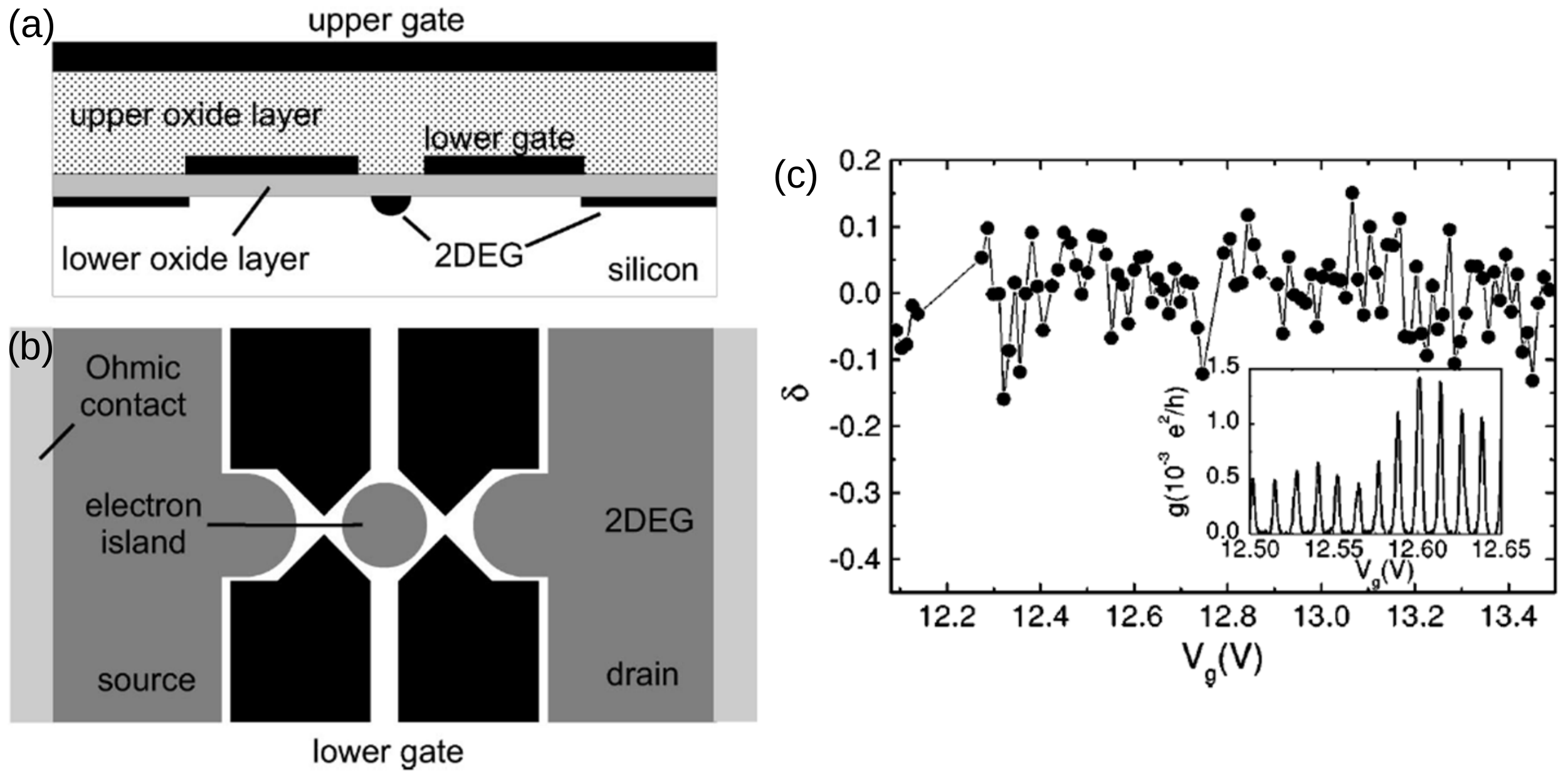}
\caption{\label{MOS_fig2} \textbf{Si MOS quantum dot with large-area top gate.} (a) Cross-sectional schematic, showing two oxide and two gate layers, formed on a silicon substrate. The lower SiO$_2$ layer is thermally grown, while the upper oxide layer is formed using plasma deposition. The large-area upper gate induces a 2DEG at the Si/SiO$_2$ interface, while the lower gates locally deplete the 2DEG to form a quantum dot. (b) Top-view schematic, showing lower depletion gates (black) and induced electron layer (grey). (c) Normalized spacings $\delta$ between Coulomb peaks in dot conductance as a function of upper gate voltage. Inset: Raw Coulomb oscillations in dot conductance as a function of upper gate voltage. Data reproduced from \textcite{simmelPRB1999}.}
\end{figure}

One of the first well controlled MOS quantum dots was demonstrated by Simmel and co-workers \cite{simmelPRB1999}, see Fig.~\ref{MOS_fig2}. In this structure a continuous upper gate was used to induce a 2DEG over a large area, while four lower gates were used to confine the dot and form tunnel barriers. The resulting lower gate structure mimics those used to confine GaAs/AlGaAs quantum dots, although in the latter case the 2DEG is created by modulation doping. The resulting Coulomb oscillations in this MOS device were quite regular (Fig.~\ref{MOS_fig2}c) and provided promise for future MOS quantum dot studies. The lower gates of the device in Fig.~\ref{MOS_fig2} were made using refractory metal, since a high-temperature process was used to deposit the upper oxide isolation layer (Fig.~\ref{MOS_fig2}a).

This type of architecture, employing a large-area upper gate, has since been used by a number of groups to construct MOS quantum dots. A group at Sandia National Laboratory has demonstrated a range of quantum dot devices in which etched polycrystalline silicon (poly-Si) is used for the lower gates, and a large area upper metal gate is used to induce the 2DEG layer \cite{Nordberg:2009p115331, tracyAPL10}. The use of poly-Si gates is appealing from the perspective of future manufacturing, since it opens the way towards the use of CMOS process technologies. Similar MOS quantum dots also have been used to confine single electrons, enabling direct measurement of electron spin relaxation times \cite{xiaoPRL10}.

\begin{figure}[h]
\includegraphics[width=0.49\textwidth]{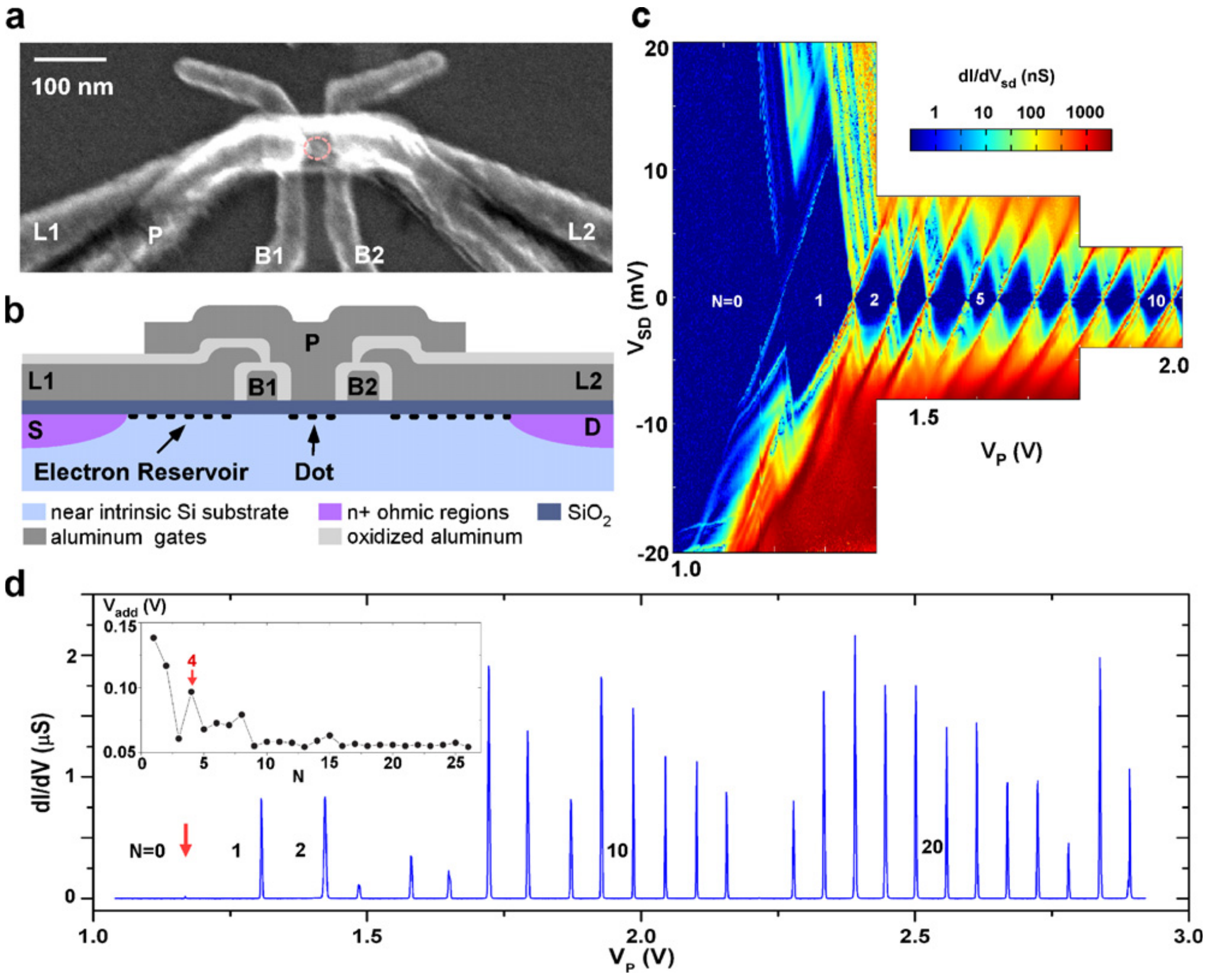}
\caption{\label{MOS_fig5}  (Color online) \textbf{Si MOS quantum dot with compact multi-layer gate stack.} (a) Scanning electron microscope image of device. (b) Cross-sectional schematic, showing three oxide layers and three Al gate layers, formed on a silicon substrate. The SiO$_2$ layer is thermally grown in a high-temperature process, while the thin Al$_2$O$_3$ layers between the gates are formed by low-temperature oxidation of the aluminum. (c) Stability map obtained by plotting differential conductance through the device as a function of source-drain bias $V_{SD}$ and plunger (P) gate voltage $V_P$. The first diamond opens up completely, indicating that the dot has been fully depleted of electrons. (d) Coulomb oscillations as a function of plunger gate voltage $V_P$ for the first 23 electrons in the dot. Data reproduced from \textcite{LimNanotech11}.}
\end{figure}

By reducing the upper MOSFET gate to nano-scale dimensions, a group at the University of New South Wales developed a highly compact multi-gate MOS architecture \cite{angusNL2007} that has since been used to construct a wide range of single \cite{limAPL09-2,LimNanotech11} and double \cite{Lim:2009:p173502,laiOther11} quantum dot structures. This architecture uses aluminum (Al) upper and lower gates, with a thin (3-5 nm) Al$_2$O$_3$ insulating layer between the gates, formed by thermally oxidizing the lower gates at the relatively low temperature of 150 $C$. Despite being very thin, the Al$_2$O$_3$ insulator can maintain inter-gate voltage differentials of up to 4 volts, allowing for high gate tunability and the formation of very small (sub-50 nm) multi-dot structures. Figure \ref{MOS_fig5} shows a quantum dot device based on this technology, in which a third layer of gate metal is used. This allows one upper gate to be used as a `plunger', to control the dot's electron occupancy, while separate upper gates are used to induce the source and drain electron reservoirs -- see Fig.~\ref{MOS_fig5}(b). In this way the dot occupancy can be reduced to the single electron level, as confirmed by the bias spectroscopy measurements in Fig.~\ref{MOS_fig5}(c), while maintaining a high density of states in the reservoirs. Such independent tuning of the dot occupancy and the reservoir electron density is not possible when a large-area upper gate is employed.

\begin{figure}[h]
\includegraphics[width=0.45\textwidth]{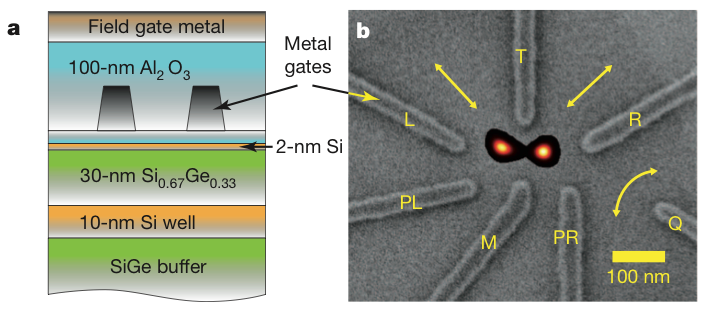}
\caption{ (Color online) \label{fig:maune} Gated quantum dot formed from a Si/SiGe heterostructure with a global accumulation gate. (a) Cross-sectional view of the heterostructure and the two layers of gates.  (b) Top-view SEM image of the gates with a numerical simulation of the electron density superimposed. Figure from \textcite{mauneNature12}.}
\end{figure}

The metal-oxide-semiconductor techniques just discussed can be applied to Si/SiGe heterostructures, yielding extremely stable and tunable quantum dots~\cite{hayesArxiv09,borselliAPL2011}.  The device design, as shown in Fig.~\ref{fig:maune}, uses a Si quantum well surrounded by epitaxial SiGe barriers to provide a clean environment for the electrons in the device.  Those electrons are induced by an accumulation gate at the top of the structure.  Depletion gates in between the accumulation gate and the heterostructure surface are used to control size and shape of the dot.

MOS-based quantum dots, using architectures like those in Figs. \ref{MOS_fig2} and \ref{MOS_fig5}, have since been used in a range of advanced measurements, including single-spin measurement, and spin- and valley-state spectroscopy, as will be discussed in Sections \ref{spinsdots} and \ref{doubledots}.

\subsubsection{Quantum dots in etched silicon nanowires}\label{etchedsilicon}

As discussed in Section \ref{dotsearlywork}, some of the earliest silicon-based single-electron devices (e.g., \textcite{takahashiIEDM1994,takahashiEL1995} were based upon narrow nanowires, patterned using traditional top-down lithographic techniques, and etched from thin (typically $< 50$ nm) silicon layers that form the upper layer of silicon-on-insulator (SOI) wafers. These early devices used the pattern-dependent oxidation (PADOX) technique to create additional confinement along the length of the nanowire, but in subsequent structures researchers have incorporated `wrap-around' gates, positioned along the wire to provide additional confinement.

\begin{figure}[h]
\includegraphics[width=0.45\textwidth]{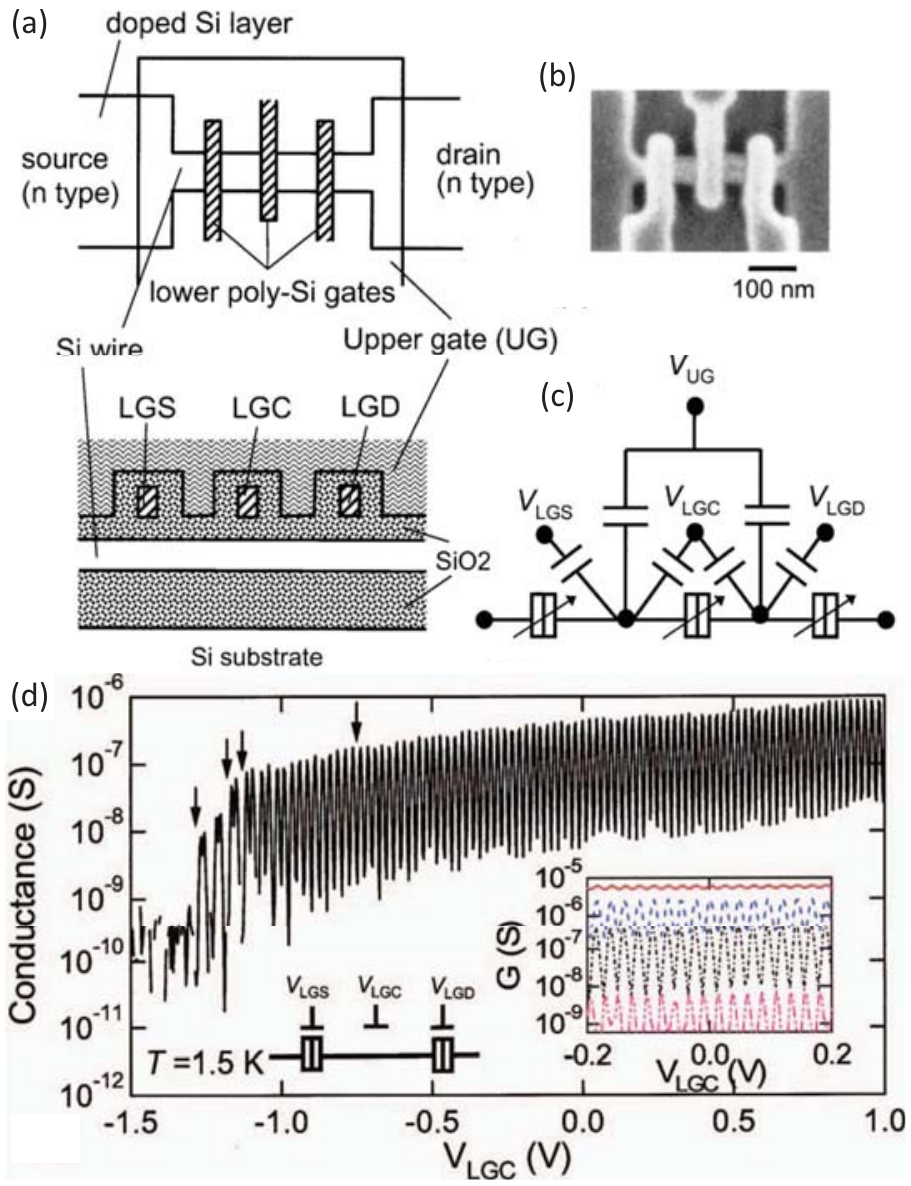}
\caption{\label{MOS_fig8}  (Color online) \textbf{Multi-gated quantum dot in etched silicon nanowire.} (a) Schematic top-view and cross-sectional view of the device. Three lower `wrap-around' gates (LGS, LGC, LGD) are used to form tunnel barriers in an etched silicon nanowire. (b) Top-view scanning electron microscope image of the device before the upper gate is deposited. (c) Equivalent circuit of the device. (d) Coulomb blockade oscillations in device conductance as a function of central gate voltage $V_{LGC}$ when the two outer gates (LGS, LGD) are biased to set each tunnel barrier to $G=1$ $\mu$S. Inset: Coulomb oscillations for a range of values of barrier conductance from 20 nS to 8 $\mu$S. Data reproduced from \textcite{fujiwaraAPL2006}.}
\end{figure}

One of the first examples of this type of gated silicon nanowire was demonstrated by a group at NTT in Japan \cite{fujiwaraAPL2006} -- see Fig.~\ref{MOS_fig8}. Here, confinement in the y and z directions was provided by the narrow wire, of width 20 nm and thickness 20 nm. Confinement along the wire was created by wrap-around lower gates, which in this case were made from poly-Si. Finally, a large-area poly-Si upper gate, isolated from the gates below using SiO$_2$, was patterned above the entire structure to induce carriers in the nominally un-doped nanowire. The resulting structure is entirely CMOS compatible, making it convenient for production using well established manufacturing processes, and also utilizing the high-quality thermally grown SiO$_2$ insulator, which is known for having very low charge noise. In subsequent measurements on these devices it was found that they exhibited extremely high charge stability, with a drift of less than 0.01$e$ over several days \cite{zimmermanAPL2007}.

As seen in Fig.~\ref{MOS_fig8}(d), a quantum dot could be formed by using the outer gates LGS and LGD to create tunnel barriers, with the central gate LGC acting as a `plunger' to control the dot occupancy. The Coulomb oscillations (Fig.~\ref{MOS_fig8}d) were highly periodic over a large gate voltage range ($-0.5 V  < V_{LGC} < 1.0 V$), with a deviation of less than 1 percent, although the dot occupancy $N_e$ in this case was relatively large, with $N_e \sim$ 200 electrons at $V_{LGC}  = 0 V$. The peak conductance could also be tuned over more than three orders of magnitude by varying the barrier gate voltages. For central gate voltages $V_{LGC}  < -1.0 V$, an additional tunnel barrier was formed, breaking the quantum dot into two dots in series. Using similar device structures this group could therefore operate double quantum dots, demonstrating effects such as Pauli spin blockade \cite{liuPRB08} -- discussed further in Section VI.C.4. 

\begin{figure}[h]
\includegraphics[width=0.44\textwidth]{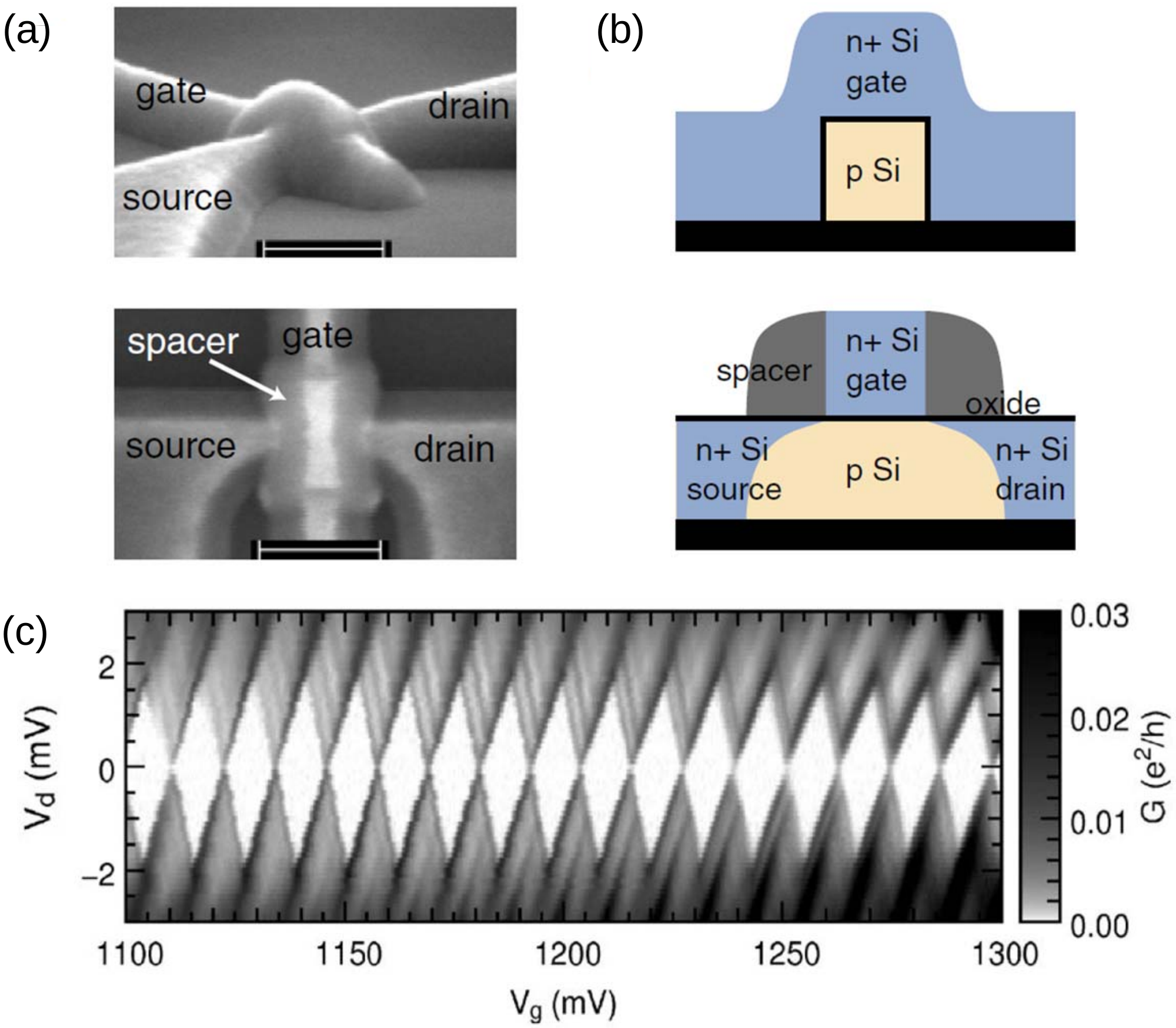}
\caption{\label{MOS_fig7}  (Color online) \textbf{Single-gated quantum dot in etched silicon nanowire.} (a) SEM images and (b) cross-sectional schematics taken perpendicular to the nanowire (upper) and along the nanowire (lower). (c) Stability map (Coulomb diamonds) obtained by plotting differential current through the device as a function of source-drain bias $V_d$ and wrap-gate voltage $V_g$. Data reproduced from \textcite{SellierPRL06} and from \textcite{hofheinzAPL06}.}
\end{figure}

It is also possible to form a quantum dot in a silicon nanowire using just a single gate, by making use of technology that has been developed for the manufacture of FinFET-type MOSFETs. FinFETs are considered likely replacements for planar CMOS technology, due to their ability to operate as FETs with good ON/OFF ratios at much shorter channel lengths. Figure \ref{MOS_fig7}(a,b) shows a FinFET structure, which is based upon a nanowire (the `fin') that is etched from a SOI wafer, as previously described. A single wrap-around poly-Si gate is encapsulated on either side by an insulating `spacer', made from either SiO$_2$ or Si$_3$N$_4$. The gate and spacer act as a mask for subsequent ion implantation of the n+ source and drain regions, which is a standard `self-aligned' gate process used in CMOS production. By applying a positive voltage to the poly-Si gate electrons can be induced below, to form a quantum dot, isolated from the source and drain due to the natural barrier created by the spacer regions -- see Fig.~\ref{MOS_fig7}(b). Such quantum dots can be extremely stable in the many-electron regime, as shown in Fig.~\ref{MOS_fig7}(c), which demonstrates bias spectroscopy (`Coulomb diamonds') taken over a wide range of electron occupancy, with high stability and almost constant charging energy \cite{hofheinzAPL06}. Similar FinFET structures have also been used for single dopant tunneling studies -- see Section \ref{dFET}.

\subsection{Charge sensing techniques}\label{chargesensing}

The non-invasive sensing of charge displacements in quantum nanostructures was first demonstrated in a GaAs/AlGaAs heterostructure device \cite{FieldPRL93}, when a quantum point contact (QPC) was used to detect the change in occupancy of a quantum dot. Here, the QPC is biased close to pinch-off, where its transconductance $dI/dV_G$ can be very large. Any small charge displacement in the vicinity of the QPC channel can then lead to a significant change in QPC current, via its capacitive coupling. This technique has since been applied widely, enabling the direct probing of single electron charges and the indirect probing of single spins in nanostructures based on a variety of materials systems, including silicon.

\begin{figure}[h]
\includegraphics[width=0.48\textwidth]{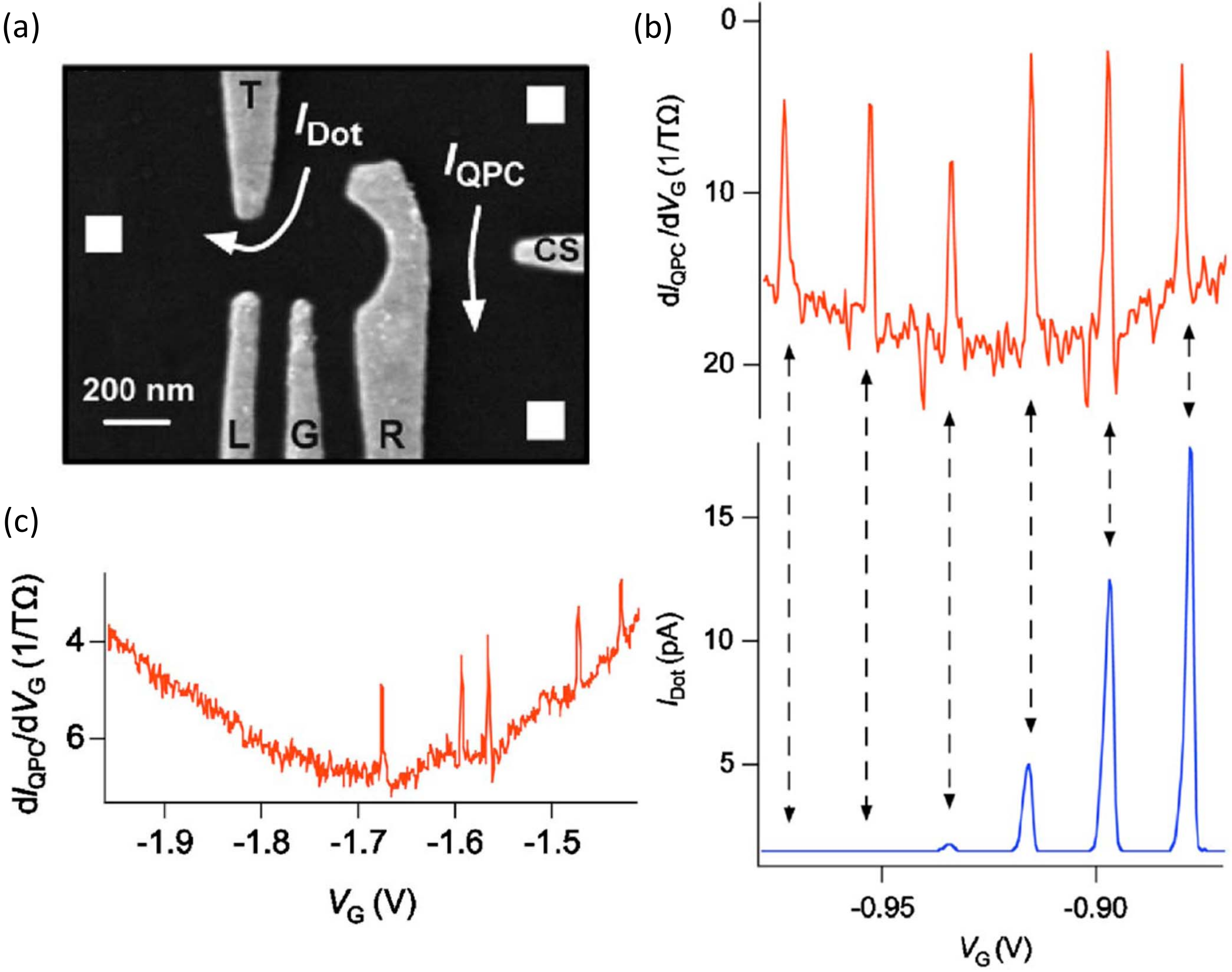}
\caption{\label{chargesensing_fig1}  (Color online) \textbf{Non-invasive charge sensing of a Si/SiGe quantum dot using a quantum point contact (QPC) sensor.} (a)  SEM device image. (b) (Top) Derivative of the QPC current $dI_{QPC}/dV_G$ as a function of gate voltage $V_G$. The peaks correspond to changes in the number of electrons in the dot. (Bottom) Current $I_{Dot}$ through the quantum dot as a function of $V_G$. (c) QPC sensor output in the few-electron limit. No further transitions occur for $V_G < 1.68$ V, indicating an empty quantum dot. From \textcite{simmonsAPL07}.}
\end{figure}

\textcite{sakrAPL2005} fabricated a QPC adjacent to a quantum dot in a Si/SiGe heterostructure using a combination of isolation etching and metal gates aligned to the etched trenches. While this structure enabled sensing of the dot's electron occupancy in the many-electron regime, it did not have sufficient sensitivity to probe down to the last electron. \textcite{simmonsAPL07} used Pd metal surface depletion gates on a Si/SiGe heterostructure to define a similar geometry -- see Fig.~\ref{chargesensing_fig1}a. By monitoring the differential conductance of the QPC sensor they were able to accurately probe the depopulation of electrons in the adjacent quantum dot, even when the transport current $I_{\textrm{Dot}}$ through the dot had fallen below the noise level (Fig.~\ref{chargesensing_fig1}b). In this way they were able to track the occupancy of the dot down to the final electron, as shown in Fig.~\ref{chargesensing_fig1}(c).

More recently, QPC sensors have been used with great versatility in both Si/SiGe and Si MOS quantum dots systems for measurement of both charge \cite{Nordberg:2009p202102} and spin \cite{hayesArxiv09,xiaoPRL10,simmonsPRL11} states.  A technique developed to measure the spin state of a single electron in a GaAs/AlGaAs quantum dot \cite{elzermanNature04} has been successfully applied to dots in silicon. This involves loading an electron (of indeterminate spin) into an empty quantum dot and positioning the Fermi level so that only a spin-up electron is able to tunnel out, with the charge displacement monitored by a QPC sensor. The technique has been used to measure the spin lifetime of single electrons loaded into Si/SiGe \cite{hayesArxiv09,simmonsPRL11}and Si MOS \cite{xiaoPRL10} quantum dots. These experiments are discussed in more detail in Section \ref{timeresolved}.

Single electron transistors (SETs) can also been used as highly sensitive electrometers in nanostructure devices. The most sensitive such electrometers employ Al metal islands, with Al$_2$O$_3$ tunnel barriers, which can be integrated with both MOS~\cite{Andresen:2007p2000} and Si/SiGe-based quantum dots~\cite{yuanAPL11}.  Integrating such SETs into a radio-frequency (rf) tank circuit forms an rf-SET \cite{schoelkopfScience98}, which can operate at frequencies above 100 MHz with charge sensitivities approaching $\sim$10$^{-6}$ $e/\sqrt{\textrm{Hz}}$. \textcite{Andresen:2007p2000} fabricated such an Al-Al$_2$O$_3$ rf-SET on the surface of a phosphorus-doped silicon (Si:P) device to study the gate-controlled transfer of an electron between two implanted phosphorus donors, with a measurement bandwidth exceeding 1 MHz. They were able to study the charge relaxation rate as a function of gate-induced detuning between the two donor levels, measuring an oscillating relaxation rate consistent with acoustic phonon emission in silicon.

\begin{figure}[h]
\includegraphics[width=0.48\textwidth]{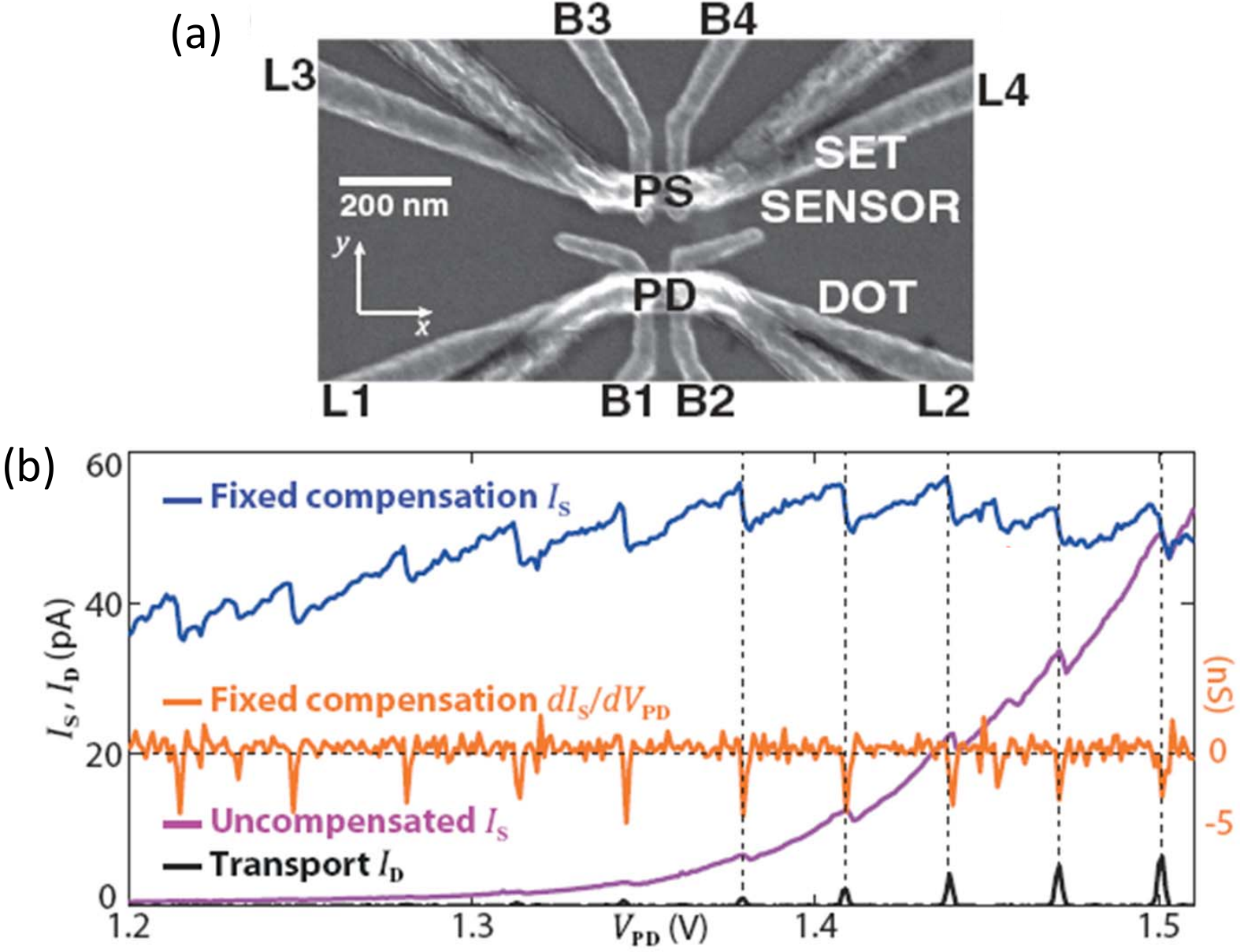}
\caption{\label{chargesensing_fig2} \textbf{Non-invasive charge sensing of a Si MOS quantum dot using a single electron transistor (SET) sensor.} (a) SEM device image, showing a Si MOS SET sensor (upper device) that is capacitively coupled to a Si MOS quantum dot (lower device). (b) Transport current $I_D$ through the quantum dot shows Coulomb peaks as a function of dot plunger gate voltage $V_{PD}$. The changing potential on the dot is detected by monitoring the uncompensated current $I_S$ through the SET sensor, which shows charge transfer events superimposed on a rising background, due to the coupling of the SET to $V_{PD}$. This background can be largely removed by adding a linear correction (fixed compensation) to the SET gate voltage $V_{PS}$, and then further enhanced by plotting the derivative $dI_S/dV_{PD}$.   Data reproduced from \textcite{yangAIP11}.}
\end{figure}

While Al-Al$_2$O$_3$  rf-SETs are well established as fast charge sensors, it is advantageous to integrate the SET sensor into the silicon device itself, as has been done with silicon-based QPC sensors, since this can improve the capacitive coupling to the system being measured and can also simplify fabrication. Furthermore, the larger charging energies that can be obtained with silicon quantum dots, compared with Al metal islands, provides the potential for increased sensitivity and higher operating temperature. Figure \ref{chargesensing_fig2} shows an example of a silicon SET integrated adjacent to a Si MOS quantum dot \cite{yangAIP11}. In this experiment, Yang and co-workers also employed a dynamic feedback technique to keep the SET sensor at a point of constant sensitivity, allowing for more robust measurements that can tolerate random charge displacement events. Podd and co-workers in Cambridge also demonstrated a capacitively coupled pair of Si MOS quantum dots, in which one of the dots could be used to sense the potential of the other \cite{PoddAPL2010}.

\textcite{angusNL2007} configured a silicon-based rf-SET by using a double-gate structure to induce a Si-MOS quantum dot and connecting this within a radiofrequency tank-circuit. They demonstrated a charge sensitivity of better than 10$^{-5}$  $e/\sqrt{\textrm{Hz}}$ at a bandwidth up to 2 MHz, which compares well with metallic rf-SETs. In their device the bandwidth was limited by a high gate resistance, but there is no reason why such a structure could not be designed to operate at bandwidths above 100 MHz. One advantage of a Si-MOS SET compared with its Al-Al$_2$O$_3$ counterpart is that the tunnel barriers of the Si-MOS device are gate controlled, meaning that the resonant frequency of the tank circuit can be easily tuned to optimize its operation.

For studies of spin dynamics, which can be orders of magnitude slower than charge dynamics in silicon, the need for high-frequency sensing becomes less critical and standard low-frequency (sub-MHz) SET operation can be used \cite{HofheinzEPJ06}.  Most notably, \textcite{morelloNature10} used a Si-MOS SET, similar to the structure used by \textcite{angusNL2007}, to detect charge motion between the SET island and implanted phosphorus dopants, thus enabling single-shot spin readout of an electron bound to a phosphorus donor. This experiment is discussed further in Section VI.C.3. 

\subsection{Few-electron quantum dots}\label{fewelectron}

\begin{figure}[h]
\includegraphics[width=0.3\textwidth]{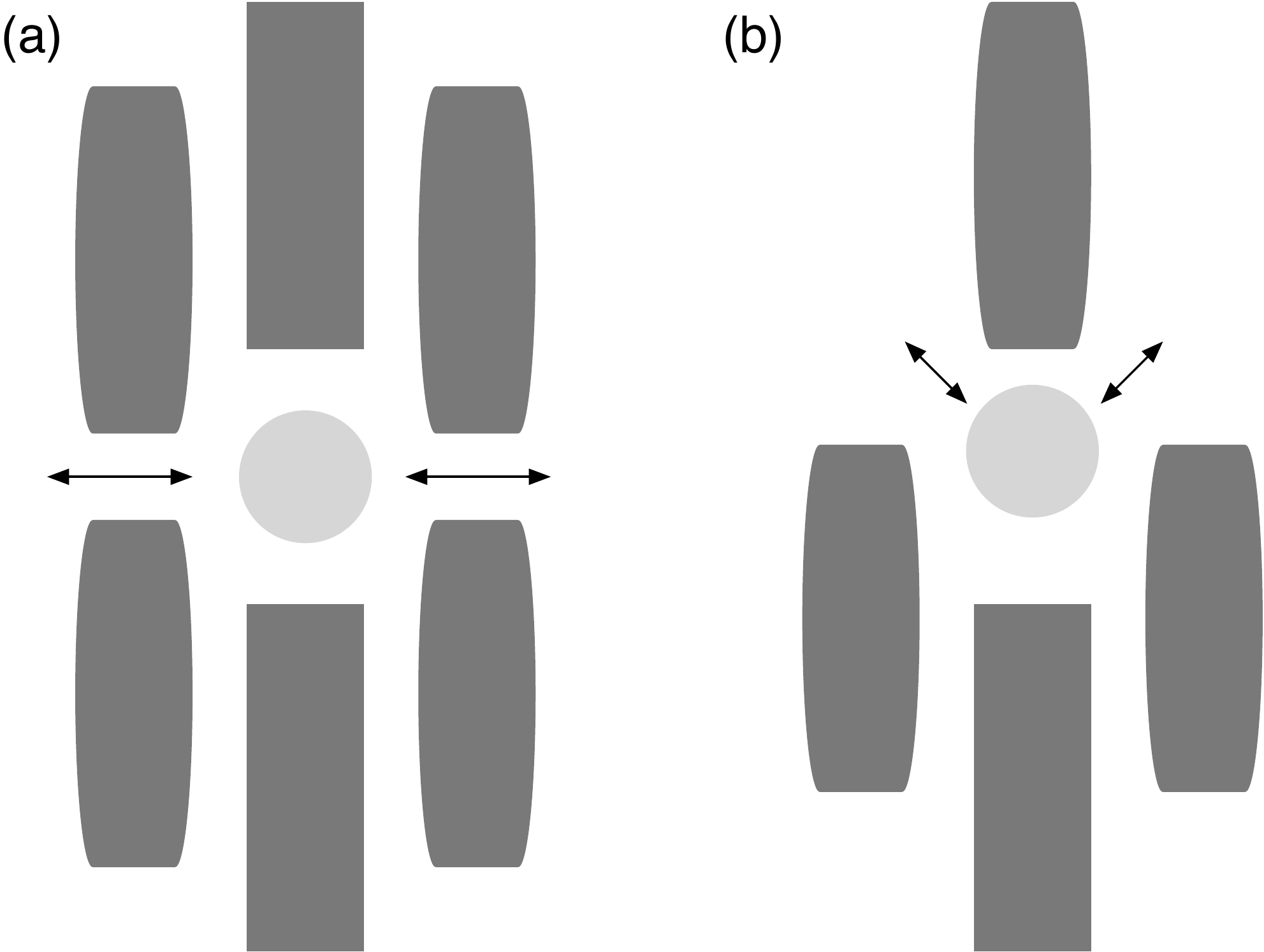}
\caption{\label{fig:few-electron-schematic} \textbf{Gate design
    enabling few-electron occupation.} The gate design in (a) is a
  natural way to form a quantum dot tunnel-coupled to two reservoirs,
  as shown by the arrows.  As the dot becomes smaller, however, it is
  very difficult to maintain a high tunnel rate to both reservoirs.
  The gate design in panel (b), based on Fig.~1 of
  \textcite{ciorgaPRB2000}, enables a small dot to be coupled to both
  reservoirs.}
\end{figure}

For many years it was difficult to achieve single-electron occupation
in gated quantum dots, in spite of the tunability of such dots.  The
fundamental problem was the difficulty maintaining reasonably fast
tunnel rates between a quantum and nearby charge reservoirs.  A common
gate design (see, e.g.,~\textcite{waughPRL1995}), is shown schematically
in Fig.~\ref{fig:few-electron-schematic}(a).  As the quantum dot is
made smaller, by making the gate voltages more negative, the tunnel
barriers to one or both reservoirs must become wider.

Fig.~\ref{fig:few-electron-schematic}(b) shows an alternative approach
for the formation of few-electron quantum dots in GaAs, developed by
the group in Ottawa~\cite{ciorgaPRB2000}.  The advantage of this gate
design is that it enables strong tunnel coupling to both reservoirs
even when the quantum dot is small.  This gate design is equally
useful for gated dots in Si, and it was first implemented in a Si/SiGe
heterostructure in~\textcite{sakrAPL2005}, enabling observation of both
Coulomb blockade and charge sensing, but not single-electron
occupation.

The challenge to achieving single electron occupation in both single
and double one-electron dots in Si/SiGe has been to bring under
control instability in the background offset charge of the quantum
dots.  In 2007 \textcite{simmonsAPL07} demonstrated single-electron
occupation in a top-gated, Si/SiGe quantum dot. In
that work, care was taken to ensure that the doping of phosphorous in
the modulation doping layer was not larger than necessary; limiting
the doping in this layer appears to improve the stability of devices.
The primary evidence for single-electron occupation was the absence of
additional charge transitions, as shown in
Fig.~\ref{chargesensing_fig1}, for a change in gate voltage more than
3.5 times as large as that required to add the last observed electron.

Metal-oxide-semiconductor quantum dots can also approximate the few-electron regime \cite{PratiAPL11}. In the approach of \textcite{xiaoAPL10}, the depletion gates underneath a global accumulation gate form the quantum dot. Using an approach analogous to this type of MOS Si structure, Borselli and collaborators have shown that single-electron occupation can be achieved in very stable Si/SiGe quantum dots when the doping is
removed from the structure~\cite{borselliAPL2011}, see section \ref{planarmos}.

\begin{figure}[h]
\includegraphics[width=0.4\textwidth]{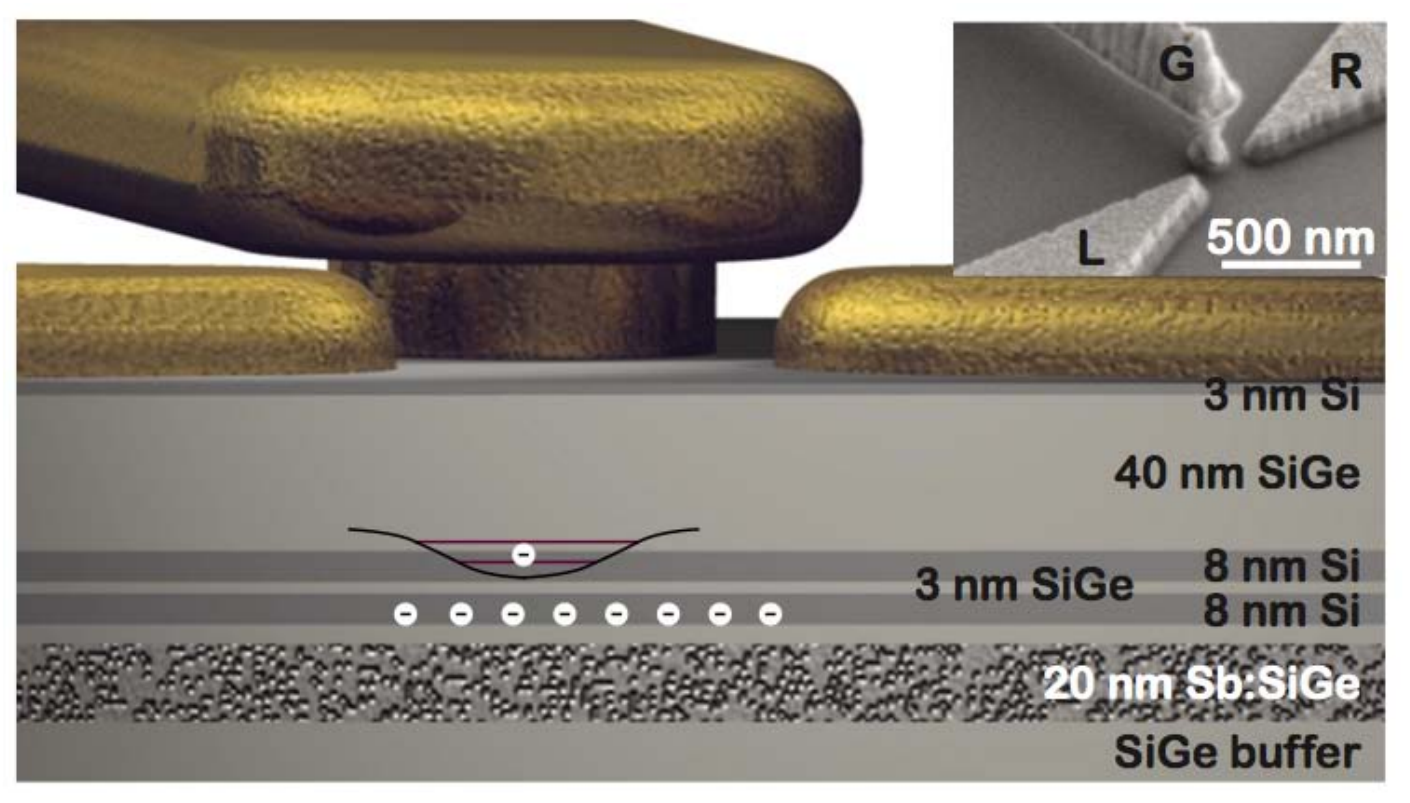}
\caption{\label{fig:few-electron-hrl-double-qw}  (Color online) Schematic diagram of a
  few-electron quantum dot formed from a Si/SiGe heterostructure with
  a double quantum well and an accumulation gate contacted by an air
  bridge.  Inset: SEM micrograph of the gate region of a corresponding
  device.  Figure from \textcite{borselliAPL11}.}
\end{figure}

A novel approach to achieving single-electron occupation was
demonstrated by Borselli and colleagues at HRL
Laboratories~\cite{borselliAPL11}.  As shown in
Fig.~\ref{fig:few-electron-hrl-double-qw}, the device structure uses
two quantum wells, the lower of which is doped.  An air bridge is used
to apply a positive voltage to an isolated, circular surface gate,
pulling electrons into the upper quantum well.  Nearby surface gates
are negatively biased, enabling the formation of a charge-sensing
channel in the lower electron layer.  Such a device forms an extremely
symmetric quantum dot that is easily tuned to the one-electron charge
state.

\begin{figure}[h]
\includegraphics[width=0.48\textwidth]{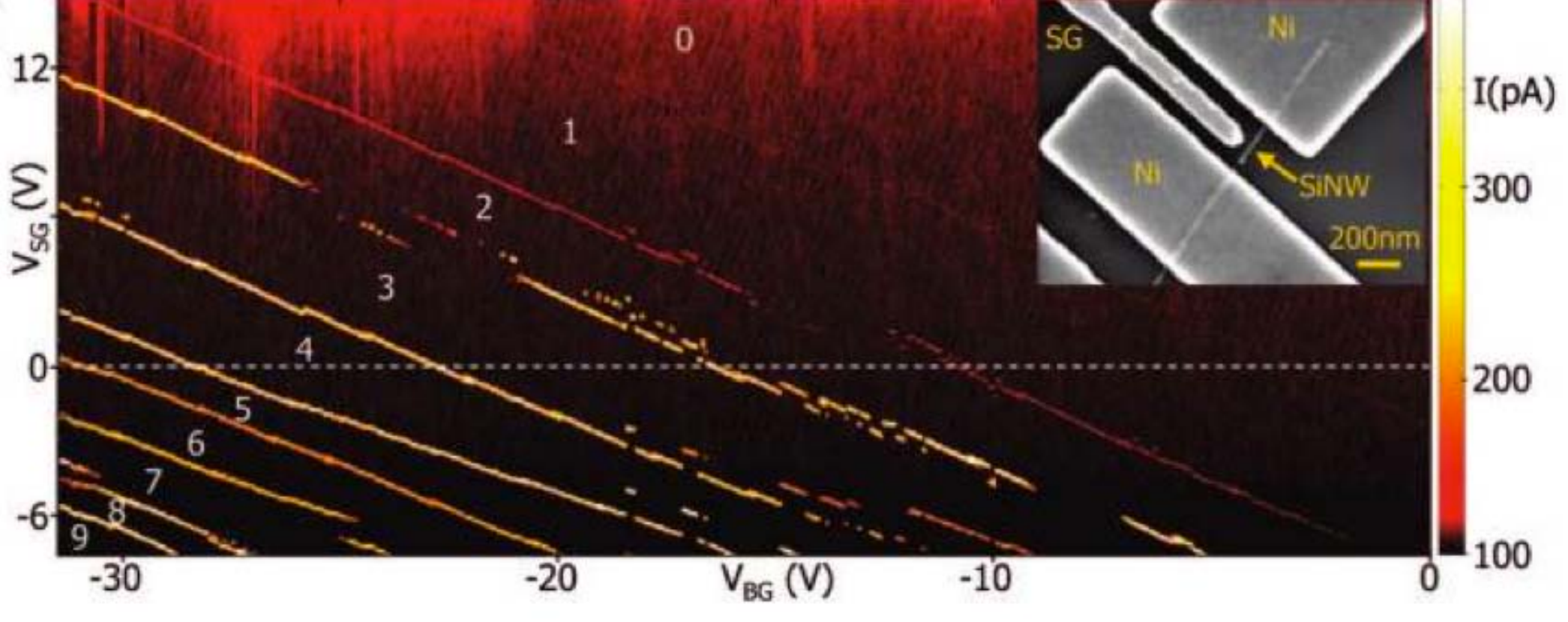}
\caption{\label{fig:few-electron-zwanenburg}  (Color online) Transport data showing
  the last hole in a Si nanowire based quantum dot.  Inset: SEM image
  of the device showing the NiSi contacts and a Cr/Au side gate.
  device.  Figure from \textcite{zwanenburgNL09}.}
\end{figure}

Few-carrier occupation can be accomplished even in the absence of
charge sensing, as demonstrated in nanowire-based hole quantum dots
for which the Coulomb diamonds open to very large gate voltages at
sufficiently positive gate voltage~\cite{zhongNL05}.  Zwanenburg and
collaborators have reached the one-hole state in a very small Si
quantum dot in a nanowire, enabling them to perform spin
spectroscopy~\cite{zwanenburgNL09}.  The device made use of NiSi
contacts, in which a Schottky barrier defines the quantum dot, as
shown in Fig.~\ref{fig:few-electron-zwanenburg}.  The few-electron
regime was also observed without charge sensing in planar MOS Si
quantum dots, thanks to the high degree of tunability of these devices
\cite{limAPL09-2,LimNanotech11}, and in MOSFETs built within a pre-industrial Fully Depleted Silicon
On Insulator technology \cite{pratiNanotech12}.

\subsection{Spins in single quantum dots}\label{spinsdots}

In the previous sections we have established the evolution in recent years from the observation of simple localization and coulomb blockade to few-electron quantum dots in silicon. With the understanding and control of the charge side of electrons one can also probe their spins. In this section we first discuss experiments on ground-state and excited-state magnetospectroscopy in silicon quantum dots. The existence of valleys in silicon make the spin filling non-trivial: the configuration and mixing of valleys and orbits determines how electrons will consecutively occupy the available spin-up or down states.

\subsubsection{Spin-state spectroscopy}\label{spinstate}
\begin{figure}[htb]
\includegraphics[width=0.48\textwidth]{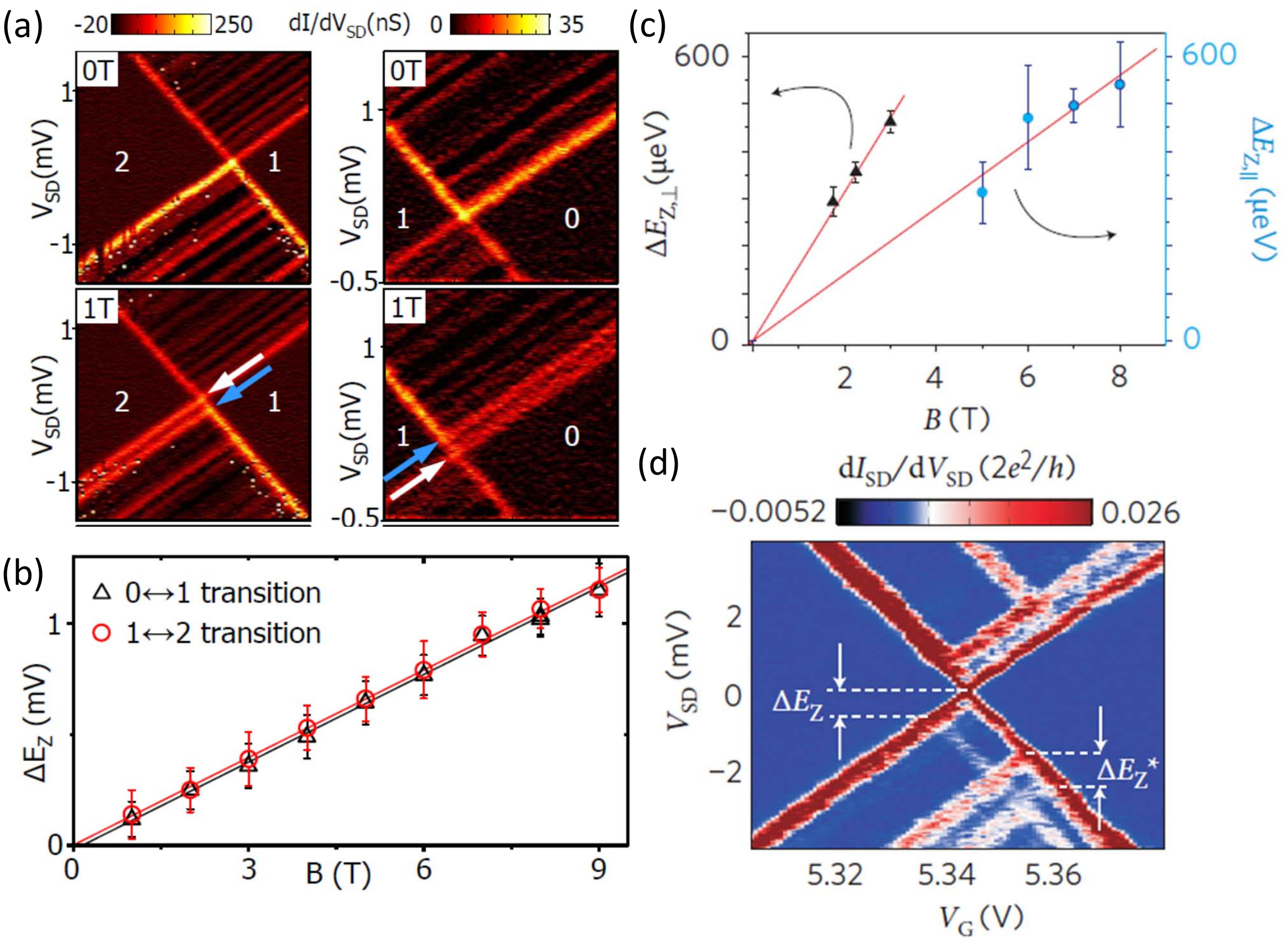}
\caption{\label{spinsdots_fig2}  (Color online) \textbf{Excited-state magnetospectroscopy in Si quantum dots.} (a) Zeeman splitting at the 0-1 and 1-2 transition in a few-hole Si nanowire quantum dot and (b) the corresponding magnetic field dependence of the Zeeman energy, data from \textcite{zwanenburgNL09} (c), Anisotropic g-factors in SiGe  nanocrystals, and (d) the corresponding excited-state magnetospectroscopy, data from \textcite{KatsarosNNano10}.}
\end{figure}

The most straightforward methods of measuring electron spin states in quantum dots are ground-state and excited-state magnetospectroscopy \cite{hansonrmp07}. Excited-state magnetospectroscopy allows observation of spin excited states at a fixed magnetic field \cite{cobdenPRL98}, as long as the Zeeman energy can be resolved. Four experimental demonstrations in silicon systems are bottom-up Si and SiGe nanowires \cite{zwanenburgNL09, roddaroPRL08, huNNano11} and SiGe nanocrystals \cite{KatsarosNNano10}, see Fig.~\ref{spinsdots_fig2}. When the spin-excited state is measured at different magnetic fields, one can extract the g-factor by plotting the Zeeman energy versus magnetic field, see Fig.~\ref{spinsdots_fig2}(b). The first two holes in a Si nanowire quantum dot were found to have a g-factor of $2.3\pm 0.2$ in perpendicular magnetic field. In SiGe nanocrystals and nanowires the g-factor is anisotropic: the results in Fig.~\ref{spinsdots_fig2}(c) show g-factors of $g_{\parallel} = 1.21$ and $g_{\perp} = 2.71$ for respectively parallel and perpendicular field.

In case of ground-state magnetospectroscopy, the spin filling is investigated by measuring the magnetic field dependence of the electrochemical potential $\mu_N$, which is by definition the energy required for adding the $N^{th}$ electron to the dot. The slope of $\mu_N(B)$ is given by  $\frac{\partial \mu_N}{\partial B} = -g\mu_B \Delta S_{\textrm{tot}}(N)$, where $g$ is the g-factor, the Bohr magneton $\mu_B = 58$ $\mu$eV/T and $\Delta S_{\textrm{tot}}(N)$ is the change in total spin of the dot when the $N^{\textrm{th}}$ electron is added~\cite{hadaPRB03}. The electrochemical potential has a slope of $+g\mu_B/2$ when a spin-up electron is added, whereas addition of a spin-down electron results in a slope of $-g\mu_B/2$. The rate at which $\mu_N$ changes with magnetic field thus reveals the sign of the added spin.

\textcite{rokhinsonPRB00} were the first to observe the theoretically expected slopes in multiples of $g\mu_B/2$ in an n-type Si quantum dot. They show the peak shift with magnetic field of 29 electrons entering the dot, and more detailed measurements on two sets of Coulomb peaks with slopes of $\pm1/2g\mu_B$ and $\pm3/2g\mu_B$. The charge transitions display an unexpected large number of kinks at which the slope changes sign, and thus the spin state as well. They conclude that the spin filling is inconsistent with a simple picture of non-interacting electrons in four single-particle levels. Later reports are more straightforward to interpret and will be discussed below.

\begin{figure}[htb]
\includegraphics[width=0.48\textwidth]{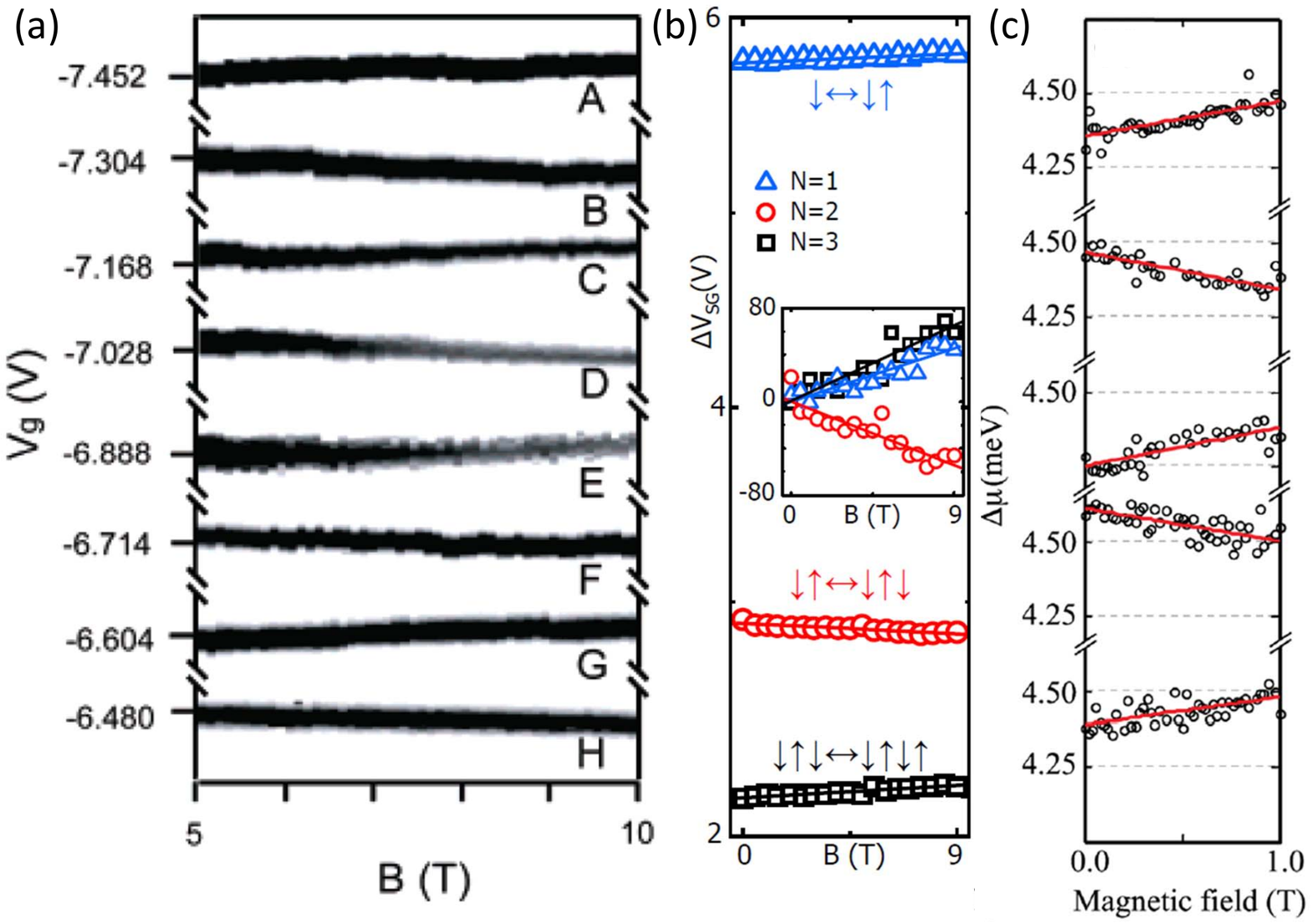}
\caption{\label{spinsdots_fig1}  (Color online) \textbf{Ground-state magnetospectroscopy.} Three examples of even-odd hole spin filling in (a) a many-hole Si nanowire quantum dot \cite{zhongNL05} (b), a few-hole Si nanowire quantum dot \cite{zwanenburgNL09} and (c), a many-hole Ge/Si nanowire quantum dot \cite{roddaroPRL08}.}
\end{figure}

The spin filling of holes has been investigated in nanowire quantum dots. In 2005, \textcite{zhongNL05} found alternating spin-up and spin-down holes in a many-hole quantum dot. The magnetic field evolution of the positions of eight consecutive Coulomb peaks in Fig.~\ref{spinsdots_fig1}(a) reveals alternating slopes of $\pm g\mu_B/2$, with an extracted g-factor of 2$\pm0.2$. The few-hole regime displayed similar spin filling of the first four holes in an empty dot \cite{zwanenburgNL09}, see Fig.~\ref{spinsdots_fig1}(b). The even-odd filling suggests that the degeneracy of heavy and light holes is lifted due to strain and confinement effects; see, for example, calculations based on density functional theory \cite{leuPRB06, sorokinPRB08} and tight-binding models \cite{niquetPRB06, buinNL08}. SiGe nanowires have been shown to exhibit the same spin filling, see \cite{roddaroPRL08} and Fig.~\ref{spinsdots_fig1}(c).

\subsubsection{Spin filling in valleys and orbits}\label{spinvalleys}
The even-odd spin filling as observed in p-type silicon quantum dots (see Section \ref{spinstate}) is not very different from similar devices in other material systems. However, the valleys in the silicon conduction band make the spin filling of \textsl{electrons} non-trivial. Valley physics in silicon has been studied extensively both theoretically~\cite{Culcer:2010p155312,Friesen:2010p115324,Saraiva:2009p081305,Culcer:2010p205315,saraivaArxiv10} and experimentally~\cite{kohlerOther79, nicholasOther80, pudalovOther85, koesterSST97, Takashina:2006p236801, goswamiNPhys06, McGuireNJP10, FuechsleNnano10}.

\begin{figure}[htb]
\includegraphics[width=0.48\textwidth]{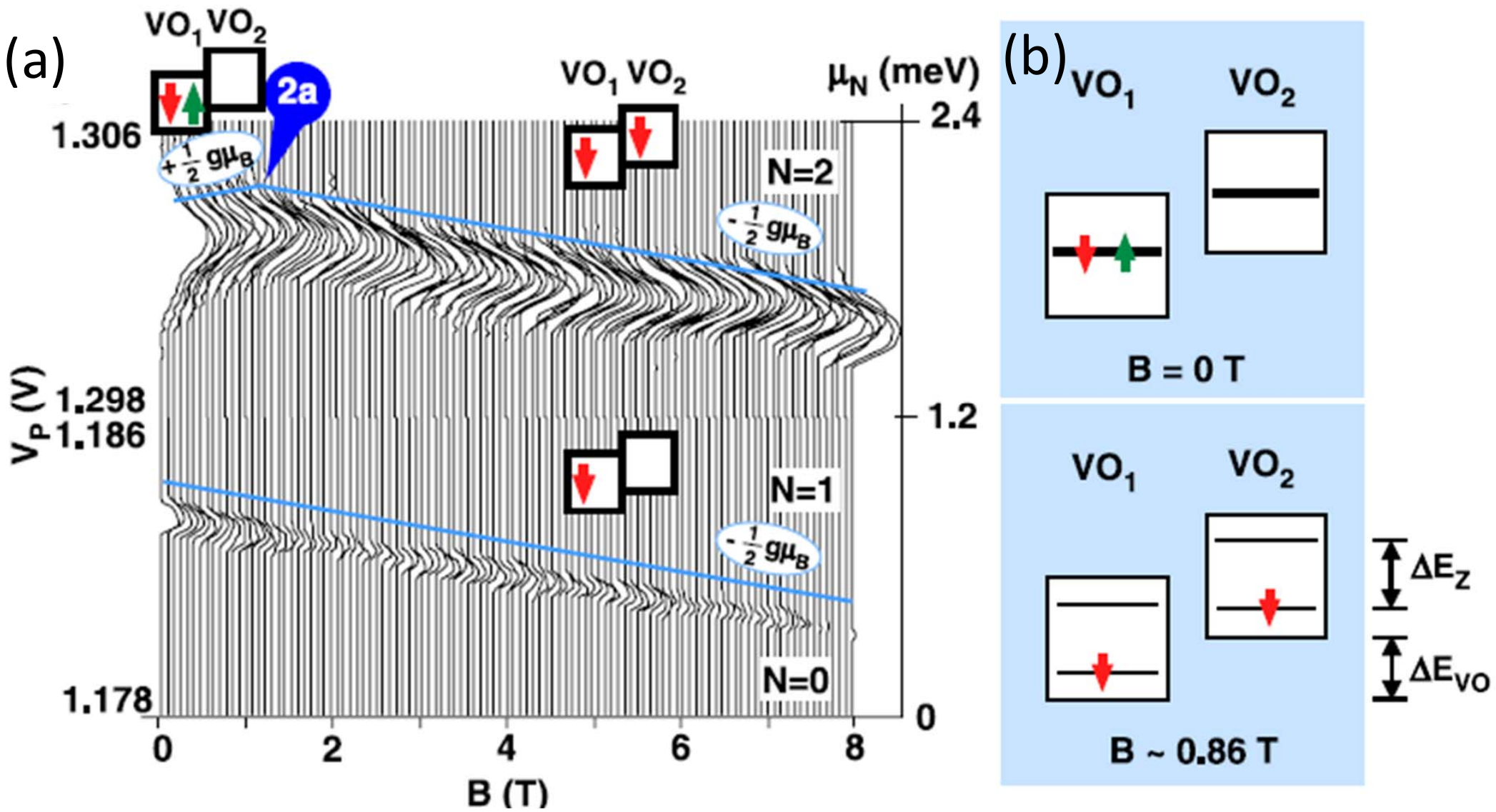}
\caption{\label{spinsdots_fig3}  (Color online) \textbf{Spin filling in valleys in a planar MOS Si quantum dot.} (a) Magnetospectroscopy of the first two electrons entering the quantum dot. The circle 2a marks a kink in the second Coulomb peak at $\sim$0.86 T. The arrows in the boxes (VO$_1$ for valley-orbit 1 and VO$_2$ for valley-orbit 2) represent the spin filling of electrons in the quantum dot. (b) For $B<0.86$ T, the first two electrons fill with opposite spins in the same valley-orbit level (left panel). The Zeeman energy at the kink is equal to the valley-orbit splitting (0.10 meV). Data reproduced from \textcite{LimNanotech11}}
\end{figure}

As discussed in Section \ref{phys-valleyphysics} a 2-dimensional electron gas has two $\Gamma$-valleys, separated by the valley splitting $E_V$, see Fig.~\ref{valleys-bulk-2D-dopants-alternative}. A finite valley splitting influences the spin filling as observed in ground-state magnetospectroscopy~\cite{hadaPRB03}: the first electron is always a spin-down, yielding a slope of the corresponding Coulomb peak of $-g\mu_B/2$, see the experiment by \textcite{LimNanotech11} in Fig.~\ref{spinsdots_fig3}b. The kink in the second Coulomb peak (marked 2a) at $\sim$0.86 T is caused by a sign change of the $N=2$ ground-state spin: at low magnetic field (before the kink), the second electron fills the quantum dot with a spin-up. As the magnetic field is increased, the sign of the second electron spin changes from up to down at $B\sim0.86$ T.

When the valleys and orbits are mixed (section \ref{phys-valleyphysics}), there are no pure valleys or pure orbits, and the lowest available levels are referred to as valley-orbits. The sign change can then be explained with a simple model where the two lowest valley-orbit levels are separated by the valley-orbit splitting $\Delta E_{\textrm{VO}}$, see Fig.~\ref{spinsdots_fig3}(b). At zero magnetic field, the first two electrons fill with opposite spins in valley-orbit level 1. When a magnetic field is applied, the spin-down and spin-up states are split by the Zeeman energy $E_Z$. Above 0.86 T the spin-up state of valley-orbit level 1 (VO$_1$) is higher in energy than the spin-down state of valley-orbit level 2 (VO$_2$) and it becomes energetically favored for the second electron to occupy the latter, i.e. VO$_2$. At the kink the valley-orbit splitting equals the Zeeman energy, which is 0.10 meV at 0.86 T. Comparable kinks were reported simultaneously in accumulation mode Si/SiGe quantum dots, yielding valley splittings of 0.12 and 0.27 meV~\cite{borselliAPL11}. In 2010, the absence of kinks in the ground-state magnetospectroscopy of a planar MOS Si quantum dot was explained as a result of a large exchange energy and an unusually large valley splitting of 0.77 meV \cite{xiaoAPL10}.

\subsection{Double quantum dots}\label{doubledots}

Like their counterparts in the Ga-AlGaAs material system, double quantum dots in silicon represent the natural extension from a semiconductor `artificial atom'ù to an `artificial molecule'. As outlined in the previous sections, it took until around 2006 for low-disorder silicon-based quantum dots to be produced with reasonable repeatability. Correspondingly, this is also when the first demonstrations of double quantum dots in silicon began to be reported.

\subsubsection{Charge-state control}\label{doubledotscharge}

One of the earliest reports of silicon double dot operation was by \textcite{gormanPRL05}, who formed an isolated double dot by etching a thin (35 nm) layer of bulk-phosphorus-doped silicon (Si:P) in a SOI substrate. They also integrated a nearby SET, again formed by etching the Si:P layer, which they used to monitor charge transfer in the double dot. By rapidly pulsing a nearby control gate they observed oscillations in the charge state of the double dot, as a function of pulse length, which they interpreted as coherent oscillations between the (n, m) and (n-1, m+1) charge states of the double dot. Because of the high electron numbers in the dots resulting from the degenerative doping, and the difficulty of controlling the dots size via the etching process, this type of dot structure has not progressed significantly since this time, and most studies of silicon quantum dots are now based on dots induced in undoped silicon layers.

\begin{figure}[h]
\includegraphics[width=0.4\textwidth]{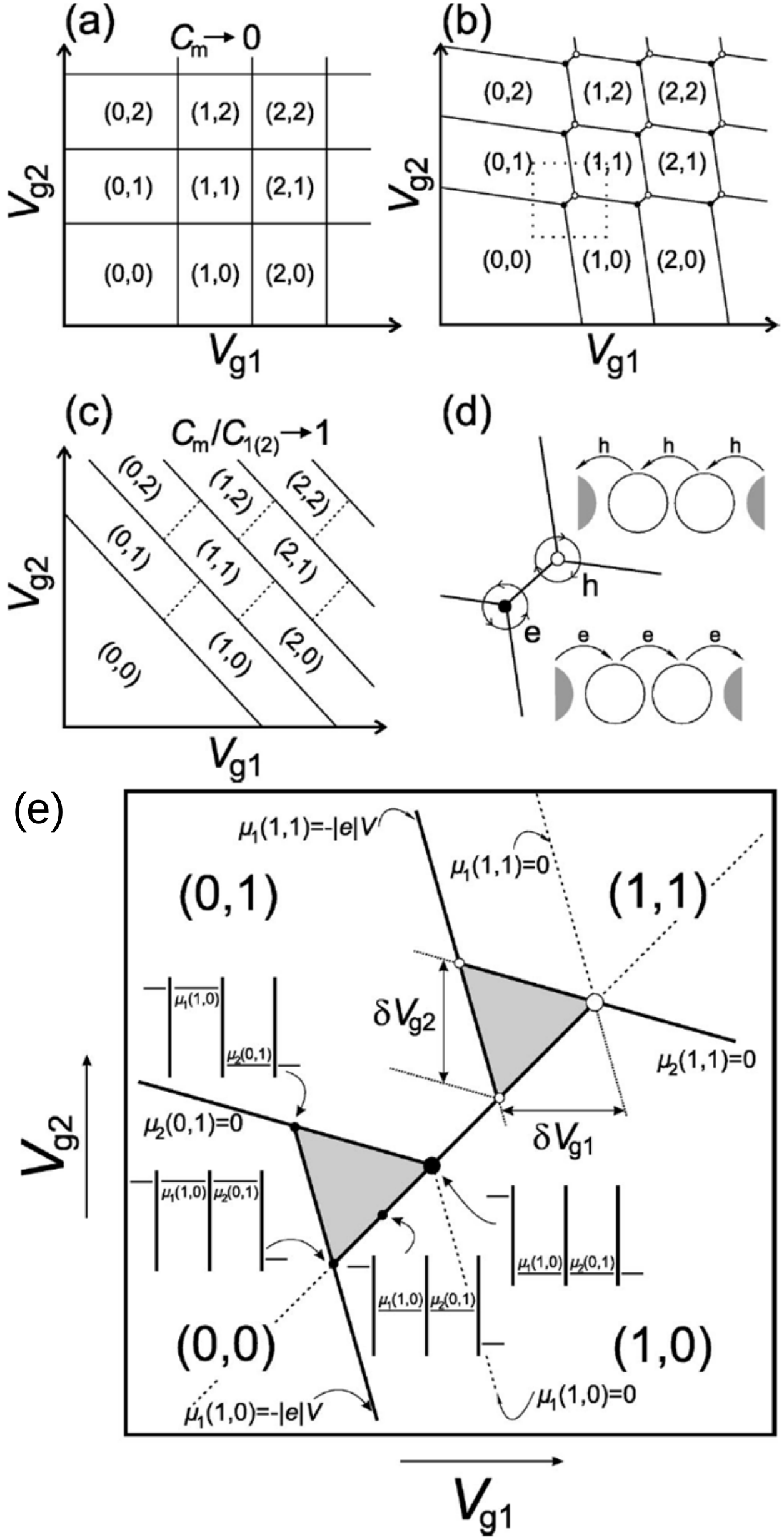}
\caption{\label{doubledots_fig1} \textbf{Schematic stability diagrams for a double dot system.} Maps are shown for (a) small, (b) intermediate, and (c) large inter-dot coupling. The equilibrium charge on each dot in each domain is denoted by (N1 ,N2). (e) Region within the dotted square of (b), corresponding to the unit cell of the double dot stability diagram, at finite bias voltage. The solid lines separate the charge domains. Classically, the regions of the stability diagram where current flows are given by the gray triangles. Reproduced from \textcite{VanDerWiel:2003p1382}.}
\end{figure}

The starting point for any experimental study of a double quantum dot is the determination of its charge state ($N_1$, $N_2$) as a function of at least two gate voltages $V_{g1}$ and $V_{g2}$ controlling the system. Here $N_1$, ($N_2$) is the electron occupancy of dot-1 (dot-2). By directly measuring the transport current $I$ through the dot, or by measuring the local electrostatic potential of the system using a nearby QPC or SET charge sensor, one can plot a charge stability map as a function of $V_{g1}$ and $V_{g2}$. \textcite{VanDerWiel:2003p1382} provide an excellent review of semiconductor double quantum dots  and Fig.~\ref{doubledots_fig1}(a-c) taken from this review depicts the charge stability maps expected for different inter-dot coupling strengths. Figure \ref{doubledots_fig1}(b) shows a map at intermediate inter-dot coupling, where one observes a characteristic `honeycomb' like structure. The points on the map where three different charge states are degenerate in energy are known as `triple points' and it is only at these points where a transport current $I$ can flow through the system, from source to drain (Fig.~\ref{doubledots_fig1}d). If we then apply a source-drain bias $V_{SD}$ across the double dot these triple points expand to take on a triangular shape, as shown in Fig.~\ref{doubledots_fig1}(e). It should be noted that while in a transport measurement only the triple points (or bias triangles) can be observed, when charge sensing is employed one can also directly detect the transition lines between charge states. \textcite{sarmaPRB11} have recently applied a Hubbard model approach to determine the charge stability diagrams for silicon double dots, showing excellent agreement with experiments \cite{simmonsNL09,laiOther11}.
\begin{figure}[h]
\includegraphics[width=0.4\textwidth]{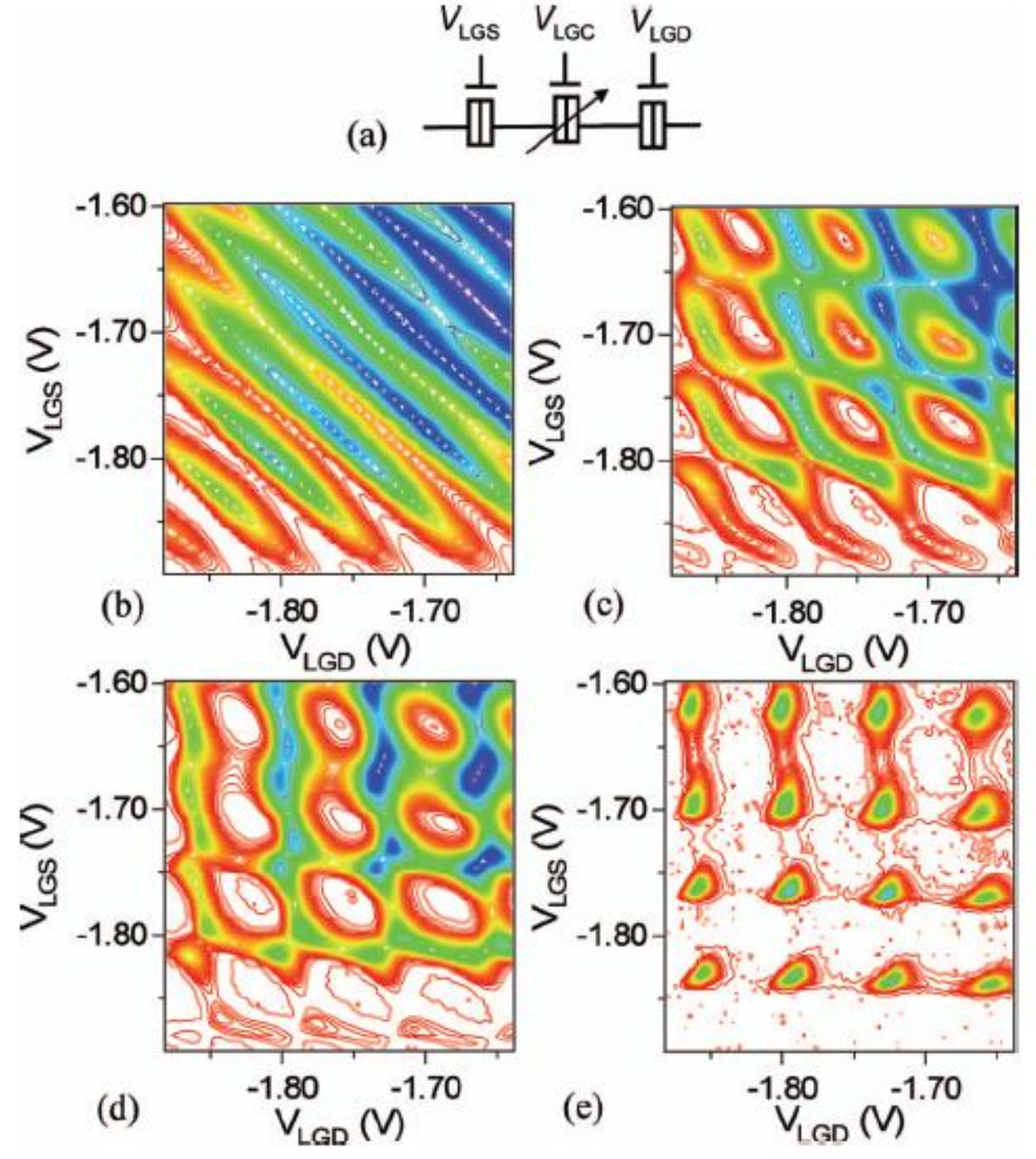}
\caption{\label{doubledots_fig2}  (Color online) \textbf{Evolution from a single dot to a double quantum dot in a gated silicon nanowire device.} (a) Equivalent circuit. (b-e) Contour plots of the drain current as a function of the outer barrier gate voltages $V_{LGS}$ and $V_{LGD}$. The central barrier gate voltages used were $V_{LGC} = -0.75$ (b), $-1.13$ (c), $-1.18$ (d), and $-1.284$ V (e). Data reproduced from \textcite{fujiwaraAPL2006}.}
\end{figure}

The first clear demonstration of a double quantum dot in nominally undoped silicon \cite{fujiwaraAPL2006} used a double-gated silicon nanowire formed from a SOI substrate, with three independently controlled barrier gates. This device structure was previous described in Section \ref{etchedsilicon} and is depicted in Fig.~\ref{MOS_fig8}(a,b). By varying the voltage $V_{LGC}$ on the central barrier gate, Fujiwara and co-workers were able to gradually tune the system from one large dot to two well-isolated dots, as evidenced by the charge stability plots shown in Fig.~\ref{doubledots_fig2}(b-e). In Fig.~\ref{doubledots_fig2}(b) the central gate voltage $V_{LGC}$ is tuned to minimize the tunnel barrier between the dots, forming one large dot. As the outer barrier gate voltages $V_{LGS}$ and $V_{LGD}$ are varied, the transport current I plotted in Fig.~\ref{doubledots_fig2}(b) shows Coulomb blockade oscillations as a function of the addition voltage ($V_{LGS} + V_{LGD}$), which can be compared with Fig.~\ref{doubledots_fig1}(c), indicating that one large dot is present. However, as the tunnel barrier height is increased by tuning $V_{LGC}$, two separate dots form, as revealed in Figs. \ref{doubledots_fig2}(c-e) with the gradual formation of a honeycomb-shaped map of $I(V_{LGS},V_{LGD}$), consistent with that in Fig.~\ref{doubledots_fig1}(b). Gate-tuneable double quantum dots based on etched silicon nanowires have since also been reported by other groups (e.g.\ \textcite{PierreNNano10}).

\begin{figure}[h]
\includegraphics[width=0.48\textwidth]{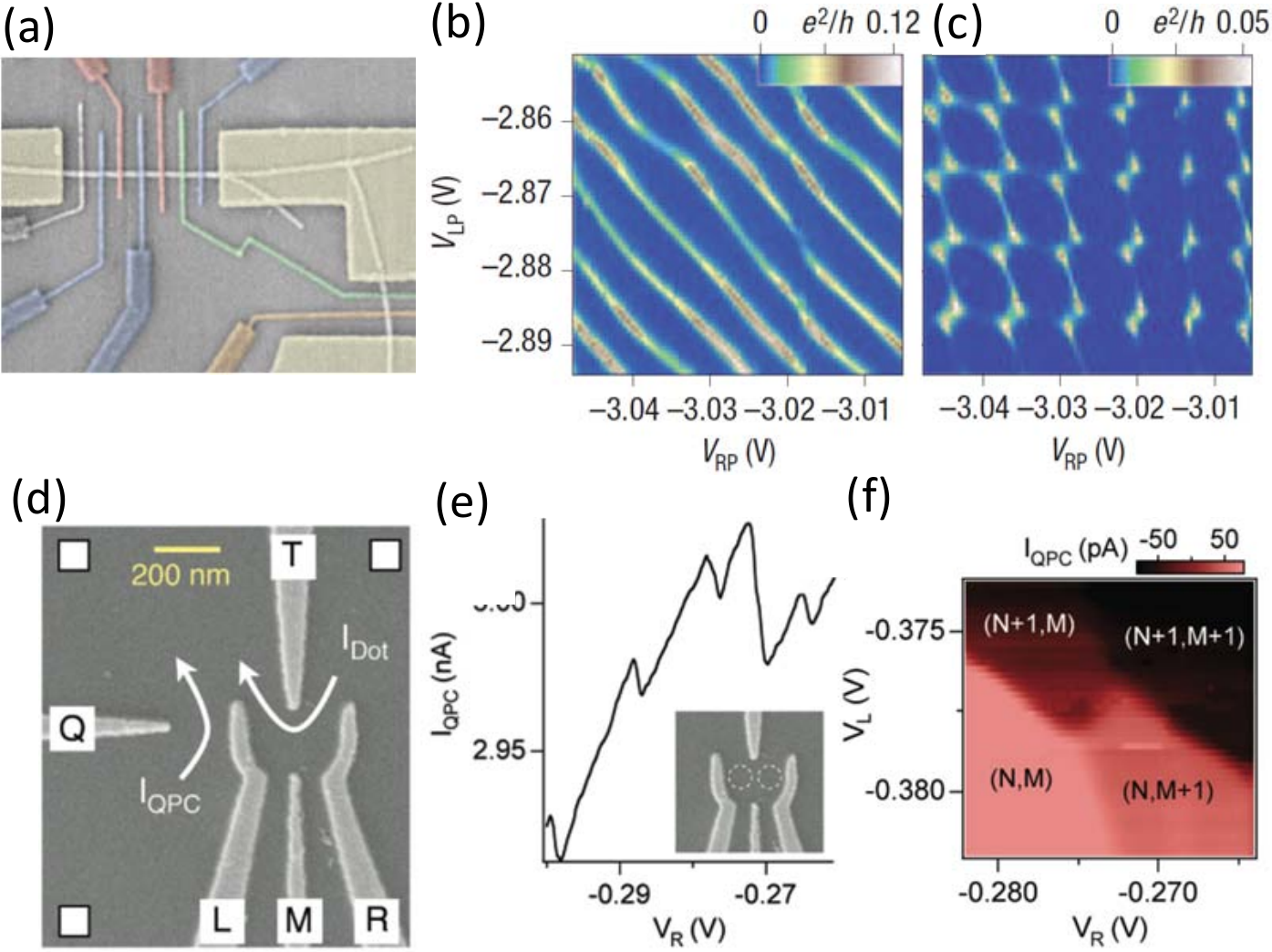}
\caption{\label{doubledots_fig4x}  (Color online) \textbf{Gate tunable double quantum dots.} (a) SEM image of a Ge/Si nanowire-based hole quantum dot. The Ge/Si nanowire at top (white in image) is gated by metal gates to form a double dot. (b,c) Charge stability maps of the conductance as a function of plunger gate voltages. (d) SEM image of an electron quantum dot defined by electrostatic top gates in a Si/SiGe heterostructure. (e) Charge sensing measurement showing the difference in the charge detection signal from the dot furthest from the QPC (4 small steps in $I_\mathrm{QPC}$) and the dot closest to the QPC (single large step) as a function of gate voltage.  (e) Two-dimensional plot of the charge sensing current showing the sequential addition of electrons to the left and right dots.  Data in (a-c) reproduced from \textcite{huNNano07}; data in (d-f) reproduced from \textcite{simmonsNL09}.}
\end{figure}

Epitaxially grown nanowires have also been configured as double quantum dots. Hu and co-workers from Harvard \cite{huNNano07} used a Ge/Si core/shell nanowire, as described in Section IV.B.1, with a number of surface metal gates to demonstrate a highly-tunable double dot device, as shown in Fig.~\ref{doubledots_fig4x}. Figs. \ref{doubledots_fig4x}(b, c) show the evolution of the charge stability map from a strongly coupled single dot to a double dot as the central barrier gate is tuned.

\begin{figure}[h]
\includegraphics[width=0.48\textwidth]{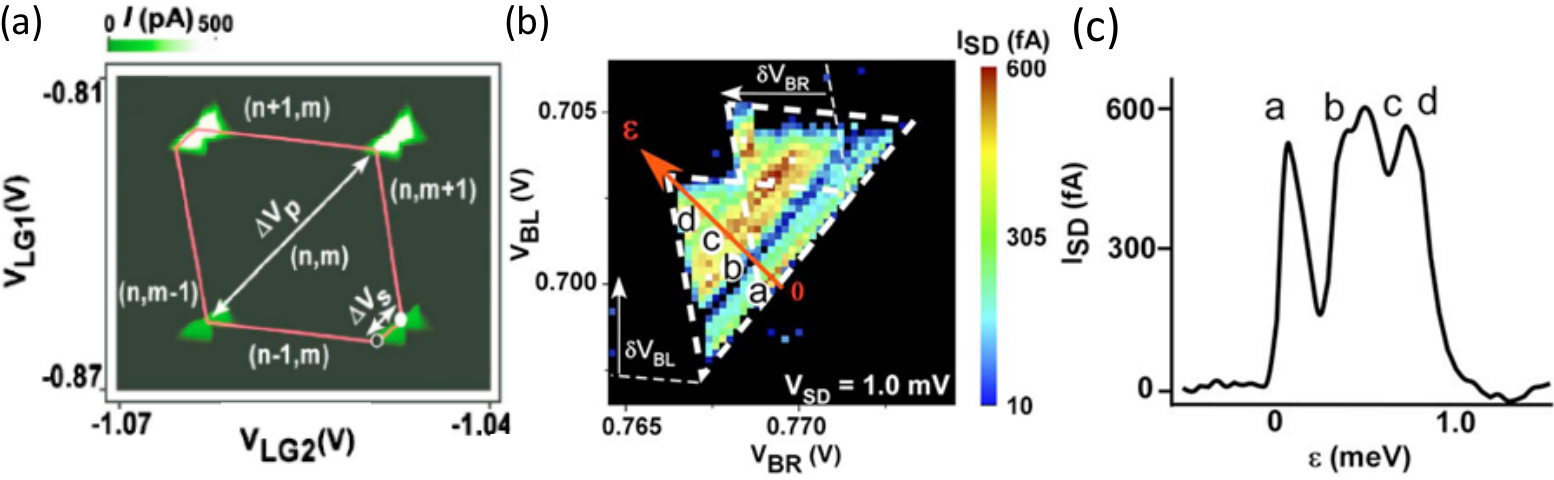}
\caption{\label{doubledots_fig3}  (Color online) \textbf{Bias spectroscopy of silicon double quantum dots.} (a) Stability map with a source-drain bias $V_{SD}=1$ mV for a silicon nanowire double dot, depicted in Fig.~\ref{MOS_fig8}, obtained by plotting source-drain current $I$ as a function of two barrier gate voltages. The triple points have clearly evolved into bias triangles. (b) Bias triangles for two triple points at $V_{SD}=1$ mV, obtained in a Si MOS double dot. (c) Line trace of $I_{SD}$, taken along arrow in (b), showing resonances corresponding to excited states in the double dot. Data reproduced from \textcite{LiuAPL08} and \textcite{Lim:2009:p173502}.}
\end{figure}

Using a gated etched nanowire device, \textcite{LiuAPL08} were able to explore the excited state energy levels within a double quantum dot, which in this case was formed using two barrier gates and the presence of a third barrier created by local disorder. Figure \ref{doubledots_fig3}(a) shows the charge stability diagram for this double dot in the presence of a source-drain bias, which transforms each triple point into a `bias triangle', as described in Fig.~\ref{doubledots_fig1}(e). By mapping the bias triangle in more detail, Liu and co-workers were able to observe resonant tunnelling through excited states of the double dots. Using a planar MOS structure, similar to that in Fig.~\ref{MOS_fig5}(a,b), \textcite{Lim:2009:p173502} were also able to observe excited state transport through a double quantum dot, this time controlled using three independent barrier gates. Figure \ref{doubledots_fig3}(b) shows a pair of bias triangles for two triple points, each showing structure in the source-drain current $I_{SD}$, that is further revealed in Fig.~\ref{doubledots_fig3}(c), which shows a line trace of $I_{SD}$ along the detuning axis $\epsilon$.

\begin{figure}[h]
\includegraphics[width=0.48\textwidth]{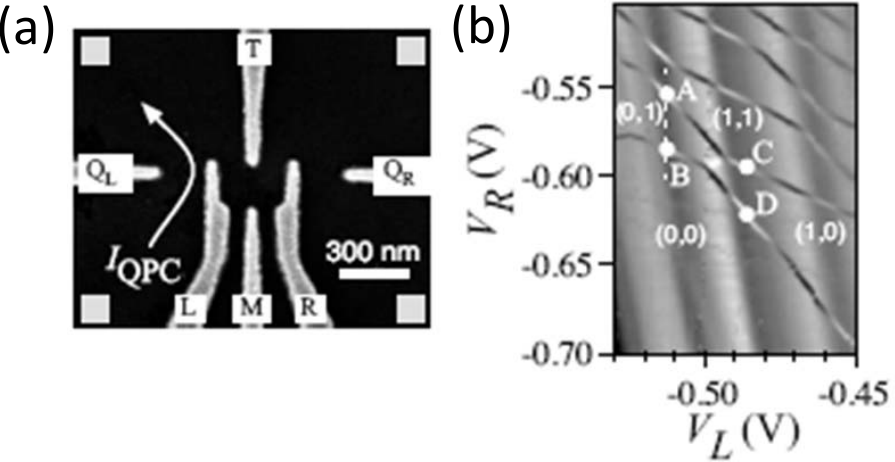}
\caption{\label{doubledots_fig4xx} \textbf{Single-electron occupancy in a Si/SiGe double quantum dot.} (a) SEM of the device. (b) Charge stability map of the double dot, obtained by plotting the QPC charge sensor output as a function of the control gate voltages $V_L$ and $V_R$. The charge configurations $(n, m)$ are marked, showing depletion to the $(0,0)$ state. Data reproduced from \textcite{thalakulamAPL10}.}
\end{figure}

\textcite{simmonsNL09} demonstrated one of the first highly tunable double quantum dots in a gated Si/SiGe device. The device structure shown in Fig.~\ref{doubledots_fig4x}(d) also incorporated a QPC charge sensor, which enabled mapping of the double dot charge stability as a function of the gate voltages $V_L$ and $V_R$ controlling the two dots (see Fig.~\ref{doubledots_fig4x}(e,f)). They were able to tune the inter-dot coupling by control of the central gate voltage $V_M$. The same group were able to demonstrate depletion of a double quantum dot to the single electron level \cite{thalakulamAPL10}. The data in Fig.~\ref{doubledots_fig4xx}(b) shows a charge stability map of the double dot, measured using the QPC sensor. The lack of charge transitions in the lower left quadrant of this map demonstrates control of electron number down to the $(0,0)$ charge state.

Occupation down to the (0,0) charge state in a double dot has also been recently demonstrated in an undoped Si/SiGe heterostructure device \cite{borselliAPL11}, in which a two-layer gate structure analogous to that used for Si MOS dots (see Fig.~\ref{MOS_fig2}a,b) was used, resulting in a very stable system.  Very recently, a few-electron double dot has been demonstrated in an isotopically-enriched $^{28}$Si quantum well~\cite{wildAPL12}.

\subsubsection{Spin transport in double quantum dots}\label{doubledotsspin}

As discussed in Section V.D, in tightly-confined quantum dots the singlet-triplet exchange energy for an electron pair can become appreciable. In a double quantum dot system this can lead to an effect known as Pauli spin blockade, where transport through the double dot is dependent upon the spin state of the electron. This phenomenon was first observed in 2002 in a GaAs/AlGaAs double quantum dot \cite{onoScience02}. In 2008 Liu and co-workers observed Pauli blockade in a silicon nanowire device similar to that shown in Fig.~\ref{MOS_fig8}(a,b), in which a double quantum dot was formed using two barrier gates and a third barrier created by local disorder \cite{liuPRB08}.

\begin{figure}[h]
\includegraphics[width=0.48\textwidth]{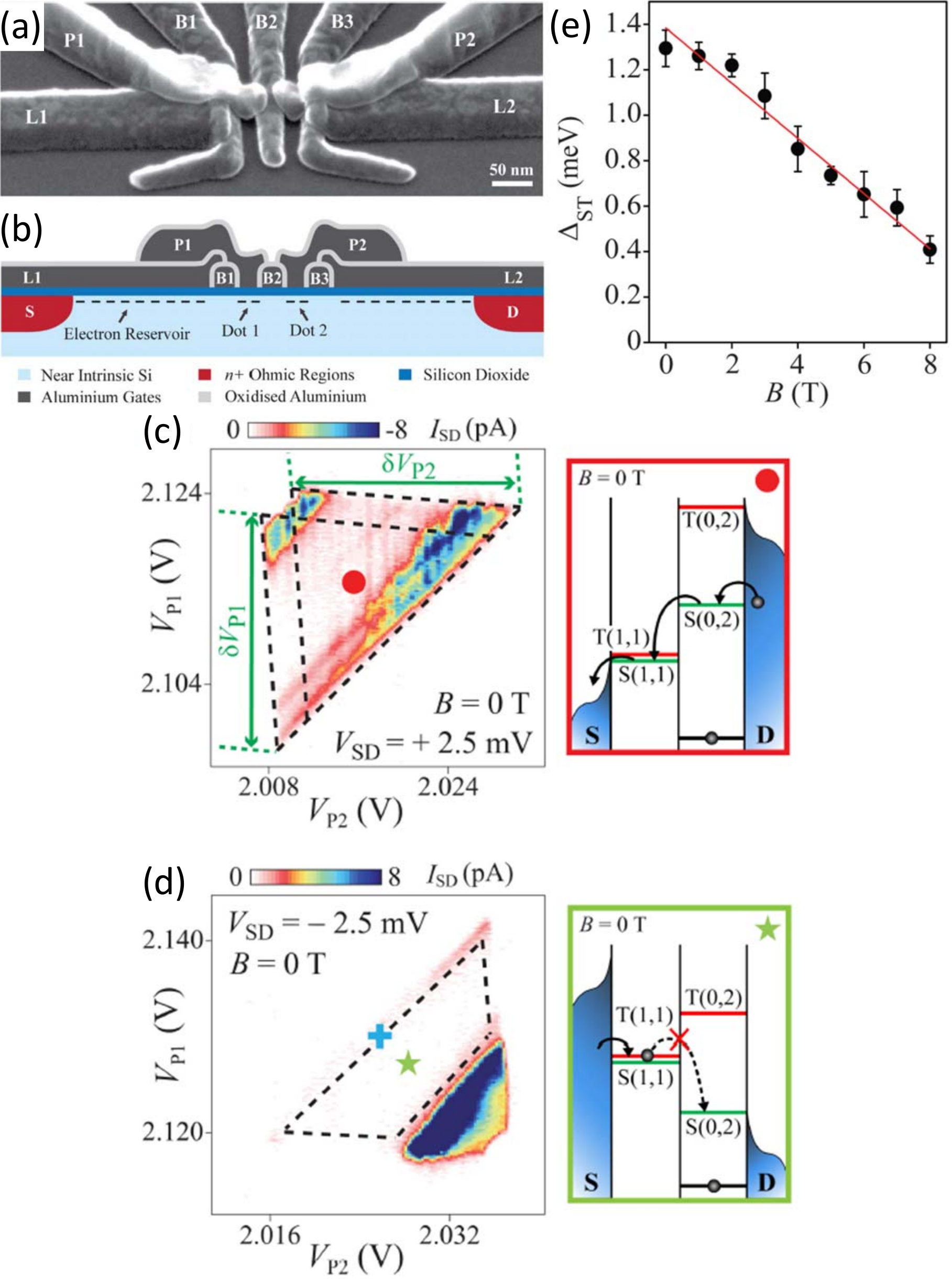}
\caption{\label{doubledots_fig5}  (Color online) \textbf{Pauli spin blockade in a silicon MOS double quantum dot.} (a) SEM image and (b) cross-sectional schematic of the Si MOS device. Gates L1 and L2 induce electron reservoirs at the Si/SiO$_2$ interface, while barrier gates B1-B3 define the double dot potential. Plunger gates P1 and P2 control the occupancy of each dot. (c,d) Current $I_{SD}$ as a function of $V_{P1}$ and $V_{P2}$ for $B=0$ T. (c) For $V_{SD}=+2.5$ mV, the ground state and excited states of a full bias triangle are observed. The current flows freely at the S(0,2)--S(1,1) transition, as illustrated in the box marked by the dot. (b) The same configuration at $V_{SD}=-2.5$ mV. Here the current between the singlet and triplet states is fully suppressed by spin blockade (box marked by star). (e) The measured singlet-triplet splitting $\Delta_{ST}$, plotted as a function of magnetic field $B$. Figure from \textcite{laiOther11}.}
\end{figure}

\textcite{laiOther11} demonstrated Pauli blockade in a Si MOS double quantum dot formed using an Al-Al$_2$O$_3$ multi-layer gate stack, similar to that discussed in Section \ref{planarmos}. Their device structure, shown in Fig.~\ref{doubledots_fig5}(a,b), incorporated three barrier gates (B1, B2, B3) and two `plunger'ù gates (P1, P2) which controlled the occupancy of the two dots. Pauli blockade occurs in one bias direction only, as depicted in Fig.~\ref{doubledots_fig5}. The phenomenon is revealed experimentally via the difference between the bias triangles for positive and negative source-drain bias $V_{SD}$, observed in the charge stability map as a function of the two plunger gate voltages.

For $V_{SD} > 0 $ (Fig.~\ref{doubledots_fig5}c) a pair of overlapping complete bias triangles are observed. Resonant transport through the ground and the excited states in the double dot occurs when the states within the dots are aligned, leading to current peaks that appear as lines parallel to the triangle base. The non-resonant background current in the triangle is due to inelastic tunneling. The non-zero current throughout the triangular region indicates that electrons from the reservoir can tunnel freely from the $S(0,2)$ singlet state to the $S(1,1)$ singlet state, as depicted in the cartoon (red box in Fig.~\ref{doubledots_fig5}). For $V_{SD} < 0 $  the current is suppressed in the region bounded by the dashed lines in Fig.~\ref{doubledots_fig5}(d). The suppression arises because the transition from $T(1,1)$ to $S(0,2)$ is forbidden by spin conservation during electron tunneling. Once the $T(1,1)$ triplet state is occupied, further current flow is blocked until the electron spin on one dot reverses its orientation via a relaxation process (green box in Fig.~\ref{doubledots_fig5}).

In Fig.~\ref{doubledots_fig5}(d) it is possible to discern some non-zero current at the bottom of the bias triangle. This `leakage current" in the spin-blockade regions has been identified as resulting from a spin-flip co-tunneling mechanism \cite{qassemiPRL09, laiOther11, coishPRB11}, where a spin-up electron from one of the reservoirs swaps with a spin-down electron in one of the dots. This effect has also been observed in a double dot formed from an etched silicon nanowire device \cite{yamahataOther11}.

Note that in this experiment \cite{laiOther11}, and also that performed in the nanowire device \cite{liuPRB08}, the electron occupancy in each dot was of order 10 or more, and so the labels (1,1) and (0,2) refer to the effective electron occupancy, whereas the true electron occupancy is $(m+m_0, n+n_0)$. Pauli blockade for two-electron singlet and triplet states therefore occurs when the total electron spin of each dot is zero in the $(m_0, n_0)$ state. More recently, \textcite{borselliAPL2011} have demonstrated a Si/SiGe double dot that exhibits Pauli blockade in the true (1,1)-(0,2) limit. The same group have since used this structure to demonstrate coherent oscillations between singlet and triplet states of the double dot system \cite{mauneNature12}, as discussed further in Section \ref{ST_readout_coherence}.

By applying a magnetic field $B$, it is possible to modify the singlet-triplet splitting $\Delta_{ST}$, defined as the energy difference between the blockaded ground state $S(0,2)$ and the excited state $T_-(0,2)$. In a magnetic field there are four accessible spin states: the singlet S; and three triplets $T_-$, $T_0$ and $T_+$, corresponding to $S_z$ = -1, 0, +1. \textcite{laiOther11} studied the singlet-triplet splitting by mapping the bias triangles in the spin blockade regime at increasing magnetic fields $B = 0-8 T$. They found that the splitting $\Delta_{ST}$ decreased linearly with increasing $B$ [Fig.~\ref{doubledots_fig5}(e)], as expected, since the triplet states split linearly by the Zeeman energy, $E_Z = \pm S_z g\mu_BB$, where $\mu_B$ is the Bohr magneton and $S_Z$ is -1, 0, +1. Here, a linear fit through $\Delta_{ST}(B)$ yielded a Land$\acute{\textrm{e}}$ g-factor of $2.1 \pm 0.2$, consistent with electrons in silicon.

\begin{figure}[h]
\includegraphics[width=0.48\textwidth]{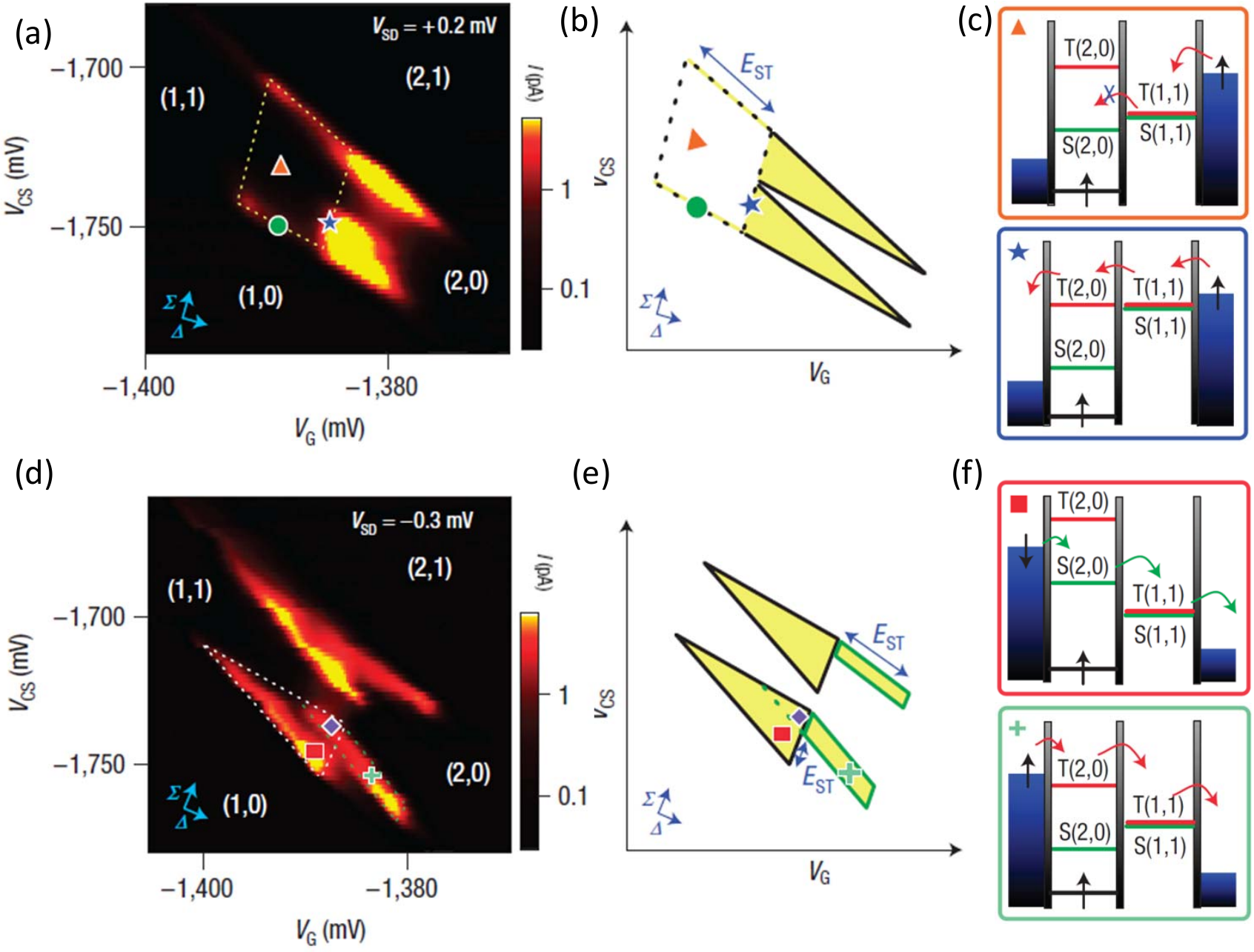}
\caption{\label{doubledots_fig7}   (Color online) \textbf{Spin blockade and lifetime enhanced transport in a Si/SiGe double quantum dot.} (a) Measured, and (b) schematic, charge stability map of current $I$ through the double dot, with a source-drain bias of $V_{SD}=+0.2$ mV. The dotted trapezoids in (a) and (b) mark the zero current regions due to spin blockade, as depicted in the schematics in (c). (d) Measured, and (e) schematic, charge stability map of current $I$ with a source-drain bias of $V_{SD}=-0.3$ mV. In this bias direction there is no blockade and current flows throughout the entire bias triangle, however, additional ìtailsî are observed due to lifetime enhanced transport, as depicted schematically in (f) and described in the text. Figure from \textcite{ShajiNPhys08}.}
\end{figure}

Spin transport in silicon double quantum dots can also be strongly affected by the relative values of the spin lifetimes and the various tunnel rates of the system. For example, it is sometimes possible to observe transport through a double dot in gate-space regions where current would normally be blockaded, a phenomenon that has been termed lifetime enhanced transport (LET), observed in a Si/SiGe quantum dot structure \cite{ShajiNPhys08}. Shaji and co-workers observed this effect in a device similar to that depicted in Fig.~\ref{chargesensing_fig1}, which was originally configured to operate as a single quantum dot. Under certain gate bias conditions a double quantum dot could be formed that exhibited a charge stability map consistent with occupancies down to the single electron level, in particular providing access to the $(1,1)$ and $(0,2)$ states. In the vicinity of the $(1,1) - (0,2)$ charge transition, a positive source-drain bias ($V_{SD}$ = +0.2 mV) led to bias triangles that exhibited a zero current region due to Pauli blockade (see Figs. \ref{doubledots_fig7}(a,b)), similar to that observed in Fig.~\ref{doubledots_fig5}.

Under negative bias ($V_{SD}$ = -0.3 mV) the entire bias triangle exhibited a transport current, as expected, but outside the bias triangles additional ``tail'' regions of non-zero current were also observed (see Figs. \ref{doubledots_fig7}(d,e)). These can be understood with reference to the green box in the schematic diagram in Fig.~\ref{doubledots_fig7}(f) and by considering the tunnel rates to the relevant states in the double dot~\cite{simmonsPRB10}. Typically, once the $S(2,0)$ state is loaded from the source reservoir (with rate $\Gamma_{LS}$), it would remained locked in position due to Coulomb blockade, and an electron could only pass to the drain via a co-tunneling process (with unloading rate $\Gamma_S$). If, however, the $S(2,0)$ loading rate is much slower than the unloading rate ($\Gamma_{LS} << \Gamma_S$) and the triplet-singlet relaxation rate $\Gamma_{TS}$ from $T(2,0)$ to $S(2,0)$ is also much slower than $\Gamma_S$, then the most likely transport pathway is via the $T(2,0)$ and $T(1,1)$ triplets, as shown in Fig.~\ref{doubledots_fig7}(f), leading to a non-zero current.  The upper triangle in the data and schematic diagram, known as the hole triangle, also shows LET behavior.  The direction of the ``tail'' in gate voltage space for the hole triangle is different than that for the electron triangle, a phenomenon that can be explained by a spin-flip co-tunneling process~\cite{KohPRL11}.  This type of spin-lifetime enhanced transport can occur in silicon quantum dots in part because of the very long spin lifetimes present, as discussed further in Section VI.A.

%% file: dopants_v11.tex
\subsection{Dopants in silicon transistors}\label{dopans_in_transistors}

\subsubsection{Early work: mesoscopic silicon transistors}\label{dopantsearlywork}

\begin{figure}[h!]
\centering
 \includegraphics[width=0.48\textwidth]{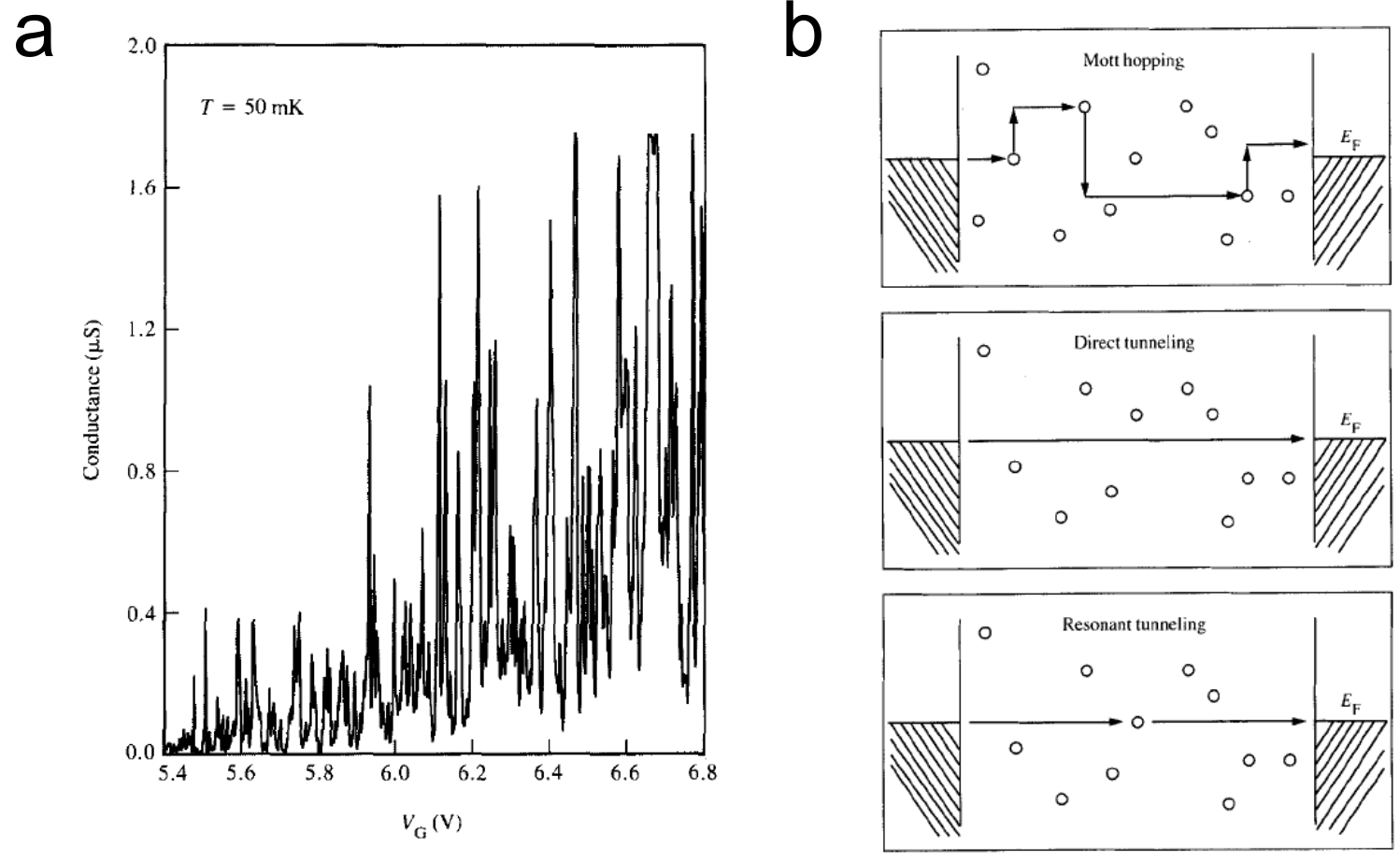}
 \caption{\textbf{Conductance in micron-scale silicon MOSFETs.} (a) Typical low-temperature conductance pattern of a 1980's generation MOSFET around the threshold regime. The strongly oscillating but chaotic pattern that appears at low temperature is associated with localized states in the channel region. (b) Schematic representation of the three major conduction mechanisms through the channel. From \onlinecite{FowlerIBM88}.} \label{dEarlyFowler}
\end{figure}

Low temperature transport experiments in silicon transistors have been used since the 1980's to perform spectroscopic measurements of dopants and defect states. Devices reached a sufficiently small length scale about 25 years ago to observe mesoscopic transport phenomenon, see the review by \textcite{FowlerIBM88}. In these devices the conductance, $G$, was found to fluctuate as a function of the gate voltage, $V_G$ around the threshold, as shown in Fig.~\ref{dEarlyFowler}a. The strongly fluctuating pattern, reproducible within the same cooling cycle, did not originate from electrical noise but from the presence of a finite number of dopant or defect states in the channel. The important length scales needed to understand this phenomenon are the localization length, $\zeta$ and the device dimensions (channel length $L$ and channel width $W$). As device dimensions approached $\sim$100 nm in size, comparable or less than the localization length, only a limited number of defect or dopant states contributed to the current in contrast to the much larger device dimensions before the 1980's where these fluctuations were always averaged out.

Three major conduction processes are known to contribute to the conductance of such small transistors, as shown in Fig.\,\ref{dEarlyFowler}b. Firstly, there is thermally activated hopping conduction, where the transported electrons hop via several dopant states from source to drain. This type of transport decreases exponentially with temperature $T$, i.e. $G \propto \exp(- \Delta E / k_\mathrm{B}T)$, where $\Delta E$ is the energy difference between the localized states of the dominant (most resistant) hop and $k_\mathrm{B}$ is the Boltzmann constant. The second process is direct tunneling, which scales with the barrier height $E_\mathrm{B}$ and $L$ roughly as $G \propto \exp -\sqrt{ 2m^* E_\mathrm{B} L / \hbar^2}$. Although typically irrelevant in the early 1980's, his conduction mechanism is playing an increasingly important role in today's nano-scale transistor operation. The last and increasingly important conduction mechanism is tunneling through a single defect. Due to the large number of conductance fluctuations, as depicted in Fig.\,\ref{dEarlyFowler}a, the identification of each state with a particular defect or dopants in the channel region was not possible. Dopants are not the only sources of disorder that cause localized states in MOSFETs as discussed in \textcite{PetersJAP98} and \textcite{SanquerPRB00}. However, these early measurements represent the first observation of mesoscopic physics in silicon MOSFET devices, and show how low temperature transport data offer a tool to electrically access dopant states in the channel region.

\subsubsection{Nano-scale transistors}\label{d_nano_transistors}

\begin{figure}[h!]
\centering
 \includegraphics[width=8cm]{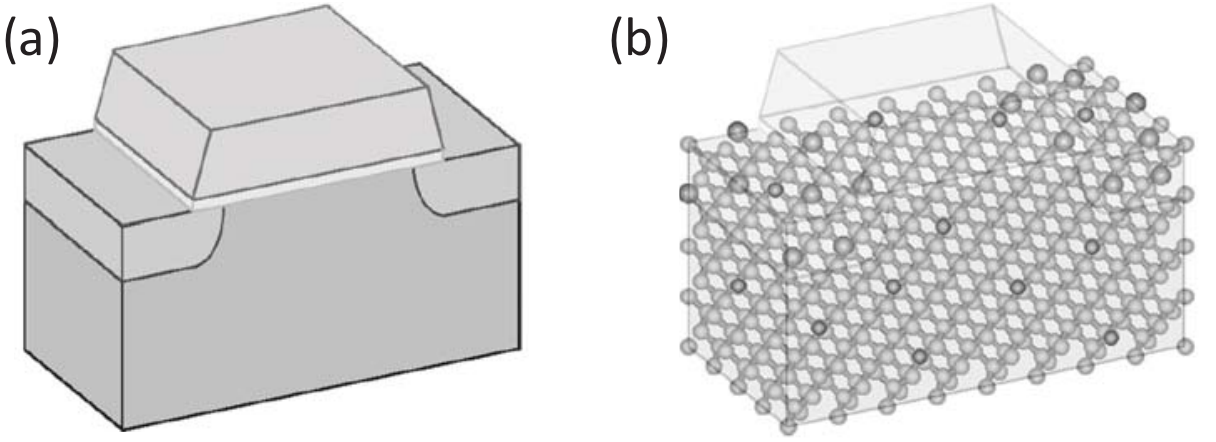}
 \caption{\textbf{The importance of discrete dopants in nano-scale MOSFETs.} (a) The transition from continuously ionized dopant charge and smooth boundaries and interfaces to (b) a 4-nm MOSFET where there are less than 10 Si atoms along the channel. From \onlinecite{AsenovIEEE03}.} \label{Asenov}
\end{figure}

Following this early work the purity of silicon MOSFETs steadily continued to improve with a concomitant decrease in device size, until the point where discrete impurities clearly started to show up in device transport properties as they reached the nano-scale \cite{MizunoIEEE94}. Here, fluctuations in the threshold voltage were observed, caused by the statistical fluctuation in the number of dopants in the channel as a result of the random Poisson distribution during doping. These results challenged the conventional understanding and modeling of micron-silicon devices where continuous ionized dopant charge with smooth boundaries and interfaces had previously been assumed. Now the granularity of the electric charge and the atomicity of matter introduced substantial variation in individual device characteristics, as shown in Fig.\,\ref{Asenov} \cite{AsenovIEEE03}. In particular the variation in number \textsl{and} position of the individual dopant atoms in the active region of MOSFETs were found to make each transistor microscopically different, introducing significant variations from device to device.

Recent advances in single ion implantation using a focussed ion beam source have shown that device to device fluctuations can be suppressed by ordering the dopants within the channel. \cite{ShinadaNature08}. The benefit of ordering the dopants was to create a homogeneous potential distribution in the channel resulting in the formation of a uniform current path. The uniformity in the channel lowers the voltage required to open the channel from the source to the drain allowing for an earlier turn on and reduction of the threshold voltage. This contrasted to devices with a random distribution of dopants where the non-homogeneous potential could lead to a block in the current path.

\subsection{Single dopant transistors}\label{d_single_dopant_transistors}

\subsubsection{The demand for single dopant architectures}\label{d_architectures}

In parallel with the increasingly important role of individual dopants in classical silicon CMOS devices has been the advent of their importance in quantum computation. In 1998 Bruce Kane introduced the concept of using the nuclear spins of individual donor atoms in doped silicon electronic devices as quantum bits or qubits \cite{KaneNature98}. Originally Kane envisioned that the quantum information could be stored in the state of the $^{31}$P nuclear spin and accessed by the electron-nuclear hyperfine coupling. Figure \ref{Kane1}(a) shows the basic two qubit unit cell of the Kane proposal based on an arrays of P donors beneath the silicon surface. The addition of a group V phosphorus donor to the silicon crystal results in electron states close in energy to the conduction band but weakly bound to the donor site at low temperatures. The electron has spin $S = 1/2$, whilst the host silicon has stable $I=0$ isotopes. It is important to isolate the qubits from any degrees of freedom that may lead to decoherence. Recent results have shown that $^{28}$Si can be isotopically refined to a level of $99.98\%$ \cite{TezukaPRB10} making it a good choice for the host material.

Any proposal for a quantum computer must meet the so-called DiVincenzo criteria \cite{DiVincenzoFDP00}, including the ability to operate on individual qubits, couple qubits into quantum logic gates, read-out the information encoded on the qubit, and to be free of environmental effects that destroy qubit entanglement. In Kane's architecture control of the qubit states is achieved by a combination of gates and globally applied a.c. magnetic field. The gates above the donors, labeled $A$ gates control the strength of the hyperfine interaction and therefore the resonance frequency of the nuclear spins beneath them. The $J$ gates are used to turn on and off the electron-mediated coupling between the nuclear spins. Finally the nuclear spins can be flipped by a resonant globally applied a.c. magnetic field. Read-out of the final spin state is achieved by spin-to-charge conversion and detection through spin-dependent tunneling to a doubly occupied donor state. This donor state is a singlet with a second electron binding energy of $1.7$ meV \cite{LarsenPRB92, LarsenPRB81}. Consequently by applying a differential voltage between the $A$ gates charge motion between the donors can only occur if the electrons are in a singlet state. This charge motion can then be detected using a sensitive electrometer nearby, such as a single-electron transistor (SET).

Subsequently a number of proposals for encoding and manipulating quantum information based on donor spin \cite{HollenbergPRB06, HillPRB05, stonehamJPCM03, skinnerPRL03, larionovNanotech01, vrijenPRA00, desousaPRA04} or charge degrees of freedom \cite{HollenbergPRB04} have been put forward. Electron spins bound to donor nuclei are particularly attractive since they have exceptionally long coherence times and relaxation times relative to the time scales for the control of the quantum state \cite{HillPRB05}. The electron spin coherence time of a P donor is $T_{2}> 60$ ms at $T=6.9$ K in isotopically pure $^{28}$Si \cite{TyryshkinPRB03}. These times are currently limited by the presence of $^{29}$Si which causes spectral diffusion due to the dipolar fluctuations of nuclear spins \cite{WitzelPRB05}. Dipolar fluctuations in the nuclear spins give rise to a temporally  random effective magnetic field at the localized electron spin leading to irreversible decoherence (i.e a T$_{2}$ process). Isotopic purification in silicon systems can in principle overcome this limiting process.

These developments have lead to the proposal of a complete 2D donor-based architecture, incorporating transport, for a truly scalable design in silicon \cite{HollenbergPRB06}. This design has considered the limitations and constraints posed by the sensitivity of the exchange interaction due to donor placement \cite{KoillerPRL02}, high gate densities required \cite{CopseyIEEE03}, spin read-out based on spin-charge conversion \cite{KaneNature98} and the communication bottleneck for linear nearest neighbor qubit arrays. A buried array of ionized donors provide pathways for coherent transport of electron spins for in-plane horizontal and vertical shuttling (dashed border sections) of qubit states into and out of the interaction zones. The overall gate density is low compared to the original Kane version since coherent spin transport is achieved adiabatically, lowering the barriers between donors in a well defined sequence to effect coherent transport by adiabatic passage (CTAP) without populating the intervening donors\cite{GreentreePRB04,rahmanPRB10}. Logic gates are carried out in interaction zones with the A and J gates for electron spin qubit control and these are distinct from the qubit storage regions. The design allows space for local B-field antennae and SET read-out devices. The introduction of coherent spin transport to donor quantum computing provides a means to consider scalable, fault tolerant architectures.

The use of single donor atoms in silicon as qubits has demanded tremendous advances in single atom fabrication and engineering. Donor separations of order $\sim10-20$ nm are required to ensure significant coupling between neighboring spins. Currently this is at the limit of what is technologically achievable. Indeed in the original critique of Kane's paper by \textcite{DiVincenzoNature98} he recognized that the fundamental and engineering obstacles to implementing the scheme were vast stating \textsl{``At the time no existing materials-preparation technology will place an array of individual phosphorus atoms at desired spots in the interior of a perfect crystal, let alone systems free from defects in the semiconductor and the overlying oxide layer''}. Despite these concerns there have been concentrated efforts internationally to realize a donor-based qubit architecture resulting in a plethora of experiments of transport in nano-scale doped quantum dots and donors. Significant to these results has been the different technologies developed to fabricate donor-based devices.

\subsubsection{Single dopants in MOS-based architectures}\label{dFET}

In this sub-section we discuss single dopant transport in ultra-scaled MOSFET structures based on randomly and deterministically doped devices. The \textcite{KaneNature98} proposal sparked the interest in single dopants and small MOSFETs were quickly identified as devices that should easily allow observation of single dopant transport. \textcite{TabePRL10} confirmed the impact of dopants on the potential landscape of a FET and discussed the evolution from many dopant to single dopant transport. In the newer generation of CMOS devices the issue of random device fluctuations is circumvented in a more straightforward approach, namely by the use of undoped channel FETs. However even in these newest generation of prototype FETs, fluctuations in device characteristics are still evident due to the presence of only a few down to a single \textsl{unintentional} dopant(s) in the channel region \cite{ColingeBook07}. By now, several groups have reported transport through a random, single dopant in a three terminal configuration \cite{SellierPRL06, CalvetPRB07, CalvetPRL07, PierreNNano10}, including microwave assisted transport \cite{PratiPRB09}. Recently this work has been extended to double gate structures and $^28$Si devices \cite{LoSST09,RochePRL12}. \onlinecite{RochePRL12} even demonstrated controlled sequential tunneling through two donors. In all these experiments, the number of electrons bound to the dopant atom could be controlled by the gate electrode. Such spectroscopic transport experiments reveal vital information on the orbital levels, the charging energy and the binding energy of the dopant atom and the spin configuration of the bound electron(s) \cite{LansbergenNphys08,LansbergenPRL11}. They thus form a powerful characterization tool in the development of single-dopant structures in parallel with the development of precision controlled single dopant devices \cite{fuechsleNatureN12}.

\begin{figure}[h!]
\centering
 \includegraphics[width=6.5cm]{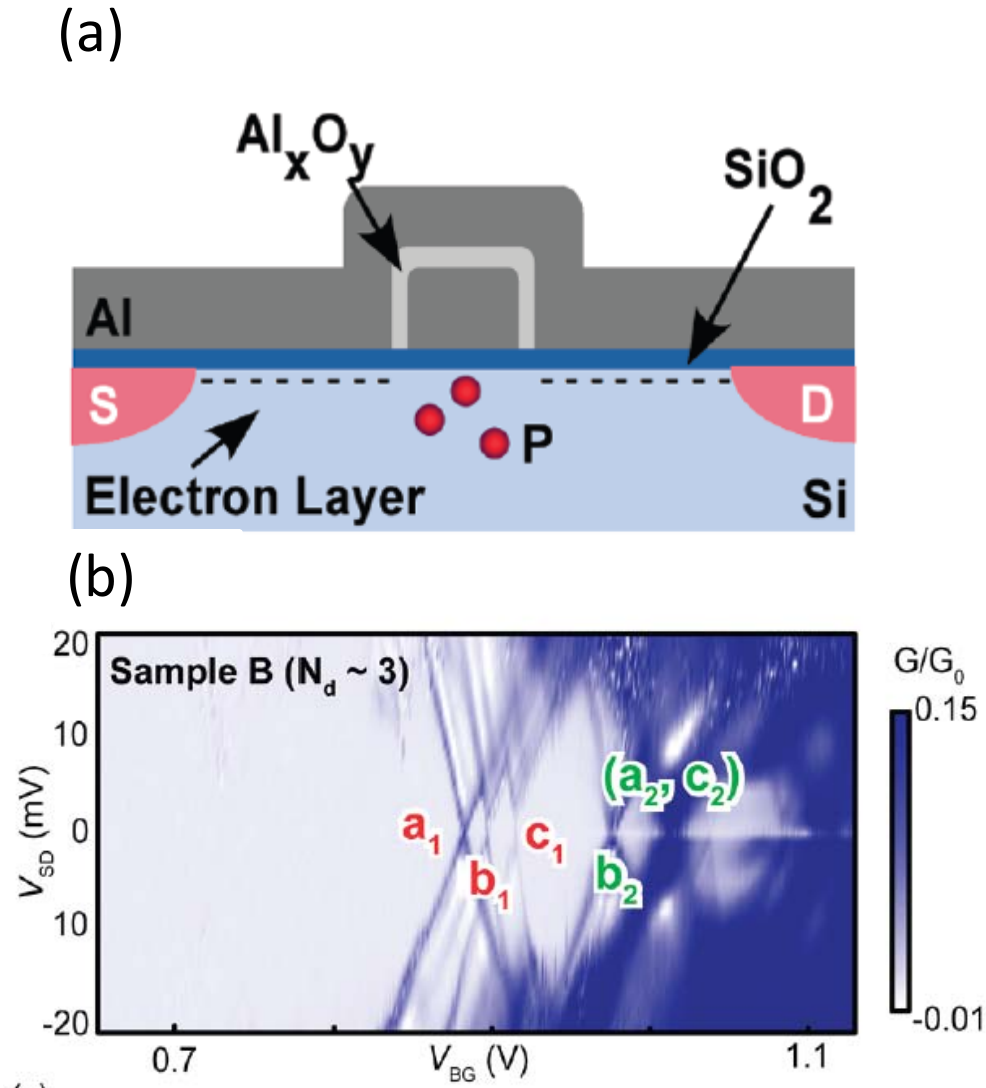}
 \caption{ (Color online) \textbf{Transport through dopants ion-implanted in a nanoFET.} (a) Schematic of a nanoFET where roughly 3 donors have been implanted into the $50 \times 30$ nm active area of the device. (b) The stability diagram showing the differential conductance as a function of the barrier gate and dc source-drain bias, highlighting the resonant tunneling peaks $a_1$, $b_1$ and $c_1$ of the three donors. From \onlinecite{TanNL10}.}\label{Tan}
\end{figure}

These experiments all relied on the in-diffusion of dopant atoms into nano-scale transistors from the source and drain regions. However several groups have concentrated on adapting the industry standard technique of ion implantation to implant single dopants into a silicon device in a controlled manner \cite{SchenkelJAP03,JamiesonAPL05,BatraAPL07,WeisJVSTB08,ShinadaNano08, seamonsAPL08,bielejecNanotech10}. There are three main limitations to applying the technique to scalable single atom architectures: the ability to register individual ion strikes, the overall spatial registration of the ion implant site and subsequent straggling of the ion due to the statistical nature of the stopping process. Once the dopants are implanted a rapid thermal anneal at high temperature is needed to repair the damage and activate the donors, causing diffusion and segregation of dopants \cite{ParkME04}. Despite these concerns silicon nanoscale transistors have been fabricated with a low density of local area implants where it has been possible to tune individual dopants into resonance and observe transport spectroscopy through a single dopant, such as a P donor in Fig.\,\ref{Tan} \cite{TanNL10}. More recently single dopant implantation has been demonstrated into the channel of a silicon nanoscale metal-oxide-semiconductor field-effect-transistor \cite{JohnsonAPL10, PratiNnano12}. In the work by \textcite{JohnsonAPL10} FinFET devices were fabricated using SOI with 20nm of Si on a 145nm thick buried oxide, giving nominal channel dimensions of $25 \times 70 \times 20$ nm. Using 14 keV there was a $57\%$ chance of a $P^{+}$ ion stopping within the channel region. The implant resulted in an increase in charge in the buried oxide, causing a shift in threshold voltage and an increase in series resistance consistent with the introduction of Frenkel pairs in the channel. In the more recent work of \textcite{PratiNnano12} donors were placed in a 1D array allowing the regime between single electron tunneling and Hubbard band formation due to inter-dot coupling to be investigated.

\begin{figure}[h!]
\centering
\includegraphics[width=6cm]{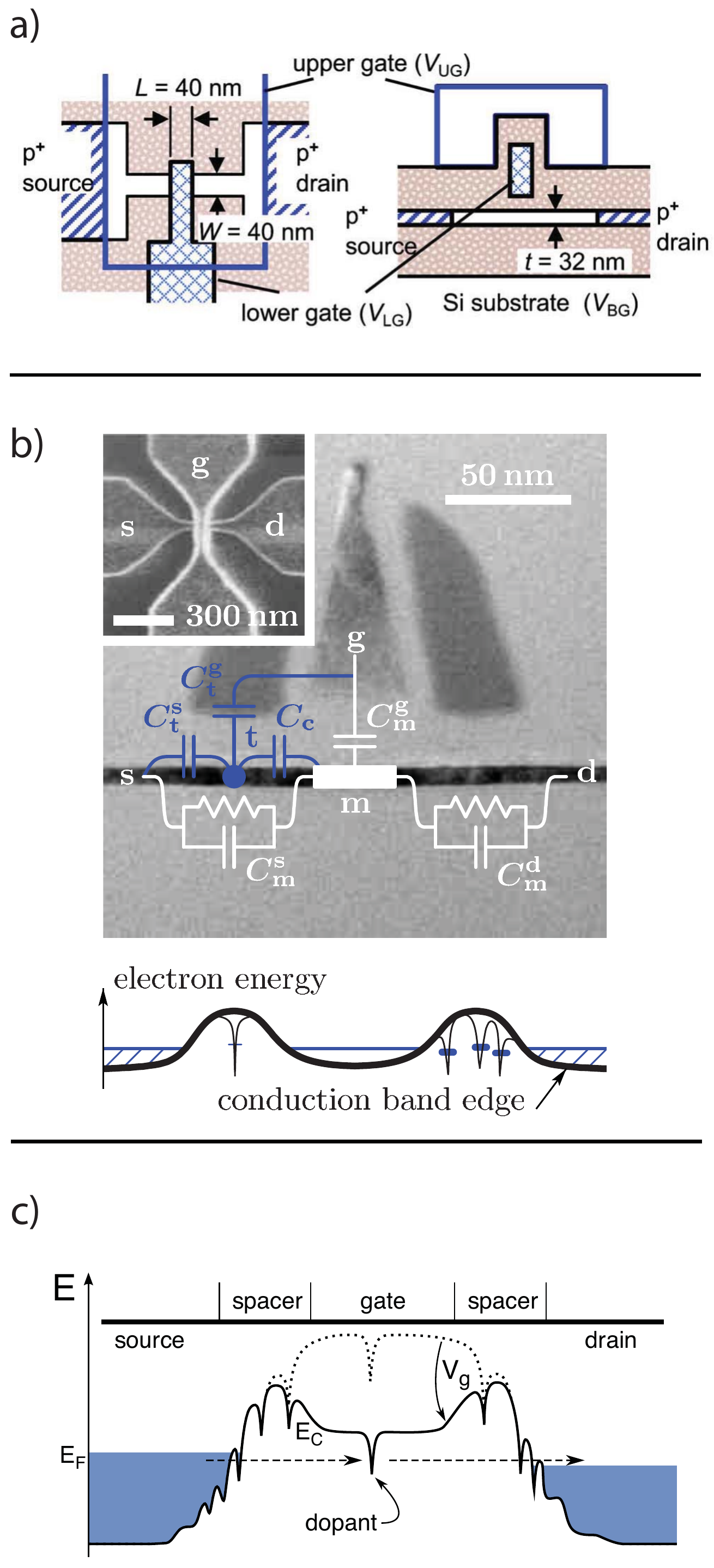}
\caption{ (Color online) \textbf{Three examples of device layouts that illustrate different transport regimes for the detection of a single dopant.} (a) Capacitive coupling to the channel which leads to a modification of the channel current due to the charge state of a dopant. (b) Tunneling through a dopant in the access region in series with transport through the channel. (c) Direct tunneling through a dopant in the channel in the sub-threshold regime. From (a) \onlinecite{OnoAPL07}, (b) \onlinecite{HofheinzEPJ06}, (c) \onlinecite{SellierPRL06}.} \label{dRegimes}
\end{figure}

\begin{figure}[h!]
\centering
\includegraphics[width=8cm]{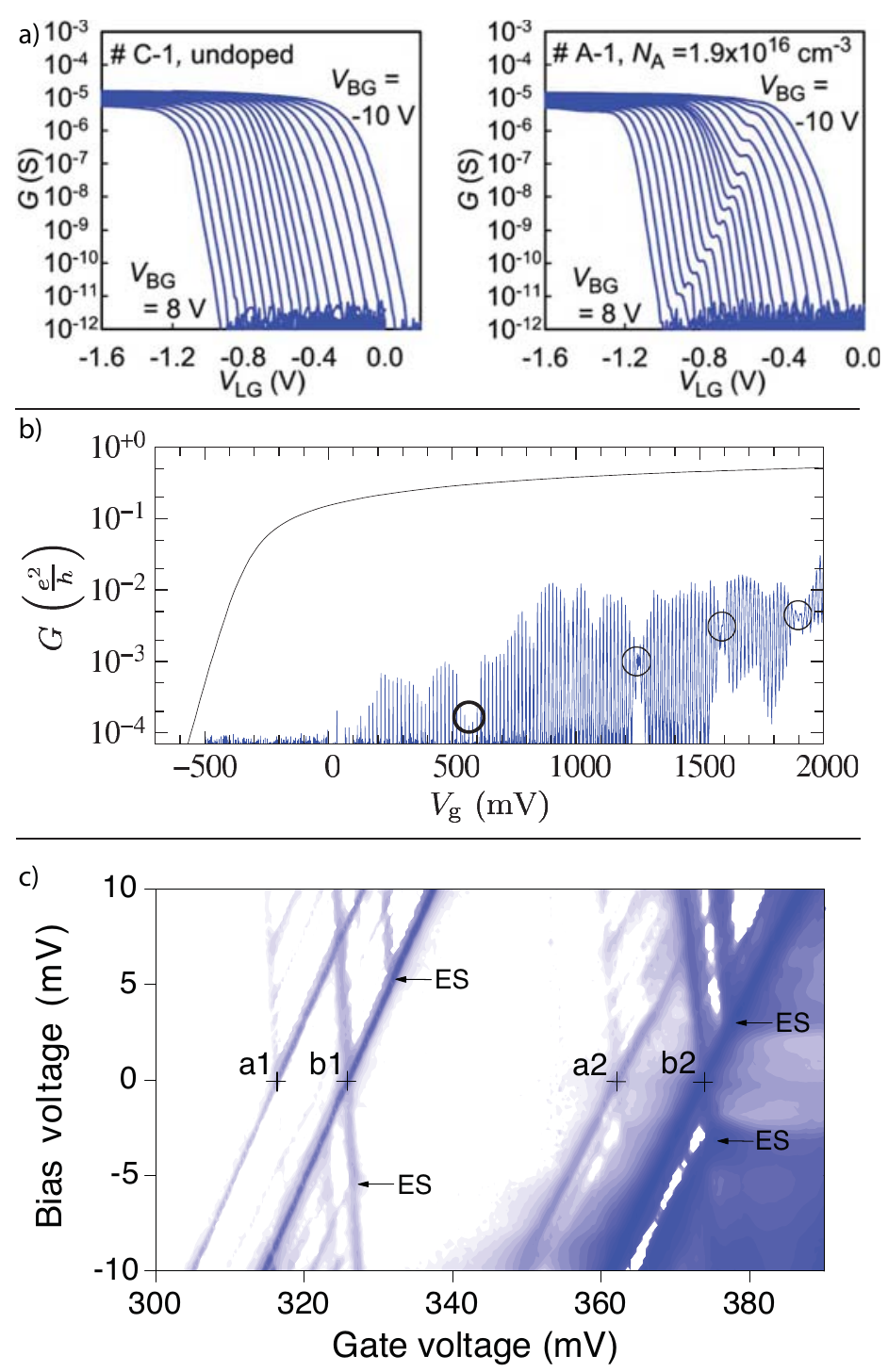}
\caption{ (Color online) \textbf{Three-dopant transport regimes in a transistor geometry.} (a) An example of the dopant detection regime based on the capacitive coupling the the channel for an undoped (left) and doped (right) double gate sample. The signature of a single acceptor charging event is evident in the doped sample. From \textcite{OnoAPL07}. (b) An example of the second regime where the dopant is in the barrier of the access region in series with a quantum dot, reproduced from \onlinecite{HofheinzEPJ06} with kind permission of The European Physical Journal. Top line represents the room temperature FET characteristics and the line below the low-temperature Coulomb peaks. (c) The third regime with direct transport through a dopant in the sub-threshold limit, taken from \cite{SellierPRL06}. }\label{dResults}
\end{figure}

There are \textsl{3 different regimes} for single dopant transport experiments, as shown in Fig.\,\ref{dResults}. \textsl{In the first} the channel current is influenced by the presence of a neutral or charged dopant. \textcite{OnoAPL07} identified a single acceptor that modified the current through a FET depending on it's charge state and refined this technique with a dual gate device to realize acceptor mapping \cite{OnoOther08,KhalafallaAPL09}. \textsl{In the second} regime there is direct transport through a dopant in the access region to a FET channel \cite{HofheinzEPJ06, CalvetPRL07, CalvetPRB07}. This allowed the study of electric field and strain effects on the acceptor \cite{CalvetPRL07, CalvetPRB07} and the Zeeman splitting \cite{HofheinzEPJ06} of the donor ground state. \textsl{The third regime}, representing the most direct way to access information about the properties of a dopant and its environment, is direct transport through the dopant states in the sub-threshold extreme of a transistor. Due to the progress in device scaling this was recently realized in ultra-scaled MOSFET devices by \cite{SellierPRL06, TanNL10, PierreNNano10, PratiPRB09}. All three transport regimes are illustrated in Fig.\,\ref{dRegimes} and will be discussed in more detail in the remaining part of the sub-section.

\begin{figure}[h!]
\centering
\includegraphics[width=6.5cm]{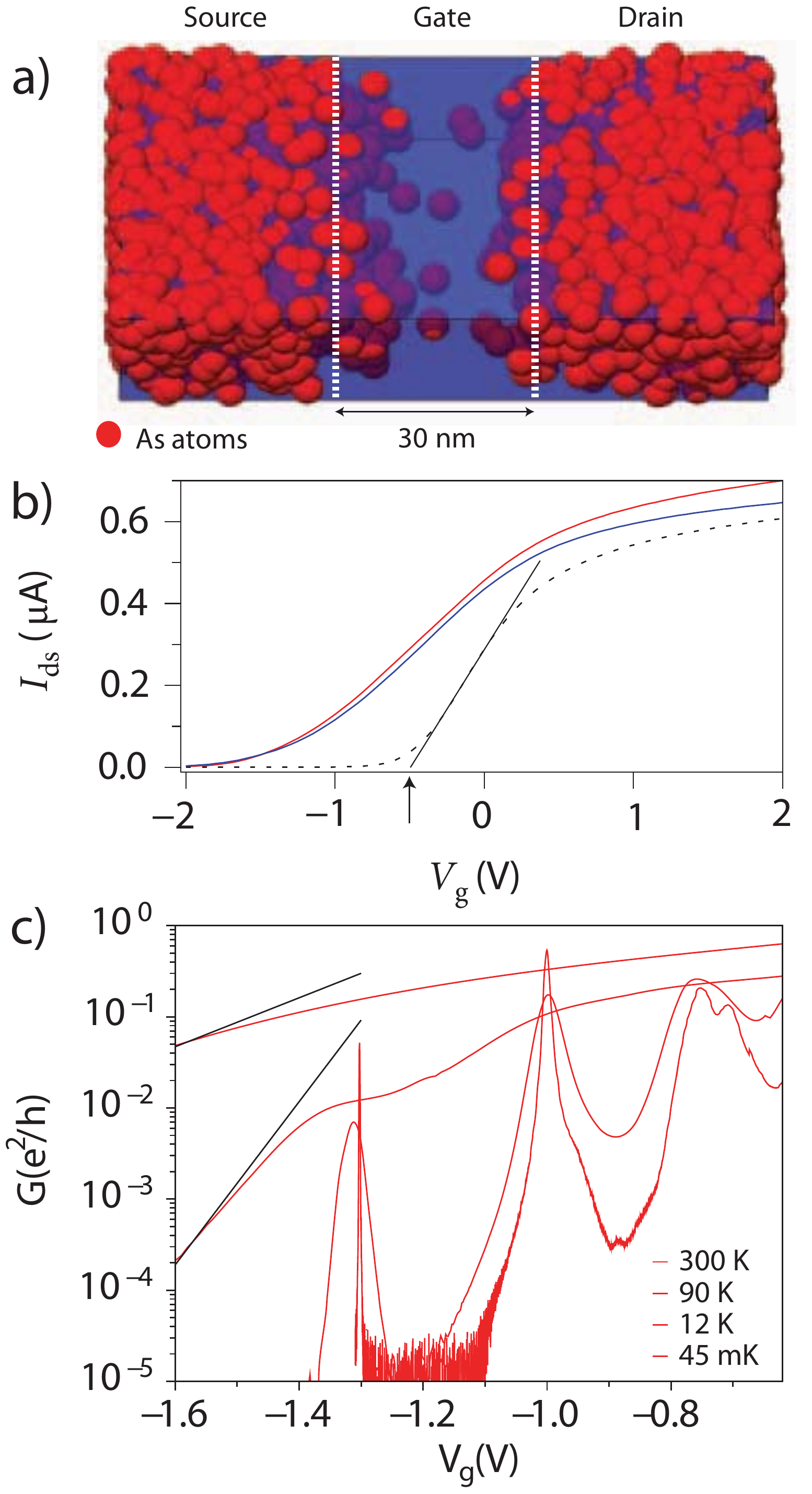}
\caption{ (Color online) \textbf{Direct tunneling through a dopant in a short-channel FET.} (a) Illustration of a Monte Carlo simulation of the doping profile in a 20nm channel where some dopants diffused into the channel region from the source and drain. (b) The dashed curve shows the current averaged over many devices where the black line indicates the threshold. Two devices show a drastically lower threshold linked to resonant transport at low temperature as indicated in (c) for the device with the lowest $V_\mathrm{th}$. From \onlinecite{PierreNNano10}. This data shows the clear connection between the low threshold of these devices at room temperature and the resonant transport at low temperature, both mediated by a single dopant.} \label{dPierre}
\end{figure}

\textcite{SellierPRL06} demonstrated transport through a donor in a nano-FET and identified excited states as well as the doubly occupied $D^-$ state confirmed by magneto transport (regime 3). The ionization energy observed in this work was consistent with As as expected from indiffusion from the source and drain region. In contrast to the ionization energy of the $D^0$ state, which was similar to bulk, the ionization energy of the $D^-$ state was enhanced. This reduction in the \textsl{charging energy}, i.e. an increase in $D^-$ ionization energy at an unchanged $D^0$ ionization energy, is discussed in comparison to several experiments in Sec.\,\ref{dCHAR}. \textcite{LansbergenNphys08} analyzed the orbital spectrum based on nonlinear transport and studied the impact of the electric field and gate interface on the donor state. They analyzed the data in comparison to a large scale atomistic model and found good agreement even though the spectrum is strongly altered from the bulk. \textcite{TanNL10} demonstrated transport spectroscopy on a transistor that was implanted with 3 donors based on a timed exposure, thus the first experiment on an intentionally placed group of donors. The intentional placement is confirmed by a sample that was not implanted and did not show any resonances. As expected the donor peaks show a paramagnetic Zeeman shift of the $D^0$ state and the opposite for the $D^-$ state. Spectroscopic measurements combined with device based modeling led to a detailed understanding of the valley-orbit coupling of donor or dot states \cite{RahmanPRB11vo} and the two electron state of a donor \cite{LansbergenPRL11, RahmanPRB11} which are discussed in detail in the next section. Recently, \textcite{PierreNNano10} linked low temperature resonant transport through donor states to the room temperature performance of a MOSFET. They performed a statistical analysis of the threshold voltage of ultra-short channel FinFETs (less than 20\,nm). Transistors with a threshold voltage far below the average display resonant transport at low temperature due to a donor in the middle of the channel, as shown in Fig.\,\ref{dPierre}. This direct link between the room temperature variability and the low temperature spectroscopy represents an important contribution to device engineering \cite{WacquezVLSI10}.

\textcite{HofheinzEPJ06} reported transport through a donor in the access region of a small MOSFET (regime 2). The access region is part of the channel that has a weaker coupling to the gate which leads to a barrier in the band structure, as shown in Fig.\,\ref{dRegimes}b. They observed sequential transport through the donor and through a localized state between the two barriers of the access region which has a much lower charging energy. Magnetotransport revealed a Zeeman shift of the resonance consistent with a paramagnetic trap. The complex interaction between a single electron transistor and a dopant was only recently explained in detail \cite{GolovachPRB11}. Calvet \textsl{et al.} investigated acceptors in the barrier of a Schottky FET. They observed a Zeeman shift of the acceptor and analyzed the impact of the electric field on the acceptor \cite{CalvetPRL07} and the effect of local strain \cite{CalvetPRB07}.

As mentioned before, single dopants can not only be detected by passing current through them but also by their electrostatic coupling to free electrons in the channel (regime 1). \textcite{KhalafallaAPL09} developed earlier single gate experiments \cite{OnoAPL07} to a multi-gate configuration. Two overlapping top-gates allowed the modulation of the channel potential profile. This made it possible to define the position of a donor along the length of the channel which led to the observation that an acceptor close to the source has the largest impact on transport. This device geometry has been further developed to allow for charge pumping based on a small number of dopants which is attractive due to the large charging energy of the dopant and its potential for high accuracy as current standard \cite{LansbergenNL12}. A double gate study of n-type channels revealed a clear difference between doped and undoped devices. In undoped devices the back gate simply shifts the threshold voltage of the device. Devices of the same geometry with $2 \times 10^{18}$P\,cm$^{-3}$ show many sub-threshold resonances with a large charging energy and a different coupling to the top gate. Furthermore, coupling between some of these is observed. These observations close the loop between the randomly- and deterministically-doped devices.

Until recently single donor transport was focused on sequential tunneling to probe the spectrum of a donor. Recently, several coherent transport experiments with single donor and double donor systems have been reported. Strong coupling to the contacts leads to the formation of a Kondo ground state where the parametric donor state is screened leading to transport in the traditional Coulomb blocked one electron region \cite{LansbergenNL10}. The valleys play a key role in Si Kondo physics as theoretically predicted \cite{ShiauPRB07,ShiauPRB07b} and experimentally confirmed for donors by the presence of spin and orbital Kondo and their combination \cite{LansbergenNL10,TettamanziPRL12}. Furthermore, quantum interference between two tunneling paths has been discussed in the context of a Fano resonance \cite{VerduijnAPL10, CalvetPRB11}. Dopants present an ideal platform to investigate the crossover from scale-less mesoscopic effects to atomistic transport.

\subsubsection{Single dopants in crystalline silicon}\label{d_dopant_STM}

An alternative technology has been developed to place dopants in silicon with atomic precision using a scanning probe microscope. Ever since its invention in 1981 by Binnig and Rohrer, the scanning tunnelling microscope (STM) has gained international recognition by not only its capability to image surfaces with unprecedented resolution but also by its potential to modify and pattern crystalline surfaces at the atomic scale. Amongst the most notable examples are the formation of the letters IBM with individual xenon atoms on a nickel surface by Eigler's group \cite{EiglerN90}.

The translation of this technology to manipulate atoms in silicon was, however, not simple due to the strong, covalent nature of silicon bonds. To position atoms in silicon it was necessary to use a lithographic process, analogous to conventional optical/e-beam lithography. Here a monolayer of hydrogen resist is patterned using the tip of the STM to create an atomic-scale template. In the early 1990s \textcite{LyoAIP90} and \textcite{LydingAPL94} proposed the use of such a template to create a pattern of highly reactive dangling bonds sites on the silicon surface which could subsequently be functionalized with various atomic and molecular species.

The process of STM hydrogen lithography has since been adapted to realize a complete fabrication strategy for atomic-scale silicon device fabrication \cite{RuessNL04,RuessN05,SimmonsMS05}. Here dopants are placed in the silicon crystal with atomic precision laterally using scanning probe techniques and atomic precision vertically with molecular beam epitaxy (MBE). The important feature of these devices is that, in contrast to modern CMOS devices and almost all quantum semiconductor devices, which use many materials and have heterogeneous interfaces, STM-patterned devices are formed in single crystal silicon. Confinement of electrons is thus achieved by atomically abrupt changes in the density of dopant atoms within the silicon crystal. The doped regions can have very high planar electron densities in the range $n_{2D}\approx 2.5 \times 10^{14}$ cm$^{-2}$. In three dimensions this density corresponds to a value $\approx 10^{21}$ cm$^{-2}$, three orders of magnitude above the Mott metal-insulator transition. At these high carrier densities one in every four silicon atoms is substituted with a phosphorus atom, so that the average separation of phosphorus atoms is $<1$ nm, much smaller than the single dopant Bohr radii ($a_B \approx 2.5$ nm). As a consequence the doped regions are highly conducting and behave like a disordered metal. Electron transport has been studied in these highly doped single crystal donor-based quantum wires \cite{RuessS07,RuessAPL08,RuessPRB07} and recently this technique has been used to realize conducting wires in silicon with Ohmic behavior down to the atomic-scale \cite{weberScience12}.

\begin{figure}[h!]
\centering
\includegraphics[width=7.5cm]{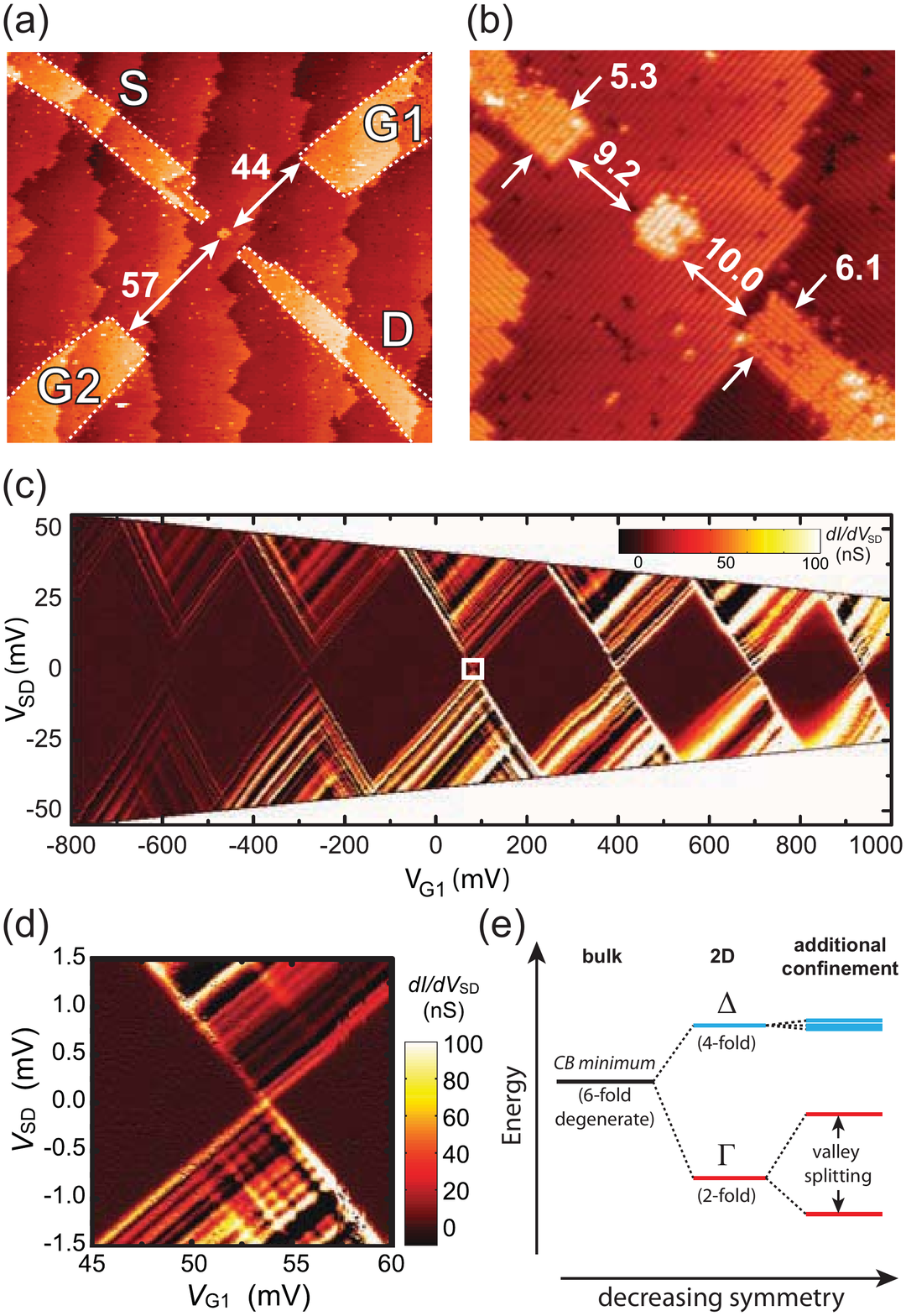}
\caption{ (Color online) \textbf{Few-electron quantum dot.} (a) An STM image of the central device region of a few-electron single-crystal quantum dot acquired during hydrogen lithography, showing a four terminal device with source (S), drain (D) and two in-plane gates (G1,G2). The bright regions correspond to areas where phosphorus donors will be incorporated. (b) A close-up showing the central quantum dot containing $6\pm3$ donors. (c) Stability diagram showing the conductance, $dI/dV_{SD}$ through the dot as a function of gate voltage, $V_{G}$ and bias voltage $V_{SD}$. (d) A close-up of the transition (white square in (c)) reveals a high density of conduction resonances with an average energy spacing of $\approx 100 \mu eV$. (e) The 6-fold degeneracy of the conduction band minima of bulk silicon is lifted by confining the electrons vertically to two dimensions and is then split again by abrupt, lateral confinement. From \onlinecite{FuechsleNnano10}.}\label{7atom}
\end{figure}

It is also possible to pattern more complex devices, such atomically abrupt, epitaxial quantum dots with 1D source and drain leads \cite{FuhrerNL09, FuechsleNnano10}. These studies have allowed the impact of vertical $\it{and}$ lateral confinement on silicon quantum dots to be investigated. In such abruptly confined quantum dots very small energy level splittings of $\approx 100 \mu eV$ have been observed in electron transport and attributed to transport through the valley states of a few-electron quantum dot, as shown in Fig.\,\ref{7atom} \cite{FuechsleNnano10}. STM images of the device in Fig.\,\ref{7atom}(a,b) show the central region of the device into which $6\pm 3$ P atoms are laterally confined with the STM. These dopants are also strongly vertically confined by low temperature silicon molecular beam epitaxy. The corresponding stability diagram shown in Fig.\,\ref{7atom}(c,d) reveals spacing in the energy spectrum of order $\approx 100\mu$eV. This very small energy level splitting was surprising given the ultra-small size ($\sim 4$ nm$^{2}$) of the quantum dot. However it is well known that strong lateral and vertical confinement breaks the degeneracy of silicon valley states.

This is illustrated in Fig.\,\ref{7atom}(e) where electrons strongly confined in a two-dimensional plane result in splitting of the 6 bulk valleys into 4 degenerate $\Delta$-pockets as well as two $\Gamma$-pockets at $k$ = 0. The remaining degeneracies can be broken in the presence of sharp lateral or vertical confinement. In these highly doped $\delta$-layers strong, abrupt quantum confinement in z-direction splits the degeneracy of the out-of-plane $\Gamma$-bands to give the lower energy $\Gamma_{1}$ and $\Gamma_{2}$ bands. The four $\Delta$-valleys, two each in the x and y directions, are usually degenerate in 2D devices. However, since the quantum dot device is also confined laterally on the nm-scale valley splitting of these states is also observed. It is this valley splitting that gives rise to the $\approx 100\mu$eV energy level separation observed experimentally in STM-patterned few-electron quantum dots. This contrasts to studies of other few-electron quantum dot systems, where it is the large size of the dot itself that gives rise to $\approx 100\mu$eV \cite{BeenakkerPRB91}.

The source and drain leads to these quantum dots are not 2D reservoirs but 1D leads. This results in the formation of resonant tunneling features due to the presence of 1D subbands. The energy separation of these subbands depends on the width of the 1D leads and for this device was found to be $\sim$10 meV. Such 1D states have also been observed in other silicon-based quantum dots \cite{limAPL09-2,mottonenPRB10} and a recent review discusses how to distinguish these in electron transport \cite{EscottN10}.

\begin{figure}[h!]
\centering
\includegraphics[width=8cm]{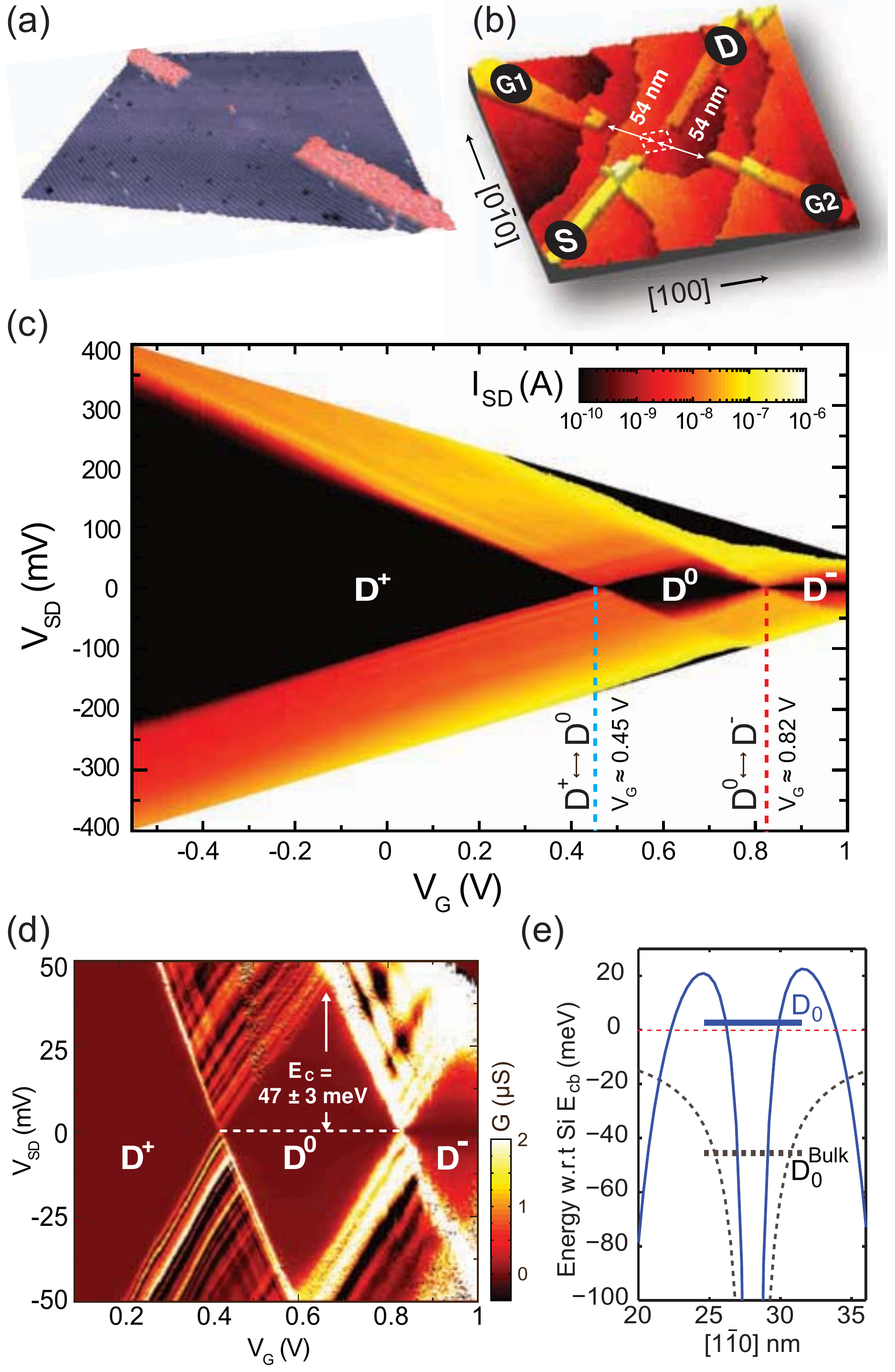}
\caption{ (Color online) \textbf{A single-atom transistor.} (a) 3D perspective STM image of a hydrogenated silicon surface. Phosphorus will incorporate in the red shaded regions selectively desorbed with an STM tip to form electrical leads to a single phosphorus atom patterned precisely in the centre. (b) the source (S), drain (D) and two gate leads (G1, G2) to the central donor, which is incorporated into the dotted square region. (c) The electronic spectrum of the single-atom transistor, showing the drain current $I_{SD}$ as a function of source-drain bias $V_{SD}$ and gate voltage $V_{G}$ applied to both gates. (d) The differential conductance $dI_{SD}/dV_{SD}$ as a function of $V_{SD}$ and $V_{G}$ in the region of the $D^{0}$ diamond shown in (c). (e) A comparison of the potential profile between the source and drain electrodes in this device (straight line) to an isolated bulk phosphorus donor (dashed line), where the $D^{0}$ state resides 45.6 meV below $E_{\textrm{cb}}$. In contrast, the $D^{0}$ state in the single-atom transistor resides closer to the top of the potential barrier. From \onlinecite{fuechsleNatureN12}.}\label{Single_atom}
\end{figure}

Using such a technology it is also possible to pattern individual dopants in silicon with atomic precision\cite{SchofieldPRL03,WilsonPRL04} to realize single dopant atom transistors, as shown in Fig.\,\ref{Single_atom}(a) \cite{fuechsleNatureN12}. The critical features of this device are that the dopant, the source-drain leads and the control gates are crystalline and all exist within one plane of the silicon crystal, as shown in Fig.\,\ref{Single_atom}(b). The encapsulation of this device in epitaxial silicon removes the confined dopant states away from the influence of surfaces and interfaces. However transport devices by definition also contain electrodes, and these electrodes are known to have profound effects on the energetics of the single dopant atom. In these epitaxial architectures the electrostatic potential at the dopant could be tuned using two in-plane gates G1 and G2 patterned either side of the transport channel defined by the S and D leads.

Fig.\,\ref{Single_atom}(c) presents the measured stability diagram of the single donor, in which the three charge states of the donor can easily be identified: the ionized $D^{+}$ state, the neutral $D^{0}$ state and the negatively charged $D^{-}$ state. The diamond below $V_G \approx 0.45 V$ does not close, as expected for the ionized $D^{+}$ state, because a donor cannot lose more than its one valence electrons. The conductance remains high (on the order of microsiemens) down to the lower end of the gating range, making the possibility of additional charge transitions unlikely. The $D^{+}\rightarrow D^{0}$ charge transition occurred reproducibly at $V_{G}\sim 0.45\pm 0.03V$, as shown in Fig.\,\ref{Single_atom}(d) for multiple cool-downs and is attributed to the high stability of the device and the inherent influence of the nearby electrodes on the position of the donor eigenstates relative to the Fermi level of the leads.

To understand quantitatively how the nearby transport electrodes affected the electronic properties of the donor, the electrostatic potential landscape of the innermost part of the device was calculated, treating the heavily doped gate regions in a self-consistent atomistic approach using a Thomas-Fermi approximation. Having established the electrostatic potential of the device, the donor electronic states were then calculated using a tight-binding approach \cite{LansbergenNphys08}. The position of the resulting one-electron ground state $D^{0}$ for the solitary phosphorus dopant is depicted in Fig.\,\ref{Single_atom}(e) (blue line). As expected, due to the electrostatic environment, the energy levels of the device are raised significantly from the bulk case (dashed grey line), where the unperturbed Coulombic donor potential asymptotically approaches the silicon conduction band minimum $E_{cb}$ (red dashed line) and $D^{0}$ has a binding energy of $E_{B}\approx -45.6$ meV. In contrast, $D^{0}$ in the effective donor potential of the single donor transport device resides much closer to the top of the barrier (solid line) along the S-D transport direction. Despite this, the charging energy $E_{C}$ could be extracted from the transport data and was found to be $47\pm 3$ meV, remarkably similar to the value expected for isolated phosphorus donors in bulk silicon ($\approx 44$ meV)\cite{RamdasRPP1981}.

These results are in sharp contrast to previous experiments on single dopant in silicon transport devices, which have reported charging energies that significantly differ from the bulk case \cite{LansbergenNphys08, PierreNNano10, RahmanPRB11}. There, the difference was attributed either to screening effects resulting from strong capacitive coupling to a nearby gate \cite{LansbergenNphys08} or strong electric fields \cite{RahmanPRB11}, or to an enhanced donor ionization energy in the proximity of a dielectric interface \cite{PierreNNano10}. Importantly, these effects are small for a single phosphorus dopant, which is symmetrically positioned between two gates, encapsulated deep within an epitaxial silicon environment.

\subsection{Discussion}\label{d_discussion}

A dopant in a semiconductor represents the ultimate limit of a quantum dot. In contrast to a quantum dot the confinement potential is given by the three dimensional Coulomb potential of the dopant ion (see section \ref{confinement}) and not by external gates. In quantum dots the orbital energy is small and mainly probed as the energy difference between the triplet and single sates in the two electron problem. The dopant has hydrogenic level spectrum with splittings in excess of 10meV. The valleys in Si lead to interesting corrections due to the restricted momentum space of these states. Again in strong contrast to a quantum dot, the confinement potential of a dopant is strongly altered by the amount of charge on the dopant. A shallow impurity can only bind two electrons where the 2nd charging energy for the 2nd electron is almost the ionization energy of the first, i.e. $H^-$ like the two electron state is very close in energy to the continuum. Here we compare the physical properties of the dopant confined states for the different devices. The orbital spectrum of a gated donor is illustrated followed by a discussion of the charging energy between the one and two electron state. Finally, we look at the interaction between the donor and the leads and the interaction between two donors.

\subsubsection{Orbital structure of a dopant in a nanostructure}\label{dORB}

Section \ref{physicsdopants} discussed the spectrum and orbital structure of a bulk dopant. A dopant in a nanostructure possesses not necessarily these properties as already calculated by \textcite{MacmillenPRB84}. The environment in a nanodevice has a large impact on the orbital spectrum of a donor. An electric field will lift degeneracies \cite{SmitPRB04, FriesenPRL05} and a triangular well, e.g.\ due to a gate, lowers the excited states of the dopant due to the interaction with the interface well \cite{martinsPRB04, CalderonPRB07}. These theoretical predictions have been experimentally confirmed in the interface \cite{LansbergenNphys08} as well as the bulk regime \cite{fuechsleNatureN12}. Critically important for the ability to model the devices and obtain metrology data with respect to impurity type and depth, was the ability to compute the excited state spectra as well as the ground state spectra in NEMO3D, see section III of the supplemental material. The effective mass models offer key physical insight into the problem and go hand in hand with the tight-binding work which generates accurate predictions to interpret the experimental data.

\begin{figure}[h!]
\centering
\includegraphics[width=8cm]{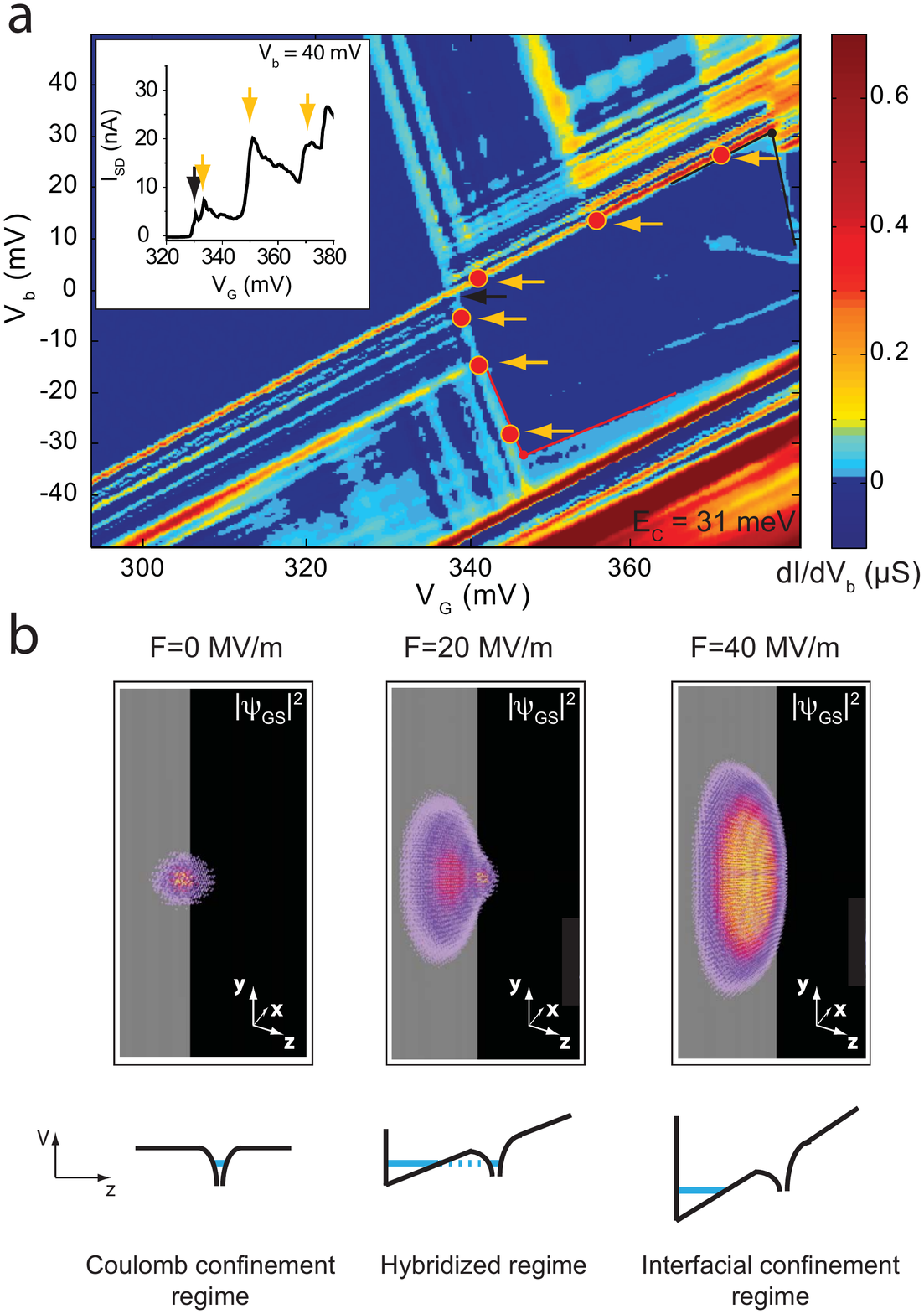}
\caption{ (Color online) \textbf{Excited-state spectroscopy of single gated donors.} (a) Differential conductance of a dopant in a FET. Excited states are indicated by the dots and arrows. Inset in (a) shows current $I_{\textrm{SD}}$ as a function of gate voltage at V$_{\mathrm{b}}$ = 40\,mV where each plateau indicates addition of a quantum channel due to an orbital. (b) Simulations of the gated donors eigenstates: wave function density of the $D^{0}$-ground state ($|\Psi_\mathrm{GS}|^2$) located 4.3\,nm below the interface in three different electric field regimes: Coulomb confinement regime, 0\,MV\,m$^{-1}$ (left), hybridized regime, 20\,MV\,m$^{-1}$ (middle) and interfacial confinement regime, 40\,MV\,m$^{-1}$ (right). The grey plane indicates the Si/SiO$_2$ interface. From \onlinecite{LansbergenNphys08}.} \label{figORB}
\end{figure}

The environment, i.e. the leads, an interface to a gate, or an electrical field, can drastically alter the orbitals of a dopant in comparison to the unperturbed bulk condition. \textcite{LansbergenNphys08} measured the energy spectrum of single donors, located in the channel of FinFETs by transport spectroscopy, as shown in Fig.\,\ref{figORB}(a). They were not bulk like but agreed well with a multimillion atom simulations of the complete system. In conjunction with the data, the theoretical analysis allowed the authors to identify the species of the donors (As) and furthermore provided an explicit determination of the degree of gate-controlled quantum confinement in each device. Figure\,\ref{figORB}(b) shows the 3 confinement regimes that can be distinguished: Coulomb, hybridized, and interfacial confinement with the charge density and schematic potential landscape for these three regimes. At low electric fields the electron is located at the donor site and its ground state corresponds to a donor in bulk (thus full lattice symmetry). At high electric fields the electron is pulled inside the triangular potential well at the interface reducing the symmetry of the system. The electron is still localized near the donor site in the lateral directions though, in correspondence with the results of \textcite{CalderonPRB07}. At the crossover between these regimes, the electron is delocalized over the donor- and well-sites.

Donors in devices fabricated with the STM, as discussed in sub-section \ref{d_dopant_STM}, exhibit a more bulk like orbital spectrum \cite{fuechsleNatureN12}. This is due to the fact that in these devices the donor is far away from a dielectric interface and the gate does not create large fields in comparison with MOSFET structures \cite{LansbergenNphys08}. \textcite{fuechsleNatureN12} also showed that the charging energy of a dopant in an STM fabricated device is comparable to the bulk value. This is to be expected for an environment that is close to bulk besides the source and drain electrodes with a cross-section of only a few square nanometers.

\subsubsection{Charging energy of a dopant in a nanostructure}\label{dCHAR}
In the constant-interaction model \cite{BeenakkerPRB91} the charging energy of a Coulomb island is independent of the number of electrons $N$ localized on the charge island. This assumption is valid as long as the confinement potential is not affected by $N$ which is not at all the case for isolated donors. For shallow donors, only a single charge transition ($N=1 \rightarrow N=2$) plays a role since it is not possible to bind a 3rd electron. The addition of a single electron to an ionized donor site will screen the positive nucleus and thus strongly alter the confinement potential for an additional electron. The Coulomb interaction between an electron on the donor and all other electrons in its environment can still be parameterized by a single capacitance $C$, which is specific to $N$ and the donor environment. The charging energy, represented by $e^2 / 2 C$, of donors close to a gate interface is modified due to the screening at the interface as well as the applied electric field which was experimentally demonstrated by \textcite{LansbergenNphys08}. \textcite{fuechsleNatureN12} showed that an STM fabricated single dopant device displays an unaltered charging energy consistent with the bulk-like environment. Recent theoretical work addresses this problem and progress has been made using effective mass \cite{FangPRB02, HollenbergPRB04, CalderonPRB10, HaoJOP11} and self-consistent field tight-binding \cite{RahmanPRB11} treatments.

\subsubsection{Interactions between donors}\label{dInter}

The interaction between donors plays a central role in quantum information science. The goal is to achieve tunable interaction that preserves coherence \cite{KaneNature98}. This has not been achieved yet but it is within reach. Initial experiments have focused on the study of capacitive coupling as well as the coherent coupling between dopants. A detailed understanding of tunnel coupling as well as capacitive coupling between a donor and a SET is a key issue since this is the central read-out mechanism for qubits \cite{MorelloPRB09}. This complex coupling between a dopant and a quantum dot, i.e. a semiconductor SET, has been analyzed in detail in an experimental and theoretical effort by \textcite{GolovachPRB11}. Coherent coupling between dopants has been achieved in the limit of weakly coupled dopants \cite{VerduijnAPL10} and strongly coupled dopants \cite{CalvetPRB11}. Both rely on the interference between two coherent transport channels which leads to a specific line shape \cite{FanoPR61} that is sensitive to the phase difference between the two transport paths. This phase difference can be modified by changing the magnetic flux that is enclosed in the loop of the transport paths. \textcite{VerduijnAPL10} studied two As atoms in a nano MOSFET and showed that the distance between the dopants is about 30 nm based on the magnetic field dependence. \textcite{CalvetPRB11} studied acceptors in a Schottky FET and also observed a Fano resonance which proofs coherent exchange of electrons. The lack of magnetic field dependence confirms their expectation that the acceptors studied are strongly coupled in these devices.

\subsection{Double dopant quantum dots}\label{d_DD}

\begin{figure}[h!]
\centering
\includegraphics[width=8cm]{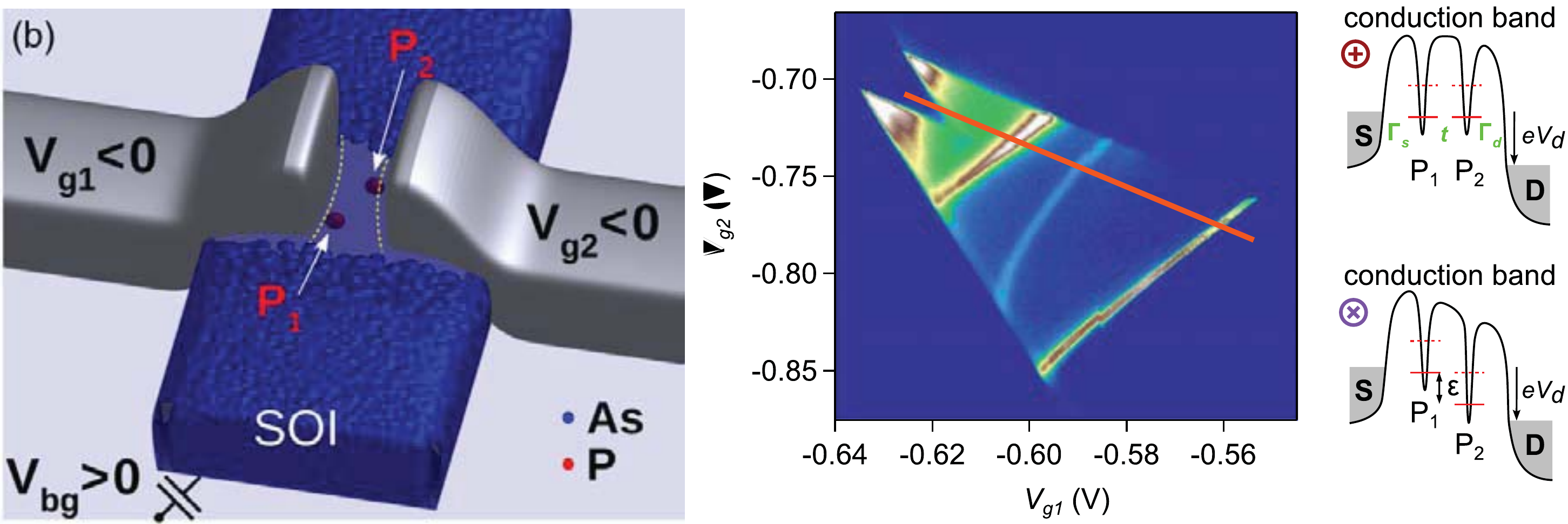}
\caption{ (Color online) \textbf{Sequential transport through a double donor device with independent gate control.} The left panel shows the two opposing gates similar to a conventional FinFET geometry but with a split gate. The channel received a background doping of $10^{18}$ P/cm$^{3}$ and this device demonstrates independent gate control of two dopants. The right panel shows a finite bias stability diagram revealing bulk-like excited states of the dopant. From \onlinecite{RochePRL12}.}\label{DoubleDonor12}
\end{figure}

The study of transport and interactions in donor-based double quantum dots has been motivated by their potential for solid state quantum computing applications \cite{LossPRA98,TaylorNP05}. Initial studies concentrated on ion-implanted devices, where both independent gate control \cite{HudsonN08} on the dot occupancies and charge detection using surface aluminum SETs were demonstrated \cite{MiticN08} in multi-donor devices which contained hundreds of dopants in each dot. Characteristic honeycomb structures (see section \ref{doubledots}) were observed in the charge stability maps. However difficulty was encountered going to smaller dot sizes due to the inherent straggling in the ion implantation process. Non the less, sequential transport through a stochastically doped FinFET structure has been demonstrated by \cite{RochePRL12}. They used a split gate geometry to independently control the chemical potential of two dopants and probe the excited states by tunneling spectroscopy in a similar manner to a double quantum dot, as shown in Fig.\,\ref{DoubleDonor12}.

\begin{figure}[h!]
\centering
\includegraphics[width=8cm]{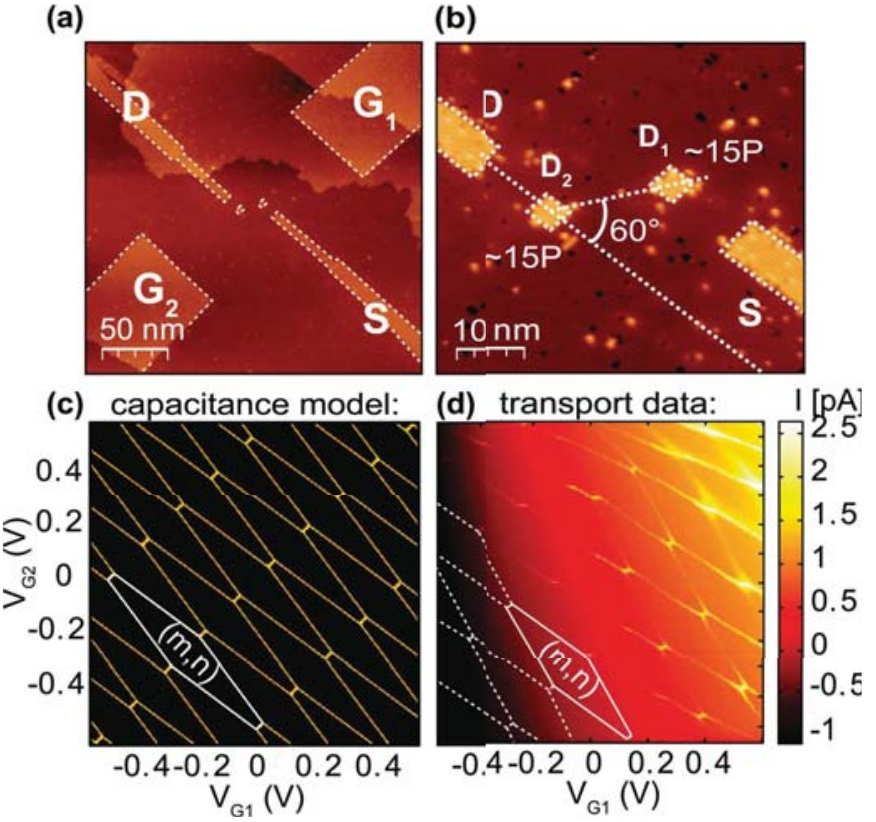}
\caption{ (Color online) \textbf{A donor-based double quantum dot in silicon.} (a) An overview STM-image of the device showing the two quantum dots, tunnel-coupled to the source and drain (S/D) leads and capacitively coupled to the gates G$_{1(2)}$. (b) Close-up of the two quantum dots, $\sim$4 nm in diameter. The DQD angle $\alpha $=$60\pm3^{\textrm{o}}$ has been optimized for maximum electrostatic control whilst suppressing parallel leakage through the dots. The modeled (c) and measured (d) charge stability diagrams show excellent agreement, demonstrating independent electrostatic control of the individual dots. From \textcite{WeberNL12}.}\label{Few_Electron_double_dot2}
\end{figure}

Few-electron single-crystal quantum dots have recently been realized using STM-patterned devices, as shown in Fig.~\ref{Few_Electron_double_dot2}. Here independent electrostatic control of the ultra-small dots was achieved by careful modeling and optimization using the capacitance modeling tool FASTCAP~\cite{NaborsIEEE91} and a single-electronics modeling tool (SIMON) \cite{WasshuberIEEE97}. This is quite remarkable given the small physical size of the dots ($\sim$4nm in diameter) and their close spacing ($\sim$10 nm). At such small dimensions cross-capacitances between the quantum dots become considerable and the dots need to be positioned at an angle, $\alpha\sim 60^{o}$ with respect to one another to achieve independent electrostatic control. One of the advantages with donor-based quantum dots is that this combination of device modeling and precision lithography using scanning probe microscopy allows reliable predictive device design, an important tool as devices scale to the single donor level.

\subsection{Charge sensing in few-electron dopants}\label{d_charge_sensing}

Following the principal work by \textcite{FieldPRL93} on remote sensing of charge using a quantum point contact in a AlGaAs/GaAs system, \textcite{elzermanNature04} adapted this technique to perform single shot detection of spin dependent single electron tunneling events in a single electron quantum dot. Real time sensing of single electron tunneling is fundamental to electrical read-out of qubit states in spin quantum computing. In these experiments \cite{elzermanNature04} a quantum point contact was capacitively coupled to the qubit. However the visibility, and therefore fidelity of spin read-out, of these charge detectors is greatly enhanced when a single electron transistor rather then quantum point contact is used as a charge detector \cite{MorelloPRB09}. Here the SET is additionally tunnel-coupled to the qubit and electrons can be loaded from the SET-island itself, thus eliminating the need for a separate electron reservoir.

\begin{figure}[h!]
\centering
\includegraphics[width=7.5cm]{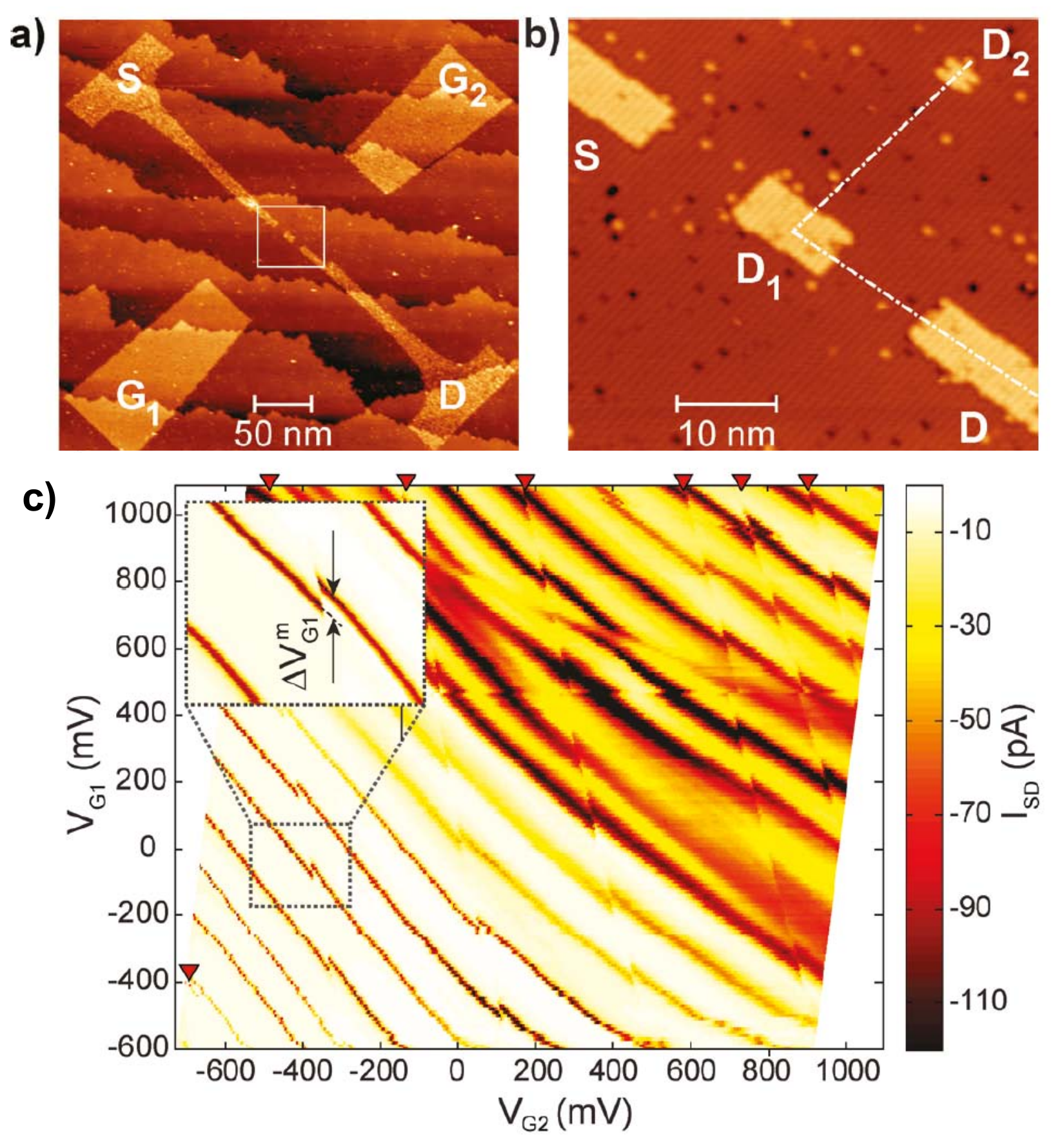}
\caption{ (Color online) \textbf{Charge sensing using a donor-based single electron transistor coupled to a small donor dot.} (a) Filled-state STM image of the overall device pattern, showing (in lighter contrast) the regions where the hydrogen resist monolayer has been desorbed to create the source (S) and drain (D) contacts of the single-electron transistor, and the two gates (G$_1$, G$_2$). (b) High-resolution image of the device pattern within the white box in (a), showing the SET-island (D$_1$) and the quantum dot (D$_2$) (c) Charge stability plot showing the dependence of $I_{SD}$ on the gate voltages ($V_{G1}$, $V_{G2}$), for a constant $V_{SD}$ = -50 $\mu$V. The high current lines correspond to the Coulomb peaks of the SET. Inset: High resolution map of a small section of (c) showing discontinuity of a current line, due to a particular charge transition of D2. The triangles in the main map indicate a total of 7 such transitions of D2. From \onlinecite{MahapatraNL11}.} \label{Charge_sensing}
\end{figure}

High-fidelity spin read-out of a P-donor-bound electron in Si has recently been demonstrated within this architecture \cite{morelloNature10} by implanting a small number of P donors in the vicinity of an electrostatically induced SET, at the Si-SiO$_2$ interface. This has established the feasibility of fiducial detection of P-donor-based spins qubits, however the uncertainty in the number and position of the donors relative to the SET is of concern for scaling up to a practical 2D Si:P quantum computer with multiple donor arrays. To reproducibly achieve sufficient charge sensitivity and electron tunnel rates, it is important to precisely situate the SET with respect to deterministically positioned array of Si:P qubits. Towards this end STM-lithography has recently developed a charge detection device lay-out, wherein a SET charge sensor and an ultra-small quantum dot are both patterned on the same plane of a Si crystal, at an atomically-precise separation, as shown in Fig.~\ref{Charge_sensing}.

In this way two of the key design parameters, i.e., the sensitivity of charge detection and electron tunnel rates can be made sufficiently high to enable projective spin readout of individual P donors in Si. Since the fabrication technique is essentially identical to the established approach for deterministic placement of P donors in Si, this charge sensing layout can be readily integrated in scalable Si:P spin quantum computing architectures. Recent results demonstrate that single electron tunneling between the quantum dot and the SET island occurred on a time-scale ($\tau \sim$ ms) two-orders-of-magnitude faster than the spin-lattice relaxation time of a P donor in Si and suitable for projective readout of Si:P spin qubits \cite{MahapatraNL11}. Another work by \textcite{mazzeoAPL12} reported on the charge dynamics of a single donor coupled to a few-electron silicon quantum dot. Single-spin sensitivity is discussed in more detail in section VI.A. 

%% file: timeresolved.tex
In the previous sections we have reviewed the quantum electronic
properties of silicon, and how such properties can be harnessed in
nanoscale structures. That field of research is fairly mature,
partly thanks to its technological links with classical silicon
nanoelectronics.

Once the confinement of a single electron in silicon has been
achieved, it is possible to start exploring the properties of
the electron spin. An exciting possible application is
quantum computing \cite{LaddNature10}, where the electron spin is
used as a quantum bit \cite{mortonNature11}. Another vast
field of research is spin transport
\cite{appelbaumNature07,dashNature09} and spintronics
\cite{zuticRMP04,awschalomNphys07,jansenNMat12}, which we will not
discuss in this review.

In this section we present a theoretical introduction to the
behavior of electron spins confined in silicon, and a snapshot of
the current state of experimental progress. The discussion below is meant to
provide the reader with an outlook on the future direction of the
field. A thorough review of spin control in silicon will only be
possible several years from now.

\subsection{Spin relaxation and decoherence}\label{spinrelaxation}

The suitability of a physical system to encode and preserve quantum
information is quantified by parameters such as the relaxation and
coherence times. The first, called $T_1$, describes the timescale
over which an energy-excited state decays to the ground state. The
second, called $T_2$, describes the timescale over which the phase
coherence between different branches of a quantum superposition can
be preserved.

A spin $S = 1/2$, such as an electron spin confined in a Si
structure by a natural or artificial potential, can be described by
a $2\times 2$ density matrix $\rho = |\psi\rangle\langle\psi|$. The
state vector $|\psi\rangle$ belongs to a 2-dimensional Hilbert space
with basis vectors $| {\downarrow} \rangle$, $| {\uparrow} \rangle$,
normally chosen as the eigenstates of the spin operator $S_z$ if a
static magnetic field $B_0 \parallel \hat{z}$ is applied. At thermal
equilibrium, the diagonal elements of the density matrix are related
by a Boltzmann factor $\rho_{11}^{th} = \exp(-E_Z/k_B T)
\rho_{22}^{th}$, where $E_Z = g \mu_B B_0$ is the Zeeman splitting.
Therefore, the diagonal elements of $\rho$ are related to the degree
of spin polarization. After a perturbation involving exchange of
energy with an external field or reservoir, the spin returns to
equilibrium in a typical time scale $T_1$, e.g.\ $\rho_{11}(t) -
\rho_{11}^{th} \propto \exp(-t/T_1)$. $T_1^{-1}$ represents the
spin-lattice relaxation rate. $T_1$ measurements in bulk samples are
performed by observing the timescale over which the thermal
equilibrium is recovered after either an `inversion pulse', which
swaps the populations of the ground and excited spin states, or a
`saturation comb', which equalizes the populations. Section
\ref{SS_singlespin} discusses how to obtain the $T_1$ of a single
spin in a nanostructure from a measurement of the probability of
detecting the spin excited state as a function of the waiting time
after the excited state preparation.

A coherent superposition of the
$| {\uparrow} \rangle,| {\downarrow} \rangle$ basis states results in
nonzero off-diagonal elements (`coherences')
$\rho_{12}=\rho_{21}^*$. The preparation and manipulation of such
coherent superpositions is at the heart of quantum information
technology \cite{nielsenQCQI}, and relies on well-established
techniques that belong to the vast field of magnetic resonance
\cite{slichter}. Even in the absence of energy exchange with the
environment, the coherence may decay in time like $\rho_{12}(t)
\propto \exp[-(t/T_2)^{\alpha}]$, where $T_2^{-1}$ is the
decoherence rate, and $\alpha$ is an exponent that depends on the
details and the dynamics of the environment coupled to the spin. For
electron spins in solid state, a major contribution to decoherence
is given by the hyperfine coupling between the electron and the
surrounding nuclear spins. The nuclear spins exhibit complex
dynamics, driven by the interplay of their mutual interactions and
the coupling with the electron. The time fluctuations of the
hyperfine field randomize the electron spin precession frequency and
destroy its coherence.

Even in the presence of a perfectly static nuclear spin bath, a
macroscopic ensemble of spins would exhibit a spread of precession
frequencies as a consequence of inhomogeneity in the local magnetic
field, caused e.g.\ by the difference in the instantaneous value of
the local hyperfine field at every electron site.  The resulting
dephasing time $T_2^{*}$ represents the timescale over which a free
induction decay occurs, i.e. the vector sum of all the spins in the
ensemble averages to zero. For a single spin, the free precession
cannot be observed in a single experiment, and must be obtained
through repetition averaging. Therefore, a $T_2^{*}$-process arises
when the quasi-static value of the local magnetic field changes from
one repetition to the next. The `true' decoherence time $T_2$ is
obtained in the experiments by `refocusing' the quasi-static
inhomogeneity (in space, for a spin ensemble, or in time, for a
single spin) of the magnetic field through a Hahn echo technique
\cite{slichter}.

The definitions and discussion above can be readapted to the case
where a two-level system is obtained from the truncation of the
Hilbert space of two exchange-coupled spins
\cite{levyPRL02,pettaScience05}. The basis states then become the
singlet and triplet states, $|S\rangle = (| {\uparrow\downarrow} \rangle
- | {\downarrow\uparrow} \rangle)/\sqrt{2}$ and $|T_0\rangle =
(| {\uparrow\downarrow} \rangle + | {\downarrow\uparrow} \rangle)/\sqrt{2}$,
and the energy splitting caused by the exchange interaction $J$
replaces $E_Z$ in the expressions above. This way to define
spin-based two-level systems has been proposed to allow the control
of the qubit purely by electrical means, i.e. without resorting to
magnetic resonance techniques.

Relaxation and decoherence of spins in semiconductors has been the
subject of intense research, and an accessible review is given in
\cite{hansonrmp07}. Here we highlight the specific phenomena that
arise in silicon, in particular due to the valley degeneracy of the
conduction band.

The spin-lattice relaxation, i.e. the return of the diagonal
elements of the spin density matrix to their equilibrium value,
requires the coupling of the spin to a phonon reservoir. Silicon
lacks piezoelectric effect, which is often the dominant source of
spin-phonon coupling in III-V materials. The only type of phonons
present in Si is the ``deformation potential'', i.e. a local change
in lattice spacing which propagates with wave vector $q$. A
deformation potential phonon alters the band gap in an inhomogeneous
and time-dependent way, with repercussions on the exact mixture of
spin, valley and orbital nature of the electronic wave functions.

The relaxation rate $T_1^{-1}$ is obtained in a 'Fermi golden rule'
approach as:
\begin{equation}
T_1^{-1} \approx \frac{2\pi}{\hbar}|\langle
\uparrow|\mathcal{H}_{e-ph;SO}| {\downarrow} \rangle|^2 N(E_Z)
\label{T1_FGR}
\end{equation}
where $N(E_Z) \propto E_Z^2$ is the density of phonon states at the
energy splitting $E_Z$, and $\mathcal{H}_{e-ph;SO}$ is a Hamiltonian
term that includes the electron-phonon interaction and the
spin-orbit coupling. It should be noted that the electron-phonon
interaction does not directly couple Zeeman-split \emph{pure} spin
states. However, a nonzero coupling is obtained if the true
eigenstates contain admixtures of other orbital or valley states,
mixed in by the spin-orbit interaction.

Another way to look at the problem -- more familiar to the spin
resonance community -- is to think of the spin as being subject to
an effective magnetic field, whose magnitude and direction can be
modulated by a lattice phonon. Then $T_1^{-1}$ is proportional to
the spectral density, at frequency $\omega_e = E_Z/\hbar$, of the
component of the phonon-induced fluctuating local field
perpendicular to the spin quantization axis.

Let us recall a simple expression for the electron $g$-factor in a
semiconductor \cite{rothPR60,kittelQTS}:
\begin{equation}
g \approx 2 - \frac{m}{m^*}\left(\frac{2\Delta_{\rm SO,VB}}{3E_g +
2\Delta_{\rm SO,VB}}\right),
\end{equation}
where $\Delta_{\rm SO,VB}$ is the spin-orbit splitting of the
valence band, $E_g$ is the band gap, $m$ and $m^*$ are the free
electron and the effective mass, respectively. In Si, $\Delta_{\rm
SO,VB} \sim 40$~meV is relatively small, due to the small atomic
number. The large band gap $E_g = 1.12$~eV results in electron
$g$-factors very close to 2. Accordingly, spin relaxation in Si is
relatively slow, since the modulation of the $g$-factor due to
phonon scattering is very small.

There are four main differences in the spin relaxation behavior
between Si and III-V semiconductors such as GaAs
\cite{blakemoreJAP82}: (i) Si has no piezoelectric effect, therefore
only deformation potential phonons are present; (ii) Si has no bulk inversion asymmetry, therefore is immune from Dresselhaus spin-orbit coupling effects \cite{dresselhausPR55,hansonrmp07}; (iii) The small
atomic number and large band gap of Si produce a weak spin-orbit
coupling; (iv) The physical mechanism and the magnetic field
dependence of $T_1^{-1}$ depends on the nature (valley or orbital) and the details
of the excited states above the valley-orbit ground state.

\subsubsection{Electron spin relaxation in donors} \label{relaxation_donors}

Let us consider first the case of a shallow donor such as P, As or
Sb, where doublet and triplet valley-orbit excited states (see
section \ref{phys-valleyphysics}) lie $\sim 10-15$ meV above the
singlet (spin-degenerate) ground state. All of these 6 states share
the same hydrogenic $1s$ orbital nature, and lie well below the $2p$
orbital states. In this situation, the dominant contribution to
spin relaxation arises from valley effects, which can take two forms: (i) ``valley repopulation'' or (ii) ``one-valley'' mechanisms.

(i) ``\textsl{Valley repopulation}'': within one valley, the $g$-factor of an electron is slightly different (anisotropic) for
magnetic field parallel ($g_{\parallel}$) or perpendicular
($g_{\perp}$) to the valley axis. In the unperturbed $1s$ singlet
ground state of a donor, all 6 valleys contribute equally, and the overall
$g$-factor of the donor-bound electron is isotropic. However, the
local strain produced by a phonon has the effect of disrupting the
symmetry of the 6 valleys, lowering certain valleys with respect to
others. Now the $g$-factor may assume an
anisotropic character, which can be interpreted as a phonon-induced
modulation of the effective local field, with a component perpendicular to the spin quantization axis. The resulting relaxation
rate becomes \cite{hasegawaPR60}:
\begin{eqnarray}
T_1^{-1}(B,T) = f_{\rm
Si}(\theta,\phi)\frac{6}{5\pi}\left(\frac{g'\Xi}{3g E_{vo}}\right)^2
\times
\nonumber \\
 \times \left(\frac{1}{\rho \nu_t^5} + \frac{2}{3 \rho \nu_l^5}
\right)\left(\frac{g \mu_{\rm B}B}{\hbar}\right)^4 k_{\rm
B}T, \label{hase}\\
= K_4 B^4 T. \label{K_4}
\end{eqnarray}
Here $\Xi$ is the deformation potential parameter \cite{bardeenPR50,herringPR56},
representing the energy shift of the valleys due to a deformation of
the crystal lattice. $g' = (g_l - g_t)/3$ describes the anisotropy
of the $g$-factor along the principal axes of the effective mass
tensor for each valley, $E_{vo}$ is the energy difference between
the first excited valley-orbit state and the ground state, $\rho =
2330$~kg/m$^3$ is the density of Si, and $\nu_t = 5860$~m/s and
$\nu_l = 8480$~m/s are the transverse and longitudinal sound
velocities, respectively. $f_{\rm Si}(\theta,\phi)$ is an angular factor that goes to zero for $\theta = 0$ ($B \parallel [001]$), and is maximum for $B \parallel [111]$.

Eq.~\ref{hase} was derived in the high-$T$
limit, appropriate for typical X-band ($\sim 10$~GHz) ESR
experiments at $T
> 1$~K \cite{feherPR59b}. The full expression contains the term $(1+n_{ph}) \approx
k_B T / g \mu_B B$, where $n_{ph} = \left(\exp(g \mu_B B / k_B
T)-1\right)^{-1}$ is the Bose occupation factor of the phonon mode
at the Zeeman energy. For this reason, $T_1^{-1} \propto T$ in the
high-$T$ limit. Conversely, single-spin experiments in
nanostructures \cite{morelloNature10} are conducted in the low-$T$
limit, where $(1+n_{ph}) \approx 1$. This indicates that only
spontaneous emission of phonons can take place. Eq.~(\ref{K_4})
becomes:
\begin{eqnarray}
T_1^{-1}(B)|_{\mathrm{low-}T} = K_4 \frac{g \mu_{\rm B}B}{k_{\rm B}
T}B^4 T= K_5 B^5. \label{K5}
\end{eqnarray}
The $T_1^{-1} \propto B^5$ dependence arises from the following
factors: (i) The density of phonon states $N(E_Z)$ is proportional
to $B^2$; (ii) In the matrix element $\langle
\uparrow|\mathcal{H}_{e-ph;SO}| {\downarrow} \rangle$, a factor
proportional to $B$ accounts for the need to break time-reversal
symmetry, while another factor proportional to $\sqrt{B}$ arises
from the $\sqrt{q}$ dependence of the strain caused by a deformation
potential phonon, where $q \propto g \mu_B B$ is the wave number.
Therefore, $|\langle
\uparrow|\mathcal{H}_{e-ph;SO}| {\downarrow} \rangle|^2$ is proportional
to $B^3$.

(ii) ``\textsl{One-valley}'' mechanism: A phonon-induced strain introduces a coupling between the $\Gamma$ band and the nearest $\Delta$
band (see section \ref{phys-valleyphysics}). This yields an anisotropic modulation of the
$g$-factor even for an electron confined to a single valley. This
one-valley mechanism also yields $T_1^{-1} \propto B^5$ in
the low-$T$ limit, but has a different angular dependence, with fastest relaxation for $B \parallel [001]$ and slowest along [111]. Unlike the valley repopulation, the one-valley mechanism always produces a nonzero relaxation rate.

A detailed discussion and experimental study of these
relaxation channels for Si:P was given by \textcite{wilsonPR61}, in
the high-$T$ limit (see Fig.~\ref{spinrelaxation_fig1}(a)). Since both the valley repopulation and the one-valley mechanism are generally active at the same time and have a comparable strength, \textcite{wilsonPR61} included the analysis of the angular dependence of $T_1^{-1}$ to unravel the different contributions. The low-$T$
limit has been investigated in the single-shot spin readout
experiments of \textcite{morelloNature10}, where the $T_1^{-1}
\propto B^5$ law was verified, and the experimental values of
$T_1^{-1}$ were found to be in quantitative agreement with the
prediction of Eq.~\ref{K5} to within factors $\sim 2$
(see Fig.~\ref{spinrelaxation_fig1}(b)). The longest observed relaxation
time for a single spin was $T_1 \approx 6$~s at $B = 1.5$~T. Because
of the very strong field dependence of $T_1$, \textcite{feherPR59b}
were able to observe $T_1 \approx 5000$~s at $T = 1.25$~K and $B =
0.32$~T in a bulk sample.

\subsubsection{Electron spin relaxation in quantum dots} \label{relaxation_dots}

The spin relaxation mechanisms for an electron confined to a quantum dot obtained differ slightly from those in a donor, due to the different valley and orbital nature of the electron states (see section \ref{physicsdots} and Fig.~\ref{valleys-bulk-2D-dopants-alternative}). The ground and first excited electron wave functions are symmetric or antisymmetric combination of the $\pm z$ valleys, because of the strong vertical confinement in the quantum well from which the dot is formed. It can be shown that, under this circumstance, the ``valley repopulation'' mechanism does not contribute to spin relaxation \cite{glavinPRB03,tahan07}.

The ``one-valley'' mechanism, on the contrary, is active and yields a relaxation rate $T_1^{-1} \propto B^5$ \cite{glavinPRB03}. Notice that, unlike in the donor case, the one-valley mechanism in dots obtained from a [001] quantum well gives vanishing relaxation for $B \parallel$~[001] and [110] \cite{tahan07}.

\begin{table}[t]
\begin{tabular}{ l l l }
\ \ & Donors  & Quantum dots  \\
\hline
\hline
Valley Repopulation \ \ \ \        & $T_1^{-1} \propto B^5$ & negligible \\
\ \ & max $\parallel$  [111] \ \ \ \ & \ \ \\
\ \ & 0 $\parallel$  [001] & \ \ \\
\hline
One-valley   & $T_1^{-1} \propto B^5$  & $T_1^{-1} \propto B^5$  \\
\ \ & max $\parallel$  [001] & max $\parallel$  [100],[010] \ \ \\
\ \ & min $\parallel$  [111] & 0  $\parallel$ [001],[110]  \\
\hline
Rashba spin-orbit \ \        & negligible & $T_1^{-1} \propto B^7$  \\
coupling \ \ & \ \ & max $\parallel$  [100],[010]  \\
\ \ & \ \ & min  $\parallel$ [001]  \\
\hline
\end{tabular}
\caption{\label{tableT1} Summary of the magnetic field dependence of the spin relaxation rates $T_1^{-1}(B)$, for different mechanisms applicable to donors and quantum dots in Si. It is assumed that the dots are formed from [001] quantum wells.}
\end{table}

\begin{figure}[t] \center
\includegraphics[width=8.5cm]{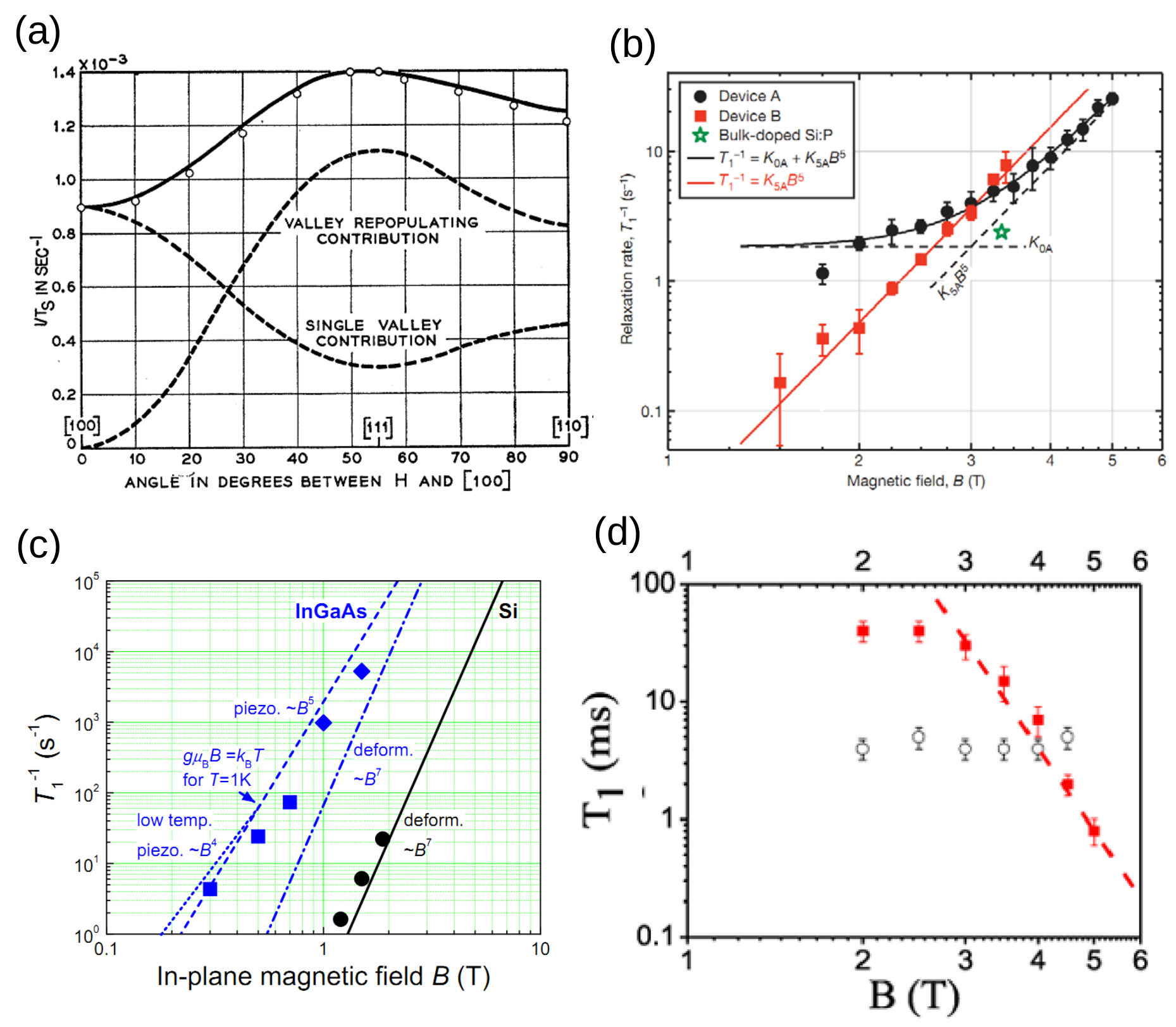}
\caption{ \label{spinrelaxation_fig1}  (Color online) (a) Spin-lattice relaxation
rate $T_1^{-1}$ of P donors in bulk Si, at $B \approx 0.3$~T and $T
= 1.2$~K, as a function of the field orientation. The angular
dependence allows the separation of ``valley repopulation'' and
``single valley'' contributions. From \textcite{wilsonPR61}. (b) $T_1^{-1}(B)$
for single P donors in two different devices. Both show a $T_1^{-1}
\propto B^5$ contribution, but Device A also exhibits a
$B$-independent plateau, attributed to dipolar flip-flops with
nearby donors. Also shown is $T_1^{-1}(3.3\mathrm{T})$ in bulk Si:P. From \textcite{morelloNature10}. (c) $T_1^{-1}(B)$ in a gate-defined Si/SiGe
dot ($\bullet$), compared to data for a InGaAs dot ($\blacksquare, \blacklozenge$). From \textcite{hayesArxiv09}. (d) $T_1(B)$ in a gate-defined Si MOS
dot, for the 1-electron ($\blacksquare$) and 2-electron ($\circ$) states. From \textcite{xiaoPRL10}.}
\end{figure}

An additional mechanism for spin relaxation in quantum dots arises from the structural inversion asymmetry of the quantum well in which the dot in confined, known as Rashba spin-orbit coupling (SOC) \cite{rashbaSPSS60,khaetskiiPRB00,tahanPRB05,hansonrmp07}. Spin relaxation due to Rashba SOC can become dominant in Si quantum dots if the dot geometry gives rise to low-lying excited states of different orbital symmetry as compared to the ground state, or when the ``one-valley'' mechanism vanishes due to $B \parallel$ [001] or [110]. The magnetic field dependence of the Rashba-SOC spin relaxation channel is $T_1^{-1} \propto B^7$ for deformation potential phonons \cite{tahan07,hansonrmp07,raithPRB11}, the only ones present in Si.  The
additional factor $B^2$ as compared to the valley-related mechanisms, arises from the linear dependence on $q$ of
the matrix element for deformation potential phonons to couple
states of different orbital nature. Two experiments
\cite{hayesArxiv09,xiaoPRL10} have indeed found a behavior
consistent with $T_1^{-1} \propto B^7$ in gate-defined Si dots
(see Fig.~\ref{spinrelaxation_fig1}(c,d)). Relaxation times as long as $T_1
\approx 2.8$~s at $B = 1.85$~T have been measured in a Si/SiGe dot
\cite{simmonsPRL11}. The relaxation rate depends on the size and shape of the dot, and is inversely proportional to the square of the orbital level spacing, producing longer $T_1$ for smaller dots. Importantly, the Rashba SOC relaxation channel gives nonzero contribution for any magnetic field direction.  A summary of the magnetic field dependencies of the spin relaxation mechanism for donors and dots in Si is given in Table \ref{tableT1}.

\subsubsection{Singlet-triplet relaxation} \label{ST_relaxation}

The Zeeman-split states of a single electron bound to a donor or a
dot constitute a natural qubit. An alternative scheme has been
proposed, where the logical qubit is obtained from the two-electron
singlet and triplet (S/T) spin states of a double quantum dot
\cite{levyPRL02}. Coherent manipulation of S/T qubits has been
pioneered in GaAs double dots \cite{pettaScience05}, and has
recently been demonstrated in Si/SiGe double dots as well
\cite{mauneNature12}. A `digital' scheme applicable to donors in Si
has also been proposed \cite{skinnerPRL03}. The first measurement of
S/T relaxation in a Si double quantum dot \cite{prancePRL12},
discussed in Section \ref{ST_readout_coherence}, gave a value of
$T_{1,\mathrm{S/T}} \sim 10$~ms at $B=0$, two orders of magnitude
longer than in GaAs dots \cite{johnsonNature05}. In an applied
magnetic field, the spin lifetime of the $T_{-}$ state grows up to
values of order 3 s at $B = 1$~T. The theory of S/T spin relaxation
in the (1,1) charge configuration of double dots in Si/SiGe was
discussed in \cite{raithArxiv12,wangJAP11}, while two groups
\cite{pradaPRB08,wangPRB10} have analyzed the relaxation mechanisms
in a two-electron single dot, which is relevant for the
understanding of lifetime-enhanced transport \cite{ShajiNPhys08} or
the direct measurement of spin relaxation in a 2-electron dot
\cite{xiaoPRL10}.

A theory for exchange-coupled donor pairs exists
\cite{borhaniPRB10}, and predicts a complicated dependence of the
triplet$\rightarrow$singlet relaxation rate on the exchange
interaction $J$, ranging from $\propto J$ to $\propto J^3$. Valley
interference plays a crucial role, since $J$ can vary over orders of
magnitude by simply changing the direction of the axis joining two
donors.

\subsubsection{Spin decoherence} \label{spin_decoherence}

The attractiveness of silicon for quantum computing applications
arises in large part because of the predicted weakness of the
decoherence mechanisms \cite{tahanPRB02,desousaPRB03,tahanPRB05}.
The main source of decoherence for electron spins in solid state is
the coupling to the bath of nuclear spins in the host material. In
the spin resonance literature this goes under the name of ``spectral
diffusion'', to indicate that the time evolution of the state of a
bath of nuclear spins coupled to an electron spin causes the
electron spin resonance frequency (the ``spectrum'') to ``diffuse''
over a certain range \cite{klauderPR62}. Natural silicon has only a
4.7\% concentration of spin-carrying ($I=1/2$) $^{29}$Si isotope,
greatly reducing the effects of nuclear fields compared to GaAs
devices~\cite{witzelPRB06}. The isotopic purification to silicon
consisting of only spinless $^{28}$Si has been demonstrated for some
time \cite{agerJEC05},  and further pursued to extreme levels in the
context of the Avogadro project \cite{andreasPRL11}, where a
$^{28}$Si sphere with less than $ 5\times 10^{-5}$  $^{29}$Si
concentration has been produced with the goal of redefining the
kilogram. Purified material originating from the Avogadro project
has been used to demonstrate exceptional electron spin coherence
times $T_2 > 10$~s \cite{tyryshkinNmat11}. Natural germanium
contains 7.7\% $^{73}$Ge ($I=9/2$), all other isotopes being
spinless. An isotopically purified Si/SiGe heterostructure has been
demonstrated \cite{sailerPSS09}. Even with unenriched Ge, the
effects of the Ge nuclear spins are greatly suppressed because in
the typical heterostructures used for Si/SiGe qubits, only about
0.6\% of the electron density resides in the SiGe
barrier~\cite{Shi:2012p140503}.

\begin{figure}[t] \center
\includegraphics[width=8.5cm]{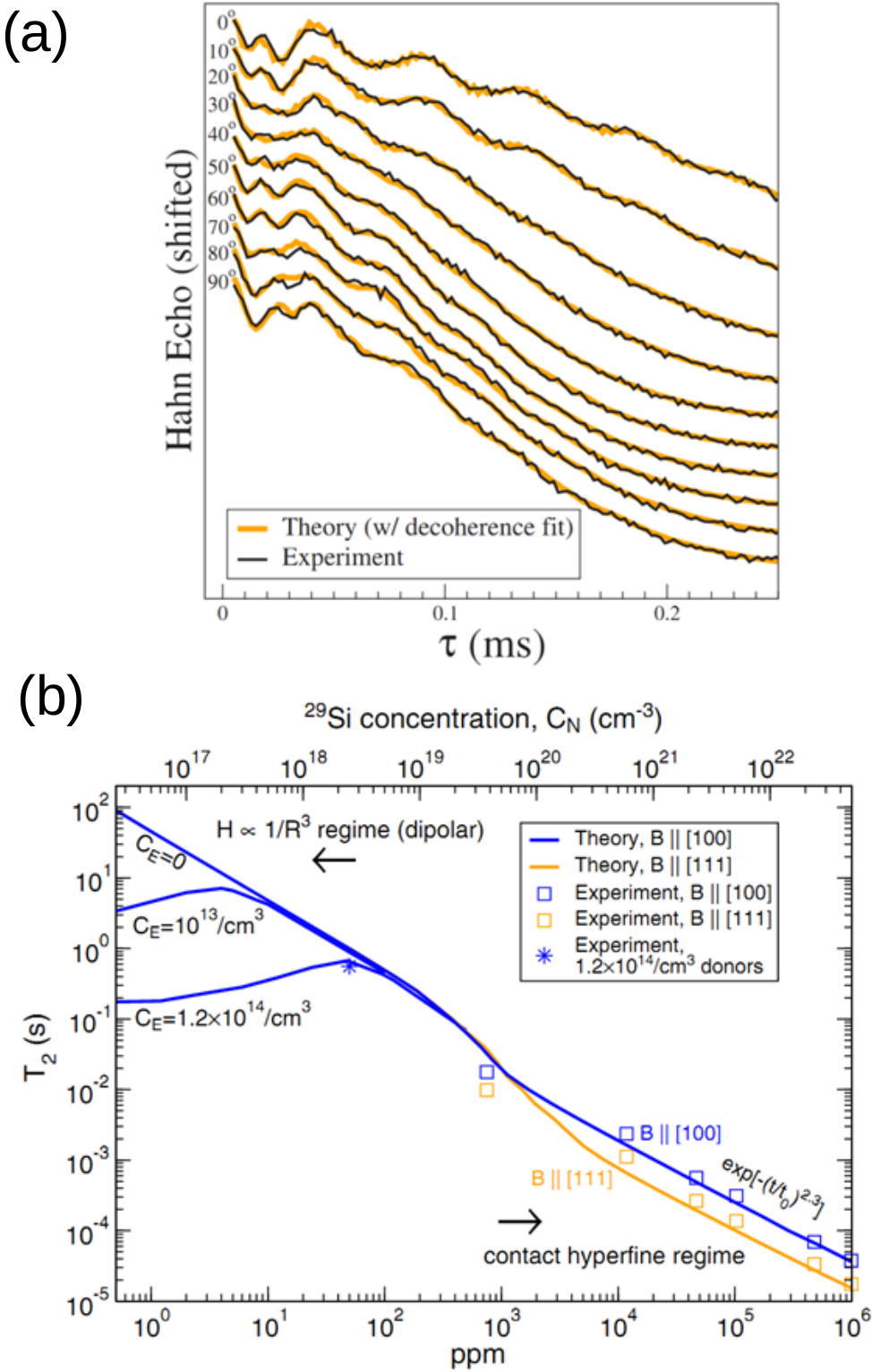}
\caption{ \label{spinrelaxation_fig2}  (Color online) (a) Experimental echo decay
(black) and cluster expansion theory (orange) for $^{\rm nat}$Si:P
at different angles of the magnetic field with respect to the
crystallographic [001] axis. Notice the echo envelope modulation
arising from anisotropic hyperfine coupling between donor electron
and $^{29}$Si nuclei. From \cite{witzelPRB07}. (b) Decoherence time
$T_2$ for Si:P as a function of $^{29}$Si concentration, $C_{\rm
N}$, for different dopant concentrations,  $C_{\rm E}$. Symbols
are experimental data points. From \cite{witzelPRL10}.}
\end{figure}

The theory of electron spin decoherence arising from nuclear spins
in Si is well understood. The correct order of magnitude of the
decoherence time $T_2$ was already predicted by early studies, where
the dynamics of the nuclear spin bath was approximated as Markovian,
i.e. neglecting time correlations \cite{desousaPRB03}. Full
agreement with experimental data in bulk Si:P \cite{TyryshkinPRB03}
requires a more sophisticated treatment of the spin bath dynamics,
where non-Markovian time correlations are taken into account. Quantum mechanical solutions of the dynamics of electron spins in a
nuclear bath included nuclear spin pair correlations
\cite{yaoPRB06}, all the way to higher-order cluster-expansion
techniques \cite{WitzelPRB05,witzelPRB06}. The echo
decay takes the form:
\begin{eqnarray}
V(2\tau) \propto \exp(-2\tau/T_{\rm R} \times
\exp[-(2\tau/T_{\rm SD})^{\alpha}]
\end{eqnarray}
where $\tau$ is the time interval before and after the refocusing
$\pi$-pulse in the Hahn-echo sequence \cite{slichter}, $T_{\rm R}$
is a relaxation time that accounts for both instantaneous diffusion
and $T_1$ processes, $T_{\rm SD}$ is the spectral diffusion time,
which depends on the internal dynamics of the nuclear bath, and the
exponent $\alpha$ takes the value 2.3 \cite{witzelPRB07}. Full matching between cluster expansion theory and experimental data \cite{TyryshkinPRB03} (see Fig.~\ref{spinrelaxation_fig2}(a)) was obtained by including the Electron Spin Echo Envelope Modulation (ESEEM) effect \cite{rowanPR65},
which arises from the anisotropic component of the hyperfine
coupling to the $^{29}$Si nuclei\cite{iveyPRB75,saikinPRB03,ParkPRL09}.

The cluster expansion technique has been extended to the study of
decoherence upon isotopic purification \cite{witzelPRL10} (Fig.
\ref{spinrelaxation_fig2}(b)). An interesting phenomenon that must
be taken into account to match Si:P bulk data is the interplay
between $^{29}$Si spectral diffusion and dipole-dipole coupling
among electron spins. The latter gives an additional contribution to
decoherence, which is  always present because any realistic Si
crystal contains some level of background doping. The dipolar
coupling between donor electron spins contains terms of the form
$S_1^+S_2^- + S_1^-S_2^+$, which allow the excitation of spin 1
and de-excitation of spin 2 (``flip-flop'' process) or vice-versa,
while conserving total energy to within the strength of the coupling
term. This process produces additional magnetic noise on a spin
qubit. However the presence of some $^{29}$Si nuclei may actually be
beneficial in this context. When the inhomogeneity of the local
hyperfine fields is stronger than the electron dipole-dipole
coupling, flip-flop processes are suppressed by energy conservation.
The concept can be extended to any source of local field
inhomogeneity, and the field inhomogeneity does not degrade the
potential implementation of exchange
gates~\cite{deSousaPRA01,HuPRL01}. On this basis,
\textcite{tyryshkinNmat11} have measured $T_2 > 10$~s by using a
highly purified $^{28}$Si:P crystal, and deliberately introducing a
magnetic field gradient across the sample to prevent neighboring
spins from undergoing energy-conserving flip-flop processes. Another
way to suppress flip-flop is lowering the temperature such that $g
\mu_{\rm B}B \gg k_{\rm B}T$, thereby polarizing the electron spins.
This would lead to an exponential suppression of the dipolar
decoherence channel \cite{morelloPRL06,witzelPRL10} because of the
scarcity of spins in an excited state.

Once the nuclear spin and dipole-dipole decoherence mechanisms have
been thoroughly suppressed, one may expect the remaining dominant
decoherence channel to be charge noise, particularly in the case
where exchange coupling is used to implement quantum logic
gates~\cite{culcerAPL09,gambleArxiv12}. Dephasing from charge noise
is expected to be more pronounced in quantum dot qubits than in
donor qubits, but less pronounced than in superconducting qubits,
because the characteristic size of quantum dot qubits is
intermediate between the sizes of impurity qubits and
superconducting qubits.

\subsection{Orbital and valley relaxation}\label{orbitalrelaxation}

So far we have discussed the relaxation processes for the electron
spin confined to a donor or a gate-defined dot, with the intention
of describing the lifetime of an excited qubit state encoded in the
spin Hilbert space. Excited orbital or valley states then act as
intermediate states for perturbations involving lattice phonons and
spin-orbit coupling to cause spin relaxation.

However, the orbital and valley excited states can also be used
actively, for instance to mediate strong interaction between nearby
donors. An early proposal suggested the use of the excited $2p$
orbital states of a deep donor to induce a superexchange interaction
between pairs of shallow donors placed on either sides of the
central one \cite{stonehamJPCM03}. This involves the coherent
manipulation of hydrogenic Rydberg states, a well established
practice in atomic physics. The $2p$ Rydberg state lifetime for P
donors in Si was found to be $T_1 \approx 200$~ps, attributed to the
spontaneous emission of phonons \cite{vinhPNAS08}. Coherent control
of the Rydberg states has also been achieved, with an orbital
coherence time $T_2 \approx 28$~ps \cite{greenlandNature10}.

Valley states are expected to have much longer lifetimes and
coherence, due to the unlikelihood of processes that cause
inter-valley transitions. A recent proposal describes the use of
singlet and triplet valley states of a double quantum dot to encode
and manipulate quantum information with reduced sensitivity to noise
\cite{CulcerPRL12}. Recent experiments have shed light on the valley
physics and its effect on electronic states. Through transport
spectroscopy measurements of donor states in FinFETs,
\textcite{LansbergenPRL11} showed that under certain conditions
relaxation of excited states into lower manifolds is suppressed due
to a combination of both spin and valley blockade. This enhanced
lifetime results in an additional transport path through the excited
state, and appears as a current step in the stability diagram. The
phenomena dubbed as `lifetime enhanced transport' (LET) was first
observed in a silicon double quantum dot \cite{ShajiNPhys08} due to
a blocked relaxation of a spin triplet into a ground state spin
singlet, arising from the long spin relaxation times in silicon (see
section \ref{doubledotsspin}). In the experiment, LET enabled
\textcite{LansbergenPRL11} to identify a blocked transition between
states that have different valley symmetries. They confirm this
observation (i) by extracting the tunnel rates in and out of the
donor states through a temperature dependent measurement and
analysis, and (ii) by computing the low-energy two-electron spectrum
of the system from a multimillion atom tight-binding method to
compare and identify the measured excited manifolds.

\subsection{Control and readout of spins in silicon}\label{singleshot}

\subsubsection{Bulk spin resonance} \label{bulkESR}

The dynamics of spins in bulk materials has been traditionally
studied by electron spin resonance (ESR) and nuclear magnetic
resonance (NMR) techniques. Pioneering experiments on ESR of
donors in Si by \textcite{feherPR59b} measured exceptionally
long electron spin-lattice relaxation times at low
temperature, with a longest measured $T_{1e} \approx 1.4$~hours at
1.25 K and 0.3 T. These experiments were crucial in the development
of the general theory of spin relaxation in semiconductors, as well
as for the understanding of the electronic structure of donors
\cite{feherPR59a}.

Bulk spin resonance has also been used to study electron gases in
modulation-doped Si/SiGe quantum wells
\cite{jantschPSS98,tyryshkinPRL05}, and  the paramagnetic defects that
occur at Si/SiO$_2$ interfaces \cite{poindexterPSS83,browerSST89}
and in amorphous silicon \cite{stutzmannPRB83,askewSSC84}. The
temperature dependence of $T_1$ in paramagnetic dangling bonds
points to a relaxation mechanism where the electron spin is coupled
to the charge fluctuations of the defect, which acts as a tunneling
2-level system \cite{askewSSC84,desousaPRB07}. Thus, ESR can be
used as a non-invasive diagnostic tool to characterize the distribution of tunneling energies of defects in or near amorphous interfaces.

The possibility of enhancing the electron spin coherence of donors
in Si by reducing the concentration of the spin-$1/2$ $^{29}$Si
isotope \cite{abePRB10} was
 demonstrated as early as 1958 \cite{gordonPRL58}. More recently, the quality of isotopic purification has been further improved \cite{agerJEC05} and reached a pinnacle with the Avogadro project \cite{beckerPSS10}, to redefine the kilogram as a sphere of pure $^{28}$Si. \textcite{TyryshkinPRB03} showed that a
 $^{28}$Si:P sample with P doping concentration $n \approx
 10^{15}$~cm$^{-3}$ exhibits a coherence time $T_{2e} = 60$~ms, by using a conventional Hahn-echo technique \cite{slichter}, but
 accounting for the effect of instantaneous diffusion. In bulk
 experiments, the dipole-dipole coupling between the spins
 introduces an artefact whereby the refocusing pulse has the effect
 of flipping the coupled spins, therefore instantaneously changing
 the local magnetic field and artificially suppressing the echo. The
 ``true'' $T_2$ must be obtained by extrapolating the echo decay
 time constant in the limit $\theta_2 \rightarrow 0$. This
 extrapolation method, however, does not eliminate the dynamical
 effect of dipole-dipole coupling \cite{witzelPRL10} during the wait time $\tau$. The
 decoherence due to dipolar interaction can be suppressed by
 introducing a magnetic field gradient across the sample, of
 magnitude larger than the spin-spin coupling strength. With this
 method, and using a bulk sample with extreme isotopic purity ($<
 50$~ppm $^{29}$Si) and low doping ($n \sim 10^{14}$~cm$^{-3}$),
 \textcite{tyryshkinNmat11} obtained a record value of $T_{2e} \approx 10$~s. The combination of narrow ESR absorption
 lines, very long spin coherence and the presence of a nuclear spin
 with  $I=1/2$, make the $^{28}$Si:P system an ideal candidate to
 explore sophisticated techniques to encode, retrieve and manipulate
 non-trivial quantum states. \textcite{mortonNature08} demonstrated the ability
 to store and retrieve an arbitrary quantum state of the P electron
 onto the $^{31}$P nucleus, obtaining a quantum memory with
 coherence time $T_{2n} > 1$~s.

 \subsubsection{Electrically-detected magnetic resonance} \label{EDMR}

\begin{figure}[t] \center
\includegraphics[width=8.5cm]{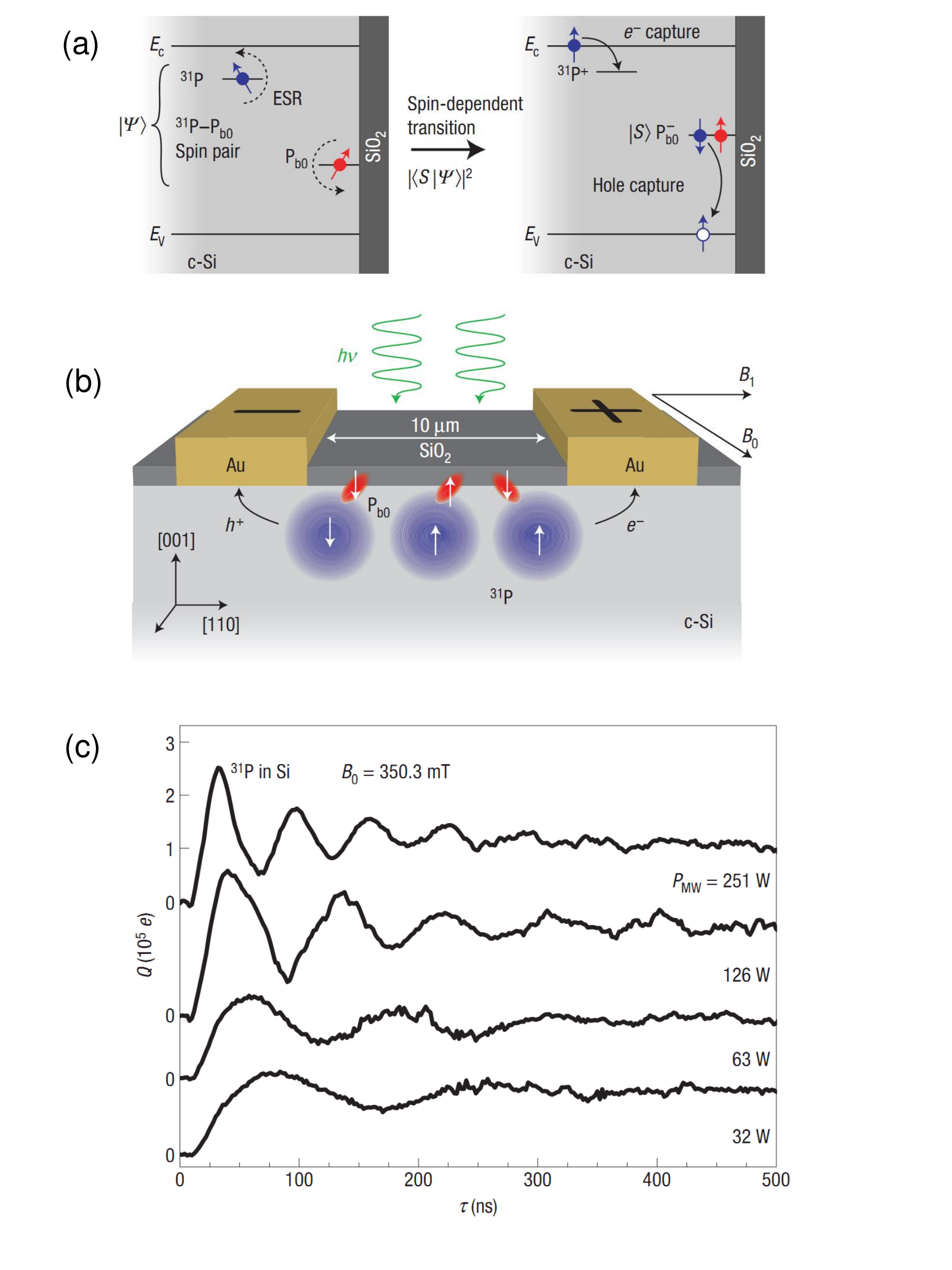}
\caption{ \label{EDMR_fig1}  (Color online) (a) Sketch of the spin-dependent
transition between a donor-bound electron and an interface trap,
following the creation of free carriers through illumination. (b)
Schematics of an EDMR device. P donors close to charge traps at the
Si/SiO$_2$ interface contribute a spin-dependent scattering
mechanism for the electrons traveling between the Au contacts. A
resonant microwave excitation alters the polarization of the
donor-bound electrons, causing a measurable change of the overall
device resistance (c) Electrically detected Rabi oscillations of
P-donor electrons at different values of the driving power. From
\textcite{stegnerNphys06}.}
\end{figure}

In a bulk spin resonance experiment, the precession of a spin
ensemble is detected through the electromotive force induced in a
cavity or pick-up coil. In this way, one can only detect a
macroscopic number of spins, typically~$> 10^{15}$.  In
semiconductors, however, it is possible to make localized spins and
mobile electrons coexist. This allows the detection of spin
resonance by electrical means, and yields a significant improvement
in detection sensitivity. Electrically-detected magnetic resonance
(EDMR) exploits spin-dependent scattering between free carriers and
localized spins. A change in the current (or the conductance) of a
suitably designed nanostructure is observed when a resonant
oscillating magnetic field alters the equilibrium magnetization of
localized spins onto which the free carriers are made to scatter \cite{desousaPRB09}.
The free carriers can be  generated either by illumination
\cite{boehmePRL03} or by electrostatically inducing an electron
layer in a MOSFET structure
\cite{ghoshPRB92,loAPL07,vanbeverenAPL08}. EDMR has been
successfully applied to the detection of spin resonance and coherent
control of $^{31}$P dopant spins in Si
\cite{stegnerNphys06,hueblPRL08,luPRB11} (Fig. \ref{EDMR_fig1}), and
to demonstrate a very long-lived classical spin memory
\cite{mccameyScience10}. The detection sensitivity has been pushed
to the level of $~100$ donors in ion-implanted nanostructures
\cite{mccameyAPL06}, and it has been proposed that reaching the
single-spin limit is possible, and would yield a quantum
nondemolition measurement of the donor nuclear spin
\cite{sarovarPRB08}.

\subsubsection{Single-shot readout of a single electron spin} \label{SS_singlespin}
\begin{figure}[t] \center
\includegraphics[width=8.5cm]{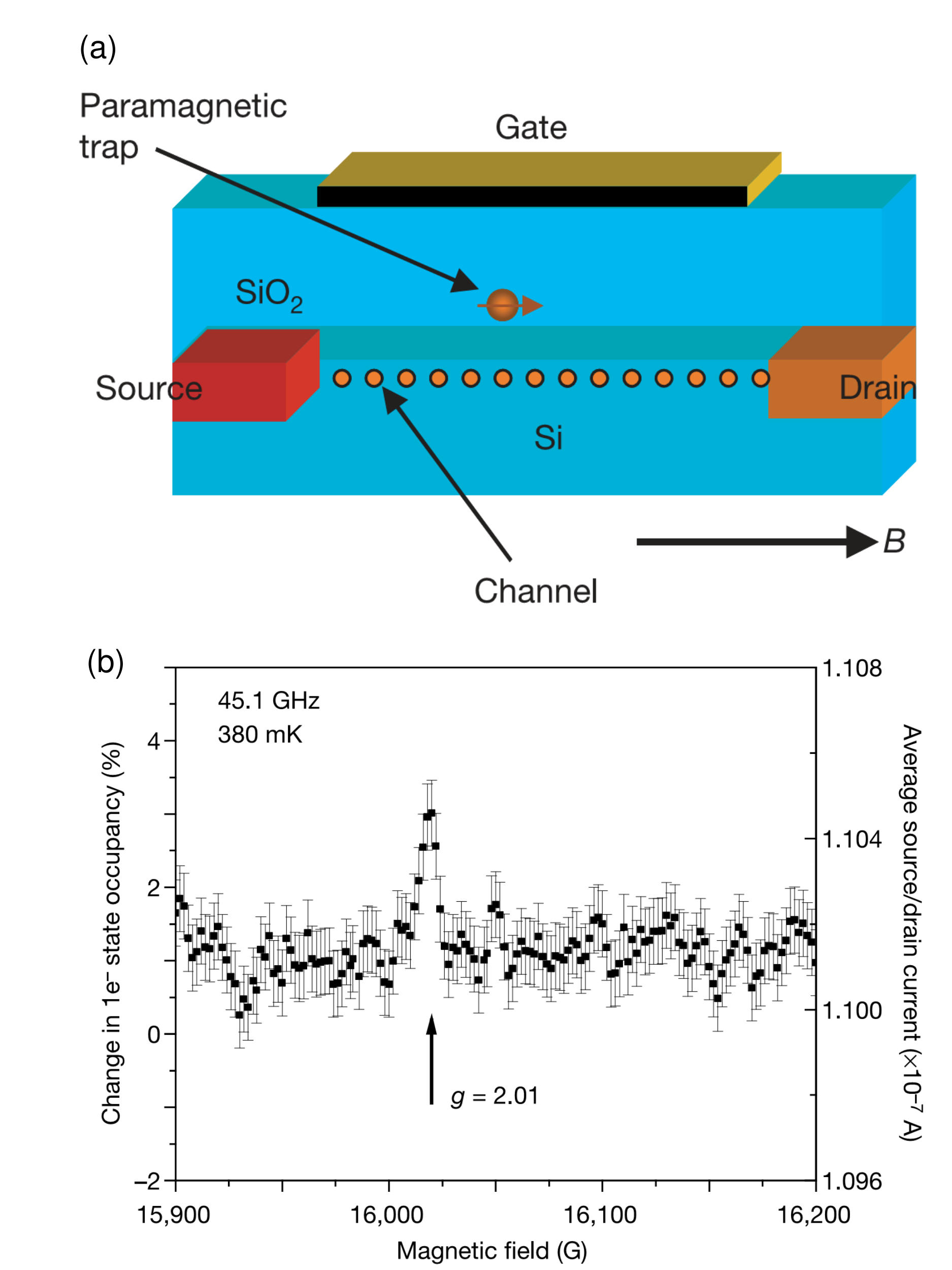}
\caption{ \label{singleshot_fig1}  (Color online) (a) Schematics of a single
charge trap coupled to the channel of a Si transistor. (b)
Single-electron spin resonance measurement, obtained by monitoring
the average current through the transistor as a function of magnetic
field, while applying a microwave excitation at 45 GHz. The excess
current at the resonance frequency arises from the change in charge
occupancy of the trap, made possible by the driven flipping of its
electron spin. From \textcite{xiaoNature04}.}
\end{figure}

To reach single-spin sensitivity, it is necessary to integrate
single-charge detection with a spin-dependent displacement of the
charge. This idea was already incorporated in the Kane proposal for
a Si:P quantum computer \cite{KaneNature98}, where the readout of
the electron spin state would take place by detecting the transfer
of an electron from the $D^0$ state of a donor to the $D^-$ state on
its neighbor. This transfer is only allowed if the two electrons
form a spin singlet state. The detection of the spin-dependent
charge transfer would occur via a single-electron transistor (SET)
on the surface of the device. Subsequent proposals pointed out that
it is possible to detect the displacement of a single charge through
the change in conductance of a small transistor \cite{vrijenPRA00}.
In addition to the spin-dependent addition of a second electron to
an already occupied donor, \textcite{martinPRL03} pointed out that,
in the presence of a large magnetic field, a charge center (not
necessarily a donor) can change its occupancy state when the excited
spin state lies above the Fermi level of a nearby electron reservoir
(e.g.\ the channel of a transistor), while the ground spin state
lies below. This process corresponds to an energy-dependent
spin-to-charge conversion. If a resonant magnetic field is applied
to drive transitions between the spin states, one expects to observe
a switching behavior in the current through the transistor, as the
system goes through the cycle: excite spin-up state $\rightarrow$
ionize $\rightarrow$ load spin-down electron. This method was
employed by \textcite{xiaoNature04} to detect the spin resonance of
a single charge trap coupled to a small Si transistor
(Fig.~\ref{singleshot_fig1}). The same type of spin-to-charge
conversion lies at the heart of the single-shot readout of a single
electron confined to a GaAs quantum dot \cite{elzermanNature04}. In
that case, the spin state of the single electron was detected in a
single-shot manner, i.e., with no need for repetition averaging,
thanks to the large electrical signal obtained by monitoring the
change in conductance of a quantum point contact with strong
electrostatic coupling to the quantum dot.

\begin{figure}[t] \center
\includegraphics[width=8.5cm]{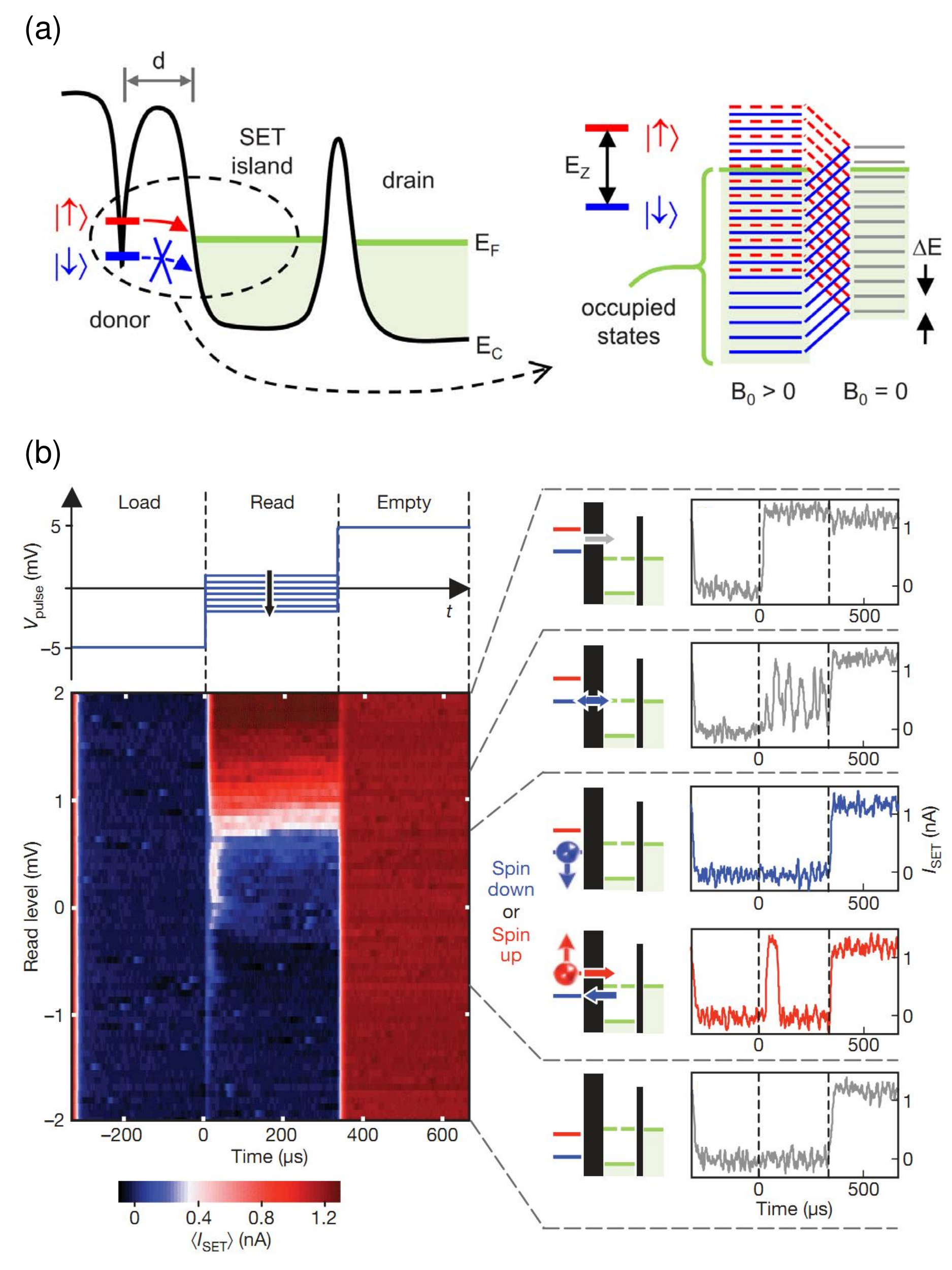}
\caption{ \label{singleshot_fig2}  (Color online) (a) Spin-to-charge conversion
scheme for a single donor tunnel-coupled to the island of an SET. The
presence of quantized states inside the SET island can be ignored if
the single-particle energy level spacing is smaller than the thermal
broadening. From \textcite{MorelloPRB09}. (b) Single-shot readout of
a donor electron spin. The individual traces show the evolution of
the readout signal as a function of the donor electrochemical
potential with respect to the Fermi level. From
\textcite{morelloNature10}.}
\end{figure}

The spin-to-charge conversion, and therefore the single-shot spin
readout, is considerably more challenging in Si than e.g in GaAs
quantum dots. This is because the large effective mass requires
tighter electron confinement, and decreases the transparency of
tunnel barriers. Averaged spin readout experiments were performed in
Si/SiGe \cite{hayesArxiv09} and Si MOS \cite{xiaoPRL10} quantum
dots, yielding the spin relaxation time $T_1$. The first successful
single-shot electron spin readout in Si was obtained by Morello et
al. \cite{morelloNature10}, where the electron was bound to a
$^{31}$P donor and tunnel-coupled to the island of an induced
Si-SET. The readout scheme is a  modification of the
energy-dependent spin-to-charge conversion used by
\textcite{elzermanNature04}. The donor and the SET island
effectively form a hybrid double quantum dot \cite{hueblPRB10}
connected ``in parallel'' \cite{hofmannPRB95}, where one dot is
coupled to source and drain leads, and the other (the donor in this
case) is only coupled to the main dot \cite{MorelloPRB09}. This
results in a very compact structure and charge transfer signals
large enough to completely switch the SET from Coulomb blockade
($I_{\rm SET} = 0$) to the top of a Coulomb peak ($I_{\rm SET} \sim
2$~nA), resulting in single-shot readout of the donor spin with $>
90\%$ visibility (Fig. \ref{singleshot_fig2}). Single-shot spin
readout has also been achieved in a gate-defined Si quantum dot
\cite{simmonsPRL11}, using a QPC as charge sensor and the 2DEG in a
Si/SiGe heterostructure as the charge reservoir. There, the weaker
(purely capacitive) coupling between sensor and dot lead to a
current signal $\sim 20$~pA upon spin-dependent displacement of a
single electron charge.

\subsubsection{Readout and control of singlet-triplet states in double quantum dots} \label{ST_readout_coherence}

Some of the most successful implementations of spin-based qubits in
semiconductors have made use of two-electron systems
\cite{levyPRL02}, where quantum information can be encoded into the
singlet and triplet (S/T) states of exchange-coupled electrons,
instead of the Zeeman split spin states of a single electron.
Coherent control \cite{pettaScience05}, single-shot readout
\cite{barthelPRL09} and dynamical decoupling methods
\cite{bluhmNphys10} for S/T qubits have been demonstrated in GaAs
double quantum dots. In the quest to implement S/T qubits in Si, the
large effective mass plays again a role in requiring very tight
electron confinement and reducing the tunnel couplings, which in
this case also have the essential role of determining the spin
exchange coupling $J$. In addition, most S/T qubit implementations
in GaAs have made use of a gradient of hyperfine field, $\Delta
B_z$, between the two dots to be able to control the qubit along two
orthogonal axes in the S/T basis \cite{folettiNphys09}. Because of
the much smaller hyperfine interaction in Si \cite{assaliPRB11} as
compared to GaAs, the two-axis control of a S/T qubit through $J$
and $\Delta B_z$ becomes more challenging. On the other hand, the
weak coupling to the nuclear spin bath allows for substantially
longer coherence times.

\begin{figure}[t] \center
\includegraphics[width=8.5cm]{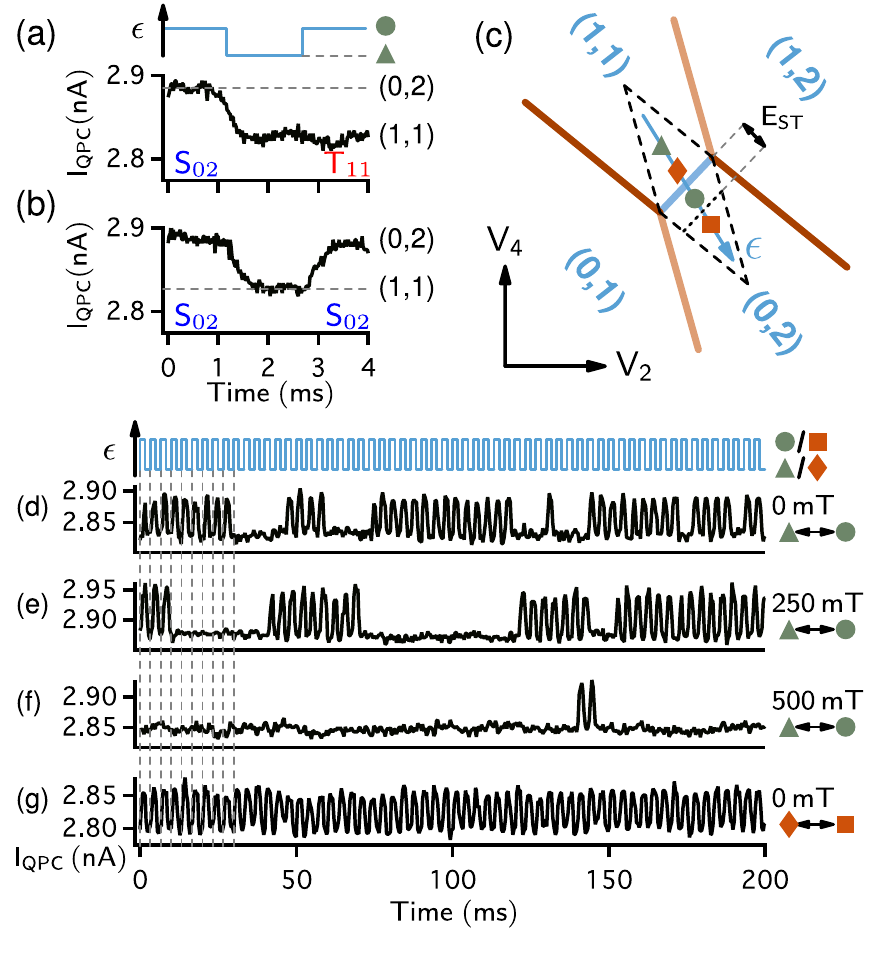}
\caption{ \label{singleshot_fig3}  (Color online) \textbf{Single-shot readout of
singlet-triplet states in a Si/SiGe double quantum dot.} (a,b) QPC
current traces, $I_{\rm QPC}$, while pulsing the detuning with a
square wave. Singlet states are identified when $I_{\rm QPC}$
returns to a high value, as in (b). (c) Charge stability diagram and
pulsing levels. (d -- f) Time traces of  $I_{\rm QPC}$ at different
magnetic fields, as indicated. Increasing $B$ extends the lifetime
of the $T_{11}$ (constant current) state. (g) Control sequence,
pulsing outside the spin blockade region. From \textcite{prancePRL12}.}
\end{figure}

As long as the valley degeneracy of the Si conduction band (Section
\ref{phys-valleyphysics}) is completely lifted, the singlet/triplet
spin states in Si double quantum dots can be detected and
manipulated in the same way as in GaAs dots. The readout mechanism
involves Pauli spin blockade (Section \ref{doubledotsspin}). Figure
\ref{singleshot_fig3} shows the single-shot readout of the singlet
and triplet states of a Si/SiGe double quantum dot
\cite{prancePRL12}. The state of the two-electron system is detected
by pulsing the detuning $\epsilon$ from negative -- where the (1,1)
charge state is stable -- to positive -- where the electrons can
occupy the (0,2) state, provided their spin state is a singlet.
Switching between the (1,1) and the (0,2) state produces a signal on
the QPC current which can be measured in single-shot. This
experiment also yields the triplet spin relaxation time $T_1$, which
is found to be $\sim 10$~ms for all triplets at $B=0$, but extends
up to $\sim 3$~s at $B=1$~T for the $T_-$ state, whose splitting
from the singlet state is reduced by the applied field.

\begin{figure}[t] \center
\includegraphics[width=0.48\textwidth]{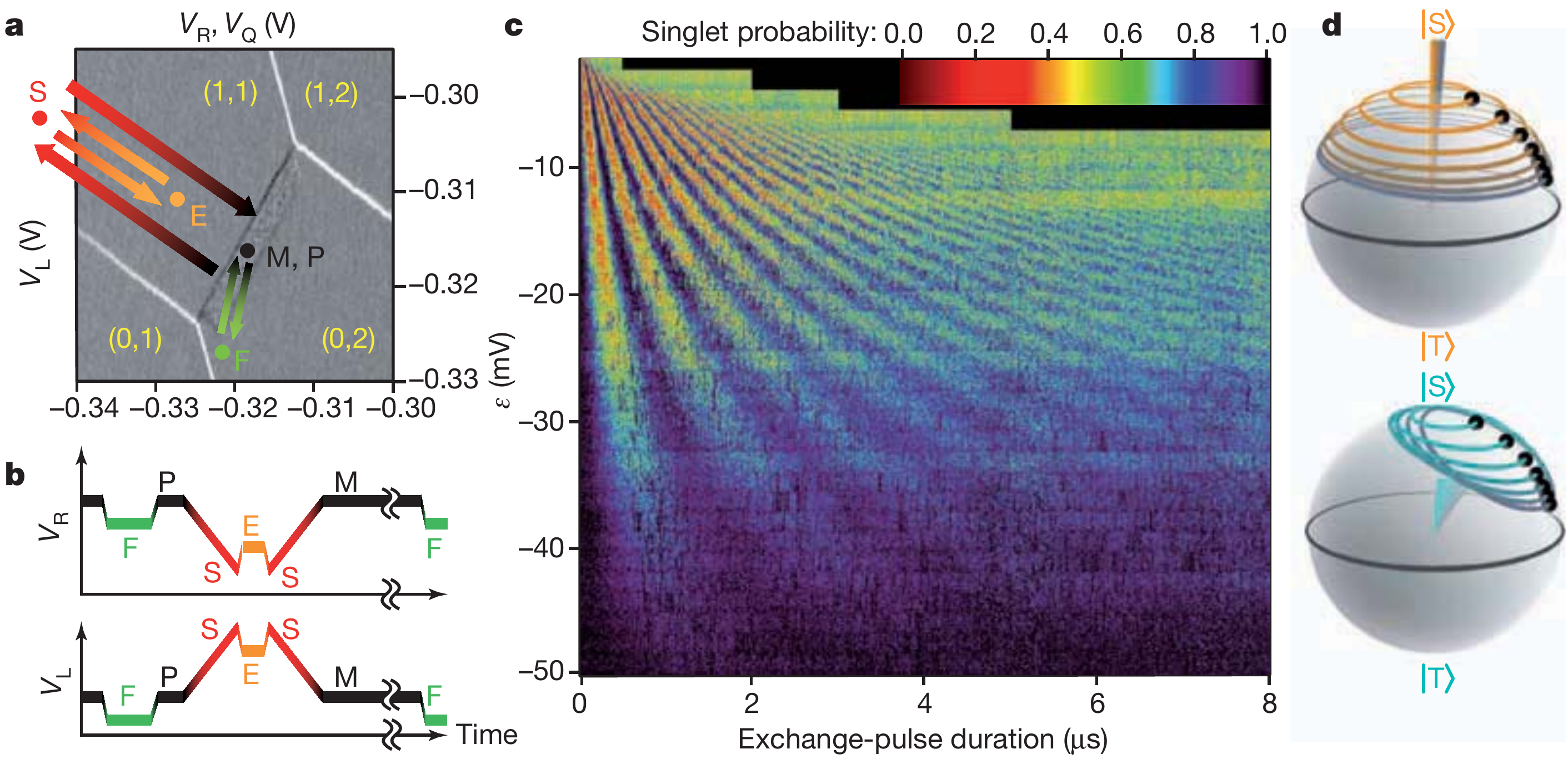}
\caption{ \label{singleshot_fig4}  (Color online) \textbf{Coherent manipulation of
singlet-triplet states in a Si/SiGe double quantum dot.} (a) Charge
stability diagram of the double dot system. Arrows describe
the trajectory in gate space during the pulsing sequence shown in
panel (b). The (0,2) singlet state is prepared at point F.
Adiabatically moving to point S, where the exchange coupling is very
weak, brings the system to the (1,1) singlet. Pulsing to point E
turns on the exchange and causes the two-spin state to oscillate
between the (1,1) singlet and triplet. M is the measurement point
where the electrons recombine in the (0,2) state if in singlet state.
(c) Rabi oscillations of the singlet probability, as a function of
the exchange pulse duration (time spent at point E) and (0,2) --
(1,1) detuning $\epsilon$. (d) Bloch sphere representation of the trajectories
of the two-spin states for different initial values of the
hyperfine fields. From \textcite{mauneNature12}.}
\end{figure}

The coherent control of singlet/triplet states has been achieved in
an accumulation-mode Si/SiGe (Section \ref{sigedots}) double quantum
dot \cite{mauneNature12}. In the S/T qubit  basis, one can represent
the singlet and triplet states as the poles of a Bloch sphere, with
the $| {\uparrow\downarrow} \rangle$, $| {\downarrow\uparrow} \rangle$
states on the equator \cite{pettaScience05}. The exchange
interaction $J$ acts equivalently to an effective field along
$\hat{z}$, while a gradient of hyperfine field $\Delta B_z$ between
the two dots acts as an effective field along $\hat{x}$. Figure
\ref{singleshot_fig4} shows the measurement of Rabi oscillations in
the S/T basis of a Si/SiGe DQD. The system is prepared in the (0,2)
singlet state by exchange with the leads. A subsequent adiabatic
pulse (i.e., slow with respect to the interdot tunnel rate, fast
with respect to the S/T mixing time arising from hyperfine
interactions) prepares a state close to the equator of the Bloch
sphere. The exchange oscillations are then initiated by pulsing
closer to the zero detuning line, causing $J$ to increase and the
spin state to undergo  a rotation around an axis that depends on the
instantaneous value of $J$ and $\Delta B_z$. A final adiabatic pulse
brings the state back to $\epsilon > 0$, where the electrons occupy
the same dot if they returned to a singlet state. The oscillations
of the singlet return probability constitute a demonstration of
coherent control of the two-electron spin states. The dephasing
introduced by the randomness of the hyperfine field can be measured
with a modified pulse sequence, where the electrons prepared in the
(0,2) singlet state are rapidly separated and left to dephase at $J
\approx 0$ before being brought back to the (0,2) region for
readout. The measured dephasing time  $T_2^{*} \approx 360$~ns
\cite{mauneNature12} represents an improvement by nearly two orders
of magnitude over the value observed in GaAs dots
\cite{pettaScience05}, as expected from the much weaker hyperfine
coupling in Si as compared to GaAs \cite{assaliPRB11}. It should be
noted, however, that the smallness of the hyperfine field poses a
challenge when attempting to reach the regime where $J < \Delta B_z$
(see Fig.~\ref{singleshot_fig4}(d)).

\subsubsection{Single-atom spin qubit} \label{single_atom_qubit}

\begin{figure}[t] \center
\includegraphics[width=8.5cm]{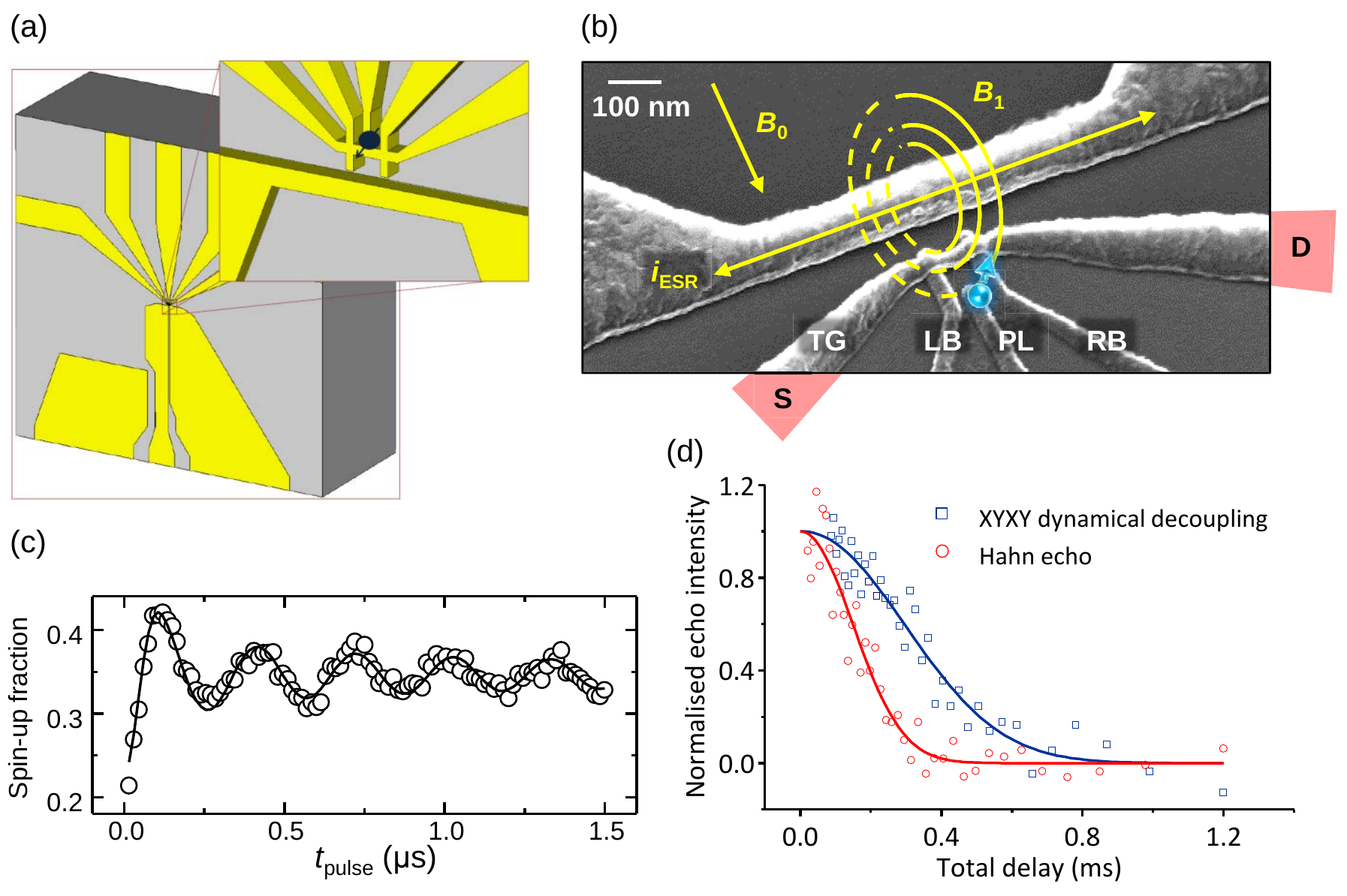}
\caption{ \label{singleshot_fig5_qubit}  (Color online) \textbf{Single-atom electron spin qubit based on an implanted $^{31}$P donor.} (a) Optimized design of an on-chip planar transmission line capable of delivering coherent microwave pulses at frequencies up to 50~GHz. From \textcite{dehollainNT13} (b). Scanning electron micrograph of the spin qubit device. (c) Rabi oscillations of the electron spin state, with 10~dBm driving power at 30~GHz. (d) Measurement of spin coherence with Hahn echo and XYXY dynamical decoupling. (b-d) from \textcite{plaNature12}.}
\end{figure}

The single-shot spin readout techniques discussed in Section \ref{SS_singlespin} have been recently combined with coherent spin control via microwave pulses, yielding the demonstration of a spin qubit based on the electron bound to a single $^{31}$P atom implanted in a silicon MOS device \cite{plaNature12} (Fig.~\ref{singleshot_fig5_qubit}(b)). The use of energy-selective spin-dependent tunneling as the readout process \cite{morelloNature10} has the consequence that the qubit must be operated in a magnetic field $B$ such that the Zeeman splitting $g\mu_B B$ is larger than the thermal broadening of the charge reservoir at electron temperature $T_{\rm el}$, $\approx 5 k_B T_{\rm el}$. For instance, $T_{\rm el} \sim 200$~mK requires $B > 1$~T, and consequently a qubit operation frequency $\nu > g\mu_B B/h \approx 20$~GHz. Careful microwave design is necessary to achieve a strong oscillating magnetic field $B_1$ at the qubit, while minimizing the electric field disturbance to the charge detector \cite{dehollainNT13} (Fig.~\ref{singleshot_fig5_qubit}(a)).

Coherent control of a donor-bound electron spin was achieved by electrically initializing the qubit in the $| {\downarrow} \rangle$ state, applying short microwave pulses to rotate the spin to $| {\uparrow}  \rangle$,  and reading out the final state in single-shot. The probability to measure a $| {\uparrow} \rangle$ electron reveals the characteristic Rabi oscillation as function of the pulse duration (Fig.~\ref{singleshot_fig5_qubit}(c)). A microwave power of 10 dBm ($\approx -20$~dBm at the chip) resulted in a Rabi frequency of 3.3~MHz, i.e. an oscillating field $B_1 \approx 0.12$~mT. The coherence time of the electron spin qubit was measured with a Hahn echo technique, yielding $T_{2} \approx 210$~$\mu$s, close to the value obtained in bulk ESR experiments in natural Si \cite{gordonPRL58}. Coherent qubit rotations around two orthogonal axes were also demonstrated, with an XYXY sequence yielding $T_{2}^{XYXY} \approx 410$~$\mu$s \cite{plaNature12} (Fig.~\ref{singleshot_fig5_qubit}(d)).

\section{Outlook} \label{outlook}

We have presented a comprehensive review of the progress of quantum electronic devices in silicon. This field of research has reached maturity in the theoretical understanding and experimental demonstration of the confinement, manipulation and measurement of single electrons in nanometer-scale structures. These developments have the potential to inform and support the progress of classical silicon-based nanoelectronic devices, as necessary to keep pursuing Moore's law in computer chips. Most importantly, they underpin the manipulation of quantum degrees of freedom such as those of electron and nuclear spins, and open the perspective of processing quantum information using the same technological platform that has allowed the extraordinary progress of microelectronics over the last half-century.

While the manipulation of individual spin qubits in silicon is still in its infancy, the results obtained so far \cite{mauneNature12,plaNature12} have confirmed that silicon is an excellent host for spin qubits, as expected on the basis of the weak spin-orbit coupling and the abundance of isotopes with zero nuclear spin. To build a large-scale quantum computer in silicon, the next milestones will involve harnessing the exchange interaction (Section~\ref{wavefunction}) and demonstrating 2-qubit logic gates, as well as coherent transport of quantum spin states. This is arguably the most challenging goal in the whole program, and several theoretical ideas have been put forward to achieve it \cite{GreentreePRB04,friesenPRL07}. Among solid-state qubits, great success in multi-qubit coupling has been obtained by combining superconducting qubits and microwave cavities \cite{youNature11}, and similar methods are being applied to semiconductor quantum dots \cite{freyPRL12,peterssonNature12}. Once again, silicon appears to be the ideal host to attain strong coupling between spins qubits and cavities \cite{huPRB12}, thanks to the extremely narrow resonance lines achievable with isotopic purification. The ideas and methods for silicon quantum electronics described here lay the foundations for future efforts to bring the potential of silicon to full fruition in the quantum information era.

%% file: acknowledgements.tex
\section*{Acknowledgments}
We thank Malcom Carroll, Akira Fujiwara, Mark Gyure, Xuedong Hu, Marc Sanquer, Sankar Das Sarma, Andre Saraiva, Charles Tahan and Lieven Vandersypen for commenting on the manuscript, Mark Friesen for useful discussions, and Gabri Lansbergen for sharing his thesis work on transport regimes. FAZ, AM, ASD, MYS, LCLH, GK and SR acknowledge support from the Australian Research Council Centre of Excellence for Quantum Computation and Communication Technology (project number CE110001027) and the US Army Research Office under contract number W911NF-08-1-0527. GK, SNC and MAE acknowledge support by the U.S. Army Research Office (W911NF-08-1-0482, W911NF-12-1-0607). GK acknowledges the technical support by Zhengping Jiang in the preparation of the NEMO figures and editorial aspects. GK also acknowledges the long time collaborations with Dr. Timothy B. Boykin and his former student Dr. Neerav Kharche. nanoHUB.org computational resources operated by the network for computational nanotechnology funded by the National Science Foundation have been used in this work. FAZ acknowledges support from the Foundation for Fundamental Research on Matter (FOM), which is part of the Netherlands Organization for Scientific Research (NWO), and support from the European Commission under the Marie Curie Intra-European Fellowship Programme.